\documentclass[11pt,a4paper]{article}

\usepackage[T1]{fontenc}
\usepackage[utf8]{inputenc}
\usepackage[english]{babel}

\usepackage{mathpazo}
\usepackage{tgheros}

\usepackage{geometry}
\usepackage{microtype}
\usepackage{xcolor}

\usepackage{titlesec}
\titleformat{\section}
	{\Large\sffamily\bfseries}{\thesection}{1em}{}
\titleformat{\subsection}
	{\large\sffamily\bfseries}{\thesubsection}{1em}{}
\titleformat{\paragraph}[runin]
	{\normalsize\sffamily\bfseries}{}{}{}
\titlespacing*{\section}{0pt}{1.8ex plus .4ex minus .2ex}{0.8ex plus .2ex}
\titlespacing*{\subsection}{0pt}{1.4ex plus .3ex minus .2ex}{0.5ex plus .1ex}

\usepackage{fancyhdr}
\fancypagestyle{sbarunning}{%
	\fancyhf{}%
	\fancyhead[L]{\small\sffamily\textcolor{gray}{Extended Scenario Bundle Analysis}}%
	\fancyhead[R]{\small\sffamily\textcolor{gray}{Pitz \& Ferraz (2026)}}%
	\fancyfoot[C]{\small\thepage}%
}
\fancypagestyle{plain}{%
	\fancyhf{}%
	\fancyfoot[C]{\small\thepage}%
}

\usepackage{amsmath,amssymb,amsthm}
\usepackage{mathtools}

\usepackage{array}
\usepackage{float}             
\usepackage{makecell}
\usepackage{ragged2e}
\usepackage{enumitem}
\newlist{sbaactions}{description}{1}
\setlist[sbaactions]{leftmargin=1.8em, labelsep=0.6em, style=nextline, font=\bfseries}
\usepackage[section]{placeins}
\usepackage{verbatim}

\usepackage{graphicx}
\usepackage[labelfont=bf]{caption}
\usepackage{tabularx}
\usepackage{booktabs}
\usepackage{multirow}

\usepackage{tikz}
\usetikzlibrary{arrows.meta,positioning,calc,fit,
                 decorations.pathreplacing,backgrounds,patterns,
                 shapes,arrows,shapes.geometric}

\tikzset{
  tevbox/.style={draw,rounded corners,align=center,inner sep=4pt},
  tevsmall/.style={font=\small},
  tevarrow/.style={-{Latex[length=2mm]},thick},
  tevdash/.style={-{Latex[length=2mm]},thick,dashed}
}

\usepackage{pgfplots}
\pgfplotsset{compat=1.18}

\definecolor{accentBlue}{HTML}{2B6CB0}
\definecolor{accentBlueLight}{HTML}{BEE3F8}
\definecolor{accentOrange}{HTML}{C05621}
\definecolor{accentOrangeLight}{HTML}{FEEBC8}
\definecolor{accentRed}{HTML}{9B2C2C}
\definecolor{accentRedLight}{HTML}{FED7D7}
\definecolor{highlightRed}{HTML}{FEB2B2}
\definecolor{bgBlue}{HTML}{EBF4FA}
\definecolor{bgOrange}{HTML}{FDF2E9}
\definecolor{captionGray}{HTML}{718096}

\usepackage[authoryear,sort&compress]{natbib}

\PassOptionsToPackage{hyphens}{url}
\usepackage{hyperref}
\hypersetup{
	colorlinks=true,
	linkcolor={blue!55!black},
	citecolor={blue!55!black},
	urlcolor={blue!55!black},
	pdftitle={Extended Scenario Bundle Analysis: A Formal Framework for Strategic Scenario Modeling},
	pdfauthor={Thomas Pitz, Vinicius Ferraz},
}

\usepackage{cleveref}

\usepackage{imakeidx}
\makeindex

\usepackage{xparse}
\NewDocumentCommand{\idx}{ O{} m }{%
  \IfValueTF{#1}{#2\index{#1}}{#2\index{#2}}%
}

\newcommand{\Coalition}{Coalition\index{Coalitions}}

\newcommand{\Pow}{\mathcal{P}}

\title{Extended Scenario Bundle Analysis\index{Scenario Bundle Analysis}\\[0.4em]
\large A Formal Framework for Strategic Scenario Modeling}
\author{%
\begin{tabular}{c}
Thomas Pitz\textsuperscript{1} \qquad Vinicius Ferraz\textsuperscript{2,3}\\[0.6em]
\small
\textsuperscript{1}Faculty of Society and Economics,
Hochschule Rhein-Waal - Kleve, Germany\\
\small\texttt{thomas.pitz@hochschule-rhein-waal.de}\\[4pt]
\small
\textsuperscript{2}Institute of Management, Karlsruhe Institute of Technology (KIT)\\
\small Karlsruhe, Germany\\
\small
\textsuperscript{3}Singularity AI Research, Singularity.inc\\
\small Vienna, Austria\\
\small\texttt{vinicius@singularity.inc}
\end{tabular}%
}
\date{April 2026}

\begin{document}
	\thispagestyle{plain}
	\pagenumbering{arabic}
	\pagestyle{plain}

	\noindent\rule{\textwidth}{1.2pt}

	\vspace{1.2em}
	\begin{center}
		{\LARGE\sffamily\bfseries
		Extended Scenario Bundle Analysis\index{Scenario Bundle Analysis}\par}
		\vspace{0.4em}
		{\large\sffamily
		A Formal Framework for Strategic Scenario Modeling\par}
		\vspace{1.2em}
		\noindent\rule{\textwidth}{1.2pt}
		\vspace{1.0em}

		{\large Thomas Pitz\textsuperscript{1} \qquad
		Vinicius Ferraz\textsuperscript{2,3}\par}

		\vspace{0.6em}

		{\small
		\textsuperscript{1}Faculty of Society and Economics,
		Hochschule Rhein-Waal --- Kleve, Germany\\
		\texttt{thomas.pitz@hochschule-rhein-waal.de}\\[4pt]
		\textsuperscript{2}Institute of Management,
		Karlsruhe Institute of Technology (KIT), Karlsruhe, Germany\\[2pt]
		\textsuperscript{3}Singularity AI Research, Singularity.inc, Vienna, Austria\\
		\texttt{vinicius@singularity.inc}\par}

		\vspace{0.8em}

		{\sffamily April 2026}%
		\footnote{Preprint. Comments welcome.}
	\end{center}

	\vspace{0.4em}

	\begin{abstract}
		Strategic crisis analysis needs representations that combine qualitative expert judgement, explicit interdependence, and auditable update rules without requiring fully specified payoffs or probabilities.
		Scenario Bundle Analysis (SBA), developed by Amos Perlmutter and Reinhard Selten, provides such a starting point, but the original formulation leaves several database, topology, and update interfaces implicit.
		This paper presents a formal refinement and extension of the original SBA framework,
		introducing a two-layer architecture that separates a static scenario database from a
		dynamic scenario tree system.
		The extended framework incorporates a richer attitude vocabulary: beliefs, desires, intentions, fears, and coalitional commitments, with expectations treated as doxastic attitudes.
		It also adds a domain/modifier layer for contextual framing, a topology on admissible scenario spaces\index{Scenario space}, typed assessment-state updates, and multi-criteria evaluation.
		Mathematical definitions are stated with sufficient precision to support
		computational implementation.
	\end{abstract}

	\bigskip
	\noindent\textbf{Keywords:} Scenario Bundle Analysis, game theory, bounded rationality,
	crisis analysis, conflict modeling, scenario planning, strategic interaction,
	expert elicitation, semiformal methods, decision support.

	\bigskip
	\noindent\textbf{JEL Classification:} C70, C72, D81, F51.

	\newpage

	\setcounter{tocdepth}{3}
	\setcounter{secnumdepth}{3}
	\tableofcontents
	\listoffigures
	\listoftables
	
	\newpage
	
	\begin{small}
		\textit{``I did several types of work on several investigations of the application of game theory to international relations. And this one thing which is of special importance I think, is the scenario bundle method which I developed together with Amos Perlmutter and that was, I think, quite interesting for you.''} \citep{selten2004interview}  
	\end{small}
	
\pagestyle{sbarunning}

	\section{Introduction and Motivation}
	\label{sec:introduction}

	Strategic decision making in crises\index{Applications!Crisis} and
	conflicts\index{Applications!Conflict} is shaped by interdependent choices, incomplete
	information\index{Decision Theory}, and heterogeneous actor\index{Actors} motivations.
	Policy makers and analysts must anticipate stakeholder\index{Methodology!Stakeholders}
	reactions under uncertainty\index{Decision Theory!Uncertainty} while documenting the
	assumptions that drive such anticipations.
	\emph{Scenario Bundle Analysis}
	(SBA\index{Scenario Bundle Analysis}) provides a structured representation of plausible
	strategic developments that remains interpretable for practitioners.

	SBA is a game-theoretic method developed by Amos Perlmutter\index{References!Perlmutter} and Reinhard
	Selten\index{References!Selten}\index{References} \citeyearpar{selten1999sbm}.
	Applications of the original framework are reported in
	\citep{selten2004kosovo, selten2004bosnien}.
	The method is primarily qualitative in character: its workflow targets expert elicitation
	and does not presuppose formal training in game theory, supporting use in diplomatic,
	institutional, and interdisciplinary settings.

	SBA operationalizes transparent reasoning by encoding assumptions and dependency claims
	in an explicit representational structure.
	Formal models in economics or political science rely on strict rationality assumptions,
	precise payoffs, and complete information; qualitative case studies capture motivation,
	discourse, and ambiguity but often leave dependency claims implicit and therefore hard to
	audit for consistency\index{Logical Foundations!Consistency}\index{Logical Foundations}.
	SBA occupies a \emph{semiformal}\index{Methodology!Semiformal approach}\index{Methodology} position between
	these paradigms by combining qualitative content with checkable structural constraints.
	It extends the structuralist tradition\index{Theoretical Foundations!Structuralism} of
	model-based science\index{Theoretical Foundations!Model-based science} \citep{balzer1987}
	and aligns with Simon's account of bounded rationality\index{Theoretical Foundations!Bounded rationality}\index{Theoretical Foundations}
	\citeyearpar{simon1982}.

	\paragraph{Pragmatic motivation.}
	Applied settings---diplomatic negotiation, conflict prevention\index{Events!Prevention},
	and strategic foresight---require models that remain usable under sparse and contested
	information.
	SBA supports early identification of conflict dynamics and escalation pathways, comparison
	of alternative courses of action and their preconditions\index{Options!Preconditions},
	integration of quantitative indicators with expert judgment, and structured
	communication\index{Relations!Information / Communication} among analysts and decision
	makers.
	Explicit data dependencies also enable computational pipelines that assist elicitation
	and consistency checking, including NLP-based and large language model approaches\footnote{NLP: Natural Language Processing, the field concerned with computational analysis and generation of human language. LLM: Large Language Model, a class of neural language models trained on broad text corpora and used for tasks such as entity extraction, relation extraction, and stance detection.} that
	extract implicit regularities from text under coherence constraints.

	\paragraph{Software motivation.}
	The present paper serves as the scientific reference document for a software
	implementation of extended SBA.
	The implementation covers the full pipeline from structured expert elicitation through
	database encoding, scenario tree generation, consistency checking, and multi-criteria
	evaluation.
	Mathematical definitions in this paper are stated with sufficient precision to drive
	data models and algorithmic components directly.
	Where the original SBA relied on manual expert workshops, the software layer supports
	high-throughput elicitation, machine-assisted extraction, and iterative model updating.

	\paragraph{Contributions.}
	This paper contributes four formal refinements of SBA. First, it separates the static Scenario Database from the dynamic Scenario Tree System. Second, it gives typed definitions for actors, coalitions, attributes, attitudes, relations, options, and events. Third, it adds assessment-state updates and topology-based robustness analysis. Fourth, it connects the framework to empirical validation and software implementation through explicit provenance and coding interfaces.

	\paragraph{Core ontology.}
	The SBA core ontology\index{Logical Foundations!Ontology} comprises
	actors\index{Actors}, coalitions\index{Coalitions}, attributes\index{Attributes},
	attitudes\index{Attitudes}, relations\index{Relations}, options\index{Options}, and
	events\index{Events}.
	These categories encode, respectively, who participates (actors), how they form
	alliances (coalitions), what they are like (attributes), what they want or believe
	(attitudes), how they interact (relations), what actions are available (options), and
	what occurs exogenously (events).
	Section~\ref{sec:database} defines each category formally.
	Explicit encoding supports communicable reconstruction of strategic
	knowledge\index{Attitudes!Knowledge} and disciplined comparison of alternative scenario
	developments.

	\paragraph{Terminological note.}
	SBA draws on game-theoretic structure while avoiding the terms
	\emph{player}\index{Theoretical Foundations!Game Theory!Player / Actor},
	\emph{strategy}\index{Theoretical Foundations!Game Theory!Strategy}, and \emph{game}.
	The neutral vocabulary of actors, options, and scenarios accommodates social,
	institutional, and algorithmic agents and avoids colloquial connotations of ``games'' in
	crisis contexts.
	The underlying structure remains game-theoretic in the formal sense: interdependent
	choices under bounded rationality with explicitly represented preference orderings.

	\paragraph{Paper organization.}
	Section~\ref{sec:database} defines the SBA database layer comprising the seven entity
	categories.
	Section~\ref{sec:scenariotrees} introduces scenario trees and bundles.
	Sections~\ref{sec:topology} and~\ref{sec:dynamics} develop the topology of the scenario
	space and dynamic transformation rules.
	Section~\ref{sec:evaluation} covers scenario evaluation and comparison.
	Section~\ref{sec:empirical} addresses empirical grounding and validation.
	Section~\ref{sec:reflection} discusses applications, limitations, and the research
	outlook.

\subsection{Two-Layer Architecture}\index{Database Layer!Two-Layer Architecture}
\label{subsec:two_layer}

	In Selten\index{References!Selten}'s original formulation, actors, their aims and fears,
	and the state-transition structure of the scenario trees\index{Scenario Trees} are treated
	jointly within one model.
	A two-layer representation makes this dual structure explicit:
	\begin{enumerate}
		\item the \emph{Scenario Database} (\(\mathrm{DB}_{\mathrm{SBA}}\)), containing the
		non-temporal entities of analysis: actors\index{Actors}, coalitions\index{Coalitions},
		attributes\index{Attributes}, attitudes\index{Attitudes}, relations\index{Relations},
		options\index{Options}, and events\index{Events}; and
		\item the \emph{Scenario Tree System} (\(\mathbb{T}\)), representing the temporal and causal
		structure of possible state sequences.
	\end{enumerate}
	The complete extended SBA model is the pair
	\[
	\mathcal{M}_{\mathrm{SBA}} = \langle \mathrm{DB}_{\mathrm{SBA}}, \mathbb{T} \rangle,
	\]
	where \(\mathrm{DB}_{\mathrm{SBA}}\) provides the initial structure from which the
	dynamic tree system \(\mathbb{T}\) is generated and evaluated.

	Selten's early applications \citep{selten1976, selten2004kosovo, selten2004bosnien} consider only
	two internal attitudes---\emph{aims} and \emph{fears}---as the intentional basis of
	scenario evaluation.
	The extended formulation generalizes this layer by introducing additional
	attitudes\index{Attitudes} such as beliefs, desires, intentions, fears, and coalitional commitments, with expectations treated as doxastic attitudes, and
	by adding a finite set of \emph{domains}\index{Domains} that frame the contextual
	environment (see Section~\ref{subsec:domains}).
	Dynamic components such as parallel database updates and tree expansion are feasible at
	scale only with computational support; explicit coherence constraints ensure that
	machine-generated updates remain auditable.

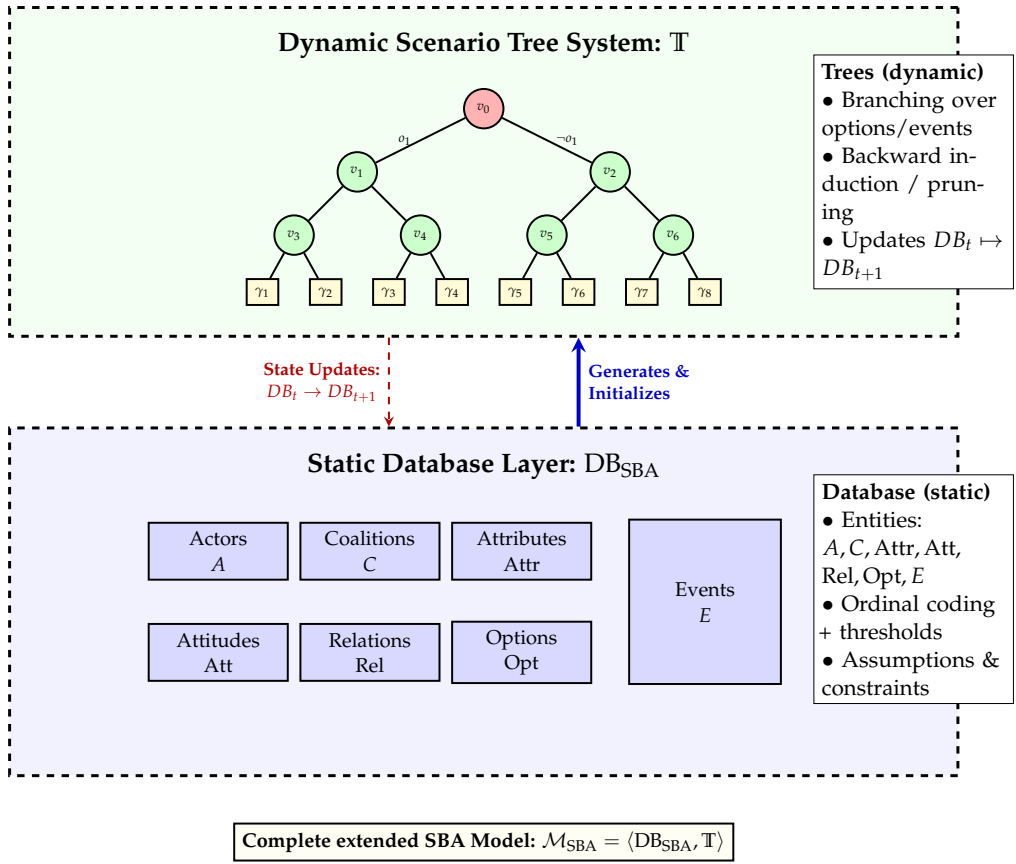
\begin{figure}[htbp]
	\centering
	\resizebox{\textwidth}{!}{%
\begin{tikzpicture}[
		scale=0.9,
		transform shape,
		db_entity/.style={rectangle, draw, thick, fill=blue!15,
			minimum width=2.2cm, minimum height=0.9cm,
			align=center, font=\footnotesize},
		db_entity_big/.style={db_entity, minimum width=2.4cm, minimum height=2.6cm},
		layer/.style={rectangle, draw, very thick, dashed, inner sep=10pt},
		db_layer/.style={layer, fill=blue!5},
		tree_layer/.style={layer, fill=green!5},
		tree_node/.style={circle, draw, thick, fill=green!20,
			minimum size=0.6cm, font=\tiny},
		terminal/.style={rectangle, draw, thick, fill=yellow!20,
			minimum width=0.5cm, minimum height=0.4cm, font=\tiny},
		generates/.style={->, ultra thick, >=stealth, blue!80!black},
		feedback/.style={->, thick, dashed, >=stealth, red!70!black},
		db_link/.style={thick, gray!30},
		annobox/.style={draw, fill=white, font=\small, text width=3.2cm},
		modelbox/.style={draw, thick, fill=yellow!5, font=\footnotesize, minimum width=6cm}
		]

		\def\LayerW{15cm}
		\def\DBH{5.5cm}
		\def\TreeH{5.2cm}
		\def\TreeY{6.8}

		\node[db_layer, minimum width=\LayerW, minimum height=\DBH] (db_container) at (0, 0) {};
		\node[font=\bfseries\large, anchor=north] at (db_container.north) [yshift=-8pt]
		{Static Database Layer: $\mathrm{DB}_{\mathrm{SBA}}$};

		\foreach \name/\title/\sym/\x/\y in {
			actors/Actors/$A$/-4.2/0.8,
			coalitions/Coalitions/$C$/-1.8/0.8,
			attributes/Attributes/$\mathrm{Attr}$/0.6/0.8,
			attitudes/Attitudes/$\mathrm{Att}$/-4.2/-0.8,
			relations/Relations/$\mathrm{Rel}$/-1.8/-0.8,
			options/Options/$\mathrm{Opt}$/0.6/-0.8
		}{
			\node[db_entity] (\name) at (\x,\y) {\title\\\sym};
		}
		\node[db_entity_big] (events) at (3.5, 0.0) {Events\\$E$};

		\begin{scope}[on background layer]
			\foreach \a/\b in {actors/coalitions,coalitions/attributes,actors/attitudes,coalitions/relations,attributes/attitudes,relations/options}{
				\draw[db_link] (\a) -- (\b);
			}
			\draw[db_link] (actors.south east) -- (relations.north west);
			\draw[db_link] (coalitions.south east) -- (options.north west);
		\end{scope}

		\node[tree_layer, minimum width=\LayerW, minimum height=\TreeH] (tree_container) at (0, \TreeY) {};
		\node[font=\bfseries\large, anchor=north] at (tree_container.north) [yshift=-8pt]
		{Dynamic Scenario Tree System: $\mathbb{T}$};

		\begin{scope}[shift={(0, \TreeY)}]
			\node[tree_node, fill=red!30] (root) at (0, 1.0) {$v_0$};

			\foreach \name/\x/\y/\lab in {
				n1/-2/0.0/$v_1$,
				n2/ 2/0.0/$v_2$,
				n3/-3/-1.0/$v_3$,
				n4/-1/-1.0/$v_4$,
				n5/ 1/-1.0/$v_5$,
				n6/ 3/-1.0/$v_6$
			}{
				\node[tree_node] (\name) at (\x,\y) {\lab};
			}

			\foreach \i [evaluate=\i as \x using (-4.5+\i*1.0)] in {1,...,8}{
				\node[terminal] (t\i) at (\x, -1.9) {$\gamma_{\i}$};
			}

			\draw[thick] (root) -- (n1) node[midway, left,  font=\tiny] {$o_1$};
			\draw[thick] (root) -- (n2) node[midway, right, font=\tiny] {$\neg o_1$};

			\foreach \p/\c in {n1/n3,n1/n4,n2/n5,n2/n6,n3/t1,n3/t2,n4/t3,n4/t4,n5/t5,n5/t6,n6/t7,n6/t8}{
				\draw[thick] (\p) -- (\c);
			}
		\end{scope}

		\def\GenX{1.5}
		\def\FbX{-1.5}

		\coordinate (gen_from) at ($(db_container.north)+(\GenX,0)$);
		\coordinate (gen_to)   at ($(tree_container.south)+(\GenX,0)$);
		\coordinate (fb_from)  at ($(tree_container.south)+(\FbX,0)$);
		\coordinate (fb_to)    at ($(db_container.north)+(\FbX,0)$);

		\draw[generates] (gen_from) -- (gen_to)
		node[midway, right, align=left, font=\scriptsize\bfseries] {Generates \&\\Initializes};

		\draw[feedback] (fb_from) -- (fb_to)
		node[midway, left, align=center, font=\scriptsize\bfseries] {State Updates:\\$DB_t \to DB_{t+1}$};

		\node[annobox, text width=2.9cm] at (6.8, 0.2) {
			\textbf{Database (static)}\\
			$\bullet$ Entities: $A,C,\mathrm{Attr},\mathrm{Att}$,
			$\mathrm{Rel},\mathrm{Opt},E$\\
			$\bullet$ Ordinal coding + thresholds\\
			$\bullet$ Assumptions \& constraints
		};

		\node[annobox, text width=2.9cm] at (6.8, 6.8) {
			\textbf{Trees (dynamic)}\\
			$\bullet$ Branching over options/events\\
			$\bullet$ Backward induction / pruning\\
			$\bullet$ Updates $DB_t \mapsto DB_{t+1}$
		};

		\node[modelbox] at (0, -3.8) {
			\textbf{Complete extended SBA Model:}
			$\mathcal{M}_{\mathrm{SBA}} = \langle \mathrm{DB}_{\mathrm{SBA}}, \mathbb{T} \rangle$
		};

	\end{tikzpicture}%
}%
	\caption{Two-layer architecture of extended SBA: the static database
	\(\mathrm{DB}_{\mathrm{SBA}}\) generates and initializes the dynamic scenario tree
	system \(\mathbb{T}\); scenario evaluations feed back as state updates.}
	\label{fig:sba_overview}
\end{figure}

	Figure~\ref{fig:sba_overview} depicts the relationship between the static database layer
	and the dynamic scenario tree system.

A minimal notation convention fixes the carrier sets used throughout.
Let \(A\) be the set of actors and let \(C\subseteq \Pow(A)\) be the set of coalitions.
Let \(E\) be the universe of event labels and let \(\mathrm{Attr}\), \(\mathrm{Att}\), \(\mathrm{Rel}\), and \(\mathrm{Opt}\) denote the attribute, attitude, relation, and option universes.
Let \(t\in\mathbb{N}\) index database stages; \(DB_t\) denotes the assessment state at stage \(t\), with \(DB_0\coloneqq \mathrm{DB}_{\mathrm{SBA}}\).

\subsection{Relationship to Related Analytical Methods}
\label{subsec:related_frameworks}

	SBA combines explicit strategic interdependence with qualitative elicitation and
	auditability.
	Classical game theory, scenario planning, and structured analytic techniques provide
	useful contrasts because they prioritize, respectively, mathematical solution concepts,
	narrative exploration of futures, and bias-aware evidence assessment.
	Table~\ref{tab:framework_comparison} summarizes the comparison at the level of
	methodological features.

	\paragraph{Classical Game Theory.}
	SBA inherits core structure from non-cooperative game theory, particularly extensive-form
	games with perfect information \citep{vonneumann1944, selten1975}: sequential decision
	structures (game trees), backward induction for equilibrium\index{Theoretical Foundations!Game Theory!Equilibrium} selection\index{Tree Analysis!Backward Induction}, and preference
	orderings over outcomes.
	It departs from standard game theory in four respects.
	First, domain-neutral terminology (``actor,'' ``option,'' ``preference ranking'') supports
	expert collaboration without game-theoretic training
	(\emph{epistemic accessibility}).
	Second, ordinal scales and qualitative likelihoods replace cardinal\index{Measurement} utilities and precise
	probabilities when such inputs are unavailable (\emph{qualitative representation}).
	Third, aims and fears are represented explicitly instead of being subsumed in utility
	functions (\emph{motivational transparency}).
	Fourth, cognitive constraints and evolving beliefs are treated as normal modeling
	conditions rather than as deviations (\emph{bounded rationality}).

	\paragraph{Scenario Planning.}
	Scenario planning \citep{kahn1967, schwartz1991, schoemaker1995} and SBA both explore
	multiple futures under uncertainty and support decision-making without prediction claims.
	SBA complements narrative scenario construction with explicit entities and dependency
	constraints---actors, options, events, and preference orderings---which improves
	auditability and consistency checking.
	Strategic interaction is modeled directly: each actor's evaluation depends on anticipated
	reactions by others, producing feedback effects typical of crises, negotiations, and
	conflicts.

	\paragraph{Analysis of Competing Hypotheses and Related Methods.}
	Analysis of Competing Hypotheses (ACH) \citep{heuer1999}, Key Assumptions Check,
	Delphi, and related Structured Analytic Techniques (SATs)
	\citep{heuer2010, phersonheuer2021sat} structure expert judgment and reduce cognitive bias.
	SBA differs by treating anticipated behavior under strategic interdependence as the
	primary analytical object and by providing backward induction as a normative benchmark
	given explicit preference rankings.
	ACH hypotheses about attitudes or causal beliefs can populate the SBA database directly,
	making the two approaches complementary.

\begin{table}[htbp]
	\centering
	\caption{Comparison of SBA with related analytical frameworks.}
	\label{tab:framework_comparison}
	\setlength{\tabcolsep}{3.5pt}
	\footnotesize
	\renewcommand{\arraystretch}{1.4}
	\begin{tabularx}{\textwidth}{@{} >{\RaggedRight\arraybackslash\bfseries}X
		>{\Centering\arraybackslash}p{1.8cm}
		>{\Centering\arraybackslash}p{1.8cm}
		>{\Centering\arraybackslash}p{1.8cm}
		>{\Centering\arraybackslash}p{1.8cm} @{}}
		\toprule
		Feature & \textbf{Game Theory} & \textbf{Scenario Planning} & \textbf{ACH/\allowbreak SATs} & \textbf{SBA} \\
		\midrule
		Formal structure        & High      & Low        & Medium     & Medium--\allowbreak High \\
		Quantitative precision  & Required  & Optional   & Optional   & Optional \\
		Strategic interaction   & Central   & Peripheral & Peripheral & Central \\
		Epistemic accessibility & Low       & High       & High       & High \\
		Equilibrium concepts    & Yes       & No         & No         & Yes (qual.) \\
		Evidence evaluation     & No        & No         & Yes        & Compatible \\
		\bottomrule
	\end{tabularx}
	\vspace{0.2cm}
	\begin{flushleft}
		\small\textit{Note: SBA combines qualitative elicitation with explicit structural
		constraints, occupying a semiformal position between game theory and scenario planning.}
	\end{flushleft}
\end{table}

\subsection{Linguistic Foundations}
\label{subsec:linguistic_foundations}

	The entities of the Scenario Database\index{Database Layer} correspond to linguistic and
	cognitive categories used in strategic descriptions of crises and conflicts.
	SBA\index{Scenario Bundle Analysis} treats formal objects as representations of
	conceptual structure encoded in natural language.

	\paragraph{Predication and description.}
	Properties and relations\index{Relations} are expressed through predicates assigning
	features or values to individuals\index{Actors}
	\citep{frege1892, carnap1947, quine1960, montague1970}.
	In SBA, attributes\index{Attributes} and relations\index{Relations} provide the predicate
	vocabulary, and actors\index{Actors} provide the individuals to which predicates apply.

	\paragraph{Entities and reference.}
	Reference to an actor\index{Actors}, a coalition\index{Coalitions}, or an
	event\index{Events} presupposes a domain of individuals about which statements can be
	made.
	In the SBA ontology\index{Logical Foundations!Ontology}, these domains are given by the
	carrier sets \(A\), \(C\), \(E\), and related constructs.
	Actors serve as nominal referents, while attributes and relations function as predicative
	structure.

	\paragraph{Intensional structure.}
	Attitudes\index{Attitudes} are modeled as intensional relations between an
	actor\index{Actors} and a proposition, in line with the intensional tradition in
	semantics \citep{carnap1947, montague1970, jackendoff1983}.
	An attitude such as \(\text{believe}(a,p)\) encodes a context-sensitive link between an
	agent's informational state and a propositional content, supporting qualitative
	representation of uncertainty without numerical probability assignments.

	\section[Database of the SBA]{Database of the SBA\index{Database Layer}\index{Scenario Bundle Analysis}}
	\label{sec:database}

	The Scenario Database (\(\mathrm{DB}_{\mathrm{SBA}}\)\index{Database Layer}) constitutes
	the static input layer of Scenario Bundle Analysis\index{Scenario Bundle Analysis}.
	It encodes all components that must be fixed before scenario trees\index{Scenario Trees}
	can be constructed: the actor\index{Actors} set~\(A\) and coalition\index{Coalitions}
	set~\(C\); their attributes\index{Attributes}~(\(\mathrm{Attr}\)), attitudes\index{Attitudes}~(\(\mathrm{Att}\)),
	and initial options\index{Options}~(\(\mathrm{Opt}\)); and the relevant exogenous
	events\index{Events}~\(E\).
	Formally, the database represents the initial state at database stage~\(t = 0\):
	\[
		\mathrm{DB}_{\mathrm{SBA}} \;=\;
		\langle A,\, C,\, \mathrm{Attr},\, \mathrm{Att},\, \mathrm{Rel},\,
		\mathrm{Opt},\, E \rangle.
	\]

\paragraph{Category roles and dependencies.}
The database categories play distinct roles in tree construction and evaluation.
Actors\index{Actors} and coalitions\index{Coalitions} provide the carriers of agency and aggregation.
Attributes\index{Attributes}, attitudes\index{Attitudes}, and relations\index{Relations} encode the assessment state that constrains what is plausible and what is desirable.
Options\index{Options} encode feasible moves (by actors or coalitions) under the currently declared constraints.
Events\index{Events} encode exogenous disturbances and observable triggers that can constrain, enable, or force transitions along scenario branches.

\paragraph{Interface discipline.}
Well-formed databases respect typing interfaces that later sections make explicit.
Attribute assignments are evaluated on the unified carrier domain \(\Omega := A\cup C\).
Relations range over \(\Omega\times\Omega\), allowing dyadic, coalition-level, and mixed actor--coalition ties.
Attitudes bind an agent \(x\in\Omega\) to propositional content \(p\in P\).
Option availability is assessed relative to an acting label \(x\in\Omega\) and the declared constraints at stage \(t\).
Event labels \(e\in E\) admit outcome labels \(e^\circ\in E^\circ\) used as edge annotations in scenario trees.
	Dynamic change of these components---coalition\index{Coalitions} formation or dissolution,
	attitude\index{Attitudes} shifts, or evolving attribute\index{Attributes} values---is
	conceptually important but is not modeled within the database itself.
	In the original SBA framework \citep{selten2004kosovo}, the database served solely as a fixed
	input layer for tree construction.
	The extended framework introduces controlled update operators
	\(\mathcal{U}_t : DB_t \to DB_{t+1}\) that govern transitions between
	assessment states; these are treated in Section~\ref{sec:dynamics}.
	Stability diagnostics---dispersion or heterogeneity measures for coalition
	attributes\index{Attributes} and attitudes\index{Attitudes}---may be used as
	\emph{initial selection criteria} to identify internally coherent entities at database stage~\(t=0\).

	\subsection{Domains of SBA\index{Scenario Bundle Analysis}}
	\label{subsec:domains}
	
	SBA employs a two-axis typology that separates interaction mechanisms from constraint regimes.
	A finite set of \emph{core domains} captures the dominant mechanism and typical actor\index{Actors} constellation of a crisis (Table~\ref{tab:crisis_domains}).
	A finite set of \emph{cross-domain modifiers}\index{Domains!Cross-Domain Modifiers} captures systemic stressors that alter resource, informational, or capacity constraints across domains (Table~\ref{tab:crisis_modifiers}).
	The event set of SBA remains denoted by \(E\); modifier symbols are kept distinct to avoid notational collision.
	
	Let
	\[
	\mathcal{D}=\{\text{Pol}\index{Domains!Political},\text{Geo}\index{Domains!Geopolitical},\text{Econ}\index{Domains!Economic},\text{Soc}\index{Domains!Social},\text{Org}\index{Domains!Organizational}\}
	\quad\text{and}\quad
	\mathcal{M}=\{\mathrm{Env}\index{Domains!Cross-Domain Modifiers!Env},\mathrm{Tech}\index{Domains!Cross-Domain Modifiers!Tech},\mathrm{Health}\index{Domains!Cross-Domain Modifiers!Health}\}.
	\]
	A concrete case is represented as \(\text{Crisis}(D,M)\) with \(D\subseteq\mathcal{D}\) and \(M\subseteq\mathcal{M}\).

	
	\subsubsection{Vertical structure: Core domains of conflict and crisis}
	\label{subsubsec:domains_vertical}
	
	The domains in Table~\ref{tab:crisis_domains} result from mapping SBA\index{Scenario Bundle Analysis}'s carrier sets onto empirically distinct mechanisms of interaction.
	Each domain is characterized by:
	\begin{enumerate}
		\item a typical \emph{actor constellation}: which types of entities (states, firms, communities, individuals) predominantly interact,
		\item a dominant \emph{interaction mechanism}: the causal process driving escalation or resolution,
		\item characteristic \emph{escalation pathways}: typical intensification and de-escalation patterns.
	\end{enumerate}
	
	The typology is intended as a compact mid-level classification for SBA applications.
	It also supports model-theoretic comparability, as a domain label \(d\in\mathcal{D}\) functions as a type parameter selecting domain-specific constraints \(\mathcal{C}_d\) (e.g., credibility requirements in geopolitical crises, liquidity constraints in economic crises).
	
	\begin{table}[h!]
		\centering
		\small
		\renewcommand{\arraystretch}{1.3}
		\setlength{\tabcolsep}{4pt}
		\begin{tabularx}{\textwidth}{
				>{\RaggedRight\arraybackslash}p{3.1cm}
				>{\RaggedRight\arraybackslash}X
				>{\RaggedRight\arraybackslash}p{2.7cm}
				>{\RaggedRight\arraybackslash}X}
			\toprule
			\textbf{Domain} & \textbf{Subtypes} & \textbf{Typical Actors} & \textbf{Characteristic Mechanisms} \\
			\midrule
			\textbf{Political (Pol)} \newline \textit{Institutional} &
			Power transition; leadership struggle; regime crisis; institutional deadlock &
			Governments, parties, parliaments, leaders &
			Contestation over authority\index{Relations!Authority / Obligation}; escalation via signaling\index{Relations!Information / Communication!Signaling}, polarization, legitimacy loss \\
			\midrule
			\textbf{Geopolitical (Geo)} \newline \textit{Security} &
			Territorial dispute; alliance instability; arms race; deterrence failure &
			States, alliances, defense institutions &
			Perceived threats and signaling\index{Relations!Information / Communication!Signaling} of intentions; escalation ladders and credibility dynamics \\
			\midrule
			\textbf{Economic (Econ)} \newline \textit{Socioeconomic} &
			Financial instability; trade or resource conflict; distributional inequality &
			Firms, unions, governments, regional blocs &
			Material scarcity; incentive misalignment; market failures \\
			\midrule
			\textbf{Social (Soc)} \newline \textit{Intergroup} &
			Ethnic or cultural tension; migration or identity crisis; protest and mobilization &
			Communities, NGOs, civil society, populations &
			Identity-based threat perception; narrative-driven polarization; collective action dynamics \\
			\midrule
			\textbf{Organizational (Org)} \newline \textit{Interpersonal} &
			Workplace or team conflict; negotiation breakdown; trust or moral dilemma &
			Individuals, teams, management units &
			Emotional or perceptual asymmetries; cooperation breakdowns; coordination failures \\
			\bottomrule
		\end{tabularx}
		\caption[Core crisis domains]{Core domains of conflict and crisis and their characteristic mechanisms. Domain abbreviations (Pol, Geo, Econ, Soc, Org) are used in formal notation.}
		\label{tab:crisis_domains}
	\end{table}
	
	\paragraph{Interpretation.}
	Pol and Geo both involve strategic signaling, but Pol foregrounds authority and legitimacy within institutions, whereas Geo foregrounds security dilemmas and credibility across states and alliances.
	Econ centers on scarcity and coordination under incentives and constraints, Soc centers on identity and mobilization dynamics, and Org centers on intra-organizational coordination and trust under short horizons.
	Domain labels support comparison of scenarios that share a dominant mechanism while allowing domain-specific constraint packages \(\mathcal{C}_D\) (e.g., credibility constraints in Geo versus liquidity and market-coordination constraints in Econ).

	\FloatBarrier
	
	\subsubsection{Horizontal structure: Cross-domain modifiers\index{Domains!Cross-Domain Modifiers}}
	\label{subsubsec:domains_horizontal}
	
	Cross-domain modifiers\index{Domains!Cross-Domain Modifiers} parameterize the conditions under which domain-specific interactions unfold.
	They modify temporal, informational, and resource constraints without introducing an additional interaction mechanism.
	
	Domains classify \emph{how actors interact} (the mechanism), while cross-domain modifiers classify \emph{under what conditions} they interact (the constraints).
	A territorial dispute in the Geo domain follows geopolitical interaction logic; \(\mathrm{Env}\) conditions can shift the relevant resource base or impose time pressure without changing the underlying strategic mechanism.
	
	Formally, any concrete case is denoted by
	\begin{equation}
		\text{Crisis}(D, M), \quad
		D \subseteq \mathcal{D}, \quad
		M \subseteq \mathcal{M}.
	\end{equation}
	
	\begin{table}[h!]
		\centering
		\small
		\renewcommand{\arraystretch}{1.3}
		\setlength{\tabcolsep}{4pt}
		\begin{tabularx}{\textwidth}{
				>{\RaggedRight\arraybackslash}p{3.2cm}
				>{\RaggedRight\arraybackslash}X
				>{\RaggedRight\arraybackslash}X
				>{\RaggedRight\arraybackslash}X}
			\toprule
			\textbf{Modifier} & \textbf{Stressors / Subtypes} & \textbf{Typical Manifestations} & \textbf{Primary Constraint Effects} \\
			\midrule
			\textbf{Environmental (\(\mathrm{Env}\))} \newline \textit{Biophysical} &
			Climate change; natural disasters; resource depletion; ecological degradation &
			Exogenous shocks\index{Events!Shocks}; drought/flood cycles; displacement pressures; resource scarcity &
			Shifts resource endowments and option preconditions\index{Options!Preconditions}; alters event likelihoods; increases time pressure and urgency \\
			
			\midrule
			\textbf{Technological (\(\mathrm{Tech}\))} \newline \textit{Informational} &
			Cyber conflict; misinformation; platform dynamics; AI-mediated decision-making; surveillance &
			Information manipulation; accelerated communication; reduced transparency; contested observability &
			Modifies information structure (observability, speed); affects epistemic trust and asymmetries\index{Modal Logics!Doxastic / Epistemic}\index{Modal Logics}; introduces new option types (e.g., cyber) \\
			
			\midrule
			\textbf{Public Health (\(\mathrm{Health}\))} \newline \textit{Humanitarian} &
			Pandemics; humanitarian emergencies; health system collapse; mass displacement &
			Cascading multi-sector stress; overload of institutions; emergency governance; refugee flows &
			Imposes hard capacity constraints; introduces humanitarian obligations as salient attitudes; increases cascade intensity of event sequences \\
			
			\bottomrule
		\end{tabularx}
		\caption{Cross-domain modifiers (\(\mathrm{Env}\), \(\mathrm{Tech}\), \(\mathrm{Health}\)) and typical manifestations. Modifiers alter constraints without introducing distinct interaction mechanisms.}
		\label{tab:crisis_modifiers}
	\end{table}
	
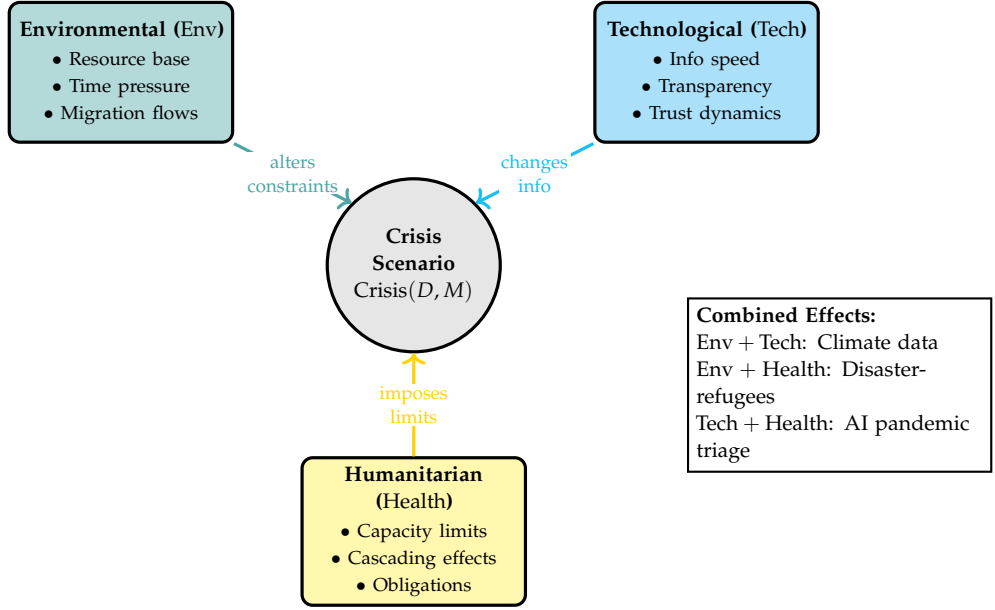
\begin{figure}[htbp]
		\centering
		\begin{tikzpicture}[
			scale=0.88,
			transform shape,
			every node/.style={font=\footnotesize},
			crisis/.style={circle, draw, very thick, fill=gray!20,
				minimum size=2.6cm, align=center, font=\footnotesize\bfseries},
			modbox/.style={rectangle, rounded corners, draw, very thick,
				fill=#1, minimum width=3.2cm, minimum height=2.1cm,
				text width=3.1cm, align=center},
			arrowlbl/.style={font=\scriptsize, align=center, fill=white, inner sep=1pt}
			]
			
			\node[crisis] (crisis) at (0, 0) {Crisis\\Scenario\\$\mathrm{Crisis}(D, M)$};
			
			\node[modbox=teal!30] (env) at (-4.4, 2.9) {
				\textbf{Environmental (\(\mathrm{Env}\))}\\[1pt]
				{\scriptsize $\bullet$ Resource base}\\
				{\scriptsize $\bullet$ Time pressure}\\
				{\scriptsize $\bullet$ Migration flows}
			};
			
			\node[modbox=cyan!30] (tech) at (4.4, 2.9) {
				\textbf{Technological (\(\mathrm{Tech}\))}\\[1pt]
				{\scriptsize $\bullet$ Info speed}\\
				{\scriptsize $\bullet$ Transparency}\\
				{\scriptsize $\bullet$ Trust dynamics}
			};
			
			\node[modbox=yellow!40] (health) at (0, -4.0) {
				\textbf{Humanitarian (\(\mathrm{Health}\))}\\[1pt]
				{\scriptsize $\bullet$ Capacity limits}\\
				{\scriptsize $\bullet$ Cascading effects}\\
				{\scriptsize $\bullet$ Obligations}
			};
			
			\draw[->, very thick, teal!70] (env.south east) -- (crisis.north west)
			node[midway, arrowlbl] {alters\\constraints};
			
			\draw[->, very thick, cyan!70] (tech.south west) -- (crisis.north east)
			node[midway, arrowlbl] {changes\\info};
			
			\draw[->, very thick, yellow!70!orange] (health.north) -- (crisis.south)
			node[midway, arrowlbl] {imposes\\limits};
			
			\node[draw, thick, fill=white, text width=4.3cm, align=left, font=\footnotesize] (info) at (6.4, -1.8) {
				\textbf{Combined Effects:}\\
				\(\mathrm{Env}+\mathrm{Tech}\): Climate data\\
				\(\mathrm{Env}+\mathrm{Health}\): Disaster-refugees\\
				\(\mathrm{Tech}+\mathrm{Health}\): AI pandemic triage
			};
			
		\end{tikzpicture}
		\caption{Cross-domain modifiers as constraint regimes on resource, information, and capacity conditions.}
		\label{fig:crossdomain_impact}
	\end{figure}

	\paragraph{Interpretation.}
	\(\mathrm{Env}\) modifies resource availability and time pressure; \(\mathrm{Tech}\) modifies information structure and epistemic trust; \(\mathrm{Health}\) modifies capacity limits and humanitarian obligations.
	Modifiers can be active in any domain configuration \(D\) and are recorded separately to preserve comparability of mechanism labels across different constraint regimes.
	Figure~\ref{fig:crossdomain_impact} illustrates the constraint pathways of all three modifiers.
	

A crisis instance is represented as \(\mathrm{Crisis}(D,M)\), where \(D\) is a finite set of active domains and \(M\) a finite set of cross-domain modifiers.

	\subsubsection{Examples of composite crisis types}
	\label{subsubsec:domains_examples}
	
	The notation \(\text{Crisis}(D, M)\) specifies composite scenarios by separating dominant mechanisms from modifier conditions:
	
	\begin{itemize}[leftmargin=1.5em]
		
		\item \textbf{Pandemic governance crisis}:\\
		\(\text{Crisis}(\{\text{Pol},\text{Econ}\},\{\mathrm{Health}\})\) --- a public-health emergency
		disrupts political decision-making and economic activity, with spillovers across sectors.
		Relevant actors include governments, health agencies, firms, and labor organizations.
		The interaction combines political contestation and economic coordination under
		extraordinary capacity constraints.
		
		\item \textbf{Climate-induced migration and protest}:\\
		\(\text{Crisis}(\{\text{Soc}\},\{\mathrm{Env}\})\) --- environmental degradation and resource scarcity
		drive displacement and migration, triggering social unrest and identity-based mobilization
		(\(\text{Soc}\)). The core mechanism is polarization and collective action, accelerated by
		environmental stress.
		
		\item \textbf{Supply-chain breakdown with misinformation}:\\
		\(\text{Crisis}(\{\text{Econ}\},\{\mathrm{Tech}\})\) --- cyber attacks, misinformation campaigns, and algorithmic
		failures undermine supply-chain coordination and trust in economic institutions
		(\(\text{Econ}\)). The mechanism is coordination breakdown and market failure, amplified by
		information manipulation.
		
		\item \textbf{AI-managed disaster response failure}:\\
		\(\text{Crisis}(\{\text{Org}\},\{\mathrm{Env},\mathrm{Tech},\mathrm{Health}\})\) --- algorithmic decision systems mismanage disaster
		response, producing organizational coordination failures (\(\text{Org}\)) under simultaneous
		environmental, technological, and humanitarian stressors.
		
		\item \textbf{Resource conflict with regime instability}:\\
		\(\text{Crisis}(\{\text{Geo},\text{Pol}\},\{\mathrm{Env}\})\) --- a territorial dispute over water or energy
		resources (\(\text{Geo}\)) coincides with domestic political instability (\(\text{Pol}\)),
		intensified by climate-driven scarcity. Adequate modeling combines geopolitical
		signaling with domestic legitimacy dynamics under resource constraints.
		
	\end{itemize}
	
	The examples instantiate compositional analysis: \(D\) identifies dominant interaction mechanisms, while \(M\) records constraint regimes that can recur across otherwise distinct domain configurations.
	
	\subsubsection{Formal implications}
	\label{subsubsec:domains_formal}
	
	The domain typology translates into structural constraints on SBA models.
	Each domain \(d \in \mathcal{D}\) imposes characteristic constraints \(\mathcal{C}_d\):
	\textbf{Geo} requires credibility mechanisms and escalation thresholds;
	\textbf{Econ} imposes budget and liquidity constraints;
	\textbf{Soc} emphasizes identity attitudes and collective-action thresholds;
	\textbf{Pol} requires legitimacy relations and institutional veto points;
	\textbf{Org} focuses on trust dynamics and short temporal horizons.
	Cross-domain modifiers alter these constraints without changing interaction mechanisms:
	\(\mathrm{Env}\) shifts resource endowments and time pressure;
	\(\mathrm{Tech}\) modifies information structure and epistemic trust;
	\(\mathrm{Health}\) imposes capacity limits and humanitarian obligations.
	
	\paragraph{Compositional semantics.}
	A crisis scenario \(\text{Crisis}(D, M)\) inherits constraints from all active domains and modifiers:
	\begin{equation}
		\mathcal{C}_{\text{Crisis}(D,M)} = \bigcup_{d \in D} \mathcal{C}_d \cup \bigcup_{m \in M} \mathcal{C}_m
	\end{equation}
	This supports modular construction of compound scenarios and systematic comparison across cases that share mechanisms but differ in constraint regimes.
	
	\subsubsection{Scope and limitations}
	\label{subsubsec:domains_scope}
	
	The distinction between domains and cross-domain modifiers\index{Domains!Cross-Domain Modifiers} does not claim to exhaust all conceivable forms of conflict or crisis.
	It delineates constellations that can be represented within the logical and structural framework of SBA\index{Scenario Bundle Analysis} at a mid-level of abstraction intended for applied modeling. Several caveats apply:
	
	\begin{enumerate}
		\item \textbf{Domain boundaries}: Empirical cases often overlap. A trade war (\(\text{Econ}\)) may involve geopolitical signaling (\(\text{Geo}\)) and domestic political contestation (\(\text{Pol}\)). The notation \(\text{Crisis}(D, M)\) with \(|D|>1\) accommodates such cases, while requiring explicit specification of the intended dominant mechanism.
		\item \textbf{Cultural and ideational factors}: The typology includes no separate ideational or cultural domain. Cultural factors are treated as attributes (e.g., cultural identity) or attitudes (e.g., normative beliefs) within the Social domain. This reflects an action-theoretic commitment: culture primarily shapes interpretation and evaluation rather than constituting an independent interaction mechanism.
		\item \textbf{Emerging mechanisms}: New crisis forms may require extension of the typology. Autonomous AI agents could motivate a distinct domain if their interaction mechanisms become irreducible to human organizational dynamics.
		\item \textbf{Granularity trade-offs}: The five-domain structure fixes a middle level of abstraction. Finer partitions increase descriptive precision but reduce comparability; coarser partitions increase comparability but blur mechanisms.
	\end{enumerate}
	
	Nevertheless, the formal architecture---\(\text{Crisis}(D, M)\) and constraints ---supports refinement by adding domain or modifier labels.

	\subsubsection{Theoretical foundation}
	\label{subsubsec:domains_theory}
	
	Multi-axis typologies in conflict and security research justify separating mechanism categories from orthogonal conditioning dimensions.
	Frameworks such as Waltz's levels of analysis \citep{waltz1959}, Buzan et al.'s sectoral analysis \citep{buzan1998}, and Galtung's distinctions between direct, structural, and cultural violence \citep{galtung1996} support the methodological claim that crises admit more than one organizing axis without collapsing heterogeneous drivers into a single taxonomy.
	
	Environmental degradation, technological interdependence, and global health crises plausibly function as cross-cutting stressors rather than standalone interaction mechanisms \citep{beck1992, homer1999, castells1996, helbing2013, fidler2004, who2022}.
	This motivates a separation between (i) domain labels that classify the dominant strategic mechanism and (ii) modifier labels that parameterize the constraints under which that mechanism unfolds.
	
	The vertical-horizontal distinction serves two analytical purposes.
	First, it constrains category proliferation: environmental, technological, and health-related stressors are treated as modifiers of time, information, and resources rather than as additional domains with distinct interaction logics.
	Second, it supports compositional comparison: a crisis can be analyzed as a primary domain mechanism (e.g., political contestation) under specific modifier conditions (e.g., pandemic constraints), enabling systematic comparison across structurally similar scenarios.

	\subsection{Actor\index{Actors} Categories}
	\label{subsec:actors}
	
	Actors\index{Actors} (\(A\)) constitute the central carrier set\index{Logical Foundations!Carrier Sets} of SBA\index{Scenario Bundle Analysis} and anchor the attitudinal and relational structure of any scenario.
	Five generic actor categories capture recurring differences in scale, institutional embedding, and autonomy (Table~\ref{tab:actor_categories}).
	
	The categories are analytical distinctions rather than rigid ontological classes.
	An entity can shift category across scenarios or levels of granularity: a government ministry functions as an institutional actor in geopolitical analysis but can be decomposed into collective or individual actors when internal organizational dynamics become decision-relevant.

	\subsubsection{Typology of actors\index{Actors}}
	\label{subsubsec:actors_typology}
	
	\begin{table}[h!]
		\centering
		\small
		\renewcommand{\arraystretch}{1.3}
		\setlength{\tabcolsep}{4pt}
		\begin{tabularx}{\textwidth}{
				>{\RaggedRight\arraybackslash}p{2.7cm}
				>{\RaggedRight\arraybackslash}X
				>{\RaggedRight\arraybackslash}X
				>{\RaggedRight\arraybackslash}p{3.0cm}}
			\toprule
			\textbf{Category} & \textbf{Description} & \textbf{Examples} & \textbf{Typical SBA contexts} \\
			\midrule
			\textbf{Individuals} &
			Human agents with bounded rationality, preferences, and intentions. &
			Citizens, employees, leaders, experts. &
			\(\text{Org}\), \(\text{Soc}\), \(\text{Pol}\). \\
			
			\midrule
			\textbf{Collective actors} &
			Groups with joint aims, internal coordination, and a comparatively stable collective identity. &
			Parties, movements, firms, NGOs. &
			\(\text{Pol}\), \(\text{Econ}\), \(\text{Soc}\). \\
			
			\midrule
			\textbf{Institutional actors} &
			Formal entities endowed with authority\index{Relations!Authority / Obligation}, resources, and legal standing. &
			States, ministries, corporations, international organizations. &
			\(\text{Pol}\), \(\text{Geo}\), \(\text{Econ}\). \\
			
			\midrule
			\textbf{Hybrid or networked constellations} &
			Loosely coupled constellations of human and technical components with distributed agency. &
			Platforms, alliances, AI--human teams. &
			Across domains; often under \(\mathrm{Tech}\) modifier. \\
			
			\midrule
			\textbf{Artificial / algorithmic actors} &
			Non-human autonomous systems with decision procedures. &
			Algorithms, LLMs, autonomous agents, bots. &
			Across domains; often under \(\mathrm{Tech}\) modifier. \\
			
			\bottomrule
		\end{tabularx}
		\caption{Generic actor\index{Actors} categories in Scenario Bundle Analysis\index{Scenario Bundle Analysis} (SBA\index{Scenario Bundle Analysis}).}
		\label{tab:actor_categories}
	\end{table}
	
	Distinct categories correspond to typical differences in (i) how aims, fears, and beliefs are attributed, (ii) how relations\index{Relations} are stabilized, and (iii) how option sets\index{Options} are generated and constrained.
	Individuals are a paradigmatic carrier of attitudes and choice under bounded rationality; collective and institutional actors encode coordinated agency under internal procedures and authority constraints; hybrid and algorithmic actors encode distributed or software-mediated decision processes that frequently arise under technological/informational conditions.
	
	\subsubsection{Formal representation}
	\label{subsubsec:actors_formal}
	
	Actor categories can be represented as subsets of \(A\):
	\begin{equation}
		A_{\mathrm{ind}},\ A_{\mathrm{coll}},\ A_{\mathrm{inst}},\ A_{\mathrm{hyb}},\ A_{\mathrm{alg}} \subseteq A,
		\qquad
		A = \bigcup_{x \in \{\mathrm{ind},\mathrm{coll},\mathrm{inst},\mathrm{hyb},\mathrm{alg}\}} A_x,
	\end{equation}
	where the subsets are not necessarily disjoint.
	Overlap reflects granularity choices and hybrid constructions (e.g., an AI--human team can be modeled as hybrid, while its algorithmic component can be modeled separately when its autonomy is decision-relevant).
	
	Each category induces characteristic constraints on attitudes, relations, and options.
	Institutional actors require explicit authority relations and procedural option constraints; collective actors require internal coordination assumptions; hybrid and algorithmic actors often require explicit modeling of observability, information flows, and software-mediated action channels.
	
	\subsubsection{Theoretical foundation}
	\label{subsubsec:actors_theory}
	
	The actor typology reflects standard distinctions in social and institutional theory.
	Frameworks in rational-choice sociology and institutional economics emphasize the interplay between individual choice, organizational structure, and institutional constraint \citep{coleman1990, north1991, march1984}.
	These distinctions support separating individual, collective, and institutional actors when they display different coordination structures and constraint profiles.
	
	Network-oriented perspectives extend the actor concept to socio-technical systems and distributed agency \citep{castells1996, latour2005}.
	Actor-Network Theory motivates treating technical artifacts and organizational infrastructures as components that can materially shape strategic outcomes, even when they do not instantiate human-like intentionality.
	
	Research on algorithmic and artificial agency analyzes decision capacity in digital environments and motivates representing autonomous systems as actors when their outputs generate stable, consequential choice patterns \citep{floridi2019, taeihagh2021}.
	This stance is pragmatic: actorhood in SBA is a modeling commitment about explanatory and predictive relevance, not a claim about consciousness or moral status.
	
	\paragraph{The homogeneity assumption.}
	Actors\index{Actors} are treated as internally coherent units in the sense that their encoded attitudes, relations, and option sets remain sufficiently stable to support a consistent scenario representation.
	Homogeneity is a modeling simplification rather than an empirical claim: real organizations, states, and individuals can be internally inconsistent, but SBA requires a stable interface for assigning attitudes and options to decision-making units.
	Homogeneity is relative to the chosen granularity: a state may be treated as homogeneous in a geopolitical model, but decomposed into ministries or factions when internal bargaining drives the scenario dynamics.
	
	When internal conflicts within an entity become decision-relevant, the entity should be decomposed into multiple actors or represented as a coalition\index{Coalitions} (Section~\ref{subsec:coalitions}) with explicit coordination assumptions.
	
	\subsubsection{Operationalization and practical guidelines}
	\label{subsubsec:actors_guidelines}
	
	Actor identification is operationally guided by decision relevance, homogeneity at the chosen granularity, and autonomy of choice.
	
	\paragraph{Step 1: Identify decision-relevant entities.}
	Collect entities that make consequential choices in the scenario:
	\begin{itemize}
		\item named individuals (political leaders, CEOs),
		\item organizations (governments, corporations, NGOs),
		\item informal groups (protest movements, groups of investors),
		\item technical systems (algorithms, platforms).
	\end{itemize}
	
	\paragraph{Step 2: Assess coherence at the intended granularity.}
	Treat an entity as a single actor if its aims, fears, and option set can be represented as sufficiently coherent for the scenario:
	\begin{itemize}
		\item stable aims and salient fears,
		\item internal conflicts negligible for the focal decisions,
		\item capacity to commit to options without continuous internal renegotiation.
	\end{itemize}
	If these conditions fail, decompose the entity into multiple actors or model it as a coalition (see Section~\ref{subsec:coalitions}).
	
	\paragraph{Step 3: Assign a working category.}
	Assign the category that best captures the entity's strategic role:
	\begin{itemize}
		\item scale: individual vs.\ collective vs.\ institutional,
		\item structure: formal vs.\ informal vs.\ networked/hybrid,
		\item nature: human vs.\ algorithmic vs.\ mixed.
	\end{itemize}
	
	\paragraph{Step 4: Record category-specific constraints.}
	Institutional actors: legal mandate, authority structure, resource base.\\
	Collective actors: membership structure, coordination mechanisms.\\
	Hybrid constellations: human and technical components, interaction channels, observability constraints.\\
	Algorithmic actors: decision procedure, objective proxies, hard constraints.
	
	\paragraph{Example: Climate negotiation scenario.}
	\begin{itemize}
		\item \textbf{Institutional actors}: nation-states (USA, China, EU), international organizations (UNFCCC, World Bank),
		\item \textbf{Collective actors}: environmental NGOs (Greenpeace, WWF), industry associations,
		\item \textbf{Individuals}: key negotiators, scientific advisors,
		\item \textbf{Hybrid constellations}: modeling consortia combining institutions and computational infrastructure,
		\item \textbf{Algorithmic actors}: automated market mechanisms (emissions trading), platform systems shaping communication and coordination.
	\end{itemize}
	
	\subsubsection{Scope and limitations}
	\label{subsubsec:actors_scope}
	
	The five-category typology is a compact mid-level classification for SBA applications rather than an exhaustive ontology.
	Boundary cases arise when legal form, autonomy, and coordination structures point to different categorizations:
	\begin{itemize}
		\item a political party may be modeled as collective or institutional depending on its formal legal status and decision procedures,
		\item a human-supervised AI system may be modeled as algorithmic or hybrid depending on autonomy and control points,
		\item a decentralized movement may be modeled as collective or hybrid depending on coordination mechanisms.
	\end{itemize}
	
	New actor forms continue to emerge, including decentralized autonomous organizations (DAOs), multi-agent AI systems with emergent collective behavior, and transnational epistemic communities coordinating via digital platforms.
	The typology can accommodate such entities as hybrid or algorithmic actors, while domain-specific constraints may require refinement as empirical understanding develops.
	
	Actor categories depend on institutional and cultural context.
	Non-Western governance forms (e.g., clan-based organizations, religious authorities, traditional leadership structures) can be represented in SBA, but the typology may need adaptation to preserve explanatory adequacy.
	
	Classifying an entity as an actor with aims and options encodes a modeling stance about decision relevance and strategic impact.
	This stance is unambiguous for humans and human organizations but remains contested for algorithmic systems; SBA treats actorhood pragmatically as a representational choice rather than a claim about moral responsibility.


	\subsection{Coalitions\index{Coalitions}}
	\label{subsec:coalitions}
	
	Coalitions\index{Coalitions} are temporary or enduring configurations of actors\index{Actors} that coordinate behavior within a scenario.
	Formally, the set of coalitions \(C\) is a selected family of actor subsets:
	\begin{equation}
		C \subseteq \{X \subseteq A \mid |X| \geq 2\}.
	\end{equation}
	Each \(X \in C\) denotes a concrete coordination unit in the model, specified by its member set and by the coalition-relevant constraints encoded in the scenario database.
	
	Coalitions derive their strategic capacity from member coordination rather than from a unified internal decision procedure.
	Coalition formation, stability, and dissolution therefore act as central mechanisms in crises and conflicts, especially in settings where institutional enforcement is weak or contested.

	\subsubsection{Conceptual distinction: Coalitions vs.\ collective actors}
	\label{subsubsec:coalitions_distinction}
	
	Coalitions\index{Coalitions} differ from collective actors\index{Actors} (\(A_{\mathrm{coll}}\)) in representational status, persistence, and decision structure.
	Collective actors are elements of the carrier set \(A\) with an internal coordination interface treated as part of the actor description.
	Coalitions are derived configurations of actors, represented as subsets \(X \subseteq A\) whose coordination is contingent and defeasible.
	Table~\ref{tab:collective_vs_coalition} summarises the key representational differences between coalitions and collective actors.
	
	\begin{table}[h!]
		\centering
		\small
		\renewcommand{\arraystretch}{1.4} 
		\setlength{\tabcolsep}{4pt}
		
		\begin{tabularx}{\textwidth}{@{}
				>{\bfseries\raggedright\arraybackslash}p{2.8cm} 
				>{\RaggedRight\arraybackslash}X       
				>{\RaggedRight\arraybackslash}X       
				@{}}
			\toprule
			Feature & \textbf{Collective actor (\(a \in A_{\mathrm{coll}}\))} & \textbf{Coalition (\(X \in C \subseteq \mathcal{P}(A)\))} \\
			\midrule
			
			Standing / identity &
			Persistent identity with an internal coordination interface (e.g., organization, party). &
			Relational configuration of actors whose identity is exhausted by coordinated membership in the scenario. \\
			\addlinespace 
			
			Persistence &
			Typically persists across scenarios and retains resources, roles, and personnel. &
			Situational and contingent; dissolves when coordination loses relevance or feasibility. \\
			\addlinespace
			
			Decision procedure &
			Internal procedures can generate binding choices (bylaws, voting rules, executive authority). &
			Coordination is negotiated; members retain autonomy and can defect. \\
			\addlinespace
			
			Defection possibility &
			Internal dissent exists but is mediated by institutional procedures. &
			Defection is a first-order strategic possibility and a central stability concern. \\
			\bottomrule
		\end{tabularx}
		\caption{Collective actors versus coalitions: representational status and stability-relevant differences.}
		\label{tab:collective_vs_coalition}
	\end{table}

	The boundary between coalitions and collective actors is permeable.
	Persistent coalitions can evolve into collective actors when stable coordination mechanisms and formal representation become part of the actor description.
	Conversely, collective actors can fragment into competing coalitions under internal crises.
	
	\textbf{Example.}
	The European Coal and Steel Community (1951) began as a coalition of states coordinating industrial policy and gradually institutionalized into the European Union, an actor with supranational authority.
	Conversely, the Soviet Union fragmented into independent states and informal coalitions after 1991.
	
	\subsubsection{Formal representation}
	\label{subsubsec:coalitions_formal}
	
	\paragraph{Basic structure.}
	A coalition \(X \in C\) is represented as a finite member set
	\begin{equation}
		X = \{a_1, a_2, \ldots, a_n\} \subseteq A, \quad n \geq 2,
	\end{equation}
	excluding single-actor ``coalitions'' by definition.
	
	\paragraph{Coordination requirements.}
	Coalitions require (i) aim alignment, (ii) coordination capacity, and (iii) stability mechanisms against defection:
	\begin{itemize}
		\item \emph{Aim alignment}: members share or negotiate a sufficiently aligned set of aims\index{Aim} for joint action,
		\item \emph{Coordination capacity}: members can communicate and synchronize options\index{Options},
		\item \emph{Stability mechanisms}: incentives or constraints deter unilateral deviation (treaties, sanctions, repetition, reputational costs).
	\end{itemize}
	Sections~\ref{subsec:attitudes}, \ref{subsec:relations}, and \ref{subsec:options} provide the representational machinery that specifies these conditions in terms of attitudes, relations, and feasible option sets.
	
	\paragraph{Coalition stability.}
	A coalition \(X\) is stable at a scenario node when each member expects weakly better ranked terminal outcomes under continued coalition coordination than under unilateral deviation, given the encoded beliefs/expectations and the scenario tree continuation.
	This stability notion is conceptually related to core stability in cooperative game theory \citep{gillies1959}, but it is evaluated within scenario trees rather than as a static allocation problem.
	
	\subsubsection{Cross-domain coalitions\index{Coalitions}}
	\label{subsubsec:coalitions_crossdomain}
	
	Coalitions may link actors from heterogeneous vertical domains (Section~\ref{subsec:domains}).
	A coalition \(X \in C\) is \emph{cross-domain} if its members carry at least two distinct analyst-assigned domain labels from \(\{\text{Pol}, \text{Geo}, \text{Econ}, \text{Soc}, \text{Org}\}\).
	Using the (database-level) label function \(\text{domain}:A \to \{\text{Pol}, \text{Geo}, \text{Econ}, \text{Soc}, \text{Org}\}\), interpreted as an analyst-assigned primary label for the current model granularity, cross-domain membership is expressed by:
	\begin{equation}
		X \text{ is cross-domain} \;\Leftrightarrow\; \exists a_i, a_j \in X:\ \text{domain}(a_i) \neq \text{domain}(a_j).
	\end{equation}
	
	\paragraph{Examples of cross-domain coalitions.}
	\begin{itemize}
		\item \textbf{Political--Economic alliances:} governments (\(\text{Pol}\)) coordinate with firms and financial institutions (\(\text{Econ}\)) on infrastructure development or trade policy (often under \(\mathrm{Tech}\) when digital infrastructure is central).
		\item \textbf{Public--private partnerships:} state agencies (\(\text{Pol}\)), NGOs (\(\text{Soc}\)), and corporations (\(\text{Econ}\)) collaborate on public goods provision; pandemic-response coalitions are frequently catalyzed under \(\mathrm{Health}\).
		\item \textbf{Scientific and regulatory consortia:} research institutions and organizational units (\(\text{Org}\)), industry (\(\text{Econ}\)), and regulators (\(\text{Pol}\)) coordinate on standards and joint projects; coordination often depends on \(\mathrm{Tech}\).
		\item \textbf{Geopolitical--Social alignments:} states (\(\text{Geo}\)) align with diaspora communities or transnational movements (\(\text{Soc}\)); climate-justice coalitions are frequently catalyzed under \(\mathrm{Env}\).
	\end{itemize}
	
	\paragraph{Analytical significance.}
	Cross-domain coalitions matter because they (i) pool complementary resources across institutional boundaries, (ii) couple vertical domain constraints with modifier-driven stressors (\(\mathrm{Env}\), \(\mathrm{Tech}\), \(\mathrm{Health}\)), and (iii) introduce accountability tensions between heterogeneous constituencies (sovereignty, profit, legitimacy, humanitarian obligations).
	
	\subsubsection{Theoretical foundation}
	\label{subsubsec:coalitions_theory}
	
	Coalition formation has long been analyzed in political science and economics, with size, composition, and stability as key determinants of strategic success \citep{riker1962, axelrod1984}.
	
	\paragraph{Size and composition.}
	Riker's theory of political coalitions \citep{riker1962} characterizes minimal winning coalitions in competitive settings.
	Non-zero-sum contexts permit larger coalitions when scale effects, legitimacy, or risk pooling dominate per-member distributive incentives.
	
	\paragraph{Stability and cooperation.}
	Axelrod's analysis of cooperation \citep{axelrod1984} explains coalition stability through repeated interaction, reciprocity, and credible punishment of defection, including reputational mechanisms and withdrawal of future support.
	
	\paragraph{Game-theoretic formalization.}
	Cooperative game theory formalizes coalition structure through stability and
	allocation concepts such as the core, Shapley value, and nucleolus
	\citep{gillies1959, greenberg1993}.
	SBA draws primarily on the stability intuition while treating coalition outcomes as tree-embedded and context-dependent rather than as a static division problem.
	
	\paragraph{Social and organizational perspectives.}
	In organizational and collective-action contexts, coalitions are temporary alignments enabling joint action without full institutional integration \citep{olson1965}.
	Olson's free-rider problem highlights coordination costs and defection incentives, motivating explicit attention to enforcement and commitment mechanisms.
	
	\subsubsection{Coalition dynamics in scenario trees}
	\label{subsubsec:coalitions_dynamics}
	
	Coalitions form, evolve, and dissolve as scenarios unfold because aim alignment, feasibility, and credibility change with events and prior moves.
	Formation corresponds to decision points where actors accept coordination constraints in exchange for anticipated improvements in ranked outcomes.
	Dissolution corresponds to states where defection becomes weakly preferred for at least one member under updated beliefs, constraints, or aims.
	Evolution includes membership change, role differentiation, and partial re-coordination as event realizations modify resources and information.
	
	In scenario tree notation, coalition dynamics appear as (i) nodes where coordination is chosen or rejected, and (ii) database updates that modify coalition membership and coordination relations along paths.
	
	\subsubsection{Practical guidelines for modeling coalitions}
	\label{subsubsec:coalitions_guidelines}
	
	Coalition modeling is guided by aim overlap, coordination feasibility, and defection incentives.
	
	\paragraph{Step 1: Identify candidate coalitions.}
	Which actors share salient aims or face common threats?
	Which combinations pool complementary resources or legitimacy?
	Which cooperation patterns have historical precedent?
	
	\paragraph{Step 2: Assess stability drivers.}
	Which actors face defection incentives under plausible event realizations?
	Which enforcement mechanisms exist (treaties, sanctions, repetition, reputational costs)?
	Which information asymmetries undermine credibility?
	
	\paragraph{Step 3: Specify collective capacity.}
	Which options become feasible only under coordination (pooled resources, joint commitments, synchronized timing)?
	Which option constraints arise from coalition membership (e.g., incompatibility with bilateral deals)?
	
	\paragraph{Step 4: Analyze cross-domain tensions.}
	Which domain-specific constraints generate internal tension (sovereignty vs.\ profit vs.\ legitimacy vs.\ humanitarian obligations)?
	Which modifiers (\(\mathrm{Env}\), \(\mathrm{Tech}\), \(\mathrm{Health}\)) amplify or relax those tensions?
	
	\paragraph{Step 5: Track coalition-relevant updates along tree paths.}
	At which nodes can coalitions plausibly form or dissolve?
	Which events shift aim alignment, feasibility, or credibility?
	
	\subsubsection{Limitations and extensions}
	\label{subsubsec:coalitions_limitations}
	
	The present coalition representation isolates a core modeling interface while leaving several refinements outside the current scope.
	
	\paragraph{Nested coalitions.}
	Hierarchical coalition structures (coalitions of coalitions) are not represented explicitly.
	Cases such as forums containing states and regional blocs can be approximated by introducing an institutional actor that represents the forum, or by modeling bloc membership at the actor level.
	
	\paragraph{Partial membership.}
	Binary membership \(a \in X\) abstracts from issue-specific participation and graded commitment.
	Issue-specific coalitions can be represented by separate coalitions indexed by topic, at the cost of increased database size.
	
	\paragraph{Internal bargaining.}
	Coalition formation is treated at the interface level rather than by modeling intra-coalition bargaining over aims, contributions, and benefit distribution.
	When internal bargaining is decision-relevant, decomposition into multiple actors or explicit coalition-internal options becomes necessary.
	
	\paragraph{Coalition competition.}
	Competing coalition structures require explicit modeling of coalition--coalition interaction, not only coalition--actor interaction.
	This interaction can be represented by treating coalitions as decision nodes in the tree when they act through a joint option interface.
	
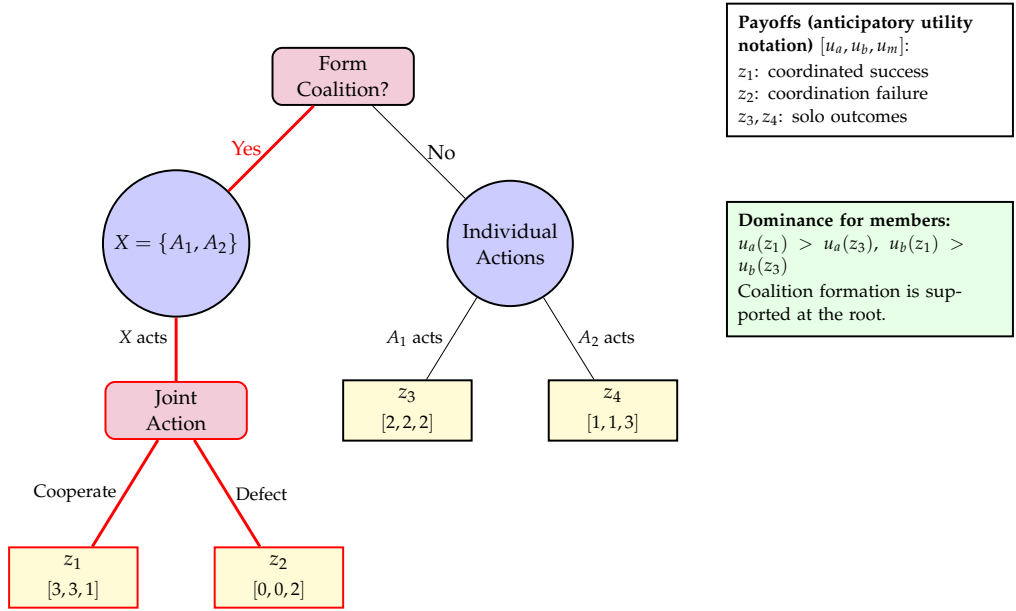
\begin{figure}[htbp]
		\centering
		\resizebox{\textwidth}{!}{%
\begin{tikzpicture}[
			scale=0.92,
			transform shape,
			level distance=2.6cm,
			sibling distance=5.2cm,
			every node/.style={font=\footnotesize},
			level 2/.style={sibling distance=3.2cm},
			decision/.style={circle, draw, thick, minimum size=0.95cm, fill=blue!20, align=center},
			coalition/.style={rectangle, draw, thick, minimum width=2.2cm, minimum height=0.85cm, fill=purple!20, align=center, rounded corners},
			terminal/.style={rectangle, draw, thick, minimum width=2.0cm, minimum height=0.75cm, fill=yellow!20, align=center},
			chosen/.style={very thick, red},
			annobox/.style={draw, thick, fill=white, text width=4.1cm, align=left, font=\scriptsize, inner sep=5pt}
			]
			
			\node[coalition] (root) {Form\\Coalition?}
			child {
				node[decision] (yes) {$X=\{A_1,A_2\}$}
				child {
					node[coalition] (joint) {Joint\\Action}
					child {
						node[terminal] (t1) {$z_1$\\[0.05cm]{\scriptsize $[3,3,1]$}}
						edge from parent node[left, font=\scriptsize] {Cooperate}
					}
					child {
						node[terminal] (t2) {$z_2$\\[0.05cm]{\scriptsize $[0,0,2]$}}
						edge from parent node[right, font=\scriptsize] {Defect}
					}
					edge from parent node[left, font=\scriptsize, pos=0.3] {$X$ acts}
				}
				edge from parent[chosen] node[left, font=\footnotesize] {Yes}
			}
			child {
				node[decision] (no) {Individual\\Actions}
				child {
					node[terminal] (t3) {$z_3$\\[0.05cm]{\scriptsize $[2,2,2]$}}
					edge from parent node[left, font=\scriptsize] {$A_1$ acts}
				}
				child {
					node[terminal] (t4) {$z_4$\\[0.05cm]{\scriptsize $[1,1,3]$}}
					edge from parent node[right, font=\scriptsize] {$A_2$ acts}
				}
				edge from parent node[right, font=\footnotesize] {No}
			};
			
			\node[annobox] at (8.2, 0.1) {
				\textbf{Payoffs (anticipatory utility notation)} $[u_a,u_b,u_m]$:\\[0.2em]
				$z_1$: coordinated success\\
				$z_2$: coordination failure\\
				$z_3,z_4$: solo outcomes
			};
			
			\node[annobox, fill=green!10] at (8.2, -3.0) {
				\textbf{Dominance for members:}\index{Decision Theory!Dominance}\\
				$u_a(z_1) > u_a(z_3)$,\;
				$u_b(z_1) > u_b(z_3)$\\[0.2em]
				Coalition formation is supported at the root.
			};
			
		\end{tikzpicture}%
}%
		\caption{Coalition formation as a coordination decision in a scenario tree.}
		\label{fig:coalition_tree_example}
	\end{figure}

	Despite these limitations, the interface-level coalition representation captures common coordination patterns that drive crisis dynamics in applied SBA models.
	Figure~\ref{fig:coalition_tree_example} shows how coalition formation appears as a coordination decision node in a scenario tree.
	

	
\begin{figure}[htbp]
		\centering
		\resizebox{\textwidth}{!}{
			\begin{tikzpicture}[
				actor/.style={circle, draw, thick, minimum size=0.75cm, font=\scriptsize},
				coalition/.style={draw, very thick, dashed, fill=purple!5, inner sep=10pt, rounded corners=15pt},
				event/.style={rectangle, draw, thick, fill=#1, text width=2.0cm, align=center, font=\scriptsize, inner sep=4pt}
				]
				
				\node[font=\footnotesize\bfseries] at (0, 5) {$z_0$: Initial state};
				\node[font=\footnotesize\bfseries] at (5.5, 5) {$z_1$: Coalition forms};
				\node[font=\footnotesize\bfseries] at (11, 5) {$z_2$: Coalition dissolves};
				
				\node[actor, fill=blue!20] (a1t0) at (-0.6, 3.2) {$A_1$};
				\node[actor, fill=green!20] (a2t0) at (0.6, 3.2) {$A_2$};
				\node[actor, fill=red!20] (a3t0) at (0, 2.2) {$A_3$};
				
				\node[font=\scriptsize, text width=2.4cm, align=center] at (0, 0.5) {Actors pursue\\individual aims};
				
				\node[coalition] (coal_bg) at (5.5, 2.7) {\phantom{\rule{2cm}{1.2cm}}};
				\node[actor, fill=blue!20] (a1t1) at (5.0, 3.1) {$A_1$};
				\node[actor, fill=green!20] (a2t1) at (6.0, 3.1) {$A_2$};
				\node[actor, fill=red!20] (a3t1) at (5.5, 2.3) {$A_3$};
				
				\node[font=\scriptsize, text width=3.2cm, align=center] at (5.5, 0.5) {
					\textbf{Coalition $X=\{A_1,A_2,A_3\}$}\\Joint action on aligned aims
				};
				
				\node[actor, fill=blue!20] (a1t2) at (10.2, 3.6) {$A_1$};
				\node[actor, fill=green!20] (a2t2) at (11.8, 3.6) {$A_2$};
				\node[actor, fill=red!20] (a3t2) at (11, 1.8) {$A_3$};
				
				\draw[very thick, dashed, orange, rounded corners=10pt] (9.7, 3.1) rectangle (12.3, 4.1);
				\node[font=\scriptsize\bfseries, orange] at (11, 4.35) {$X'=\{A_1,A_2\}$};
				
				\node[font=\scriptsize, text width=2.6cm, align=center] at (11, 0.5) {Aims diverge,\\coalition fragments};
				
				\draw[->, ultra thick, gray!60] (1.3, 2.7) -- (3.7, 2.7) node[midway, above, font=\scriptsize\bfseries, text=black] {Shared threat};
				\draw[->, ultra thick, gray!60] (7.3, 2.7) -- (9.2, 2.7) node[midway, above, font=\scriptsize\bfseries, text=black] {Divergence};
				
				\node[event=yellow!20] (e1) at (2.5, 1.2) {\textbf{Event:}\\External threat};
				\node[event=orange!20] (e2) at (8.2, 1.2) {\textbf{Event:}\\Threat removed};
				
				\draw[->, thick, dashed, gray] (e1.north) -- (4.2, 2.2);
				\draw[->, thick, dashed, gray] (e2.north) -- (9.8, 2.2);
				
			\end{tikzpicture}
		}
		\caption{Coalition formation and fragmentation as path-dependent coordination dynamics.}
		\label{fig:coalition_dynamics}
	\end{figure}
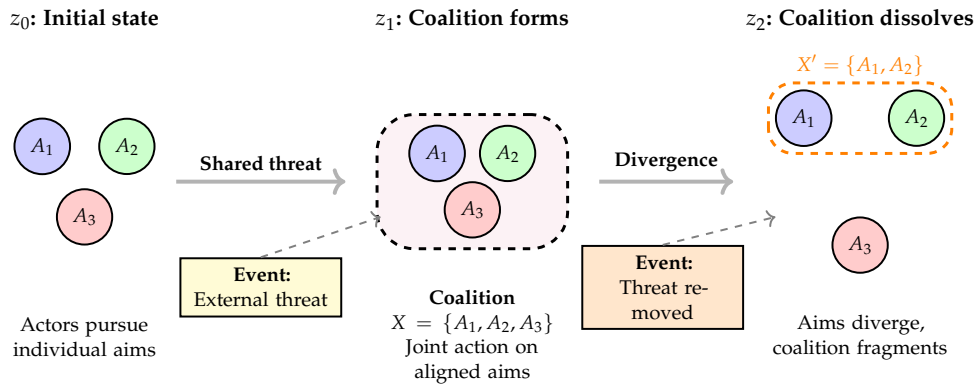

	Figure~\ref{fig:coalition_dynamics} extends this to a temporal trajectory, tracing coalition formation and fragmentation across three time periods.
	%
	
	
\subsection{Attributes\index{Attributes} of Actors\index{Actors} and Coalitions\index{Coalitions}}
\label{subsec:attributes}

Attributes\index{Attributes} are descriptive properties used to characterize
actors\index{Actors} and coalitions\index{Coalitions} in the Scenario Database\index{Database Layer}.
They specify dimensions along which entities differ (e.g., capabilities, resources, legitimacy),
thereby constraining and enabling strategic interaction.

While attitudes (Section~\ref{subsec:attitudes}) capture \emph{what actors believe and aim at},
and options (Section~\ref{subsec:options}) specify \emph{what actors can do},
attributes describe \emph{what actors (and coalitions) are like} in the scenario:
their relevant background properties at the level of analysis chosen.

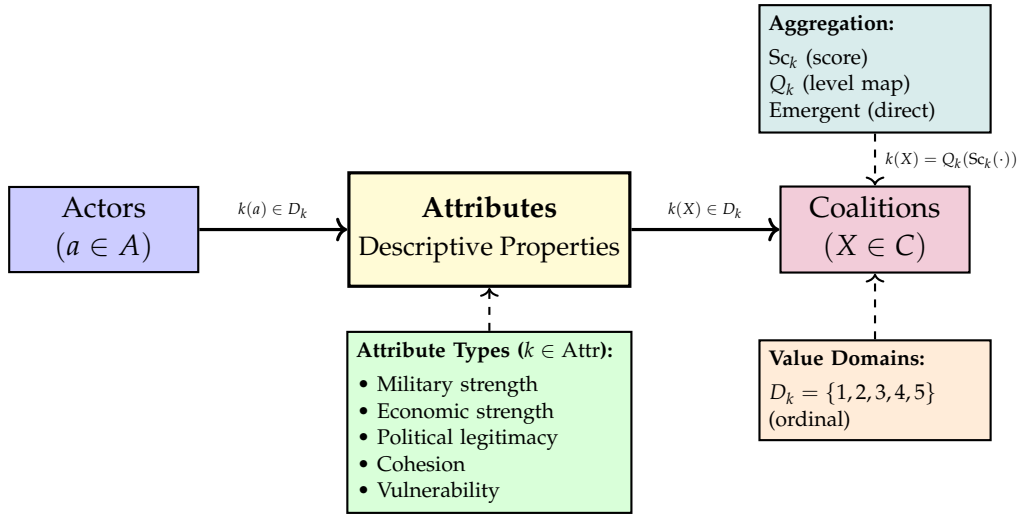
\begin{figure}[htbp]
	\centering
	\begin{tikzpicture}[scale=0.85,
		node distance=1.5cm and 2cm,
		block/.style={rectangle, draw, thick, align=center, minimum height=1cm, minimum width=2.5cm},
		main/.style={block, fill=yellow!20, very thick, minimum width=3.5cm, minimum height=1.5cm, font=\bfseries},
		actor_style/.style={block, fill=blue!20},
		coal_style/.style={block, fill=purple!20},
		info_style/.style={block, fill=#1, text width=3.5cm, font=\scriptsize, align=left}
		]
		
		\node[main] (attr) at (6, 3) {Attributes\\{\normalfont\small Descriptive Properties}};
		
		\node[actor_style] (actors) at (0, 3) {Actors\\$(a \in A)$};
		
		\node[coal_style] (coal) at (12, 3) {Coalitions\\$(X \in C)$};
		
		\draw[->, very thick] (actors) -- (attr) node[midway, above, font=\tiny] {$k(a)\in D_k$};
		\draw[->, very thick] (attr) -- (coal) node[midway, above, font=\tiny] {$k(X)\in D_k$};
		
		\node[info_style=green!15] (types) at (6, 0) {
			\textbf{Attribute Types ($k \in \mathrm{Attr}$):}\\[0.1cm]
			\textbullet\ Military strength\\
			\textbullet\ Economic strength\\
			\textbullet\ Political legitimacy\\
			\textbullet\ Cohesion\\
			\textbullet\ Vulnerability
		};
		\draw[->, thick, dashed] (types) -- (attr);
		
		\node[info_style=orange!15, text width=2.8cm] (values) at (12, 0.5) {
			\textbf{Value Domains:}\\[0.1cm]
			$D_k = \{1,2,3,4,5\}$\\
			(ordinal)
		};
		\draw[->, thick, dashed] (values) -- (coal);
		
		\node[info_style=teal!15, text width=2.8cm] (agg) at (12, 5.5) {
			\textbf{Aggregation:}\\[0.1cm]
			$\mathrm{Sc}_k$ (score)\\
			$Q_k$ (level map)\\
			Emergent (direct)
		};
		\draw[->, thick, dashed] (agg) -- (coal) node[midway, right, font=\tiny] {$k(X)=Q_k(\mathrm{Sc}_k(\cdot))$};
		
	\end{tikzpicture}
	\caption{Attribute structure in SBA. Each attribute type \(k\) is an attribute function \(k:A\cup C\to D_k\).
		Coalition values may be computed by aggregation \((\mathrm{Sc}_k,Q_k)\) or assigned directly (emergent attributes).}
	\label{fig:attribute_structure}
\end{figure}

Figure~\ref{fig:attribute_structure} sketches the attribute assignment pattern for actors and coalitions.

\subsubsection{Formal definition}
\label{subsubsec:attr-formal}

Let \(\mathrm{Attr}\) denote a finite or countable set of attribute types.
For each \(k \in \mathrm{Attr}\), let \(D_k\) be the domain of admissible values.
An attribute type \(k\) is represented by a (possibly partial) attribute function\index{Attributes!Attribute function}
\begin{equation}
	k : A \cup C \longrightarrow D_k,
	\qquad k \in \mathrm{Attr},
\end{equation}
where \(A\) is the actor set and \(C\) the coalition set (Section~\ref{subsec:coalitions}).
For \(x \in A \cup C\), the value \(k(x)\in D_k\) denotes the attribute value of \(x\) with respect to \(k\).

\paragraph{Type safety\index{Attributes!Type safety}.}
For each attribute type \(k\), admissible values are restricted to \(D_k\).
In particular, numerical codes are attribute-specific: a value ``3'' for military strength and a value ``3'' for legitimacy
need not be comparable unless an explicit mapping between the corresponding domains is introduced.

\paragraph{Coalitional attributes.}
Coalitions\index{Coalitions} may inherit attributes\index{Attributes} compositionally (via aggregation or dominance rules),
or may carry explicitly assigned coalition-level attributes. The corresponding rules are specified in
Section~\ref{subsec:attr_aggregation}.

\subsubsection{Attribute values and measurement scales}
\label{subsubsec:attr-scales}

\paragraph{Ordinal measurement as default.}
In SBA, many attributes are operationalized on finite ordinal scales\index{Measurement!Ordinal scales}:
\begin{equation}
	D_k = \{1,\ldots, n_k\},
\end{equation}
equipped with the natural order \(\preceq_k\).
For actors or coalitions \(x_i,x_j \in A \cup C\),
\begin{equation}
	x_i \preceq_k x_j
	\quad \Longleftrightarrow \quad
	k(x_i) \le k(x_j).
\end{equation}
Ordinal scales support robust comparative judgments (ranking without spurious precision) and are directly elicitable from experts.

\paragraph{Other measurement scales (raw values vs.\ operational levels).}
Nothing prevents storing attributes with categorical (nominal) or numerical (interval/ratio) raw domains in the database.
However, when attributes enter scenario comparison, evaluation, or update rules, they are \emph{operationalized}
via ordinal levels (Section~\ref{subsec:attr_aggregation} and Figure~\ref{fig:ordinalization}).
Examples of raw domains include:
\begin{itemize}
	\item \textbf{Categorical attributes:} \(D_k=\{\text{democratic},\text{authoritarian},\text{hybrid}\}\) (regime type)
	\item \textbf{Interval attributes:} temperature, dates (when meaningful distances matter)
	\item \textbf{Ratio attributes:} budgets, population size (when ratios are meaningful)
\end{itemize}
The measurement scale should reflect the epistemic status of the attribute (reliability, meaningful operations) and the scenario horizon.

\subsubsection{Examples of attribute types}
\label{subsubsec:attr-examples}

Table~\ref{tab:attribute_examples} lists representative attribute types used in SBA applications,
organized by typical relevance across domains (Section~\ref{subsec:domains}) and modifiers.
At this stage, we do not fix specific measurement scales (e.g., 1--5 vs.\ 1--7), since
operationalization depends on the scenario horizon, available data, and the coding protocol.
Where quantitative indicators exist, they can be mapped to ordinal levels via a documented
ordinalization rule (Figure~\ref{fig:ordinalization}).

\begin{table}[h!]
	\centering
	\small
	\renewcommand{\arraystretch}{1.3}
	\setlength{\tabcolsep}{5pt}
	\begin{tabularx}{\textwidth}{
			>{\RaggedRight\arraybackslash}p{3.0cm}
			>{\RaggedRight\arraybackslash}X
			>{\RaggedRight\arraybackslash}p{3.5cm}}
		\toprule
		\textbf{Attribute type} & \textbf{Description} & \textbf{Typical relevance (domains/modifiers)} \\
		\midrule
		Military strength &
		Capacity for armed coercion and credible threat projection. &
		Domains: Geo, Pol \\
		\midrule
		Economic strength &
		Material resources and production capacity relevant for sustained action. &
		Domains: Econ, Geo \\
		\midrule
		Political legitimacy &
		Acceptance of authority and perceived right to rule (domestic/international). &
		Domains: Pol, Soc \\
		\midrule
		Technological capability &
		Access to advanced technology, innovation capacity, and critical infrastructure. &
		Domain: Econ;\ Modifier: \(\mathrm{Tech}\) \\
		\midrule
		Organizational cohesion &
		Internal unity, discipline, and ability to coordinate and implement decisions. &
		Domain: Org;\ Actor category: collective actors \\
		\midrule
		Diplomatic influence &
		Capacity to shape norms, agendas, and coalition formation. &
		Domain: Geo \\
		\midrule
		Resource endowment &
		Access to natural resources (oil, minerals, water, arable land) relevant under scarcity. &
		Domain: Econ;\ Modifier: \(\mathrm{Env}\) \\
		\midrule
		Vulnerability &
		Exposure to shocks and constraints (security, economic, environmental, institutional). &
		Domains/modifiers: cross-cutting \\
		\midrule
		Information capacity &
		Ability to gather, process, and disseminate information; resilience to manipulation. &
		Domain: Pol;\ Modifier: \(\mathrm{Tech}\) \\
		\midrule
		Regime type &
		Institutional form of rule (used as a classificatory attribute when relevant). &
		Domain: Pol \\
		\bottomrule
	\end{tabularx}
	\caption[Representative attribute types]{Representative attribute types in SBA and their typical relevance. Specific measurement scales are fixed at the level of the scenario coding protocol (including any ordinalization of quantitative inputs).}
	\label{tab:attribute_examples}
\end{table}

\noindent
In most applications, the first nine entries are operationalized ordinally, whereas \emph{regime type} is treated as a categorical attribute.

\paragraph{Example: Military strength ranking.}
Let \(k=\text{mil\_strength}\) with \(D_k=\{1,2,3,4,5\}\).
A typical assessment may be (Table~\ref{tab:mil_strength_example}):

\begin{table}[h!]
	\centering
	\small
	\renewcommand{\arraystretch}{1.2}
	\setlength{\tabcolsep}{6pt}
	\begin{tabular}{
			>{\RaggedRight\arraybackslash}p{4.2cm}
			>{\centering\arraybackslash}p{3.2cm}
			>{\RaggedRight\arraybackslash}p{4.4cm}}
		\toprule
		\textbf{Actor} & \textbf{\(k(a)\)} & \textbf{Interpretation} \\
		\midrule
		\(A_1\) (State X)         & \(5\) & Very high \\
		\midrule
		\(A_2\) (State Y)         & \(3\) & Moderate \\
		\midrule
		\(A_3\) (Non-state group) & \(2\) & Low \\
		\bottomrule
	\end{tabular}
	\caption{Illustrative ordinal assessment for military strength \(k=\text{mil\_strength}\).}
	\label{tab:mil_strength_example}
\end{table}

which induces the order \(A_3 \prec_k A_2 \prec_k A_1\).
Such rankings support comparative questions (credible threats, feasible coalitions, strategic asymmetries) without requiring cardinal comparability.

\subsubsection{Conceptual foundations}
\label{subsubsec:attr-foundations}

\paragraph{What attributes represent.}
Attributes\index{Attributes} capture descriptive dimensions of actors\index{Actors} and coalitions\index{Coalitions} that are strategically relevant in the scenario, including:
\begin{itemize}
	\item \textbf{Material resources}: wealth, hardware, natural resources, infrastructure
	\item \textbf{Institutional properties}: legal authority, decision procedures, organizational structure
	\item \textbf{Positional characteristics}: geographic position, network centrality, alliance membership
	\item \textbf{Reputational dimensions}: credibility, trustworthiness, historical track record
	\item \textbf{Capacities and vulnerabilities}: mobilization ability, information processing, exposure to shocks
\end{itemize}

\paragraph{Sources of attribute values.}
Attribute values may be drawn from quantitative indicators, qualitative expert assessments, mixed indices, or computational derivations (e.g., text mining, network metrics).
For scenario-level modeling, such inputs are mapped to ordinal levels (``ordinalized'') to enforce comparability under uncertainty.

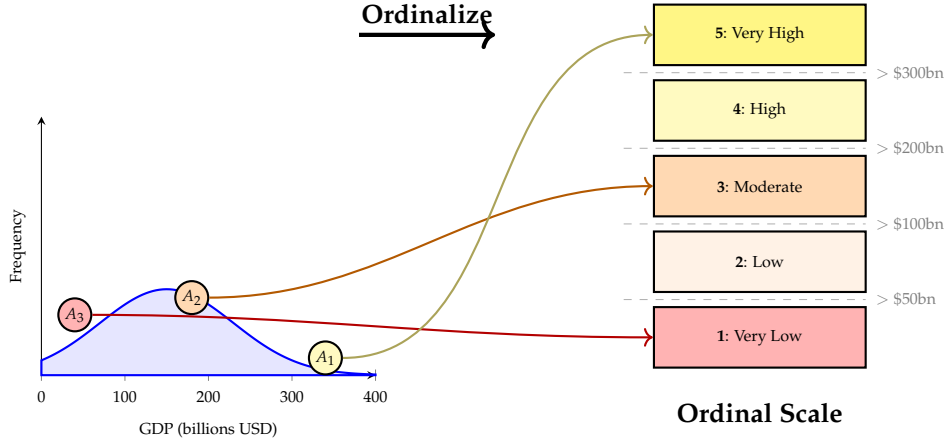
\begin{figure}[htbp]
	\centering
	\begin{tikzpicture}[scale=1.0]
		
		\begin{axis}[
			name=cont,
			at={(0,0)},
			width=6cm,
			height=5cm,
			axis lines=left,
			xlabel={GDP (billions USD)},
			ylabel={Frequency},
			ymin=0, ymax=0.015,
			xmin=0, xmax=400,
			ytick=\empty,
			samples=100,
			domain=0:400,
			font=\tiny,
			clip=false
			]
			\addplot[thick, blue, fill=blue!10] {1/(80*sqrt(2*3.14159))*exp(-((x-150)^2)/(2*80^2))} \closedcycle;
			
			\node[circle, draw, thick, fill=red!30, inner sep=1pt] (A3) at (axis cs:40, 0.0035) {$A_3$};
			\node[circle, draw, thick, fill=orange!30, inner sep=1pt] (A2) at (axis cs:180, 0.0045) {$A_2$};
			\node[circle, draw, thick, fill=yellow!30, inner sep=1pt] (A1) at (axis cs:340, 0.001) {$A_1$};
		\end{axis}
		
		\draw[->, ultra thick] (4.2, 4.5) -- (6.0, 4.5) node[midway, above, font=\small\bfseries] {Ordinalize};
		
		\begin{scope}[shift={(9.5,0)}]
			\foreach \y/\val/\lbl/\f in {
				0.5/1/Very Low/red!30,
				1.5/2/Low/orange!10,
				2.5/3/Moderate/orange!30,
				3.5/4/High/yellow!30,
				4.5/5/Very High/yellow!60} {
				
				\node[rectangle, draw, thick, fill=\f, minimum width=2.8cm, minimum height=0.8cm, align=center, font=\tiny]
				(R\val) at (0, \y) {\textbf{\val}: \lbl};
			}
			\node[font=\small\bfseries] at (0, -0.5) {Ordinal Scale};
			
			\foreach \y/\thresh in {1/50, 2/100, 3/200, 4/300} {
				\draw[dashed, gray!60] (-1.8, \y) -- (1.4, \y);
				\node[font=\tiny, gray, right] at (1.4, \y) {$>\$\thresh$bn};
			}
		\end{scope}
		
		\draw[->, thick, red!70!black] (A3) to[out=0, in=180] (R1.west);
		\draw[->, thick, orange!70!black] (A2) to[out=0, in=180] (R3.west);
		\draw[->, thick, yellow!60!black] (A1) to[out=0, in=180] (R5.west);
		
	\end{tikzpicture}
	\caption{Ordinalization: continuous quantitative inputs (left) are mapped to ordinal levels (right) via defined thresholds.}
	\label{fig:ordinalization}
\end{figure}

\paragraph{Ordinalization procedure.}
Quantitative inputs can be mapped to ordinal levels via:
\begin{itemize}
	\item \textbf{Percentile binning} (e.g., quintiles/septiles)
	\item \textbf{Expert thresholds} (domain-defined cut-points)
	\item \textbf{Theoretical benchmarks} (externally established categories)
	\item \textbf{Relative ranking} (within-scenario ordering)
\end{itemize}
Figure~\ref{fig:ordinalization} illustrates this transformation.

\paragraph{Relevance criterion\index{Attributes!Relevance criterion}.}
Only attributes that are strategically relevant for the scenario should be included. An attribute is relevant if it
(i) constrains or enables actions,
(ii) affects how actors are perceived by others,
(iii) shapes vulnerability to events,
or (iv) systematically correlates with actor roles or domain membership.
Irrelevant or weakly connected attributes are excluded to preserve coherence\index{Logical Foundations!Coherence} and avoid unnecessary parameterization.

\subsubsection{Attribute aggregation for coalitions}
\label{subsubsec:attr-aggregation}
\label{subsec:attr_aggregation}

Coalitions\index{Coalitions} require valuation rules for coalition-level attributes.
SBA distinguishes between \emph{aggregation-based} attributes and \emph{emergent} attributes:
\begin{itemize}
	\item \textbf{Aggregation-based attributes} derive the coalition value from member values via an explicit rule.
	\item \textbf{Emergent attributes} are assessed at coalition level because a simple function of member values is not an adequate operational model (e.g., coalition cohesion).
\end{itemize}
Accordingly, aggregation rules apply to a designated subset \(\mathrm{Attr}_{\mathrm{agg}} \subseteq \mathrm{Attr}\).

\paragraph{Formal aggregation rule (aggregative attributes only).}
Fix \(k \in \mathrm{Attr}_{\mathrm{agg}}\) with domain \(D_k\).
Let \(X \in C\) be a coalition with member set \(X=\{a_1,\ldots,a_n\}\).
Aggregation is modeled in two steps:
\begin{enumerate}
	\item A \emph{score function} \(\mathrm{Sc}_k\) maps the multiset of member levels to a numerical score,
	\[
	\mathrm{Sc}_k:\ \mathcal{M}(D_k)\to \mathbb{R}, \qquad
	\mathrm{Sc}_k\big(\{k(a):a\in X\}\big)\in \mathbb{R}.
	\]
	\item A \emph{level mapping} \(Q_k\) maps scores back to admissible levels,
	\[
	Q_k:\ \mathbb{R}\to D_k.
	\]
\end{enumerate}
The coalition-level attribute is then defined by
\begin{equation}
	k(X)\ \coloneqq\ Q_k\!\left(\mathrm{Sc}_k\big(\{k(a):a\in X\}\big)\right).
\end{equation}
Here \(Q_k\) implements the agreed ordinalization/threshold rule (including possible capping to \(\max D_k\)).

\paragraph{Monotonicity constraint (fixed membership).}
For aggregative attributes, monotonicity means: if coalition membership \(X\) is fixed and all member levels weakly increase,
then the coalition level does not decrease. Formally, for two profiles \(k\) and \(k'\) on the same \(X\),
\begin{equation}
	\Big(\forall a\in X:\ k'(a)\succeq_k k(a)\Big)
	\ \Rightarrow\
	k'(X)\succeq_k k(X),
\end{equation}
where \(k(X)\) and \(k'(X)\) are computed via the same \((\mathrm{Sc}_k,Q_k)\).
This constraint applies only to aggregative attributes; emergent attributes may violate monotonicity (e.g., coordination loss under expansion).

\paragraph{Common aggregation rules (scores).}
Table~\ref{tab:aggregation_functions} lists standard choices for \(\mathrm{Sc}_k\) (with coalition level \(k(X)=Q_k(\mathrm{Sc}_k(\cdot))\)).

\begin{table}[h!]
	\centering
	\small
	\renewcommand{\arraystretch}{1.3}
	\setlength{\tabcolsep}{5pt}
	\begin{tabularx}{\textwidth}{
			>{\RaggedRight\arraybackslash}p{3.1cm}
			>{\RaggedRight\arraybackslash}X
			>{\RaggedRight\arraybackslash}X}
		\toprule
		\textbf{Aggregation type} & \textbf{Score \(\mathrm{Sc}_k\)} & \textbf{Interpretation} \\
		\midrule
		Additive (sum) &
		\(\mathrm{Sc}_k = \sum_{a\in X} k(a)\) &
		Cumulative resources (before mapping \(Q_k\)) \\
		\midrule
		Averaging (mean) &
		\(\mathrm{Sc}_k = \frac{1}{|X|}\sum_{a\in X} k(a)\) &
		Average capacity/perception (before mapping \(Q_k\)) \\
		\midrule
		Dominance (max) &
		\(\mathrm{Sc}_k = \max_{a\in X} k(a)\) &
		Best-shot / strongest-member determines level \\
		\midrule
		Constraint (min) &
		\(\mathrm{Sc}_k = \min_{a\in X} k(a)\) &
		Weakest-link properties \\
		\midrule
		Weighted sum &
		\(\mathrm{Sc}_k = \sum_{a\in X} w_a\cdot k(a)\) \ (\(\sum_a w_a=1\)) &
		Heterogeneous contributions (before mapping \(Q_k\)) \\
		\bottomrule
	\end{tabularx}
	\caption{Common aggregation rules for coalitional attributes. Coalition levels are obtained by \(k(X)=Q_k(\mathrm{Sc}_k(\cdot))\).}
	\label{tab:aggregation_functions}
\end{table}

\paragraph{Selection criteria.}
The choice of aggregation rule depends on the semantics of the attribute:
additive for pooled resources, mean for averaged tendencies, max for best-shot capacities, min for weakest-link constraints, and weights for unequal contributions.

\paragraph{Emergent attributes.}
Some coalition attributes are assessed directly at coalition level because a simple function of member attribute levels is not an adequate operational model.
Examples include coalition cohesion, coordination capacity, and symbolic legitimacy (which may increase or decrease relative to member legitimacies depending on institutional context).

\paragraph{Example: Military strength aggregation.}
Let \(k=\text{mil\_strength}\) and \(X=\{A_1,A_2,A_3\}\) with
\(k(A_1)=5\), \(k(A_2)=3\), \(k(A_3)=2\).
Score-based aggregation yields:
\begin{itemize}
	\item \textbf{Sum:} \(\mathrm{Sc}_k=5+3+2=10\), then \(k(X)=Q_k(10)\). (E.g., capped/\allowbreak thresholded to ``5''.)
	\item \textbf{Mean:} \(\mathrm{Sc}_k=10/3\approx 3.3\), then \(k(X)=Q_k(3.3)\). (E.g., ``3''.)
	\item \textbf{Max:} \(\mathrm{Sc}_k=\max(5,3,2)=5\), so \(k(X)=Q_k(5)=5\)
	\item \textbf{Min:} \(\mathrm{Sc}_k=\min(5,3,2)=2\), so \(k(X)=Q_k(2)=2\)
\end{itemize}
The appropriate choice is scenario- and coalition-structure dependent (unified command vs.\ loose alliance; weakest-link tasks; etc.).

\subsubsection{Domain-specific attributes}
\label{subsubsec:attr-domain}

Different domains (Section~\ref{subsec:domains}) emphasize different attribute types. Table~\ref{tab:domain_attributes} maps typical attribute families to primary domains.

\begin{table}[h!]
	\centering
	\small
	\renewcommand{\arraystretch}{1.3}
	\setlength{\tabcolsep}{5pt}
	\begin{tabular}{
			>{\RaggedRight\arraybackslash}p{3cm}
			>{\RaggedRight\arraybackslash}p{10cm}}
		\toprule
		\textbf{Domain} & \textbf{Typical attributes} \\
		\midrule
		Political (Pol) & Legitimacy, authority, institutional stability, regime type, veto points \\
		\midrule
		Geopolitical (Geo) & Military strength, alliance membership, geographic position, deterrence capacity, diplomatic influence \\
		\midrule
		Economic (Econ) & Economic strength, resource endowment, market access, technological capability, financial reserves \\
		\midrule
		Social (Soc) & Group cohesion, identity salience, mobilization capacity, social capital, demographic composition \\
		\midrule
		Organizational (Org) & Cohesion, decision efficiency, information capacity, hierarchy, expertise \\
		\bottomrule
	\end{tabular}
	\caption{Domain-specific attribute emphases. Many attributes (e.g., legitimacy, cohesion) can matter across domains.}
	\label{tab:domain_attributes}
\end{table}

\subsubsection{Temporal dynamics of attributes}
\label{subsubsec:attr-temporal}

\paragraph{Static vs.\ dynamic attributes.}
Many attributes are treated as static within a scenario tree (fixed at the root node),
reflecting typical SBA horizons (weeks to months). Some attributes can be modeled dynamically:
reputation/credibility, resource stocks, cohesion, or vulnerability under sustained stress.

Dynamic attributes require explicit update rules at tree nodes.
A convenient notation is node-indexed attribute evaluation \(k_n(x)\), where \(n\) is a scenario-tree node.
Exogenous events may trigger attribute updates (e.g., disaster reduces economic strength; defeat reduces legitimacy).

\subsubsection{Practical guidelines for attribute specification}
\label{subsubsec:attr-guidelines}

When applying SBA to empirical scenarios, analysts may proceed as follows:

\paragraph{Step 1: Select relevant attributes.}
Based on domain and actor types, list candidate attributes and apply the relevance criterion
(keep only attributes that shape the strategic landscape).

\paragraph{Step 2: Choose operational scales.}
For each retained attribute, specify its operational scale for scenario analysis
(ordinal levels), and document the ordinalization rule for quantitative inputs.

\paragraph{Step 3: Specify coalition rules where needed.}
For attributes used at coalition level, decide whether the attribute is aggregative or emergent,
and (if aggregative) specify \((\mathrm{Sc}_k,Q_k)\).

\paragraph{Step 4: Document sources and validation.}
Record definitions, value domains, elicitation/data sources, and checks for consistency across coders/\allowbreak experts.

\subsubsection{Theoretical foundation}
\label{subsubsec:attr-theory}

The treatment of attributes as structured properties of decision-making entities connects SBA to representational approaches in decision and game theory \citep{arrow1951, sen1977, kreps1988},
and to organizational theory emphasizing how capacities and constraints shape strategic behavior \citep{simon1957, cyert1963, march1984}.
Aggregation questions for coalition attributes relate to social choice and cooperative game theory \citep{arrow1951, sen1977, gillies1959, shapley1953},
and to the micro--macro link problem in sociology \citep{coleman1990}.

\subsubsection{Limitations and extensions}
\label{subsubsec:attr-limitations}

\paragraph{Attribute interdependence.}
The present formalization treats attributes as independent dimensions. Empirically, attributes may be correlated or causally linked.
Extensions could incorporate dependence structures (e.g., graphical models).

\paragraph{Uncertainty and contestation.}
Attribute values may be uncertain or disputed. Extensions could represent intervals or distributions, or allow actor-relative perceptions of attributes (linked to belief structures in the Attitudes section).

\paragraph{Cultural variation in attribute salience.}
Attribute salience can be culturally specific. SBA can accommodate additional attributes, but analysts must justify their relevance and operationalization in context.

 \subsection{Attitudes\index{Attitudes}}
 \label{subsec:attitudes}

 Attitudes\index{Attitudes} are propositional states that encode how actors\index{Actors}
 and coalitions\index{Coalitions} perceive, evaluate, and motivate action in a scenario.
 Where attributes\index{Attributes} describe what entities \emph{are}, attitudes describe
 what they \emph{believe, desire, fear, expect, and commit to}.
 They therefore provide the cognitive and motivational interface between database-level
 representation and scenario-tree dynamics.

 \subsubsection{Foundations and Formal Types}
 \label{subsubsec:att-foundations}
 
 Attributes describe externally observable properties of actors and
 coalitions. \emph{Attitudes}, by contrast, represent internal propositional
 states: how agents perceive, evaluate, and interpret situations. This section
 fixes the primitive attitude vocabulary before adding coherence postulates,
 coalitional variants, and an optional parameter layer.
 
 Linguistically, propositional attitudes are expressed by \emph{that}-clauses, e.g.\
 \emph{``Actor~A believes that $p$''} or \emph{``Actor~B intends that $p$''}.
 Formally, such expressions relate an agent to a propositional content.
 We use the following domains: let \(A\) be the set of actors\index{Actors},
\(C \subseteq \Pow(A)\) the set of coalitions\index{Coalitions},
and let \(P\) denote the set of propositional contents.
 
 \paragraph{Primitive attitude types.}
 The SBA framework distinguishes a finite set of primitive attitude operators:
	\[
	\mathrm{Att}_{\mathrm{prim}}=\{K,B,W,I,F,\mathrm{Com}\},
	\]
	where \(K\) and \(B\) denote epistemic--doxastic states (knowledge, belief),
	\(W\) denotes conative preference, \(I\) denotes practical intention,
	\(F\) denotes affective evaluation (fear or negative anticipation), and
	\(\mathrm{Com}\) is a primitive coalitional commitment operator (indexed by coalitions \(X\in C\) as \(\mathrm{Com}_X\)).
	For each actor \(a\in A\), write \(K_a,B_a,W_a,I_a,F_a\) for the
	agent-indexed knowledge, belief, desire, intention, and fear operators
	associated with \(a\).
 
 For each individual attitude type \(k \in \mathrm{Att}_{\mathrm{prim}}\setminus\{\mathrm{Com}\}\),
 we interpret the modal notation \(k_a p\) through a binary relation
 \[
 \mathrm{att}_k \subseteq A \times P, \qquad
 (a,p) \in \mathrm{att}_k \;\text{ iff }\; k_a p.
 \]
 
 \paragraph{Aim as a definition (weak intention).}
 Following usage in scenario analysis, \emph{aiming} is introduced as a defined
 predicate (not a primitive operator):
 \[
 \mathrm{aim}(a,p)
 \;\Leftrightarrow\;
 W_a p \;\wedge\; \neg B_a \neg p.
 \]
 An actor \emph{aims} at \(p\) if \(p\) is desired and not believed impossible,
 without committing to the full deliberative structure associated with intention.
 
 \paragraph{Fear as aversive conative state (crisis-model default).}
 In the SBA model, fear is treated as an evaluative attitude whose functional role
 is captured by an associated aversion:
 \[
 F_a p \;\Rightarrow\; W_a \neg p.
 \]
 This is a modeling assumption tailored to crisis and conflict settings, not a universal
 semantic axiom.
 
 \paragraph{Coalitional attitudes.}
 Coalition-level commitments are introduced by a primitive operator
 \[
 \mathrm{Com}_X p, \qquad X \in C,
 \]
 interpreted as ``coalition \(X\) is (weakly) committed to \(p\).''
 Member-level alignment conditions for \(\mathrm{Com}_X p\) are treated as coherence postulates
 (idealized defaults that can be relaxed) in Section~\ref{subsubsec:coalitional-attitudes}.
 
 \paragraph{Role in the Scenario Database.}
 Attitudes link the Database Layer to the dynamic structure of scenario trees by
 shaping an actor's evaluation of options and the perceived feasibility or desirability
 of events. Knowledge and belief determine the informational perspective; desires and
 aims encode motivational aspects; intentions guide action selection; fear captures negative
 anticipation. Coalitional commitments represent joint stances that affect coordination and
 alignment across actors.
 
 \subsubsection{Operator Taxonomy and Semantic Families}
 \label{subsubsec:att-operator-taxonomy}
 
 Propositional attitudes in the SBA framework are represented by modal operators that
 connect an actor to a proposition. Each operator family corresponds to a semantic
 domain such as knowledge, belief, desire, intention, affective evaluation, or collective
 commitment. These families reflect established distinctions in epistemic and cognitive
 action theory \citep{hintikka1962, bratman1987, searle1983, tuomela2002, fagin1995}.
 
 The following operators constitute the core repertoire of individual and coalitional
 attitudes:
 
 \begin{table}[H]
 	\small
 	\centering
 	\renewcommand{\arraystretch}{1.15}
 	\setlength{\tabcolsep}{4pt}
 	\begin{tabular}{p{1.5cm} p{3.0cm} p{3.0cm} p{5.0cm}}
 		\hline
 		\textbf{Symbol}
 		& \textbf{Attitude Class}
 		& \textbf{Typical Verbs}
 		& \textbf{Canonical Sources} \\
 		\hline
 		
 		$K_a p$
 		& Knowledge (epistemic)
 		& know, realize, understand
 		& \citet{hintikka1962, fagin1995} \\[4pt]
 		
 		$B_a p$
 		& Belief (doxastic)
 		& believe, assume, expect
 		& \citet{hintikka1962, meyer1999} \\[4pt]
 		
 		$W_a p$
 		& Desire (conative)
 		& wish, hope, prefer
 		& \citet{searle1983, bratman1987, wooldridge2000} \\[4pt]
 		
 		$I_a p$
 		& Intention (practical)
 		& intend, plan, decide
 		& \citet{anscombe1957, bratman1987, vanderhoek2003} \\[4pt]
 		
 		$F_a p$
 		& Fear / evaluation (affective)
 		& fear, distrust, suspect
 		& \citet{cloreortonysmith1988} \\[4pt]
 		
 		$\mathrm{Com}_X p$
 		& Collective commitment (coalitional)
 		& agree, commit, resolve
 		& \citet{tuomela2002, searle1995} \\
 		\hline
 	\end{tabular}
 	\caption{Core operator families for propositional attitudes in SBA.}
 	\label{tab:attitude_operators_master}
 \end{table}
 
 \medskip
 \noindent\textit{Note.}
 \(\mathrm{aim}(a,p)\) is a defined predicate (from \(W\) and \(B\)); it is therefore omitted from
 Table~\ref{tab:attitude_operators_master}.
 
 \medskip
 Each operator abstracts a class of natural-language expressions, e.g.:
 \begin{itemize}
 	\item \textbf{Epistemic} (\(K_a p\)):
 	``Actor~A knows that the border will close.''
 	\item \textbf{Doxastic} (\(B_a p\)):
 	``Actor~A believes that tensions will rise.''
 	\item \textbf{Conative} (\(W_a p\)):
 	``Actor~A prefers that negotiations continue.''
 	\item \textbf{Intentional} (\(I_a p\)):
 	``Actor~A intends that sanctions be lifted.''
 	\item \textbf{Affective} (\(F_a p\)):
 	``Actor~A fears that the ceasefire will collapse.''
 	\item \textbf{Collective} (\(\mathrm{Com}_X p\)):
 	``Coalition~\(X\) commits to implementing~\(p\).''
 \end{itemize}
 
 The defined predicate \(\mathrm{aim}(a,p)\) is used for weaker motivational stances,
 e.g.\ ``Actor~A aims that \(p\),'' without treating \(\mathrm{aim}\) as a primitive modality.
 
 \subsubsection{Axiomatic Interrelations}
 \label{subsubsec:att-operator-relations}
 
 Propositional attitudes in the SBA framework are not independent. Their interactions
 are shown in Figure~\ref{fig:attitude_taxonomy_v4}. The coherence postulates below are idealized defaults
 that can be relaxed for bounded or inconsistent agents \citep{hintikka1962, bratman1987, fagin1995, searle1983}.
\begin{table}[htbp]
	\centering
	\small
	\renewcommand{\arraystretch}{1.5} 
	\begin{tabularx}{\textwidth}{l >{\RaggedRight\arraybackslash}X l l}
		\toprule
		\textbf{ID} & \textbf{Interpretation / Description} & \textbf{Formal Expression} & \textbf{Status} \\
		\midrule
		(A1) & Knowledge entails belief & $K_a p \Rightarrow B_a p$ & Core \\
		(A2) & Closure of knowledge under known implication & $K_a(p \rightarrow q) \wedge K_a p \Rightarrow K_a q$ & Default \\
		(A3) & Intention presupposes desire and feasibility belief & $I_a p \Rightarrow W_a p \wedge B_a(\Diamond p)$ & Coherence \\
		(A4) & Aim as weak intention (definition) & $\mathrm{aim}(a,p) \Leftrightarrow W_a p \wedge \neg B_a \neg p$ & Definition \\
		(A5) & Fear as aversive conative state & $F_a p \Rightarrow W_a \neg p$ & Default \\
		(A6) & Knowledge excludes the contrary belief & $K_a p \Rightarrow \neg B_a \neg p$ & Core \\
		(A7) & Consistency of intentions & $I_a p \wedge I_a \neg p \Rightarrow \bot$ & Coherence \\
		(A8) & Practical closure of intention & $I_a p \wedge B_a(p \rightarrow q) \Rightarrow I_a q$ & Default \\
		(A9) & Fear bias in belief updating (non-rational) & $F_a p \Rightarrow \Diamond B_a \neg p$ & Optional \\
		\bottomrule
	\end{tabularx}
	\caption{Axiomatic interrelations of the actors' core attitudes ($K, B, I, W, F$). These postulates define the internal coherence requirements for an assessment state $DB_t$.}
	\label{tab:actor_attitude_axioms}
\end{table}
 
\begin{figure}[htbp]
	\centering
	\begin{tikzpicture}[
		node distance=1.8cm and 2.0cm,
		root/.style={rectangle, rounded corners=3pt, draw=black, thick, fill=gray!20,
			minimum width=4cm, minimum height=0.7cm, font=\small\bfseries},
		family/.style={rectangle, draw=gray!40, fill=gray!2, minimum width=2.8cm, 
			minimum height=0.6cm, font=\footnotesize\scshape, rounded corners=2pt},
		op/.style={circle, draw=black, thick, minimum size=0.9cm, font=\small\bfseries, align=center},
		derived/.style={circle, draw=black, thick, dashed, minimum size=0.9cm, font=\small\itshape, align=center},
		ax_core/.style={-Stealth, blue!70, line width=1.3pt},
		ax_prac/.style={-Stealth, teal!70, line width=1.3pt},
		ax_bias/.style={-Stealth, red!60, dashed, line width=1pt},
		label_node/.style={font=\tiny\bfseries, fill=white, inner sep=1pt, opacity=1}
		]
		
		\node[root] (root) at (0, 6) {Propositional Attitudes $\mathcal{A}$};
		
		\node[family] (fam_epi) at (-3.5, 4.5) {Epistemic};
		\node[op, fill=blue!15] (K) at (-4.5, 3) {$K$};
		\node[op, fill=blue!10] (B) at (-2.5, 3) {$B$};
		\draw[gray!30] (root) -- (fam_epi);
		\draw[gray!30] (fam_epi) -- (K); \draw[gray!30] (fam_epi) -- (B);
		
		\node[family] (fam_mot) at (0, 4.5) {Motivational};
		\node[op, fill=green!15] (W) at (-1.5, 1) {$W$};
		\node[derived, fill=green!5] (aim) at (0, 1) {aim};
		\node[op, fill=orange!20] (I) at (1.5, 1) {$I$};
		\draw[gray!30] (root) -- (fam_mot);
		\draw[gray!30] (fam_mot) -- (W); \draw[gray!30] (fam_mot) -- (aim); \draw[gray!30] (fam_mot) -- (I);
		
		\node[family] (fam_aff) at (3.5, 4.5) {Affective / Coll.};
		\node[op, fill=red!15] (F) at (3, 3) {$F$};
		\node[op, fill=violet!15] (C) at (5, 3) {$\mathrm{Com}_X$};
		\draw[gray!30] (root) -- (fam_aff);
		\draw[gray!30] (fam_aff) -- (F); \draw[gray!30] (fam_aff) -- (C);
		
		
		\draw[ax_core] (K) -- node[label_node] {A1, A6} (B);
		
		\draw[ax_prac] (I.north) to[bend right=15] node[label_node, pos=0.6] {A3, A8} (B.south east);
		\draw[ax_prac] (I) to[bend left=20] node[label_node, below] {A3} (W);
		
		\draw[ax_prac, Stealth-Stealth] (aim) -- node[label_node, above] {A4} (W);
		\draw[ax_prac, Stealth-Stealth] (aim) -- node[label_node, pos=0.4, sloped] {A4} (B);
		
		\draw[ax_bias] (F) to[bend left=15] node[label_node, above] {A5} (W);
		\draw[ax_bias] (F) to[bend right=10] node[label_node, midway] {A9} (B);
		
		\node[draw=gray!30, fill=gray!2, rounded corners=3pt, 
		below=2.5cm of aim, text width=12cm, align=left, inner sep=8pt] (legend) {
			\footnotesize \textbf{Coherence Mapping (Table \ref{tab:actor_attitude_axioms}):}\\
			\tikz[baseline=-0.5ex]\draw[ax_core, line width=1pt] (0,0)--(0.4,0); \textbf{A1, A6:} Cognitive consistency (Knowledge implies belief, excludes contrary belief).\\
			\tikz[baseline=-0.5ex]\draw[ax_prac, line width=1pt] (0,0)--(0.4,0); \textbf{A3, A4, A8:} Volitional coherence (Intention requires feasibility and desire; closure properties).\\
			\tikz[baseline=-0.5ex]\draw[ax_bias, line width=1pt] (0,0)--(0.4,0); \textbf{A5, A9:} Affective constraints (Fear as aversion; potential bias in belief revision).
		};
		
	\end{tikzpicture}
	\caption{Taxonomy and logical topology of propositional attitudes. Arrows indicate the mapping of internal coherence axioms onto the assessment state ontology.}
	\label{fig:attitude_taxonomy_v4}
\end{figure}

 \subsubsection{Coalitional Attitudes: Primitive Commitments and Aggregative Variants}
 \label{subsubsec:coalitional-attitudes}
 
 Collective entities in the SBA framework can hold propositional attitudes in two
 conceptually different ways: (i) as \emph{primitive} collective commitments and
 (ii) as \emph{aggregative} attitudes derived from individual members. This
 distinction corresponds to the I-mode versus We-mode analysis of collective
 intentionality \citep{tuomela2002, searle1995}, although the SBA adopts the weaker
 I-mode as its default.
 
 \medskip
 \noindent\textbf{Primitive collective attitudes.}
 The operator
 \[
 \mathrm{Com}_X p
 \]
 denotes a basic collective commitment of coalition \(X\) toward proposition \(p\).
 In the weak interpretation used here, \(\mathrm{Com}_X p\) is primitive; member-level alignment
 is imposed only via coherence postulates. A useful idealized default is:
 
 \medskip
 \noindent\textbf{(C1) (Idealized default; relaxable) Alignment of belief and intention.}
 \[
 \mathrm{Com}_X p \;\Rightarrow\; \forall a \in X:\; B_a p \;\wedge\; I_a p.
 \]
 This can be relaxed (e.g.\ to a designated core subset \(X^\ast\subseteq X\), or to
 an aggregative rule) when full alignment is empirically implausible.
 
 \medskip
 \noindent\textbf{(C2) (Coherence) Consistency of collective commitment.}
 \[
 \mathrm{Com}_X p \;\wedge\; \mathrm{Com}_X \neg p \;\Rightarrow\; \bot.
 \]
 
 \medskip
 \noindent\textbf{(C3) (Optional) Aggregative collective belief (I-mode compatible).}
 \[
 B_X p
 \;\Leftrightarrow\;
 \mathrm{Agg}_B\!\big(\{\,B_a p : a \in X\,\}\big),
 \]
 where \(\mathrm{Agg}_B\) may be majority, unanimity, weighted rules, or other application-specific schemes.
 
 \medskip
 \noindent\textbf{(C4) (Idealized default; relaxable) Closure of collective commitments.}
 \[
 \mathrm{Com}_X(p \rightarrow q) \;\wedge\; \mathrm{Com}_X p \;\Rightarrow\; \mathrm{Com}_X q.
 \]
 
 \medskip
 \noindent\textbf{(C5) (Diagnostic) Overlapping coalitions and conflict.}
 \[
 a \in X \cap Y \;\wedge\; \mathrm{Com}_X p \;\wedge\; \mathrm{Com}_Y \neg p \;\Rightarrow\; \text{conflict}(a).
 \]
 
 \medskip
 \noindent\textbf{Aggregative collective attitudes.}
 Beyond primitive commitment, coalitional belief, desire, or intention may be defined
 from member attitudes via aggregation rules:
 \[
 \begin{aligned}
 	B_X p &\;\Leftrightarrow\;
 	\mathrm{Agg}_B\!\big(\{\,B_a p : a \in X\,\}\big),\\[4pt]
 	W_X p &\;\Leftrightarrow\;
 	\mathrm{Agg}_W\!\big(\{\,W_a p : a \in X\,\}\big),\\[4pt]
 	I_X p &\;\Leftrightarrow\;
 	\mathrm{Agg}_I\!\big(\{\,I_a p : a \in X\,\}\big).
 \end{aligned}
 \]
 Both representations are useful: primitive commitments serve as normative or institutional
 anchors for coordination; aggregative attitudes represent informational structure and
 preference heterogeneity within \(X\).

 \subsubsection{Parameterized Attitudes and Coalitional Stability}
 \label{subsubsec:att-parameterized}
 
 \emph{Optional qualification layer.}
 Binary propositional attitudes provide the logical core of the SBA framework.
 In applications where belief confidence, motivational intensity, or temporal horizon
 must be represented explicitly, the signature can be extended by a finite ordinal
 parameter space.
 
 \medskip
 \noindent\textbf{Parameter space.}
 Let
 \[
 \Theta = \Theta_\ell \times \Theta_\pi \times \Theta_{\vartheta}
 \]
 be an ordinal product space with:
 \begin{itemize}
 	\item \(\Theta_\ell\): perceived likelihood levels (e.g.\ very low, low, medium, high, very high);
 	\item \(\Theta_\pi\): preference or motivation intensity levels;
 	\item \(\Theta_{\vartheta}\): temporal horizon (short, medium, long).
 \end{itemize}
 Each coordinate set is finite and totally ordered. Parameters encode qualitative
 distinctions (e.g.\ ``high confidence'', ``long-term intention'') without cardinal measurement.
 
 \medskip
 \noindent\textbf{Parameterized attitudes.}
 For each attitude type \(K,B,W,I,F\), define a parameterized version as a ternary relation
 \[
 \mathrm{Att}^+ \subseteq A \times P \times \Theta,
 \]
 where \((a,p,\theta)\) represents an attitude of type \(\mathrm{Att}\) with parameter tuple
 \(\theta = (\ell,\pi,\vartheta) \in \Theta\). The corresponding binary attitude is obtained by projection:
 \[
 \mathrm{Att}(a,p)
 \;\Leftrightarrow\;
 \exists \theta \in \Theta:\; \mathrm{Att}^+(a,p,\theta).
 \]
 Thus, the parameterized layer extends, but does not redefine, the binary attitude logic.
 
 \medskip
 \noindent\textbf{Coalitional parameterization.}
 Coalitional attitudes are extended analogously:
 \[
 \mathrm{Att}^+_C \subseteq C \times P \times \Theta,
 \quad
 \mathrm{Com}_X(p;\ell,\pi,\vartheta).
 \]
Write \(\mathrm{Com}_X(p)\) for the corresponding unparameterized commitment, defined by
\[
\mathrm{Com}_X(p)\iff \exists(\ell,\pi,\vartheta)\in\Theta:\ \mathrm{Com}_X(p;\ell,\pi,\vartheta).
\]

 Aggregate parameters may be derived from member-level parameters via application-specific
 aggregation rules:
 \[
 (\ell_X, \pi_X, \vartheta_X)
 =
 \big(
 \Psi_\ell(\{\ell_a : a \in X\}),
 \Psi_\pi(\{\pi_a : a \in X\}),
 \Psi_{\vartheta}(\{\vartheta_a : a \in X\})
 \big).
 \]
 
 \medskip
 \noindent\textbf{Coalitional coherence conditions.}
 Fix order-embeddings \(\iota_\ell:\Theta_\ell\hookrightarrow\{1,\dots,m_\ell\}\) and
 \(\iota_\pi:\Theta_\pi\hookrightarrow\{1,\dots,m_\pi\}\), and define step distances
 \(d_\ell(\ell,\ell')\coloneqq|\iota_\ell(\ell)-\iota_\ell(\ell')|\) and
 \(d_\pi(\pi,\pi')\coloneqq|\iota_\pi(\pi)-\iota_\pi(\pi')|\).
 A parameterized coalition stance \(\mathrm{Com}_X(p;\ell_X,\pi_X,\vartheta_X)\) is coherent if its aggregate
 parameters are compatible with member-level values (within tolerances):
 \[
 \forall a \in X:\;
 d_\ell(\ell_a,\ell_X) < \varepsilon_\ell,
 \qquad
 d_\pi(\pi_a,\pi_X) < \varepsilon_\pi,
 \qquad
 \vartheta_a \approx \vartheta_X.
 \]
 
 \medskip
 \noindent\textbf{Instability condition.}
 A coalition \(X\) is \emph{attitudinally unstable} w.r.t.\ \(p\) when dispersion exceeds tolerable bounds
 or no common temporal horizon exists. Using (step-coded) ranges
 \[
 \mathrm{rng}_X(\ell)\coloneqq\max_{a\in X}\iota_\ell(\ell_a)-\min_{a\in X}\iota_\ell(\ell_a),
 \qquad
 \mathrm{rng}_X(\pi)\coloneqq\max_{a\in X}\iota_\pi(\pi_a)-\min_{a\in X}\iota_\pi(\pi_a),
 \]
 define
 \[
 \mathrm{Cons}_X(\vartheta)
 \;\Leftrightarrow\;
 \exists \vartheta^\star \in \Theta_{\vartheta}:\ \forall a\in X,\ \vartheta_a=\vartheta^\star,
 \]
 and
 \[
 \mathrm{Instab}(X,p)
 \;\Leftrightarrow\;
 \Big(
 \mathrm{rng}_X(\ell) > \varepsilon_\ell
 \;\lor\;
 \mathrm{rng}_X(\pi) > \varepsilon_\pi
 \;\lor\;
 \neg\,\mathrm{Cons}_X(\vartheta)
 \Big).
 \]
 Unstable coalitions cannot sustain a joint commitment and may be treated as decomposing into member-level attitudes.
 
\begin{figure}[htbp]
 	\centering
 	\resizebox{\textwidth}{!}{%
\begin{tikzpicture}[scale=0.72]
 		
 		\definecolor{int0}{RGB}{255,255,255}
 		\definecolor{int1}{RGB}{230,230,245}
 		\definecolor{int2}{RGB}{190,190,225}
 		\definecolor{int3}{RGB}{140,140,200}
 		\definecolor{int4}{RGB}{90,90,170}
 		\definecolor{int5}{RGB}{40,40,130}
 		
 		\def\cellsize{0.8}
 		\def\actorspacing{6.4}
 		\def\legendshift{3.9} 
 		
 		\node[font=\bfseries] at (2.5*\cellsize, 6.2) {Actor A (Government)};
 		\foreach \att [count=\i from 0] in {K, B, W, I, F} {
 			\node[font=\footnotesize] at (\i*\cellsize + 0.5*\cellsize, 5.5) {\att};
 		}
 		\node[anchor=east, font=\tiny, align=right] at (-0.2, 4.5*\cellsize) {p1: Border\\closure};
 		\node[anchor=east, font=\tiny, align=right] at (-0.2, 3.5*\cellsize) {p2: Military\\intervention};
 		\node[anchor=east, font=\tiny, align=right] at (-0.2, 2.5*\cellsize) {p3: Diplomatic\\solution};
 		\node[anchor=east, font=\tiny, align=right] at (-0.2, 1.5*\cellsize) {p4: Economic\\sanctions};
 		
 		\foreach \y/\vals in {4/{5,5,1,0,0}, 3/{2,3,0,0,4}, 2/{3,4,5,4,0}, 1/{4,4,3,2,0}} {
 			\foreach \v [count=\x from 0] in \vals {
 				\fill[int\v] (\x*\cellsize, \y*\cellsize) rectangle +(\cellsize,\cellsize);
 				\pgfmathsetmacro{\tcolor}{\v > 3 ? "white" : "black"}
 				\node[font=\tiny, text=\tcolor] at (\x*\cellsize+0.5*\cellsize, \y*\cellsize+0.5*\cellsize) {\v};
 			}
 		}
 		\draw[thick] (0, \cellsize) grid[step=\cellsize] (5*\cellsize, 5*\cellsize);
 		
 		\begin{scope}[xshift=\actorspacing cm]
 			\node[font=\bfseries] at (2.5*\cellsize, 6.2) {Actor B (Opposition)};
 			\foreach \att [count=\i from 0] in {K, B, W, I, F} {
 				\node[font=\footnotesize] at (\i*\cellsize + 0.5*\cellsize, 5.5) {\att};
 			}
 			\foreach \y/\vals in {4/{3,4,4,3,0}, 3/{4,4,5,4,0}, 2/{3,3,3,2,0}, 1/{2,3,4,3,0}} {
 				\foreach \v [count=\x from 0] in \vals {
 					\fill[int\v] (\x*\cellsize, \y*\cellsize) rectangle +(\cellsize,\cellsize);
 					\pgfmathsetmacro{\tcolor}{\v > 3 ? "white" : "black"}
 					\node[font=\tiny, text=\tcolor] at (\x*\cellsize+0.5*\cellsize, \y*\cellsize+0.5*\cellsize) {\v};
 				}
 			}
 			\draw[thick] (0, \cellsize) grid[step=\cellsize] (5*\cellsize, 5*\cellsize);
 		\end{scope}
 		
 		\begin{scope}[xshift=2*\actorspacing cm]
 			\node[font=\bfseries] at (2.5*\cellsize, 6.2) {Actor C (Civil Society)};
 			\foreach \att [count=\i from 0] in {K, B, W, I, F} {
 				\node[font=\footnotesize] at (\i*\cellsize + 0.5*\cellsize, 5.5) {\att};
 			}
 			\foreach \y/\vals in {4/{5,5,0,0,5}, 3/{1,2,0,0,3}, 2/{4,5,5,4,0}, 1/{3,3,2,1,1}} {
 				\foreach \v [count=\x from 0] in \vals {
 					\fill[int\v] (\x*\cellsize, \y*\cellsize) rectangle +(\cellsize,\cellsize);
 					\pgfmathsetmacro{\tcolor}{\v > 3 ? "white" : "black"}
 					\node[font=\tiny, text=\tcolor] at (\x*\cellsize+0.5*\cellsize, \y*\cellsize+0.5*\cellsize) {\v};
 				}
 			}
 			\draw[thick] (0, \cellsize) grid[step=\cellsize] (5*\cellsize, 5*\cellsize);
 		\end{scope}
 		
 		\begin{scope}[yshift=-1cm, xshift=\legendshift cm]
 			\node[font=\small\bfseries, anchor=east] at (0,0) {Intensity:};
 			\foreach \i in {0,...,5} {
 				\fill[int\i] (\i*0.8, -0.2) rectangle +(0.6, 0.4);
 				\draw (\i*0.8, -0.2) rectangle +(0.6, 0.4);
 				\node[font=\tiny] at (\i*0.8 + 0.3, -0.5) {\i};
 			}
 		\end{scope}
 		
 	\end{tikzpicture}%
}%
 	\caption{Attitude intensity matrix across SBA actors (optional qualification layer). Intensities range from 0 (none) to 5 (very high).}
 	\label{fig:attitude_intensity_matrix}
 \end{figure}
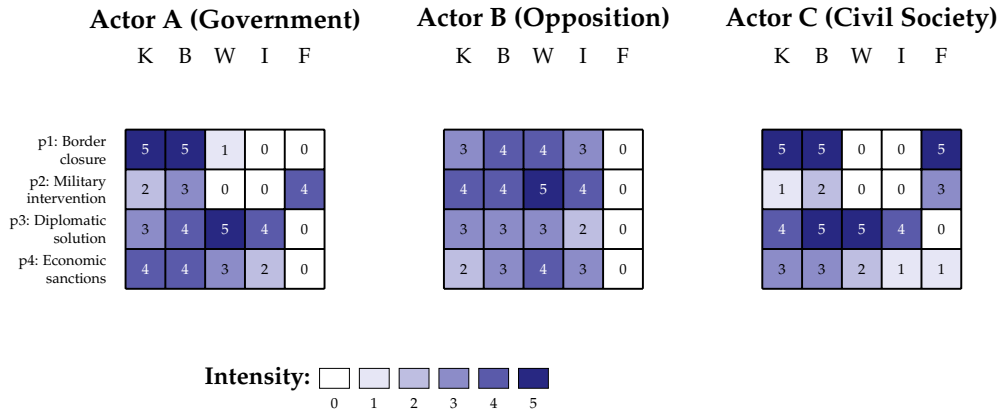

 \medskip
 \noindent\textbf{Interpretation.}
 This optional layer supports qualitative coding of confidence, intensity, and planning horizon while preserving
 the binary logical core.
 Figure~\ref{fig:attitude_intensity_matrix} shows a specimen intensity matrix for a five-actor scenario.
 
 \subsubsection{Theoretical Foundation}
 \label{subsubsec:att-theoretical-foundation}
 
 The representation of attitudes in the SBA framework is grounded in three complementary
 theoretical traditions.
 
 First, the treatment of beliefs, knowledge, desires, and intentions as relations
 between agents and propositions follows classical analyses of propositional attitudes
 \citep{frege1892, quine1956, strawson1959, anscombe1957}, in which attitude verbs are
 analyzed as intensional operators. This view is reflected in the relations \(K_a\),
 \(B_a\), \(W_a\), \(I_a\), and \(F_a\) defined over an explicit domain of propositional contents.
 
 Second, the practical interrelations among these operators draw on cognitive action
 theory and planning-based accounts of agency \citep{bratman1987, searle1983}. The
 coherence postulates for intention---its dependence on desire and feasibility belief,
 the consistency of commitments, and its closure under accepted implications---support
 scenario trees with minimally coherent motivational structure. Where indicated, these
 postulates are treated as idealized defaults and can be relaxed for bounded agents.
 
 Third, the coalitional operator \(\mathrm{Com}_X\) and its optional alignment conditions are informed
 by analyses of collective intentionality and joint commitment \citep{tuomela2002, searle1995}.
 The SBA adopts a weak I-mode interpretation as its default: collective commitment is primitive,
 and stronger We-mode conditions remain compatible as optional refinements.

\subsection{Relations\index{Relations}}
\label{subsec:relations}

Relations\index{Relations} encode directional or symmetric connections between
actors\index{Actors} and coalitions\index{Coalitions} in the Scenario Database.
Where attributes\index{Attributes} characterize entities individually and
attitudes\index{Attitudes} represent their internal states, relations represent
the \emph{structured linkages} between them: authority, alliance, conflict,
information flow, dependence, and higher-order configurations thereof.

\subsubsection{Formal definition}
\label{subsubsec:rel-formal}

Let $\mathrm{Rel}$ denote the set of relation types in the Scenario Database.
Let $\Omega \coloneqq A \cup C$ be the unified endpoint domain of actors\index{Actors} and coalitions\index{Coalitions}.
For each dyadic relation type $R \in \mathrm{Rel}$, its extensional support is represented by a binary relation
\index{Relations}
\[
R \subseteq \Omega \times \Omega,
\]
which may be directed (no symmetry assumed) or undirected (symmetric: $(x,y)\in R \Leftrightarrow (y,x)\in R$).

Dyadic relations may carry relational annotations. We treat these as \emph{partial} functions on $\Omega\times\Omega$
(defined only for $(x,y)\in R$):
\begin{align*}
	w_R       &\colon \Omega\times\Omega \rightharpoonup W, \\
	\sigma_{\mathrm{rel},R}  &\colon \Omega\times\Omega \rightharpoonup \{-1, 0, +1\}, \\
	\lambda_R &\colon \Omega\times\Omega \rightharpoonup \Lambda, \\
	\operatorname{vis}_R       &\colon \Omega\times\Omega \rightharpoonup \mathrm{Vis}.
\end{align*}
where $w_R$ is a weight (strength/intensity), $\sigma_{\mathrm{rel},R}$ a signed valence (cooperative $+1$, adversarial $-1$, neutral/unspecified $0$),
$\lambda_R$ a relational layer, and $\operatorname{vis}_R$ a visibility attribute\index{Relations!Visibility}.
The set of layers $\Lambda$ is finite (e.g.\ \textsc{Pol}, \textsc{Econ}, \textsc{Info}, \textsc{Sec}); $\mathrm{Vis}$ is an application-specific
visibility scale (e.g.\ observed vs.\ perceived vs.\ signalled).

\subsubsection{Taxonomy of relation families}
\label{subsubsec:rel-taxonomy}

Relations in SBA capture semantically distinct patterns of interaction among actors and coalitions.
They may express influence, cooperation, information flow, authority, interdependence, or hostility,
and they may operate across several domains (political, economic, informational, security).
Table~\ref{tab:relation_families} summarizes the main relation families used in the Scenario Database.

\begin{table}[tbp]
	\footnotesize
	\centering
	\renewcommand{\arraystretch}{1.15}
	\setlength{\tabcolsep}{5pt} 
	
	\begin{tabularx}{\textwidth}{@{} 
			>{\bfseries\raggedright\arraybackslash}p{3.4cm} 
			>{\RaggedRight\arraybackslash}p{4.8cm}          
			>{\RaggedRight\arraybackslash}X                   
			@{}}
		\toprule
		Relation family & \textbf{Typical instances} & \textbf{Modeling properties} \\
		\midrule
		
		Power / Influence &
		influences, coerces, protects, deters, sanctions &
		Directed, weighted; sign may indicate beneficial or harmful impact; represents the ability to alter another's feasible options. \\
		\addlinespace
		
		Alignment / Affinity &
		supports, allies with, endorses, shares ideology, cooperates &
		Often undirected or symmetric; typically positive sign; strength may encode trust, cohesion, or ideological proximity. \\
		\addlinespace
		
		Authority / Obligation &
		commands, regulates, governs, enforces, delegates &
		Directed and typically hierarchical; often acyclic within a layer; may include legitimacy or compliance attributes. \\
		\addlinespace
		
		Exchange / Interdependence &
		trades with, provides aid, shares resources, depends on &
		Directed or reciprocal; weighted by flow volume or asymmetry; reciprocity indices can indicate vulnerability or leverage. \\
		\addlinespace
		
		Information / Communication &
		informs, signals, monitors, reports to, spreads narratives &
		Directed; may be probabilistic; visibility distinguishes objectively observed ties from perceived or signalled ties. \\
		\addlinespace
		
		Adversarial (Confrontative) &
		competes, threatens, undermines, attacks &
		Negative sign; directed or reciprocal; hostility intensity and deterrence effects may vary by domain and escalation level. \\
		\addlinespace
		
		Mediative / Regulatory &
		arbitrates, mediates, guarantees, sets standards, enforces peace &
		Triadic or higher-order (arity $\ge 3$); captures stabilising roles between hostile actors; represented as hyperedges or incidence structures (Section~\ref{subsubsec:rel-higher}). \\
		\bottomrule
	\end{tabularx}
	\caption{Relation families in SBA: functional semantic classes. Relations are modeled as dyadic (arity 2) unless otherwise noted, such as in mediative roles.}
	\label{tab:relation_families}
\end{table}

\paragraph{Interpretation.}
Each relational family corresponds to a mechanism through which actors and coalitions influence one another.
Power and influence ties shape feasible options and strategic leverage; alignment and affinity relations
determine cooperation potential and the robustness of coalitions.
Authority and obligation relations impose hierarchical constraints on actions.
Exchange and interdependence ties capture resource flows and mutual dependencies, which may stabilise or
destabilise interactions depending on their symmetry.
Information and communication relations describe the pathways through which beliefs, signals, and narratives propagate,
forming an epistemic substrate for attitude updates.
Confrontative relations represent graded hostility or deterrence, while mediative and regulatory ties encode stabilising
triadic or institutional structures.
Taken together, these families \emph{shape and constrain} scenario branching and dynamic evolution in SBA in conjunction
with options\index{Options} and events\index{Events}.

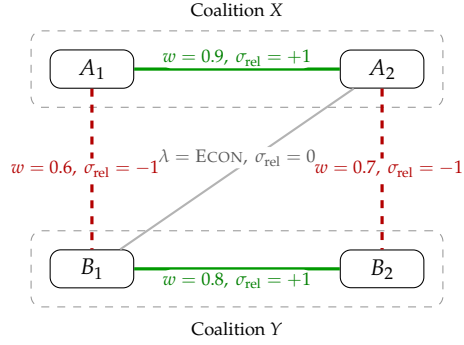
\begin{figure}[htbp]
	\centering
	\begin{tikzpicture}[
		scale=0.85, transform shape,
		node distance=25mm and 32mm,
		every node/.style={draw, rounded corners=4pt, align=center, inner sep=4pt, font=\small, minimum width=13mm},
		coop/.style={-, very thick, green!60!black},
		conf/.style={-, very thick, red!70!black, dashed},
		neutral/.style={-, thick, gray!60},
		lab/.style={font=\scriptsize, fill=white, inner sep=1pt, draw=none}
		]
		
		\node (a1) {$A_1$};
		\node[right=of a1] (a2) {$A_2$};
		\node[below=of a1] (b1) {$B_1$};
		\node[right=of b1] (b2) {$B_2$};
		
		\node[draw=gray!80, dashed, fit=(a1)(a2), inner sep=7pt, label={[font=\scriptsize, yshift=2pt]above:Coalition $X$}] {};
		\node[draw=gray!80, dashed, fit=(b1)(b2), inner sep=7pt, label={[font=\scriptsize, yshift=-2pt]below:Coalition $Y$}] {};
		
		\draw[coop] (a1) -- node[lab, midway, yshift=5pt, text=green!50!black] {$w=0.9,\ \sigma_{\mathrm{rel}}=+1$} (a2);
		\draw[coop] (b1) -- node[lab, midway, yshift=-5pt, text=green!50!black] {$w=0.8,\ \sigma_{\mathrm{rel}}=+1$} (b2);
		
		\draw[conf] (a1) -- node[lab, midway, xshift=-3pt, text=red!70!black] {$w=0.6,\ \sigma_{\mathrm{rel}}=-1$} (b1);
		\draw[conf] (a2) -- node[lab, midway, xshift=3pt, text=red!70!black] {$w=0.7,\ \sigma_{\mathrm{rel}}=-1$} (b2);
		
		\draw[neutral] (a2) -- node[lab, midway, yshift=5pt, text=gray!70!black] {$\lambda=\textsc{Econ},\ \sigma_{\mathrm{rel}}=0$} (b1);
		
	\end{tikzpicture}
	\caption{Illustrative signed and weighted relation network in SBA.
		Green edges = cooperative ties; red dashed edges = confrontative ties;
		gray edges = neutral or cross-domain links.}
	\label{fig:sba_relation_network}
\end{figure}

Figure~\ref{fig:sba_relation_network} illustrates a simplified configuration with two coalitions and several signed/weighted dyadic relation types.

\FloatBarrier

\subsubsection{Higher-order relations}
\label{subsubsec:rel-higher}

While most network representations model interactions as dyadic relations between pairs of actors,
many political, economic, and institutional dependencies inherently involve three or more entities.
Examples include defence pacts, mediation frameworks, regulatory bodies, and multilateral agreements.
SBA treats these interactions as \emph{higher-order relations} (or \emph{hyperedges}) when their meaning depends on joint participation rather than on a bundle of separable pairwise ties.

\paragraph{Formal representation (incidence view).}
For each higher-order relation type $R \in \mathrm{Rel}$, let $E_R$ be a set of relation-instances (hyperedges) and let
\[
\mathrm{inc}_R \subseteq E_R \times \Omega
\]
be the incidence relation. Each instance $e\in E_R$ determines a participant set
\[
\partial(e) \coloneqq \{\,u\in \Omega \mid (e,u)\in \mathrm{inc}_R\,\},
\qquad |\partial(e)|\ge 3.
\]
This representation supports database-level annotation on relation-instances (e.g.\ weights, layers, visibility) without
collapsing the interaction into dyadic approximations.

\paragraph{Conceptual classes of higher-order relations.}
Higher-order interactions fall into several characteristic families:
\begin{itemize}
	\item \textbf{Coalitional or collective relations:}
	Membership in a defence alliance, treaty organization, or joint venture.
	Here the existence of the group itself constitutes a relational fact.
	
	\item \textbf{Regulatory or institutional relations:}
	Structures involving governments, private actors, and oversight institutions (e.g.\ environmental compacts, trade regimes).
	These relations encode constraint-based interdependencies.
	
	\item \textbf{Mediation and arbitration configurations:}
	Interactions of the form $(\text{mediate},\{a,b,m\})$, where $m$ coordinates, stabilises, or constrains the exchange between $a$ and $b$.
	Such triadic structures are central in conflict resolution.
\end{itemize}

These classes highlight that higher-order relations express semantic content that cannot be decomposed into independent dyads.

\paragraph{Graphical representation.}
Two complementary visualizations help illustrate how higher-order relations are represented in SBA.

\begin{enumerate}
	\item \textbf{Simplicial (geometric) representation:}
	An $n$-ary relation among actors is depicted as an $(n-1)$-simplex:
	a triangle for triadic relations, a tetrahedron for tetradic relations, and so on.
	This highlights the ``all-member'' character of the interaction.
	
	\item \textbf{Bipartite incidence representation:}
	Higher-order relations are modeled as relation-instances ($R$-nodes) connected to all participating actors or coalitions.
	This makes explicit which actors co-participate in each relation-instance.
	Figure~\ref{fig:higherorder} juxtaposes the simplicial and bipartite visualisations.
\end{enumerate}

\begin{figure}[htbp]
	\centering
	\textbf{(a) Simplicial representation}\par\smallskip
	\begin{tikzpicture}[scale=1.0, every node/.style={circle,draw,minimum size=0.6cm}]
		\node (a1) at (0,0) {A};
		\node (a2) at (2,0) {B};
		\node (a3) at (1,1.7) {C};
		\fill[blue!10,opacity=0.4] (a1.center)--(a2.center)--(a3.center)--cycle;
		\node[draw=none] at (1,2.6) {\normalsize $R_1=\{A,B,C\}$};
		
		\node (b1) at (5,0) {D};
		\node (b2) at (7,0) {E};
		\node (b3) at (6,1.7) {F};
		\node (b4) at (6,0.6) {G};
		\fill[green!10,opacity=0.4] (b1.center)--(b2.center)--(b3.center)--cycle;
		\fill[green!10,opacity=0.4] (b1.center)--(b3.center)--(b4.center)--cycle;
		\fill[green!10,opacity=0.4] (b2.center)--(b3.center)--(b4.center)--cycle;
		\fill[green!10,opacity=0.4] (b1.center)--(b2.center)--(b4.center)--cycle;
		\node[draw=none] at (6,2.6) {\normalsize $R_2=\{D,E,F,G\}$};
	\end{tikzpicture}
	
	\medskip
	\textbf{(b) Bipartite incidence structure}\par\smallskip
	\begin{tikzpicture}[scale=1.0, node distance=1.6cm, every node/.style={font=\small}]
		\node[circle,draw,fill=blue!10] (a1) at (0,2) {A};
		\node[circle,draw,fill=blue!10] (a2) at (0,0) {B};
		\node[circle,draw,fill=blue!10] (a3) at (0,-2) {C};
		
		\node[rectangle,draw,fill=green!10,minimum width=1cm,minimum height=0.6cm] (r1) at (4,1) {$R_1$};
		\node[rectangle,draw,fill=green!10,minimum width=1cm,minimum height=0.6cm] (r2) at (4,-1) {$R_2$};
		
		\draw[-] (a1) -- (r1);
		\draw[-] (a2) -- (r1);
		\draw[-] (a3) -- (r1);
		\draw[-] (a2) -- (r2);
		\draw[-] (a3) -- (r2);
		
		\node[draw=none] at (2,3) {\normalsize incidence edges $\,\mathrm{inc}_R$};
	\end{tikzpicture}
	
	\caption[Higher-order relations: equivalent views]{Two equivalent visualisations of higher-order relations: (a) simplicial (geometric) depiction; (b) bipartite incidence representation suited for database and algorithmic manipulation.}
	\label{fig:higherorder}
\end{figure}

\FloatBarrier

\paragraph{Interpretation.}
The simplicial view highlights the collective nature of $n$-ary interactions: a triad or tetrad is represented as a unified geometric object rather than a set of dyads.
The bipartite view makes explicit which actors co-participate in each relation-instance and is well-suited for algorithmic manipulation.
Together, these representations show how higher-order constraints, coalitional structures, and institutional frameworks can be encoded in SBA without collapsing them into pairwise approximations.

\subsubsection{Integration with actors and coalitions}
\label{subsubsec:rel-integration}

Relations in SBA are defined on the unified domain $\Omega=A\cup C$, allowing both individual actors and coalitions to appear as relational endpoints.
The same relational formalism can therefore represent actor--actor ties, coalition--coalition ties, and cross-level links between actors and coalitions.

\paragraph{Unified relational domain.}
For any dyadic relation type $R \in \mathrm{Rel}$,
\[
R \subseteq \Omega \times \Omega,
\]
so that ties may connect two actors (e.g.\ $\text{influences}(a,b)$),
two coalitions (e.g.\ $\text{rivals}(X,Y)$), or mixed pairs (e.g.\ $\text{negotiates}(a,Y)$).
Coalitions thus participate in the network on equal footing with their members.

\paragraph{Coalitional aggregation of relations.}
Since coalitions are sets of actors, their relational attributes are derived from the ties among their members.
Let $X,Y \subseteq A$ be two coalitions and let $w_R(a,b)$ denote member-level weights for $R \in \mathrm{Rel}$.
Define the set of cross-member ties
\[
\Xi_{X,Y}^R \coloneqq \{\, (a,b)\in X\times Y \mid (a,b)\in R \,\}.
\]
The aggregate weight between \(X\) and \(Y\) is defined by an aggregation operator
\[
w_R(X,Y)=
\begin{cases}
	G_R\!\big(\{\, w_R(a,b) : (a,b)\in \Xi_{X,Y}^R \,\}\big), & \Xi_{X,Y}^R\neq\emptyset,\\
	0, & \Xi_{X,Y}^R=\emptyset,
\end{cases}
\]
where $G_R$ is selected according to the semantics of $R$ (mean, weighted mean by salience, maximum, or minimum).

Polarity and visibility attributes are aggregated analogously. In particular, to avoid treating weak and strong ties equally, the aggregate sign is weight-adjusted:
\[
\sigma_{\mathrm{rel},R}(X,Y)=
\begin{cases}
	\mathrm{sign}\!\Big(\sum_{(a,b)\in \Xi_{X,Y}^R} w_R(a,b)\cdot \sigma_{\mathrm{rel},R}(a,b)\Big), & \Xi_{X,Y}^R\neq\emptyset,\\
	0, & \Xi_{X,Y}^R=\emptyset.
\end{cases}
\]
For multiplex relations, the dominant layer is selected by mode or salience weighting:
\[
\lambda_R(X,Y)=
\begin{cases}
	\mathrm{mode}\!\big(\{\lambda_R(a,b):(a,b)\in \Xi_{X,Y}^R\}\big), & \Xi_{X,Y}^R\neq\emptyset,\\
	\text{undefined}, & \Xi_{X,Y}^R=\emptyset.
\end{cases}
\]
Visibility can be aggregated by mode (descriptive default) or unanimity (conservative default):
\[
\operatorname{vis}_R(X,Y)=
\begin{cases}
	\mathrm{mode}\!\big(\{\operatorname{vis}_R(a,b):(a,b)\in \Xi_{X,Y}^R\}\big), & \Xi_{X,Y}^R\neq\emptyset,\\
	\text{undefined}, & \Xi_{X,Y}^R=\emptyset.
\end{cases}
\]

\begin{table}[h!]
	\centering
	\renewcommand{\arraystretch}{1.15}
	\small
	\setlength{\tabcolsep}{4.5pt}
	\begin{tabularx}{\textwidth}{
			>{\RaggedRight\arraybackslash}p{3.0cm}
			>{\RaggedRight\arraybackslash}X
			>{\RaggedRight\arraybackslash}X}
		\toprule
		\textbf{Relational attribute} & \textbf{Aggregation rule} & \textbf{Interpretation} \\
		\midrule
		Weight $w_R$ &
		Mean / weighted mean / max / min &
		Strength of cross--member ties (influence, trade, information). \\[4pt]
		
		Sign $\sigma_{\mathrm{rel},R}$ &
		Weight-adjusted signed sum &
		Net cooperative vs.\ adversarial stance between groups. \\[4pt]
		
		Visibility $\operatorname{vis}_R$ &
		Mode or unanimity &
		Distinguishes observed from perceived relations. \\[4pt]
		
		Layer $\lambda_R$ &
		Mode or salience weighting &
		Dominant domain (political, economic, informational, etc.). \\
		\bottomrule
	\end{tabularx}
	\caption{Aggregation of relational attributes from actor to coalition level.}
	\label{tab:rel_aggregation}
\end{table}

\paragraph{Interpretation.}
This aggregation scheme lets coalitions inherit relational properties from member-level ties while still allowing distinct group-level structure.
It supports consistent representations of alliance cohesion, negative interdependence, hybrid actor--coalition ties, and multilayer interaction topologies.
Because coalitions appear in the same endpoint domain as individual actors, the relational network remains structurally unified and can be used directly as input for scenario-tree generation.

\subsubsection{Relational stability and coherence conditions}
\label{subsubsec:rel-stability}

Not all observed or reported ties in the relational layer are suitable for initializing a Scenario Database.
Some relations are internally inconsistent, rapidly fluctuating, or contradictory across members of a coalition.
SBA therefore uses \emph{relational stability conditions} to distinguish ties that can seed scenario-tree generation from ties that should be retained only as volatile or latent information.

\paragraph{Definition (variance-- and disagreement--based relational stability).}
Let $R \in \mathrm{Rel}$ be a dyadic relation type and let $(X,Y)\in \Omega^2$ be a pair of endpoints (actors or coalitions).
If $X$ (resp.\ $Y$) is an actor, identify it with the singleton coalition $\{X\}$ (resp.\ $\{Y\}$) for member-level aggregation.
Let
\[
\Xi_{X,Y}^R \coloneqq \{\, (a,b)\in X\times Y \mid (a,b)\in R \,\}
\]
denote the set of member-level ties generating the aggregate tie $w_R(X,Y)$ (cf.\ Table~\ref{tab:rel_aggregation}).
Define the modal sign among the member-level ties
\[
\sigma^\star_{\mathrm{rel}} \coloneqq \mathrm{mode}\!\big(\{\sigma_{\mathrm{rel},R}(a,b):(a,b)\in \Xi_{X,Y}^R\}\big),
\]
and the disagreement rate
\[
\mathrm{Disagr}_{X,Y}^R \coloneqq \Pr_{(a,b)\in \Xi_{X,Y}^R}\big[\sigma_{\mathrm{rel},R}(a,b)\neq \sigma^\star_{\mathrm{rel}}\big].
\]
The tie $R(X,Y)$ is \emph{stable} if dispersion of member-level weights is low and sign-disagreement is bounded:
\[
\mathrm{RelStab}_R(X,Y)
\;\Leftrightarrow\;
\Big(
\mathrm{Var}_{(a,b)\in \Xi_{X,Y}^R}\!\big(w_R(a,b)\big) < \varepsilon_w
\;\wedge\;
\mathrm{Disagr}_{X,Y}^R < \varepsilon_\sigma
\Big).
\]
Low variance and low disagreement indicate that members of $X$ and $Y$ maintain similar intensities and orientations toward each other.
High dispersion indicates disagreement, fragmentation, or ambiguous signaling within or between the groups.

\paragraph{Interpretation.}
Variance-based stability implements a simple coherence criterion: a coalitional tie is included in the initial relational graph only if the corresponding member-level relations point in roughly the same direction with comparable strength.
Relations violating $\mathrm{RelStab}_R$ are stored in a separate ``latent'' layer to indicate potential volatility but are excluded from the primary network used to seed scenario-tree construction.

\paragraph{Structural balance as a complementary criterion.}
For \emph{signed} relation types $R\in\mathrm{Rel}$ whose sign takes values
in $\{-1,+1\}$ on a chosen layer, an additional stability diagnostic is
given by \emph{structural balance} \citep{heider1946, cartwright1956}.
A triad $(a,b,c)$ is balanced (with respect to $R$ on that layer) if
\[
\mathrm{Bal}_R(a,b,c)
\;\Leftrightarrow\;
\sigma_{\mathrm{bal}}(a,b)\cdot\sigma_{\mathrm{bal}}(b,c)\cdot\sigma_{\mathrm{bal}}(c,a) = +1.
\]
Balanced configurations correspond to internally coherent clusters
(e.g.\ ``friend of my friend is my friend'', ``enemy of my enemy is my friend'').
Unbalanced triads indicate relational tension, potential realignment, or hidden conflict.
SBA does not enforce global balance but uses balance violations as diagnostic markers for unstable regions of the relational network.

\begin{figure}[htbp]
	\centering
	\begin{tikzpicture}[
		scale=0.9,
		actor/.style={circle, draw=black, thick, minimum size=0.8cm, font=\small\bfseries, fill=blue!10},
		positive/.style={-, very thick, green!60!black},
		neg/.style={-, very thick, red!70!black, dashed},
		strong/.style={-, ultra thick, green!40!black},
		weak/.style={-, thick, gray!60, dotted},
		directed/.style={->, thick, blue!60!black, >=stealth},
		layerlabel/.style={font=\small\bfseries, fill=white, inner sep=2pt}
		]
		
		\begin{scope}[local bounding box=layer1]
			\node[actor] (A1) at (0, 2) {A};
			\node[actor] (B1) at (2, 2) {B};
			\node[actor] (C1) at (0, 0) {C};
			\node[actor] (D1) at (2, 0) {D};
			
			\draw[positive] (A1) -- (B1);
			\draw[neg] (A1) -- (C1);
			\draw[positive] (B1) -- (D1);
			\draw[neg] (C1) -- (D1);
			\draw[neg] (A1) to[bend left=15] (D1);
			
			\node[layerlabel] at (1, -0.7) {Political};
			\node[font=\tiny, align=left, anchor=west] at (2.2, 1) {+ Alliance\\-- Rivalry};
		\end{scope}
		
		\begin{scope}[shift={(6.2,0)}, local bounding box=layer2]
			\node[actor] (A2) at (0, 2) {A};
			\node[actor] (B2) at (2, 2) {B};
			\node[actor] (C2) at (0, 0) {C};
			\node[actor] (D2) at (2, 0) {D};
			
			\draw[strong] (A2) -- (B2);
			\draw[weak] (A2) -- (C2);
			\draw[strong] (C2) -- (D2);
			\draw[positive] (B2) -- (D2);
			
			\node[layerlabel] at (1, -0.7) {Economic};
			\node[font=\tiny, align=left, anchor=west] at (2.2, 1) {$\equiv$ Strong\\$\cdots$ Weak};
		\end{scope}
		
		\begin{scope}[shift={(0,-4.5)}, local bounding box=layer3]
			\node[actor] (A3) at (0, 2) {A};
			\node[actor] (B3) at (2, 2) {B};
			\node[actor] (C3) at (0, 0) {C};
			\node[actor] (D3) at (2, 0) {D};
			
			\draw[directed] (A3) -- (B3);
			\draw[directed] (B3) -- (D3);
			\draw[directed] (D3) -- (C3);
			\draw[directed] (C3) -- (A3);
			\draw[directed] (A3) to[bend right=15] (D3);
			
			\node[layerlabel] at (1, -0.7) {Informational};
			\node[font=\tiny, align=left, anchor=west] at (2.2, 1) {$\rightarrow$ Info flow};
		\end{scope}
		
		\begin{scope}[shift={(6.2,-4.5)}, local bounding box=layer4]
			\node[actor] (A4) at (0, 2) {A};
			\node[actor] (B4) at (2, 2) {B};
			\node[actor] (C4) at (0, 0) {C};
			\node[actor] (D4) at (2, 0) {D};
			
			\draw[neg] (A4) -- (B4);
			\draw[positive] (A4) -- (C4);
			\draw[neg] (B4) to[bend left=15] (C4);
			\draw[positive] (C4) -- (D4);
			\draw[neg] (B4) -- (D4);
			
			\node[layerlabel] at (1, -0.7) {Security};
			\node[font=\tiny, align=left, anchor=west] at (2.2, 1) {+ Defense\\-- Threat};
		\end{scope}
		
		\draw[dotted, gray!30] (A1) -- (A2) (B1) -- (B2) (C1) -- (C2) (D1) -- (D2);
		\draw[dotted, gray!30] (A3) -- (A4) (B3) -- (B4) (C3) -- (C4) (D3) -- (D4);
		
		\node[draw=black, thick, rectangle, rounded corners=3pt, fill=yellow!5, align=left, font=\tiny, inner sep=5pt]
		at (4.1, -6.8) {
			\textbf{Layer Coherence Example (A \& B):}\\
			\textcolor{green!60!black}{Political (+)} / \textcolor{green!40!black}{Economic (++)} vs.\ \textcolor{red!70!black}{Security (--)}
			$\Rightarrow$ \textbf{Incoherent Layer State} (high structural tension)
		};
		
	\end{tikzpicture}
	\caption{Multiplex relational structure across four domains in SBA framework.}
	\label{fig:multilayer_relations}
\end{figure}
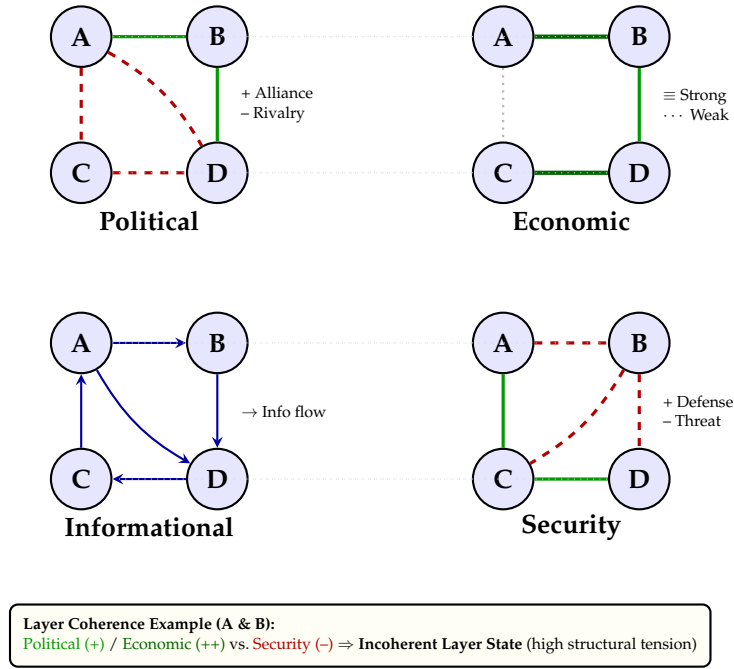

\paragraph{Coherence across layers.}
Multiplex relations may satisfy stability on one layer while violating it on another (e.g.\ strong political rivalry but simultaneous economic interdependence).
For any dyadic relation type $R\in\mathrm{Rel}$ and layer $\lambda\in\Lambda$, define the layer-slice
\[
R^\lambda \coloneqq \{\, (x,y)\in R \mid \lambda_R(x,y)=\lambda \,\}.
\]
A tie $(X,Y)$ is \emph{layer-coherent} for $R$ if, for each layer, the corresponding slice is stable or absent:
\[
\mathrm{LayCoh}_R(X,Y)
\;\Leftrightarrow\;
\forall \lambda \in \Lambda:\ 
\mathrm{RelStab}_{R^\lambda}(X,Y)
\quad\text{or}\quad
w_{R^\lambda}(X,Y)=0.
\]
Figure~\ref{fig:multilayer_relations} visualises the multiplex structure across four domains.
This prevents scenario initialization from mixing incompatible layer signals (e.g.\ a strongly confrontative political tie with a fragmented economic dependency).

\paragraph{Combined stability criterion.}
A relational tie $R(X,Y)$ is accepted into the initial SBA network if:
\[
\mathrm{Stable}(X,Y)
\;\Leftrightarrow\;
\begin{aligned}[t]
	&\mathrm{RelStab}_R(X,Y)
	\;\wedge\;
	\mathrm{LayCoh}_R(X,Y)
	\\
	&\wedge\;
	\neg\,\mathrm{StrongUnbalance}(X,Y).
\end{aligned}
\]
Here $\mathrm{StrongUnbalance}(X,Y)$ flags persistent balance violations in triads involving $X$ or $Y$ on critical signed layers
(e.g.\ political and security), and is used as a diagnostic exclusion rule rather than a global constraint.

\subsubsection{Theoretical foundation}
\label{subsubsec:rel-theory}

The relational layer of SBA ontology rests on three complementary strands of theory:
(1) network theory and structural interdependence,
(2) signed and multiplex relational structures,
and (3) higher-order and institutional dependencies.
Together, these literatures justify the use of weighted, signed, layered, and hypergraph relations in the Scenario Database.

\paragraph{(1) Network-theoretic foundations.}
Binary ties with quantitative annotations (weight, sign, layer, visibility) follow standard representations of social and institutional structures \citep{wasserman1994, jackson2008, newman2010, easley2010}.
Weights $w_R$ capture graded influence, dependence, or exposure, consistent with power--dependence theory \citep{emerson1962}.
Directed edges represent asymmetric capability or information flow, whereas undirected ties encode mutual alignment or exchange.

\paragraph{(2) Signed and multiplex relations.}
Strategic interactions frequently carry cooperative or antagonistic valence.
Signed networks \citep{heider1946, cartwright1956, szell2010, altafini2013} support the representation of polarity via $\sigma_{\mathrm{rel},R}$
and permit diagnostics such as structural balance.
Multilayer (multiplex) network theory \citep{kivela2014, boccaletti2014} justifies distinguishing relational domains (political, economic, informational, security),
each encoding distinct channels of interaction.
A visibility parameter $\operatorname{vis}_R$ links relational structure to epistemic considerations, distinguishing observed, perceived, and signalled ties.

\paragraph{(3) Higher-order and institutional dependencies.}
Many dependencies are not reducible to pairwise relations---examples include alliance pacts, mediation structures, joint regulatory regimes, and standard-setting bodies.
Higher-order network theory \citep{battiston2020, benson2016, patania2017shape} shows that such configurations require hypergraphs or incidence structures to preserve their $k$-ary semantics.
SBA adopts this insight by representing group-level commitments or institutional constraints as higher-order relations over subsets of actors or coalitions.

\paragraph{Integration with conflict and cooperation theory.}
Confrontative and cooperative ties draw on established results from strategic studies and international relations.
Negative ties represent coercion, deterrence, and hostile signalling \citep{schelling1966}, while positive ties reflect alliance, trust, or hierarchical authority \citep{morrow1994, leeds2003}.
Triadic mediation or guarantor roles \citep{bercovitch1992, greig2012} motivate representing stabilising higher-order structures explicitly.

\paragraph{Conceptual role in SBA.}
Within SBA, relations complement attributes (what actors \emph{are}) and attitudes (what actors \emph{believe, fear, or intend}).
They specify the system's \emph{interaction topology}: who influences whom, which channels transmit information,
where coalitional cohesion or fragmentation originates, and where confrontative pressures or cooperative synergies emerge.
This networked structure supports scenario branching constraints, stability diagnostics, and strategic interpretation of actor and coalition behavior.


	\subsection{Options\index{Options}}
	\label{subsec:options}
	
	Within SBA\index{Scenario Bundle Analysis}, \emph{options} represent the structured
	potential actions that actors\index{Actors} or coalitions\index{Coalitions} may perform in a given
	scenario. They form the operational interface between the descriptive
	layer of properties---attributes\index{Attributes}, attitudes\index{Attitudes}, relations\index{Relations}---
	and the generative layer of scenario development captured in
	scenario trees\index{Scenario Trees}.
	Whereas attributes describe what actors \emph{are}, and attitudes describe
	what they \emph{believe, desire, or intend}, options describe what they
	\emph{can do}.

	\subsubsection{Conceptual introduction}
	\label{subsubsec:opt-concept}
	
	In classical game theory, the elementary behavioral unit is a
	\emph{strategy}: a complete contingent plan.  
	SBA\index{Scenario Bundle Analysis} replaces this atomistic notion with the 
	structurally richer concept of \emph{options}.  
	Options represent \emph{potential actions} that are available to an acting
	entity under given structural, relational, and attitudinal conditions.
	They incorporate communicative, coercive, normative, institutional,
	and material acts without presupposing payoffs or a fixed strategy space.
	
	From an action-theoretical perspective, options are
	\emph{parametrised instances of action types}.
	An action type defines an equivalence class of behaviors that share a
	common intentional and causal structure.  
	Options are partially instantiated members of this class,
	and concrete actions are fully instantiated members:
	
	\begin{center}
		\textit{Action type\index{Options!Action Types}}
		$\;\Rightarrow\;$
		\textit{Option\index{Options} (parameters partially fixed)}
		$\;\Rightarrow\;$
		\textit{Action (fully instantiated element).}
	\end{center}
	
	\noindent
	This layered structure allows SBA to represent \emph{potentiality}---
	what could be done---without committing to realised event\index{Events} sequences.
	Options thus serve as the branching points of the scenario tree\index{Scenario Trees}.
	
	\paragraph{Acting entities.}
	Both individual actors and coalitions can perform actions.  
	The domain of potential acting entities is therefore \(A \cup C\).
	The difference between individual and collective agency is not encoded
	by separate option types but by the \emph{preconditions} of an option:
	capability, authority, legitimacy, and feasibility determine whether
	a given entity can execute a given type of act.  
	Thus, the option framework remains unified across individuals and coalitions.

	\subsubsection{Formal representation}
	\label{subsubsec:opt-formal}
	
	Let $\mathrm{ActType}$ be the finite set of action types relevant to a
	given scenario domain.  Each action type 
	$\alpha \in \mathrm{ActType}$ is defined as a structured tuple

	\[
	\alpha = \left\langle
	\begin{matrix}
		\text{Family}, \quad \text{Category}, \quad \text{Roles}, \quad \text{Content}, \quad \text{Mode}, \\
		\text{Preconditions}, \quad \text{Consequences}, \quad \text{Reversibility}, \quad \text{TargetResponse}
	\end{matrix}
	\right\rangle .
	\]
	
	
	\noindent
	The components are:
	\begin{itemize}
		\item \textbf{Family} - one of the families 
		(Assertive/\allowbreak Informative, Directive/\allowbreak Coercive,\\
		Commissive/\allowbreak Promissive,\\
		Declarative/\allowbreak Institutional, Expressive/\allowbreak Evaluative,\\
		Mediative/\allowbreak Cooperative, Procedural/\allowbreak Deliberative). See section \ref{subsubsec:actiontypes-families-categories}.
		\item \textbf{Category} - one of the fifteen granular action categories
		(e.g.\ \emph{threaten}, \emph{negotiate}, \emph{declare}, 
		\emph{praise}, \emph{reconcile}). See section \ref{subsubsec:actiontypes-families-categories}.
		\item \textbf{Roles} - the role structure of the act, specifying
		\textit{actor}, \textit{target}, and optionally 
		\textit{audience}.
		\item \textbf{Content} $p \in P$ ---\allowbreak the propositional core (``that $p$''), 
		relevant for assertive, directive, or commissive acts.
		\item \textbf{Mode} - the performative force of the act 
		(assertive, directive, commissive, declarative, expressive).
		\item \textbf{Preconditions} - capability, authority, situational 
		feasibility, and credibility requirements for executing the act.
		\item \textbf{Consequences} - epistemic, normative, relational, and 
		material effects associated with executing the act.
		\item \textbf{Reversibility} -- a binary or graded parameter 
		$\rho(\alpha)\in\{0,1\}$ indicating whether an action of type $\alpha$ 
		can be undone without producing a new compensatory action.  
		Irreversible actions (e.g.\ attack, destroy, leak) prune future 
		branches of the scenario tree;  
		reversible actions (e.g.\ threaten, promise, deploy) allow 
		withdrawal or modification.	
		\item \textbf{TargetResponse} -- a parameter 
		$\delta_{\mathrm{resp}}(\alpha)\in\{0,1\}$ specifying whether the action type 
		inherently requires a response from a target in order to be 
		meaningful.  
		For $\delta_{\mathrm{resp}}(\alpha)=1$ (offers, threats, proposals, requests),  
		the scenario tree must include downstream decision nodes for the 
		target;  
		for $\delta_{\mathrm{resp}}(\alpha)=0$ (attacks, unilateral declarations, covert 
		intelligence acts), the action proceeds without required reply.
		
	\end{itemize}
	
	\medskip
	
	\noindent
	An \emph{option} is a partially instantiated element of an
	action type.  Each option binds the acting entity and a set of 
	contextual parameters.  Formally:
	\[
	o = \langle \alpha,\, a,\, \theta \rangle ,
	\]
	where $a \in A \cup C$ is the acting entity and  
	$\theta$ is a partial assignment to the parameter space of $\alpha$.
	\medskip
	
	\noindent
	A \emph{concrete action} is obtained once all parameters of the action 
	type are bound:
	Let $\mathrm{Inst}(\alpha)$ denote the set of fully instantiated action tokens of type $\alpha$.
	\[
	\mathrm{Action}(\alpha,\theta^{\!*}) \in \mathrm{Inst}(\alpha)
	\quad\Leftrightarrow\quad
	\theta^{\!*}\ \text{assigns values to all required parameters.}
	\]
	
	\noindent
	Thus the formal hierarchy is:
	\[
\begin{aligned}
\text{Action Type} &\Rightarrow \text{Option (partially instantiated)}\\
&\Rightarrow \text{Action (fully instantiated)}.
\end{aligned}
\]
	This structure preserves the distinction between 
	general possibilities, concrete potential moves, 
	and realized actions in scenario evolution.

	\noindent
	
	\paragraph{Note on evaluative parameters.}
	Although parameters such as preference intensity ($\pi$), temporal horizon\index{Time and Dynamics!Temporal Horizon}\index{Time and Dynamics} ($\vartheta$), 
	and perceived likelihood\index{Events!Likelihood} ($\ell$) were introduced in the context of attitudes\index{Attitudes},
	they are implicitly inherited by the option\index{Options} structure of action types.
	Each parameterized action $\alpha$
	embeds evaluative components corresponding to 
	$\pi$, $\vartheta$, and $\ell$, which together determine the feasibility,
	urgency, and motivational salience of the option\index{Options} within the scenario.

\begin{figure}[htbp]
		\centering
		\begin{tikzpicture}[
			scale=0.9,
			transform shape,
			node distance=1.8cm,
			layerbox/.style={rectangle, thick, rounded corners=3pt, minimum width=9.5cm, minimum height=1.8cm, align=left, font=\small},
			typebox/.style={layerbox, draw=blue!70!black, fill=blue!5},
			optionbox/.style={layerbox, draw=green!60!black, fill=green!5},
			actionbox/.style={layerbox, draw=orange!60!black, fill=orange!5},
			thickarrow/.style={-{Stealth[scale=1.2]}, ultra thick, gray!70},
			layerlabel/.style={font=\large\bfseries, blue!70!black},
			statuslabel/.style={font=\tiny\bfseries, text=gray!60, align=center, text width=2cm}
			]
			
			\node[typebox] (type) {
				\textbf{Action Type} $\alpha \in \mathrm{ActType}$ (Abstract Template)\\[0.1cm]
				$\alpha = \langle \text{Family, Category, Roles, Content, Mode, Precond., Cons.} \rangle$\\[0.1cm]
				\textit{Status: All parameters are unbound ($\theta = \emptyset$)}
			};
			\node[layerlabel, left=0.6cm of type] {L1};
			\node[statuslabel, right=0.4cm of type] {Ontological\\Definition};
			
			\node[optionbox, below=of type] (option) {
				\textbf{Option} $o = \langle \alpha, a, \theta \rangle$ (Strategic Choice)\\[0.1cm]
				$a = \text{Acting entity (e.g., state } \alpha\text{)}$\\
				$\theta = \langle \text{target, audience, channel, scope, ...} \rangle$\\[0.1cm]
				\textit{Status: Actor bound, context partially instantiated}
			};
			\node[layerlabel, left=0.6cm of option, green!60!black] {L2};
			\node[statuslabel, right=0.4cm of option] {Potential\\Scenario Branch};
			
			\node[actionbox, below=of option] (action) {
				\textbf{Action Token} $\mathrm{Action}(\alpha, \theta^*)$ (Executed Event)\\[0.1cm]
				$\theta^* = \theta \cup \langle \text{speaker, timestamp, concrete text, ID} \rangle$\\[0.1cm]
				\textit{Status: Fully instantiated (Observable Event)}
			};
			\node[layerlabel, left=0.6cm of action, orange!60!black] {L3};
			\node[statuslabel, right=0.4cm of action] {Realized\\Node / Path};
			
			\draw[thickarrow, blue!70!black] (type.south) -- node[right, font=\tiny, align=left, xshift=0.2cm] {
				\textbf{BINDING PROCESS I:}\\
				$\bullet$ Assign acting Actor $a$\\
				$\bullet$ Set strategic parameters $\theta$
			} (option.north);
			
			\draw[thickarrow, green!60!black] (option.south) -- node[right, font=\tiny, align=left, xshift=0.2cm] {
				\textbf{BINDING PROCESS II:}\\
				$\bullet$ Fill situational data\\
				$\bullet$ Generate unique ID / Timestamp
			} (action.north);

		\end{tikzpicture}
	
		\caption{Three-layer hierarchy of SBA action ontology. The transition from L1 to L3 represents the process of \textbf{parameter binding}, transforming abstract ontological definitions into concrete scenario events.}
		\label{fig:action_hierarchy}
	\end{figure}
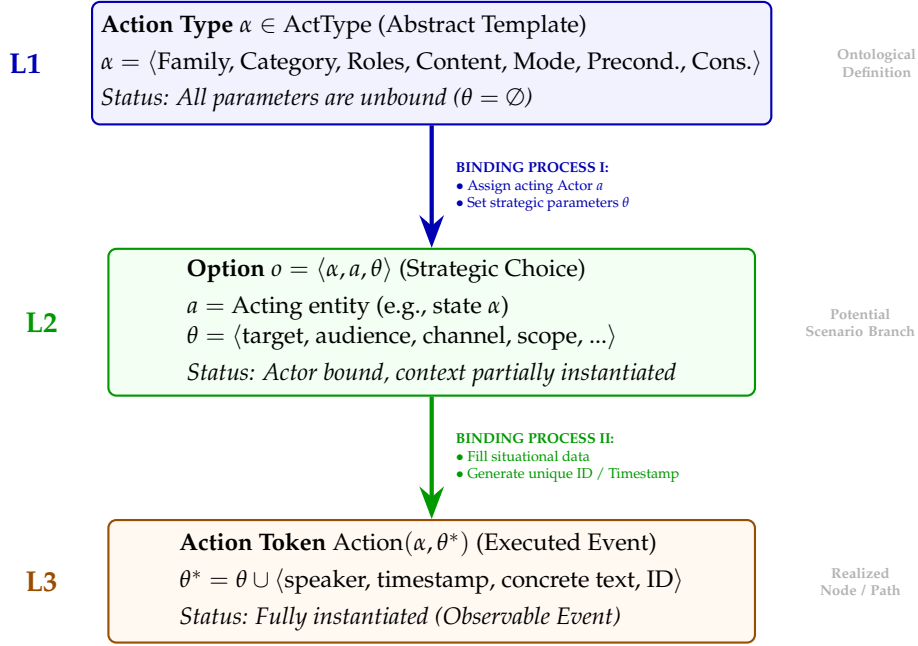

	\paragraph{Extended parameter significance.}
	The parameters \textit{Reversibility} and \textit{TargetResponse} refine 
	the generative logic of scenario construction.  
	Reversibility determines whether an option closes parts of the feasible 
	future (irreversible damage or information release) or whether it remains 
	mutable.  
	TargetResponse identifies which options necessarily induce a 
	counter--decision node in the scenario tree.  
	This distinction generalises the classical game--theoretic separation 
	between unilateral moves and interactive move pairs, but expresses it at 
	the level of action types rather than strategies.  
	Both parameters are inherited by options and instantiated actions, 
	ensuring consistent branching and pruning rules during scenario 
	generation. Figure~\ref{fig:action_hierarchy} depicts the three-layer
	hierarchy from abstract type through parameterized option to fully
	instantiated action.

	\subsubsection{Illustrative example (type \texorpdfstring{$\to$}{->} option \texorpdfstring{$\to$}{->} action)}
	\label{subsubsec:opt-example}
	
	\paragraph{Action type: Threat of sanction.}
	
	\[
	\alpha_{\text{ThreatSanction}}
	=
	\big\langle
	\begin{array}{l}
		\text{Family}=\text{Directive/\allowbreak Coercive},\\[2pt]
		\text{Category}=\text{Threatening},\\[2pt]
		\text{Roles}=\langle \text{actor},\,\text{target},\,\text{audience}\rangle,\\[2pt]
		\text{Content}=p,\\[2pt]
		\text{Mode}=\text{directive--commissive},\\[2pt]
		\text{Preconditions}=\Pi,\\[2pt]
		\text{Consequences}=\Gamma,\\[2pt]
		\text{Reversibility}=\rho(\alpha_{\text{ThreatSanction}})=1,\\[2pt]
		\text{TargetResponse}=\delta_{\mathrm{resp}}(\alpha_{\text{ThreatSanction}})=1
	\end{array}
	\big\rangle .
	\]
	
	\noindent
	The parameter $\rho=1$ indicates that the threat can be withdrawn or modified;
	$\delta_{\mathrm{resp}}=1$ indicates that the action requires an elicited response by the target.
	
	\bigskip
	
	\paragraph{Option (partially instantiated).}
	
	\[
\begin{aligned}
	o=&
	\big\langle
	\alpha_{\text{ThreatSanction}},\;
	a=\text{state } \alpha,\\
&\;
	\theta =
	\langle
	\begin{array}{l}
		\text{target}=\text{state } \beta,\\[2pt]
		\text{audience}=\text{international},\\[2pt]
		\text{channel}=\text{press conference},\\[2pt]
		\text{temporal scope}=\text{T+7},\\[2pt]
		\text{scope}=\text{trade embargo on sector S},\\[2pt]
		\rho=1,\;\delta_{\mathrm{resp}}=1
	\end{array}
	\rangle
	\big\rangle .
	\end{aligned}
\]
	
	\noindent
	The option inherits the structural parameters of the action type
	while leaving several values (speaker, timestamp, textual realisation) unspecified.
	
	\bigskip
	
	\paragraph{Concrete action (fully instantiated).}
	
	\[
	\mathrm{Action}
	\big(
	\alpha_{\text{ThreatSanction}},\,
	\theta^{\!*}
	\big),
	\qquad
	\theta^{\!*}
	=
	\theta
	\;\cup\;
	\big\langle
	\begin{array}{l}
		\text{speaker}=\text{Foreign Minister A},\\[2pt]
		\text{timestamp}=2026\text{-}02\text{-}10,\\[2pt]
		\text{text}=\text{official declaration},\\[2pt]
		\text{record ID}=\#14729
	\end{array}
	\big\rangle .
	\]
	
	\noindent
	The action type defines the structural template; the option specifies a feasible,
	partially parameterised potential move; and the concrete action instantiates all
	remaining parameters, producing a realised event in the scenario tree.

	\subsubsection{Families and Categories of Action Types}
	\label{subsubsec:actiontypes-families-categories}
	
	SBA classifies actions through a two--level semantic hierarchy derived
	from the formal representation of action types.  
	The classification separates two dimensions:
	
	\begin{itemize}[leftmargin=1.5em,itemsep=3pt]
		\item \textbf{Family} -- the high--level pragmatic orientation of an action,
		describing \emph{why} an act is performed (its illocutionary purpose).
		\item \textbf{Category} -- an empirically grounded subtype specifying
		\emph{how} an action is realised (its characteristic mode of execution).
	\end{itemize}
	
	Families provide the macro--structure of SBA action space, while
	categories refine this structure into the operational classes used for
	coding, database construction, and scenario generation.
	The two--level organization follows established distinctions in
	speech--act theory \citep{austin1962, searle1969} and the empirical
	verb taxonomy of Ballmer \& Brennenstuhl (1981), but adapts these to
	the requirements of SBA's action ontology.
	
	\paragraph{Why this hierarchy is adopted.}
	In SBA, action types need to satisfy two competing requirements.
	They must be:
	\begin{enumerate}[label=(\alph*),leftmargin=2em,itemsep=2pt]
		\item \textbf{sufficiently abstract} to function as stable equivalence
		classes across heterogeneous empirical cases, and
		\item \textbf{sufficiently specific} to support coding, inference,
		and scenario--tree branching.
	\end{enumerate}
	A single level is either too coarse (loss of empirical nuance)
	or too fine (loss of theoretical coherence).
	The family-category distinction resolves this by placing theoretical
	regularities in the family layer, and empirical differentiation in the
	category layer.
	
	\paragraph{Integrated hierarchy of families and categories.}
	Each action type belongs to exactly one family and one category.  
	The SBA action ontology distinguishes seven families of action types,
	each representing a broad illocutionary force. These families are further
	disaggregated into operational categories that link strategic intent to
	concrete scenario developments (Table~\ref{tab:action_taxonomy_comprehensive}).

	\begin{table}[htbp]
		\centering
		\caption{Systematic Taxonomy of SBA Action Types: Families and Categories}
		\label{tab:action_taxonomy_comprehensive}
		\footnotesize
		\renewcommand{\arraystretch}{1.2}
		\begin{tabularx}{\textwidth}{@{} >{\bfseries}l >{\itshape}l X @{}}
			\toprule
			Family & Category & Exemplary Action Tokens \\
			\midrule
			1. Assertive / & Epistemic-Info & assert, claim, warn, reveal, leak \\
			Informative & Cognitive-Analytic & analyze, anticipate, evaluate, plan \\
			& Intelligence & spy, surveil, monitor, investigate \\
			\midrule
			2. Directive / & Behavioral Influence & order, request, demand, advise \\
			Coercive & Threatening & threaten, coerce, deter \\
			& Escalatory & provoke, retaliate, escalate \\
			& Kinetic-Operational & attack, strike, invade, destroy \\
			& Economic-Coercive & sanction, embargo, blockade \\
			\midrule
			3. Commissive / & Commitment-Gen. & promise, pledge, assure \\
			Promissive & Offer / Exchange & offer, propose terms, guarantee support \\
			\midrule
			4. Declarative / & Status-Altering & declare, appoint, resign \\
			Institutional & Normative-Reg. & legislate, forbid, sanction, permit \\
			\midrule
			5. Expressive / & Relational-Eval. & praise, condemn, blame \\
			Evaluative & Symbolic-Comm. & justify, frame, apologize \\
			\midrule
			6. Mediative / & Negotiative & negotiate, coordinate, reconcile \\
			Cooperative & Coalition-Trans. & ally, join, defect, split \\
			\midrule
			7. Procedural / & Voting-Aggregative & vote, elect, poll, ballot \\
			Deliberative & Agenda-Setting & propose, table, adjourn, convene \\
			& Ratification & ratify, veto, approve, endorse \\
			\bottomrule
		\end{tabularx}
	\end{table}

	\medskip

	\paragraph{Role within SBA ontology.}
	In the formal schema of action types, each type~$\alpha$ contains the
	pair:
	\[
	\mathrm{Family}(\alpha)
	\qquad\text{and}\qquad
	\mathrm{Category}(\alpha),
	\]
	which jointly determine its pragmatic function and empirical subtype.
	Because options are parameterised instances of action types,
	they inherit their family and category labels from their underlying
	type.  
	This ensures that the ontology remains consistent across conceptual,
	database, and scenario-tree levels.
	
	\paragraph{Note (actors vs.\ coalitions).}
	The families and the categories apply uniformly to both
	individual actors and coalitions.  
	The acting subject \(a\) of an option or action is any
	\emph{acting entity} in \(A \cup C\);  
	differences in capability, authority, and internal coordination are
	captured by the parameter structures of the action type rather than by
	separate taxonomies.

	\subsubsection{Compositional Structure and the ``by'' Relation}
	\label{subsubsec:opt-byrelation}
	
	The family--category taxonomy (Section~\ref{subsubsec:actiontypes-families-categories})
	organizes action types at a conceptual level.
	The ``by'' relation provides a different and independent refinement:
	it captures \emph{compositional structure} among action types.
	Where families and categories group actions by pragmatic orientation,
	the ``by'' relation describes how one action type may be
	\emph{realised, implemented, or operationalised} by another.
	
	\paragraph{Conceptual purpose.}
	Following \citet{anscombe1957, vonwright1963, searle1969}, an action of type
	$\alpha_i$ is performed \emph{by} performing an action of type $\alpha_j$ if
	$\alpha_j$ constitutes the concrete means through which $\alpha_i$
	is realised:
	\[
	(\alpha_i,\alpha_j)\in\mathrm{by}
	\quad\text{iff}\quad
	\text{``one performs }\alpha_i\text{ by performing }\alpha_j\text{.''}
	\]
	
	Typical cases include:
	\begin{itemize}[leftmargin=1.6em,itemsep=1.5pt]
		\item convince \emph{by} arguing,
		\item sanction \emph{by} freezing assets,
		\item threaten \emph{by} mobilising forces,
		\item negotiate \emph{by} tabling a proposal.
	\end{itemize}
	
	Thus the ``by'' relation represents the \emph{instrumental,
		step\allowbreak-wise, or implementation\allowbreak-level} structure of actions.
	
	\paragraph{Formal properties.}
	Let $\mathcal{A}$ be the set of action types.
	The ``by'' relation is a binary relation
	\[
	\mathrm{by}\subseteq\mathcal{A}\times\mathcal{A}
	\]
	satisfying:
	\begin{itemize}[leftmargin=1.5em,itemsep=1.5pt]
		\item \textbf{Transitivity:}
		$\mathrm{by}(\alpha_1,\alpha_2)$ and $\mathrm{by}(\alpha_2,\alpha_3)$ imply
		$\mathrm{by}(\alpha_1,\alpha_3)$.
		\item \textbf{Non-symmetry:}
		in general, $\mathrm{by}(\alpha_i,\alpha_j)$ does not imply
		$\mathrm{by}(\alpha_j,\alpha_i)$.
		\item \textbf{Partial order:}
		the transitive closure of $\mathrm{by}$ induces a strict partial
		order $\prec_{\mathrm{by}}$ on $\mathcal{A}$; equivalently,
		$(\mathcal{A},\mathrm{by})$ is a directed acyclic graph,
		sometimes close to a tree but not required to be one.
	\end{itemize}
	
	\paragraph{Relation to families and categories.}
	The ``by'' relation does \emph{not} need to mirror the
	family-category taxonomy.
	It may:
	\begin{itemize}[leftmargin=1.5em,itemsep=1.5pt]
		\item cross family boundaries,
		\item connect types of the same family at different levels of
		abstraction,
		\item form deeper or denser hierarchies than the seven-family
		structure,
		\item represent implementation chains that have no analogue at the
		family/\allowbreak category level.
	\end{itemize}
	
	Thus, the ``by'' relation provides an \emph{orthogonal dimension} of
	structure: where families/categories classify actions
	\emph{taxonomically}, the ``by'' relation decomposes actions
	\emph{compositionally}.
	
	\paragraph{Optional use and computational relevance.}
	In many SBA applications, the family--category taxonomy is sufficient.
	The ``by'' relation becomes relevant when:
	\begin{itemize}[leftmargin=1.5em,itemsep=1.5pt]
		\item modeling multi-step or procedural actions,
		\item decomposing complex actions into operational components,
		\item encoding actions for algorithmic search,
		\item representing fine-grained transformations of options in
		scenario evolution.
	\end{itemize}
	
	In computational settings - for example when action types or their
	parameters are encoded as strings, bit vectors, or other structured
	representations - the ``by'' relation induces a hierarchical or
	networked neighbourhood structure.  
	This structure can be exploited by search heuristics, mutation
	operators, or evolutionary algorithms to explore the space of feasible
	actions in a controlled local manner.  
	Such applications require a more fine-grained representation than the
	family--category scheme alone can provide.
	
	\paragraph{Optional universal root type.}
	The by-relation does not in general induce a single tree; many 
	action types are incomparable and the resulting structure is a 
	directed acyclic graph.  
	If a rooted hierarchy is desired---for example in computational 
	applications that require tree-based encodings or global 
	similarity metrics---a maximally abstract action type 
	\texttt{Act} may be introduced as a formal top element such that
	$\texttt{Act} \trianglerighteq_{\mathrm{by}} \alpha$ for all 
	$\alpha \in \mathcal{A}$.  
	This construction is optional and does not affect the 
	conceptual taxonomy of families and categories; it simply 
	provides a unified root for algorithmic or hierarchical 
	representations of the action-type space.
	\subsubsection{Theoretical foundation}
	\label{subsubsec:opt-theory}
	
	\noindent
	The formal treatment of action types and their option-level
	instantiations in SBA\index{Scenario Bundle Analysis} integrates four
	complementary theoretical traditions: analytical action theory,
	speech--act theory, strategic interaction (game theory), and
	collective intentionality.  
	These traditions provide the conceptual grounding for the two-level
	taxonomy of \emph{families} and \emph{categories} introduced in
	Table~\ref{tab:action_taxonomy_comprehensive}.  
	While each tradition adopts a different perspective on what actions are
	and how they function, all converge on the view that actions are
	structured, contextualised, and parameterised entities linking internal
	attitudes to observable behavior.
	
	\paragraph{(1) Analytical action theory: types, tokens, and parameters.}
	Analytical action theory, following \citet{vonwright1963},
	\citet{davidson1963}, and \citet{anscombe1957}, treats actions as
	structured events realised under intentional and causal descriptions.
	A central distinction is drawn between \emph{action types}---abstractions
	that identify general patterns such as \emph{negotiating},
	\emph{threatening}, or \emph{promising}---and their concrete
	\emph{tokens}, which instantiate a type under full parameter
	specification.
	
	\noindent
	The revised schema for the internal structure of an action type
	(Context, Roles, Content, Mode, Preconditions, Effects) is consistent
	with this perspective: each component corresponds to an intentional or
	causal parameter whose instantiation determines when an option becomes a
	concrete action.  
	Domains are no longer treated as primitive attributes; instead, domain
	information is encoded implicitly through Context, Role assignments,
	relational structure, and feasibility preconditions.  
	This preserves expressive power without enlarging the primitive action
	signature.
	
	\paragraph{(2) Speech--act theory: pragmatic families and performative structure.}
	Classical speech--act theory, beginning with \citet{austin1962} and
	\citet{searle1969}, introduces functional classes of actions based on
	their \emph{illocutionary force}.  
	These classes---assertives, directives, commissives, expressives, and
	declaratives \citep{searle1975}---provide the conceptual origin of the seven \emph{families}
	in Table~\ref{tab:action_taxonomy_comprehensive}.  
	The categories within each family correspond to empirically grounded
	subtypes identified in the lexical taxonomy of \citet{ballmer1981}.  
	Thus SBA taxonomy is a structured operationalisation of
	speech--act-theoretic distinctions adapted to crises and conflict
	contexts.
	
	\paragraph{(3) Strategic interaction: structured moves and preconditions.}
	Game theory traditionally conceptualises behavior in terms of
	\emph{strategies}---complete plans of action.  
	SBA replaces this atomistic notion with structured \emph{options}
	tied to action types.  
	This shift reflects three insights from the theory of strategic
	interaction:
	
	\begin{enumerate}[label=(\alph*),itemsep=4pt,leftmargin=1.5em]
		\item actions are rarely atomic but consist of compositional structures
		(e.g.\ ``coerce by threatening'', ``signal by manoeuvring'');
		\item preconditions (capability, authority, feasibility) delimit the
		feasible move space independently of payoffs;
		\item actions produce epistemic, normative, and relational effects
		(belief updating, signalling credibility, altering network ties).
	\end{enumerate}
	
	\noindent
	These insights justify modeling action types as multi-parameter
	objects, treating options as feasible but partially unspecified moves,
	and viewing fully instantiated actions as realised tokens placed in the
	scenario tree.
	
	\paragraph{(4) Collective intentionality: coalitional agency and group-level actions.}
	Modern theories of collective agency
	\citep{tuomela2002, searle1995, gilbert1989, listpettit2011} distinguish
	between individual and collective intentional structures.
	Collective actors possess:
	
	\begin{itemize}[itemsep=3pt]
		\item a larger option space (jointly executable actions),
		\item internal decision procedures (role selection and parameter
		coordination),
		\item emergent capabilities (actions no single member could perform
		alone).
	\end{itemize}
	
	\noindent
	In SBA, this is captured by defining options over the domain
	\(A \cup C\) and by instantiating action types through role assignment
	and feasibility conditions.  
	Coalition-oriented categories (e.g.\ allying, defecting, coordinating)
	in Table~\ref{tab:action_taxonomy_comprehensive} correspond to these
	emergent group-level capabilities.
	Table~\ref{tab:action_theory_lineage} maps the foundational traditions to their integration into the SBA action ontology.
	
	\paragraph{Synthesis.}
	These theoretical traditions jointly justify SBA's modeling
	choices:
	
	\begin{itemize}[leftmargin=1.7em,itemsep=3pt]
		\item action types are structured templates (action theory);
		\item families and categories follow the pragmatic organization of
		illocutionary forces (speech--act theory);
		\item options encode feasible strategic moves with explicit parameter
		structure (strategic interaction);
		\item coalitional actions arise from joint attitudes and institutional
		capacities (collective intentionality).
	\end{itemize}
	
	\noindent
	The ``by''-relation, treated as an optional refinement tool, provides an
	additional structural dimension allowing complex actions to be
	decomposed or implemented through more specific subtypes; however, it is
	independent of the taxonomic family--category hierarchy.  
	Together, these elements yield an ontology that is parsimonious at the
	family level, granular at the category level, and formally precise in
	its treatment of action types, options, and instantiated actions.

	\begin{table}[h!]
		\centering
		\renewcommand{\arraystretch}{1.12}
		\small
		\setlength{\tabcolsep}{4pt}
		\begin{tabularx}{\textwidth}{
					>{\RaggedRight\arraybackslash}p{3.0cm}
					>{\RaggedRight\arraybackslash}p{3.9cm}
					>{\RaggedRight\arraybackslash}X}
				\toprule
			\textbf{Author / Tradition} &
			\textbf{Core Contribution} &
			\textbf{Role in SBA Action Ontology} \\
			\midrule
			
			\textbf{Anscombe (1957)} &
			By-relation; hierarchical realisation of actions. &
			Basis for SBA's optional hierarchical refinement of action types  
			(internal compositional structure). \\[3pt]
			
			\textbf{von Wright (1963)} &
			Logic of norms and action; deontic structure. &
			Grounds normative/regulatory action types and feasibility constraints  
			(obligation, permission, prohibition). \\[3pt]
			
			\textbf{Austin (1962)} &
			Illocutionary force; performative utterances. &
			Conceptual foundation for the \emph{families} of action types. \\[3pt]
			
			\textbf{Searle (1969, 1983)} &
			Speech-act classes; intentionality links. &
			Connects propositional attitudes (belief, desire, intention) to  
			performative structures; underpins family/\allowbreak category system. \\[3pt]
			
			\textbf{Ballmer \& Brennenstuhl (1981)} &
			Lexical taxonomy of action verbs. &
			Empirical grounding for SBA's \emph{category-level} action types. \\[3pt]
			
			\textbf{Davidson (1963, 1980)} &
			Actions as events with causal structure. &
			Defines fully instantiated actions in scenario trees; supports  
			type--token distinction. \\[3pt]
			
			\textbf{Bratman (1987)} &
			Planning theory of intention. &
			Justifies temporal coherence and role-based constraints on option  
			sequences. \\[3pt]
			
			\textbf{Schelling (1966)} &
			Threats, commitments, signalling. &
			Grounds directive/coercive families; motivates explicit preconditions  
			(capability, credibility). \\[3pt]
			
			\textbf{Lewis (1969)} &
			Conventions and coordination equilibria. &
			Supports cooperative/mediative families; links action types to  
			expectation dynamics. \\[3pt]
			
			\textbf{Tuomela (2002, 2013)} &
			Joint intentions; group agency. &
			Supports coalitional action types, emergent capabilities,  
			and options defined over \(A \cup C\). \\[3pt]
			
			\textbf{Habermas (1984)} &
			Communicative action; discourse rationality. &
			Informs integrative/mediative action types and their communicative  
			preconditions. \\
			\bottomrule
		\end{tabularx}
		\caption{Foundational traditions and their integration into SBA ontology of action types, options, and actions.}
		\label{tab:action_theory_lineage}
	\end{table}

	\subsection{Events}
	\label{subsec:events}
	
	Within SBA, \emph{events} constitute the third type of entity that 
	may occupy a node in a scenario tree, alongside actors and coalitions.
	They represent exogenous transitions that are not produced by the 
	modeled actors but that reshape attributes, relations, feasible options, 
	or the underlying system state.  
	Examples include natural hazards (earthquakes, droughts, pandemics),
	economic disruptions (currency crises, commodity shocks), and 
	geopolitical incidents (border violations, infrastructure failures).  
	Events are recorded in the initial database and determine how the scenario
	tree can branch independently of deliberate action.

	\subsubsection{Concept and Role in SBA}
	\label{subsubsec:events-concept}
	
	Events are defined as \emph{exogenous state transitions} with three 
	distinguishing properties:
	
	\begin{enumerate}[label=(\alph*), leftmargin=1.5em, itemsep=3pt]
		\item they occur independently of the choices of actors or coalitions;
		\item they introduce changes in attributes, relations, or feasible options;
		\item they generate or modify nodes in the scenario tree.
	\end{enumerate}
	
	In contrast to actions, which express intentional behavior, 
	an event $e$ is not the realisation of an option 
	but a non--agentic occurrence that alters the strategic environment.
	Thus, the event layer functions as the \emph{exogenous dynamic} of SBA, 
	supplementing the endogenous dynamics generated by actors and coalitions.
	
	Events may create:
	\begin{itemize}[itemsep=2pt]
		\item new constraints (e.g.\ infrastructure loss),
		\item new opportunities (e.g.\ ceasefire window),
		\item shifts in beliefs (e.g.\ intelligence leak, early-warning signal),
		\item structural reconfigurations (e.g.\ institutional collapse).
	\end{itemize}
	
	Their integration into scenario trees is essential for representing
	crisis dynamics in which shocks, uncertainty, and environmental pressures
	co-evolve with actor behavior.
	
	\subsubsection{Formal Representation}
	\label{subsubsec:events-formal}
	
	An event is represented as a typed tuple:
	\[
	e = \big\langle 
	\text{Name},\;
	\ell_e,\;
	\eta_e,\;
	\vartheta_e,\;
	t_e,\;
	\delta_e
	\big\rangle,
	\]
	with:
	
	\begin{itemize}[leftmargin=1.5em,itemsep=3pt]
		\item $\ell_e \in [0,1]$ - \emph{likelihood} or prior probability of occurrence,
		\item $\eta_e \in \mathbb{R}$ - \emph{impact magnitude} (direction and strength),
		\item $\vartheta_e \in \{\text{short},\text{medium},\text{long}\}$ - \emph{temporal horizon},
		\item $t_e$ - expected or realised \emph{onset time},
		\item $\delta_e$ - \emph{duration} or persistence of effects.
	\end{itemize}
	
	These parameters determine how the event enters the scenario tree, 
	how long its consequences unfold, and how it interacts with the 
	actors' epistemic attitudes or preventive actions.
	
	\paragraph{Forecast\index{Events!Forecasting}, prevention\index{Events!Prevention}, and reaction\index{Events!Reaction / Response}.}
	In empirical or policy-oriented applications,
	events\index{Events} may be subject to anticipatory modeling.
	Forecasting concerns the estimation of $\ell_e$ and $\vartheta_e$;
	prevention\index{Events!Prevention} aims to reduce either likelihood\index{Events!Likelihood} or impact magnitude;
	and reaction\index{Events!Reaction / Response} describes adaptive or compensatory actions taken once $e$ has occurred.
	The temporal parameters deliver time indices for the temporal structure of the scenario dynamics, i.e., \emph{ex ante} mitigation, \emph{in itinere} response,
	and \emph{ex post} recovery;
	Figure~\ref{fig:event_timeline} shows these three phases and their temporal parameters.

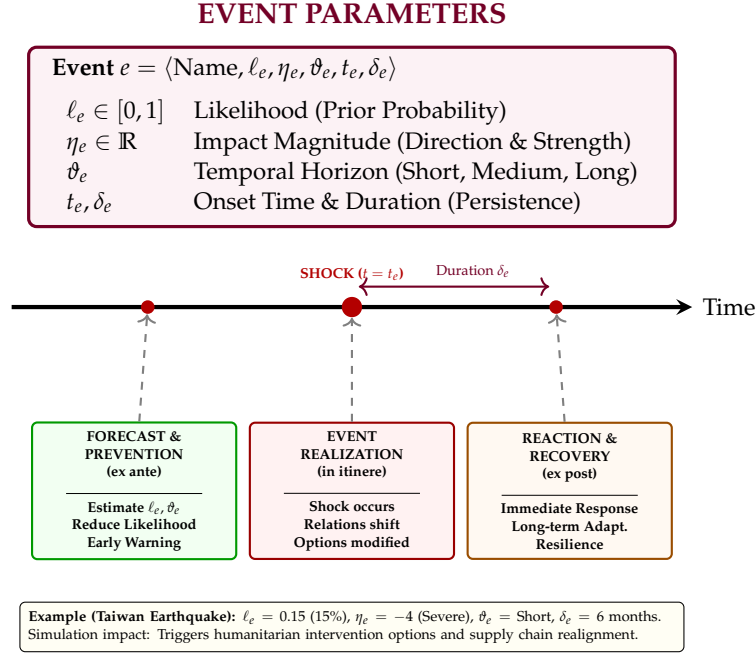
\begin{figure}[htbp]
	\centering
	\begin{tikzpicture}[
		scale=0.9,
		transform shape,
		eventbox/.style={rectangle, draw=purple!60!black, very thick, fill=purple!5, rounded corners=3pt, minimum width=9.5cm, minimum height=2.2cm, align=left, font=\small},
		phasebox/.style={rectangle, thick, rounded corners=2pt, minimum width=3cm, minimum height=2cm, align=center, font=\tiny\bfseries},
		timeline/.style={->, ultra thick, >=stealth}
		]
		
		\node[eventbox] (event) at (0, 6) {
			\textbf{Event} $e = \langle \text{Name}, \ell_e, \eta_e, \vartheta_e, t_e, \delta_e \rangle$\\[0.2cm]
			\begin{tabular}{ll}
				$\ell_e \in [0,1]$ & Likelihood (Prior Probability) \\
				$\eta_e \in \mathbb{R}$ & Impact Magnitude (Direction \& Strength) \\
				$\vartheta_e$ & Temporal Horizon (Short, Medium, Long) \\
				$t_e, \delta_e$ & Onset Time \& Duration (Persistence)
			\end{tabular}
		};
		\node[font=\large\bfseries, purple!60!black, above=0.2cm of event] {EVENT PARAMETERS};
		
		\draw[timeline] (-5, 3.5) -- (5, 3.5) node[right, font=\small] {Time};
		
		\node[circle, fill=red!70!black, inner sep=2pt] (t0) at (-3, 3.5) {};
		\node[circle, fill=red!70!black, inner sep=3pt] (te) at (0, 3.5) {};
		\node[circle, fill=red!70!black, inner sep=2pt] (t1) at (3, 3.5) {};
		
		\node[above=0.1cm of te, font=\tiny\bfseries, text=red!70!black] {SHOCK ($t = t_e$)};
		
		\draw[<->, thick, purple!60!black] (0.1, 3.8) -- node[above, pos=0.6, font=\tiny] {Duration $\delta_e$} (2.9, 3.8);
		
		\node[phasebox, fill=green!5, draw=green!60!black] (phase1) at (-3.2, 0.8) {
			\textbf{FORECAST \&}\\ \textbf{PREVENTION}\\ (ex ante)\\ \rule{2cm}{0.4pt}\\ 
			Estimate $\ell_e, \vartheta_e$\\ Reduce Likelihood\\ Early Warning
		};
		
		\node[phasebox, fill=red!5, draw=red!60!black] (phase2) at (0, 0.8) {
			\textbf{EVENT}\\ \textbf{REALIZATION}\\ (in itinere)\\ \rule{2cm}{0.4pt}\\ 
			Shock occurs\\ Relations shift\\ Options modified
		};
		
		\node[phasebox, fill=orange!5, draw=orange!60!black] (phase3) at (3.2, 0.8) {
			\textbf{REACTION \&}\\ \textbf{RECOVERY}\\ (ex post)\\ \rule{2cm}{0.4pt}\\ 
			Immediate Response\\ Long-term Adapt.\\ Resilience
		};
		
		\draw[->, thick, dashed, gray] (phase1) -- (t0);
		\draw[->, thick, dashed, gray] (phase2) -- (te);
		\draw[->, thick, dashed, gray] (phase3) -- (t1);
		
		\node[draw=black, fill=yellow!5, font=\tiny, text width=9.5cm, rounded corners=2pt] at (0, -1.2) {
			\textbf{Example (Taiwan Earthquake):} $\ell_e = 0.15$ (15\%), $\eta_e = -4$ (Severe), $\vartheta_e = \text{Short}$, $\delta_e = 6$ months. 
			Simulation impact: Triggers humanitarian intervention options and supply chain realignment.
		};
		
	\end{tikzpicture}
	\caption{Event structure and temporal phases. The model distinguishes between preventive measures (\emph{ex ante}), the realization of exogenous shocks (\emph{in itinere}), and subsequent recovery strategies (\emph{ex post}).}
	\label{fig:event_structure}
\end{figure}

	\subsubsection{Event Typology}
	\label{subsubsec:events-typology}
	
	The typology presented below serves two complementary purposes within SBA ontology. 
	First, it identifies a set of orthogonal descriptive dimensions that jointly span the 
	space of possible event types. These dimensions allow each empirical event to be mapped 
	in a structured and reproducible way to the formal tuple representation 
	\(e=\langle \text{Name},\ell_e,\eta_e,\vartheta_e,t_e,\delta_e,\dots\rangle\). 
	Second, the typology functions as an interface between heterogeneous real-world event 
	datasets and the scenario-tree structure: it provides the coding scheme through which 
	raw events are converted into attribute updates, relational changes, and branching 
	structures within SBA.

	\begin{enumerate}[label=(\arabic*), leftmargin=1.5em, itemsep=4pt]
		
		\item \textbf{Origin.}
		\begin{itemize}[leftmargin=1.5em,itemsep=2pt]
			\item \textbf{Natural}: earthquake, drought, epidemic.
			\item \textbf{Social--political}: revolution, coup, mass protest.
			\item \textbf{Technological}: cyber failure, energy blackout.
			\item \textbf{Mixed or systemic}: climate-induced migration, supply-chain shock.
		\end{itemize}
		
		\item \textbf{Scope.}  
		Local, regional, or global---indicating which actors or coalitions are affected.
		
		\item \textbf{Temporal dynamics.}
		\begin{itemize}[leftmargin=1.5em,itemsep=2pt]
			\item \textbf{Sudden shocks} (explosions, market crashes)
			\item \textbf{Slow-burning processes} (resource depletion, demographic pressure)
		\end{itemize}
		
		\item \textbf{Predictability.}  
		Unforeseen, foreseeable, or scheduled.
		
		\item \textbf{Sector of impact.}  
		Political, economic, social, technological, infrastructural, environmental.
		
	\end{enumerate}

	Each dimension plays a distinct role in the formal representation of events. 
	\emph{Origin} constrains causal interpretation and helps determine
	which effect mechanisms are plausible for the event. 
	\emph{Scope} identifies which actors or coalitions may be affected through their 
	location or domain, linking directly to the spatial footprint \(\phi_e\). 
	\emph{Temporal dynamics} determine whether the event appears as a discrete branching 
	point (sudden shock) or as a prolonged structural transformation (slow-burning 
	process), shaping both duration \(\delta_e\) and possible trajectories \(\eta_e(t)\). 
	\emph{Predictability} connects the event structure to epistemic states: 
	scheduled or foreseeable events come with more precise priors, richer 
	evidence sets \(E_e\), and in turn, more opportunities for preventive action. 
	Finally, the \emph{sector of impact} aligns the event with the relevant attribute 
	domains (political, economic, social, technological, infrastructural, environmental), 
	ensuring that the mapping from events to attribute updates is systematic.
	
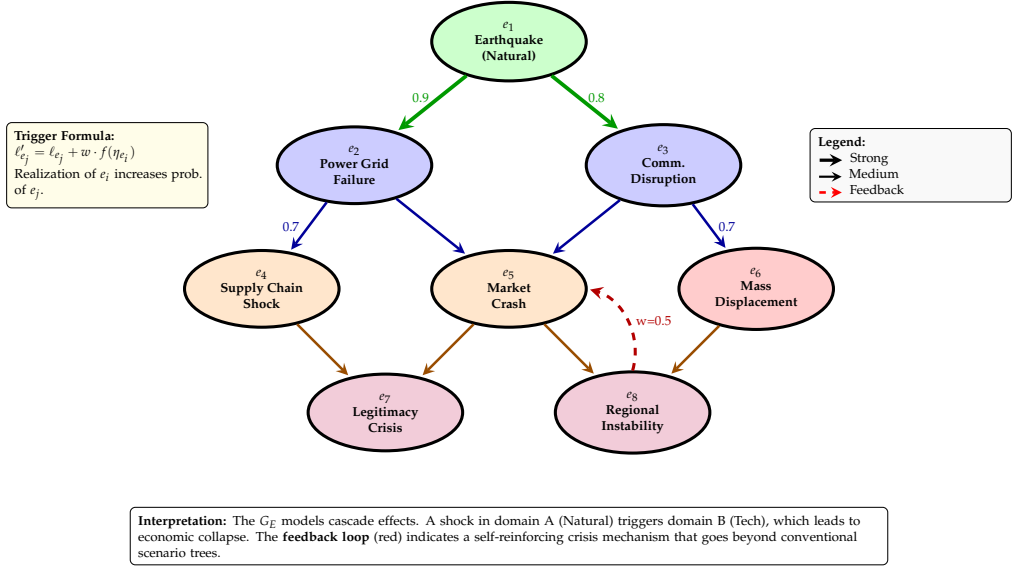
\begin{figure}[htbp]
		\centering
		\resizebox{\textwidth}{!}{%
\begin{tikzpicture}[
			scale=0.85,
			transform shape,
			eventnode/.style={ellipse, draw=black, very thick, minimum width=2.5cm, minimum height=1cm, align=center, font=\tiny\bfseries},
			natural/.style={fill=green!20},
			tech/.style={fill=blue!20},
			econ/.style={fill=orange!20},
			social/.style={fill=red!20},
			political/.style={fill=purple!20},
			primary/.style={->, line width=1.5pt, >=stealth, shorten >=1pt},
			secondary/.style={->, line width=1pt, >=stealth, shorten >=1pt},
			feedback/.style={->, line width=1.2pt, >=stealth, dashed, shorten >=1pt}
			]
			
			\node[eventnode, natural] (e1) at (0, 7) {$e_1$\\Earthquake\\(Natural)};
			
			\node[eventnode, tech] (e2) at (-2.5, 5) {$e_2$\\Power Grid\\Failure};
			\node[eventnode, tech] (e3) at (2.5, 5) {$e_3$\\Comm.\\Disruption};
			
			\node[eventnode, econ] (e4) at (-4, 3) {$e_4$\\Supply Chain\\Shock};
			\node[eventnode, econ] (e5) at (0, 3) {$e_5$\\Market\\Crash};
			\node[eventnode, social] (e6) at (4, 3) {$e_6$\\Mass\\Displacement};
			
			\node[eventnode, political] (e7) at (-2, 1) {$e_7$\\Legitimacy\\Crisis};
			\node[eventnode, political] (e8) at (2, 1) {$e_8$\\Regional\\Instability};
			
			\draw[primary, green!60!black] (e1) -- node[left, font=\tiny, pos=0.4] {0.9} (e2);
			\draw[primary, green!60!black] (e1) -- node[right, font=\tiny, pos=0.4] {0.8} (e3);
			
			\draw[secondary, blue!60!black] (e2) -- node[left, font=\tiny] {0.7} (e4);
			\draw[secondary, blue!60!black] (e2) -- (e5);
			\draw[secondary, blue!60!black] (e3) -- (e5);
			\draw[secondary, blue!60!black] (e3) -- node[right, font=\tiny] {0.7} (e6);
			
			\draw[secondary, orange!60!black] (e4) -- (e7);
			\draw[secondary, orange!60!black] (e5) -- (e7);
			\draw[secondary, orange!60!black] (e5) -- (e8);
			\draw[secondary, orange!60!black] (e6) -- (e8);
			
			\draw[feedback, red!70!black, bend right=45] (e8) to node[right, font=\tiny] {w=0.5} (e5);
			
			\node[draw, fill=yellow!10, rounded corners=2pt, text width=3cm, font=\tiny] at (-6.5, 5) {
				\textbf{Trigger Formula:}\\
				$\ell'_{e_j} = \ell_{e_j} + w \cdot f(\eta_{e_i})$\\
				Realization of $e_i$ increases prob. of $e_j$.
			};
			
			\node[draw, fill=gray!5, rounded corners=2pt, text width=3cm, font=\tiny] at (6.5, 5) {
				\textbf{Legend:}\\
				\tikz\draw[primary] (0,0)--(0.4,0); Strong\\
				\tikz\draw[secondary] (0,0)--(0.4,0); Medium\\
				\tikz\draw[feedback, red] (0,0)--(0.4,0); Feedback
			};
			
			\node[draw, fill=white, rounded corners=2pt, text width=12cm, font=\tiny, align=left] at (0, -1) {
				\textbf{Interpretation:} The $G_E$ models cascade effects. A shock in domain A (Natural) triggers domain B (Tech), which leads to economic collapse. The \textbf{feedback loop} (red) indicates a self-reinforcing crisis mechanism that goes beyond conventional scenario trees.
			};
			
		\end{tikzpicture}%
}%
		
		\caption{Event Dependency Graph $G_E$ showcasing systemic risk and cascading failures.}
		\label{fig:event_graph}
	\end{figure}

	These dimensions create a coherent classification space in which 
	empirical event data from empirical sources (Section~\ref{subsubsec:database-consistency}) can be normalised for SBA. The typology therefore underpins the consistency of scenario-tree construction: it ensures that different kinds of exogenous shocks can be formally represented, 
	compared, and integrated into scenario bundles using a shared structure.
	Figure~\ref{fig:event_graph} illustrates an event dependency graph \(G_E\) with cascading failure pathways.

	\subsubsection{Advanced event\index{Events} structure}
	\label{subsubsec:events-advanced}
	
	The elementary event\index{Events} representation
	$e=\langle \text{Name},\ell_e,\eta_e,\vartheta_e,t_e,\delta_e\rangle$
	can be enriched with additional parameters that capture
	the informational, spatial, and systemic context of exogenous changes.
	These extensions allow SBA\index{Scenario Bundle Analysis} to model how events\index{Events} are perceived,
	localized, and propagated through the actor\index{Actors} network.
	
	\paragraph{(1) Observability and confidence.}
	Not every event\index{Events} is directly observed by all actors\index{Actors}.
	Each event\index{Events} may therefore be assigned an \emph{observability} value
	$\operatorname{obs}_e\in\{\text{unobserved},\text{partial},\text{public}\}$,
	a \emph{confidence level} $c_e\in[0,1]$,
	and an optional \emph{evidence set} $E_e$
	recording the information sources or signals supporting belief\index{Attitudes!Belief} in the event\index{Events}:
	\[
	e = \langle \dots, \operatorname{obs}_e, c_e, E_e \rangle.
	\]
	These parameters enable SBA\index{Scenario Bundle Analysis} to represent asymmetric information in crises where rumor, secrecy,
	or misinformation alter actors\index{Actors}' attitudes\index{Attitudes} without a corresponding event\index{Events} having occurred.
	
	\paragraph{(2) Spatial scope.}
	Each event\index{Events} possesses a spatial or domain-specific
	\emph{footprint} $\phi_e\subseteq\text{Loc}$,
	indicating the region, sector, or network layer affected.
	This allows mapping of impacts to the relevant actors\index{Actors} or coalitions\index{Coalitions}
	through a spatial or topological distance relation\index{Relations}:
	\[
	\text{affects}(a,e)\;\Leftrightarrow\; \text{loc}(a)\in\phi_e.
	\]
	
	\paragraph{(3) Effect mapping.}
	Each event\index{Events} carries an \emph{effect map}
	\[
	\Delta_e : (A\cup C)\times\mathrm{Attr}\to E_{\mathrm{Attr}},
	\]
	specifying an attribute-level effect (a typed perturbation) rather than an absolute post-event value.
	For each attribute type \(k\in\mathrm{Attr}\), a declared update rule
	\[
	\mathrm{Upd}_k : D_k\times E_k\to D_k
	\]
	maps the prior value and the event effect to a valid post-event value:
	\[
	\mathrm{val}_{t+1}(x,k)=\mathrm{Upd}_k(\mathrm{val}_t(x,k),\Delta_e(x,k)).
	\]
	For example, if $e$ is an energy-price surge,
	$\Delta_e(a,\text{economic\_strength})=-1$
	for importers but positive for exporters.
	This function provides the quantitative link
	between exogenous shocks and endogenous attribute\index{Attributes} updates.
	
	\paragraph{(4) Event\index{Events} dependencies and cascading triggers.}
	Events\index{Events} may trigger\index{Events!Triggers} one another.
	Let $G_E=(E,\mathcal{R}_E)$ be a directed dependency graph,
	where $(e_i,e_j)\in\mathcal{R}_E$ indicates that
	the realization of $e_i$ increases the likelihood\index{Events!Likelihood} or intensity of $e_j$:
	\[
	\text{Trigger}(e_i,e_j)
	\;\Rightarrow\;
	\ell_{e_j}' = \ell_{e_j} + f(\eta_{e_i},t_{e_i}).
	\]
	This graph representation describes cascading
	and compound crises such as economic shocks induced by natural disasters.

	\paragraph{Additional dimensions.}
	Further refinements may be introduced where data permit:
	a \emph{controllability coefficient} ($\kappa_e$) expressing preventive potential;
	a \emph{reversibility index} ($\rho_e$) indicating recoverability;
	a \emph{lead time} ($\lambda_e$) for early-warning delay;
	and an \emph{impact trajectory} $\eta_e(t)$ describing evolving intensity.
	These parameters extend SBA\index{Scenario Bundle Analysis} toward dynamic risk modeling
	and integration with resilience and early-warning systems.

	\subsubsection{Theoretical foundation}
	\label{subsubsec:events-theory}
	
	The conceptual treatment of events\index{Events} within SBA\index{Scenario Bundle Analysis}
	refers to modal and temporal logic,
	decision and risk theory, and the philosophy of action.
	Events\index{Events} are formal representations of \emph{state transitions}
	whose likelihood\index{Events!Likelihood}, systemic influence, and timing affect the scenario.

	\paragraph{Forecast\index{Events!Forecasting}.}
	Forecasting potential events\index{Events} depends on epistemic\index{Modal Logics!Doxastic / Epistemic} risk perception.
	For SBA, this means estimating likelihoods (\(\ell_e\)) and temporal horizons (\(\vartheta_e\)). 
	In classical decision or game theory, this corresponds to the evaluation of
	\emph{Nature's moves} under uncertainty
	\citep{osborne1994}.
	Forecasting models in SBA\index{Scenario Bundle Analysis} can be based on
	probabilistic inference, scenario simulation, or early-warning indicators
	\citep{helbing2013}.

	\paragraph{Prevention\index{Events!Prevention}.}
	Preventive measures can change event likelihood\index{Events!Likelihood} or
	impact. These are strategic interventions that reduce vulnerability \citep{holling1973}.
	Examples of preventive options\index{Events!Prevention} chosen by actors\index{Actors} or coalitions\index{Coalitions} include infrastructure investment and pollution mitigation to lower the risk of systemic shocks.
	Such actions transform the event\index{Events} layer indirectly
	by modifying the parameters \(\ell_e\) and \(\eta_e\).
	
	\paragraph{Reaction\index{Events!Reaction / Response}.}
	Reactions are immediate or long-term responses after an event\index{Events} has materialized. 
	They instantiate endogenous adaptation after exogenous shocks
	in complex systems models \citep{arthur1999}.
	
	\paragraph{Modal and temporal semantics.}
	Formally, events\index{Events} represent transitions between
	scenario states. Therefore, each event\index{Events} $e$ defines a modal transition operator on the scenario state space:
	\[
	\square_e p \;\text{(necessarily after $e$)}, \qquad
	\Diamond_e p \;\text{(possibly after $e$)}.
	\]
	These operators correspond to the branching structure of
	the scenario tree\index{Scenario Trees}: each realized event\index{Events} generates successor states,
	while unrealized events\index{Events} define alternative branches.
	This temporal--modal interpretation connects the event\index{Events} layer
	to dynamic logic and possible--world semantics
	\citep{prior1957, rescher1971, vanbenthem1996}.
	
	\paragraph{Risk.}
	The representation of events\index{Events} refers to
	risk analysis, crisis forecasting, and resilience studies
	(e.g.\ \citealp{tverskykahneman1992, helbing2013, beck1992}).
	Event\index{Events} likelihoods (\(\ell_e\)), magnitudes (\(\eta_e\)),
	and lead times (\(\lambda_e\))
	can be determined by expert elicitation\index{Data!Expert Elicitation}\index{Data}, time--series models, or early--warning indicators.
	This view follows classical analyses
	of events\index{Events} as change--bearing particulars
	\citep{davidson1967, vonwright1963},
	extended by modern treatments in temporal and dynamic logic
	\citep{prior1957, rescher1971, vanbenthem1996}.

	\subsubsection{Database Consistency and Epistemic Coherence\index{Database Layer}\index{Logical Foundations!Consistency}\index{Modal Logics!Doxastic / Epistemic}}
	\label{subsubsec:database-consistency}
	
	The SBA\index{Scenario Bundle Analysis} database\index{Database Layer} integrates heterogeneous information about actors\index{Actors}, coalitions\index{Coalitions},
	attributes\index{Attributes}, attitudes\index{Attitudes}, relations\index{Relations}, options\index{Options}, and events\index{Events}.
	To maintain logical consistency\index{Logical Foundations!Coherence} and empirical coherence across these heterogeneous structures,
	three meta-level aspects must be addressed:
	\emph{temporal consistency}, \emph{epistemic consistency},
	and \emph{data-source coherence}.
	
	\paragraph{(1) Consistency\index{Logical Foundations!Consistency} in the temporal structure of the database.\index{Database structure}}
	All value assignments, relational links, and event\index{Events} parameters
	refer implicitly to a specific temporal layer $t$.
	Formally, let
	\[
	\mathrm{val}_t : (A\cup C)\times\mathrm{Attr}\to D_{\mathrm{Attr}}
	\]
	be the valuation function\index{Attributes!Valuation Function} for attributes\index{Attributes} at time $t$, and
	\[
	R_t \subseteq (A\cup C)^2
	\]
	the set of relations\index{Relations} instantiated at the same time.
	A minimal coherence\index{Logical Foundations!Coherence} condition requires that every element in the database\index{Database Layer}
	is indexed by a consistent temporal reference.
	This condition ensures that attribute\index{Attributes} and relation\index{Relations} data correspond to the same system snapshot and prevents contradictions between
	pre-- and post--event\index{Events} states. In the original SBA this requirement was tied mainly to the initial scenario state; here it is extended to every relevant assessment stage.
	Temporal synchronization also applies to attitudes\index{Attitudes} and perceptions,
	so that the cognitive and structural layers remain aligned when
	new events\index{Events} modify the database\index{Database Layer}.
	
	\paragraph{(2) Consistency of the epistemic\index{Modal Logics!Doxastic / Epistemic} structure of the database\index{Database structure}.}
	SBA\index{Scenario Bundle Analysis} explicitly distinguishes between \emph{objective} and \emph{subjective}
	information states.
	While the objective layer records empirically validated facts
	(e.g.\ observed alliances, measured economic indicators),
	the subjective layer represents what each actor\index{Actors} believes or knows
	about those facts.
	The epistemic\index{Modal Logics!Doxastic / Epistemic} parameters introduced throughout the database\index{Database Layer}, such as
	\[
	B_a p,\ K_a p,\ \operatorname{obs}_e,\ c_e,\ \operatorname{vis}_R
	\]
	constitute the epistemic\index{Modal Logics!Doxastic / Epistemic} structure.
	They allow modeling incomplete or asymmetric information,
	rumors, secrecy, or misperception, all of which are crucial
	for realistic crisis and conflict scenarios.
	Formally, this defines a mapping
	\[
	\mathrm{Perceive}_a : \mathrm{DB}_{\text{obj}} \to \mathrm{DB}_{\text{subj}}(a),
	\]
	which generates the actor\index{Actors}-specific perception of the global database\index{Database Layer}.
	
	\paragraph{(3) Coherence of the data source.}
	The construction of the SBA\index{Scenario Bundle Analysis} database\index{Database Layer} 
	requires integrating heterogeneous empirical inputs. 
	To ensure coherence across domains and maintain a consistent mapping into 
	attributes\index{Attributes}, relations\index{Relations}, and event structures\index{Events}, 
	the following classes of data sources and empirical methods can be combined:
	
	\begin{enumerate}[label=(\alph*), leftmargin=1.5em, itemsep=6pt]
		
		\item \textbf{Expert elicitation and Delphi surveys.}  
		Structured assessments of actor\index{Actors} attributes\index{Attributes}, 
		attitudes\index{Attitudes}, and relations\index{Relations}.  
		Suitable for strategic foresight, conflict mapping, and low-data environments.
		
		\item \textbf{Quantitative indicators.}  
		Time--series and cross-sectional data (economic, demographic, environmental, 
		financial, governance) mapped to attributes or event likelihoods.  
		Examples include the World Bank WDI \citep{wdi2024}, IMF IFS \citep{imfifs2024}, 
		UN DESA population data \citep{undesa2022}, and FAOSTAT \citep{faostat2024}.
		
		\item \textbf{Textual and discourse data.}  
		Corpus analysis, media monitoring, sentiment analysis, and speech-act coding  
		to infer attitudes\index{Attitudes}, intentions, and communicative options\index{Options}.  
		Possible sources include the GDELT Global Knowledge Graph \citep{gdelt2023}, 
		LexisNexis, parliamentary transcripts, and diplomatic statements.
		
		\item \textbf{Network and institutional datasets.}  
		Empirical foundations for relational structures\index{Relations} such as alliances, 
		trade flows, migration corridors, financial exposures, or policy coordination networks.  
		Examples include CEPII trade and migration datasets \citep{cepii2024}, 
		ICOW \citep{icow2024}, the Correlates of War alliance dataset \citep{cowalliance2024}, 
		SWIFT aggregates, and international treaty databases.
		
		\item \textbf{Event databases.}  
		Time-stamped observational records of crises, conflicts, and disasters, 
		essential for calibrating event\index{Events} likelihoods, 
		temporal patterns, and cascading dependencies.  
		Relevant repositories include GDELT \citep{gdelt2023}, ACLED \citep{acled2024}, 
		EM-DAT \citep{emdat2024}, ICEWS \citep{icews2024}, UCDP/\allowbreak PRIO \citep{ucdp2024}, 
		DesInventar \citep{desinventar2024}, and NOAA/\allowbreak ECMWF climate reanalysis datasets \citep{noaa2024, ecmwf2024}.
		
	\end{enumerate}
	
	This structured integration ensures that scenario bundles are grounded in 
	empirically coherent inputs across domains and data types.

	In practice, these sources should be triangulated.
	Expert assessments initialize qualitative parameters;
	quantitative indicators supply priors for likelihoods and impacts;
	and historical event\index{Events} data validate causal or temporal assumptions.
	The resulting database\index{Database Layer} integrates empirical evidence with explicit formal structure,
	supporting both computational processing and interpretive analysis.

	\section{Scenario Trees\index{Scenario Trees}}
	\label{sec:scenariotrees}

	Scenario trees\index{Scenario Trees} represent causally and temporally ordered
	developments of crises and conflicts as rooted branching structures generated from an
	initialized SBA database\index{Database Layer}.
	A root is selected from salient actors\index{Actors}, coalitions\index{Coalitions}, or
	initiating events\index{Events}; successive branching is generated by feasible
	options\index{Options} and admissible event realizations until terminal positions are
	reached.
	Terminal positions are evaluated by strict preference orders (per acting entity), and two
	complementary analyses are defined: a preference-based decision
	policy\index{Tree Analysis!MRP (Most Rational Path)}\index{Tree Analysis} (MRP) derived by backward induction
	on decision positions, and a likelihood-maximizing
	path\index{Tree Analysis!MLP (Most Likely Path)} (MLP) derived from conditional edge
	likelihoods.

	\paragraph{Relation to the original SBA.}
	Scenario trees in the original SBA represent fragments of extensive-form games by
	preserving sequential choice structure and backward-induction reasoning while omitting
	information sets, fixed player roles, and complete payoff functions.
	Their primary use is structural and epistemic: they trace possible evolutions under
	uncertainty, limited knowledge\index{Attitudes!Knowledge}, and dynamically changing actor
	constellations, and they admit events\index{Events} and coalitions\index{Coalitions} as
	nodes, ordinal evaluations, and context-dependent termination rules.
	The original implementation treated the database as static during tree construction; the
	initial database served as the informational basis for subsequent steps.

	A transition in a scenario tree may update any component of the SBA
	database\index{Database Layer}.
	A stage index~\(t\) is used for assessment states:
	\[
		DB_t = \langle A_t,\, C_t,\, \mathrm{Attr}_t,\, \mathrm{Att}_t,\,
		\mathrm{Rel}_t,\, \mathrm{Opt}_t,\, E_t \rangle,
	\]
	Here \(A_t\) is the actor set, \(C_t\) the coalition family, \(\mathrm{Attr}_t\) the attribute layer, \(\mathrm{Att}_t\) the attitude layer, \(\mathrm{Rel}_t\) the relation layer, \(\mathrm{Opt}_t\) the option set, and \(E_t\) the event set at stage \(t\). The index \(t\) denotes a temporal or logical stage and is not related to the symbol \(T\) for
	trees.
 
	Let \(A, C, E, \mathrm{Opt}\) be global universes, and let \(A_t\subseteq A\), \(C_t\subseteq C\), \(E_t\subseteq E\), \(\mathrm{Opt}_t\subseteq \mathrm{Opt}\) be the active stage slices recorded in \(DB_t\). The index \(t\) always refers to database stages; within-tree depth is indexed separately (e.g.\ \(h\)).
	Tree edges encode state transitions induced by executing an option or by an event
	realization.
	The update operator is
	\[
		\mathcal{U}_t : DB_t \to DB_{t+1}.
	\]

	Write \(\mathcal{U}\coloneqq(\mathcal{U}_t)_{t\ge 0}\) for the induced update family.
	\subsection{Structure of Scenario Trees\index{Scenario Trees}}
	\label{subsec:tree-structure}
	
	\paragraph{Index convention.}
	The index \(t\) refers to the assessment stage of the database \(DB_t\). Within a fixed tree generated from \(DB_t\), depth levels are indexed by \(h=0,1,\dots\) and are independent of \(t\).
	
	Each vertex \(v\in V\) is a \emph{position} in the tree. Positions are annotated by a label indicating the active entity (decision maker or exogenous event):
	\[
	\mathrm{lab}:V \to A \cup C \cup E.
	\]
	A position \(v\) is a \emph{decision position} iff \(\mathrm{lab}(v)\in A\cup C\), and an \emph{event position} iff \(\mathrm{lab}(v)\in E\). 
	Directed edges represent feasible transitions induced by executing options or by realizations of events. The edge set is \(\Gamma\subseteq V\times V\). For an edge \(e=(v,v')\in\Gamma\) we write \(\mathrm{tail}(e)\coloneqq v\) and \(\mathrm{head}(e)\coloneqq v'\). The successor set of \(v\) is
	\[
	\mathrm{succ}(v)\coloneqq\{\,e\in\Gamma:\,\mathrm{tail}(e)=v\,\}.
	\]
	Let \(L\subseteq V\) denote the set of leaf positions (terminal positions), defined by \(L\coloneqq\{v\in V:\mathrm{succ}(v)=\varnothing\}\).
	\subsubsection{Roots and initialization}
	\label{subsubsec:tree-roots}

	Scenario-tree construction begins with an initialized assessment state \(DB_0\) and a choice of a root label \(x_0\in A\cup C\cup E\). The root is represented by a distinguished position \(\rho\in V\) with \(\mathrm{lab}(\rho)=x_0\). The choice of \(x_0\) is an external modeling decision: a protocol fixes \(x_0\) given the modeling aim and the available evidence base.

	\noindent
	\textbf{Optional general selection rule.}
	If an explicit rule is preferred, define a scoring functional on labels,
	\[
	\mathrm{RootScore}: A\cup C\cup E \to \mathbb{R},
	\]
	computed from \(DB_0\) (e.g.\ weighted impact, urgency, likelihood, and network centrality features). A rule-based root label is then
	\[
	x_0 \in \arg\max_{x\in A\cup C\cup E}\ \mathrm{RootScore}(x).
	\]
	The corresponding root position \(\rho\) is then taken to be a vertex with \(\mathrm{lab}(\rho)=x_0\). If multiple positions share the same label, ties are resolved by a fixed convention (lexicographic ID order, salience priority, or analyst-declared precedence).

	Exogenous events\index{Events} (e.g.\ earthquakes, financial crashes, technological failures) frequently serve as roots because they encode an initiating shock and define the first transition in the database\index{Database Layer}. Events are not restricted to the root level: additional events may occur at deeper positions whenever their temporal horizon, likelihood, or triggering conditions (as encoded in the event-dependency structure) are satisfied. Actor\index{Actors} or coalition\index{Coalitions} roots are admissible when the scenario is driven by an endogenous initiative; in that case the initial branching reflects the execution or non-execution of salient options\index{Options} available at \(DB_0\).

	The first assessment state is fixed as
	\[
	DB_0=\langle A_0, C_0, \mathrm{Attr}_0, \mathrm{Att}_0, \mathrm{Rel}_0, \mathrm{Opt}_0, E_0\rangle,
	\]
	from which the tree is expanded by successor rules for decision positions and event positions.


\subsubsection{Edges and branching}
	\label{subsubsec:tree-edges}
	
	Outgoing edges from decision positions represent feasible option transitions, while outgoing edges from event positions represent admissible event realizations. This separation distinguishes endogenous choice from exogenous occurrence in the branching semantics.
	
	\paragraph{Auxiliary predicates.}
	Three predicates are used in the branching rules.
	\(\mathrm{Avail}(x,DB_t)\subseteq\mathrm{Opt}_t\) denotes the set of options available to entity \(x\) at assessment state \(DB_t\), determined by capability, authority, and precondition checks recorded in \(\mathrm{Opt}_t\). In binary encodings, non-execution is treated as a designated option label \(\neg o\); whenever an option \(o\) is modeled as binary at \(DB_t\), \(\mathrm{Opt}_t\) contains both \(o\) and \(\neg o\), and \(\mathrm{Avail}(x,DB_t)\) returns both whenever \(o\) is available.
	\(E^{\circ}\) denotes the finite set of event-realization labels induced by \(E_t\): for each event \(e\in E_t\) with multivalued branching, \(E^{\circ}\) contains the distinct realizations \(e^{(1)},\dots,e^{(m)}\) together with the binary outcomes \(\mathrm{occurs}(e)\) and \(\neg\mathrm{occurs}(e)\).
	\(\mathrm{valid}(e;DB_t)\) is the admissibility predicate: it holds iff the transition along edge \(e\in\Gamma\) is consistent with the constraints encoded in \(DB_t\) (option preconditions, event likelihood thresholds, and relational feasibility).

	\paragraph{Decision positions.}
	Let \(v\in V\) be a decision position with acting entity \(x=\mathrm{lab}(v)\in A\cup C\). Let \(\Sigma:\Gamma\to \mathrm{Opt}_t\cup E^\circ\) be an edge-label map. The successor set \(\mathrm{succ}(v)\) is induced by available options at the current assessment state:
	\[
	\Sigma(e)\in \mathrm{Avail}(x,DB_t)
	\quad\text{for each } e\in \mathrm{succ}(v),
	\]
	where \(\Sigma\) labels decision edges by options\index{Options} and feasibility may depend on capabilities, authority, relational constraints, and current attitudes encoded in \(DB_t\). For analytical tractability, a binary branching convention is often used: for a selected option \(o\), the two branches are
	\[
	o,\quad \neg o,
	\]
	representing execution vs.\ non-execution of the salient option at that position.
	
	\paragraph{Event positions.}
	Let \(v\in V\) be an event position with \(\mathrm{lab}(v)=e\in E\). The successor set \(\mathrm{succ}(v)\) represents admissible realizations of \(e\) at the current stage. A binary convention uses
	\[
	\mathrm{occurs}(e),\quad \neg\mathrm{occurs}(e),
	\]
	while multivalued branching refines the event into distinct realizations
	\[
	\{e^{(1)}, e^{(2)}, \dots, e^{(m)}\},
	\]
	where each \(e^{(i)}\) represents a different intensity, onset time, or partial materialization, consistent with parameters such as \(\ell_e\), \(\eta_e\), \(\vartheta_e\), and \(\delta_e\).
	
	\paragraph{Selection criteria.}
	Branch explosion is controlled by selecting only consequential or informative transitions. For options\index{Options}, fix an option-level score by assigning a preference-intensity parameter \(\pi_o\) (a scalar parameter, not a policy), a likelihood of success \(\ell_o\), a temporal horizon\index{Time and Dynamics!Temporal Horizon} \(\vartheta_o\), and an analyst-declared scoring function \(f\).
	\[
	\mathrm{salience}(o)=f\bigl(\pi_o,\ell_o,\vartheta_o,\mathrm{Rel}_t,\mathrm{Attr}_t\bigr),
	\]
	For events\index{Events}, selection is guided by realization likelihoods \(\{\ell(e^{(1)}),\dots,\ell(e^{(m)})\}\), temporal horizon \(\vartheta_e\), impact parameters \(\eta_e\), and causal relevance within the event-trigger structure.
	
	\subsubsection{Progression and termination}
	\label{subsubsec:tree-progression}
	
	Tree expansion proceeds recursively by selecting the next active position according to reactive salience and event-trigger conditions. A reactive-salience rule selects the next acting entity as the one most affected by the preceding transition:
	\[
	\text{next}(v_i)=\arg\max_{x\in A\cup C}\ \mathrm{impact}(x,v_i),
	\]
	where \(\mathrm{impact}\) quantifies the induced change in attributes\index{Attributes}, relations\index{Relations}, or attitudes\index{Attitudes}. Event progression is governed by temporal parameters and the event-dependency structure: an event position becomes eligible when its trigger conditions are satisfied and its realization likelihood exceeds an inclusion threshold.
	
	Termination occurs when further branching would be either analytically uninformative or epistemically underdetermined. A branch ends if no actor\index{Actors} or coalition\index{Coalitions} is significantly affected by the preceding transition and no event\index{Events} has sufficiently high likelihood or satisfied triggers to justify further branching, or if any extension requires speculative assumptions beyond the available information base. Terminal positions represent completed scenarios that can subsequently be evaluated and compared.
	
	\subsection{Analysis of a Single Scenario Tree\index{Scenario Trees}}
	\label{subsec:tree-analysis}
	
	Single-tree analysis separates normative choice structure from empirical plausibility by combining a preference-based decision policy (MRP) with a likelihood-based path selection (MLP). The MRP is defined on decision positions by a backward-induction recursion, while the MLP is defined as the terminal path maximizing the product of conditional edge likelihoods.
	
	\subsubsection{Backward induction\index{Tree Analysis!MRP (Most Rational Path)}
		and the Most Rational Path\index{Tree Analysis!MRP (Most Rational Path)} (MRP)}
	\label{subsubsec:tree-mpr}
	
	The MRP is a decision policy on decision positions; event positions require an additional event-realization selector to induce a single realized trajectory. Let
	\[
	V_D\coloneqq\{v\in V:\mathrm{lab}(v)\in A\cup C\},
	\qquad
	V_E\coloneqq\{v\in V:\mathrm{lab}(v)\in E\}.
	\]
	A policy is a mapping
	\[
	\mu^{\mathrm{MRP}}: V_D\to \Gamma,
	\qquad
	\mu^{\mathrm{MRP}}(v)\in \mathrm{succ}(v),
	\]
	selecting one outgoing edge at each decision position.
	
	\paragraph{Decision positions under strict preferences.}
	If \(v\in V_D\) and terminal positions carry strict preference orderings \(\succ_{\mathrm{lab}(v)}\), let \(\mathrm{Outcome}(v,e)\) denote the terminal outcome induced by choosing \(e\in\mathrm{succ}(v)\) and subsequently applying \(\mu^{\mathrm{MRP}}\) at downstream decision positions. The backward-induction step selects a maximizing continuation:
	\[
	\mu^{\mathrm{MRP}}(v)\in
	\Bigl\{ e\in\mathrm{succ}(v)\,:\,
	\begin{aligned}[t]
	&\forall e'\in\mathrm{succ}(v),\\
	&\neg\bigl(\mathrm{Outcome}(v,e')\succ_{\mathrm{lab}(v)}\mathrm{Outcome}(v,e)\bigr)
	\end{aligned}
	\Bigr\}.
	\]
	If the argmax set is non-singleton, ties are resolved by a fixed convention (lexicographic option ID, salience priority, or set-valued output).
	
	\paragraph{Decision rules under epistemic uncertainty.}
	If the acting entity at \(v\) does not know relevant propositions but only maintains a belief state encoded in \(DB_t\), the backward-induction step requires an explicit decision rule.
	Let \(\mathcal{W}(DB_t,\mathrm{lab}(v))\) denote the set of epistemically possible world-states consistent with \(\mathrm{lab}(v)\)'s information at stage \(t\).
	Let \(u_{\mathrm{lab}(v)}\) be a numerical representation of the strict terminal ordering over terminal outcomes, unique up to strictly increasing transformations.
	For \(w\in\mathcal{W}(DB_t,\mathrm{lab}(v))\), let \(\mathrm{Outcome}_w(v,e)\) denote the terminal outcome induced by choosing \(e\) when the true world-state is \(w\).
	Three standard rules are admissible choices for \(\mu^{\mathrm{MRP}}\) and determine how belief states guide choice:
	
	\begin{itemize}[leftmargin=1.5em,itemsep=3pt]
		\item \textbf{Subjective expected utility (SEU).}\index{Decision Theory!Expected Utility} A subjective probability \(\Pr_{\mathrm{lab}(v)}(\cdot\mid DB_t)\) is assumed over \(\mathcal{W}(DB_t,\mathrm{lab}(v))\), and choices maximize expected utility \citep{savage1954}:
		\[
		\mu^{\mathrm{MRP}}(v)\in
		\arg\max_{e\in\mathrm{succ}(v)}\ 
		\mathbb{E}_{\Pr_{\mathrm{lab}(v)}}\!\bigl[u_{\mathrm{lab}(v)}(\mathrm{Outcome}(v,e))\mid DB_t\bigr].
		\]
		
		\item \textbf{Maximin (Wald).}\index{Decision Theory!Maxmin} Choices maximize the worst-case utility over epistemically possible states \citep{wald1950}:
		\[
		\mu^{\mathrm{MRP}}(v)\in
		\arg\max_{e\in\mathrm{succ}(v)}\
		\min_{w\in\mathcal{W}(DB_t,\mathrm{lab}(v))}\ u_{\mathrm{lab}(v)}\!\bigl(\mathrm{Outcome}_w(v,e)\bigr).
		\]
		
		\item \textbf{Maxmin expected utility (ambiguity aversion).} A set of priors \(\Pi_{\mathrm{lab}(v)}(DB_t)\) is assumed, and choices maximize the minimum expected utility across priors \citep{gilboa1989}:
		\[
		\mu^{\mathrm{MRP}}(v)\in
		\arg\max_{e\in\mathrm{succ}(v)}\
		\min_{\Pr\in\Pi_{\mathrm{lab}(v)}(DB_t)}
		\mathbb{E}_{\Pr}\!\bigl[u_{\mathrm{lab}(v)}(\mathrm{Outcome}(v,e))\bigr].
		\]
	\end{itemize}
	
	These rules convert epistemic assessment states into choice recommendations and preserve well-defined backward induction on decision positions once a rule is fixed.

	\paragraph{Worked example: Backward induction on a four-actor tree.}
	\label{par:bi-example}
	The following example applies backward induction under strict preferences to a concrete
	scenario tree with four actors and five terminal outcomes.
	It makes the abstract recursion above fully explicit.

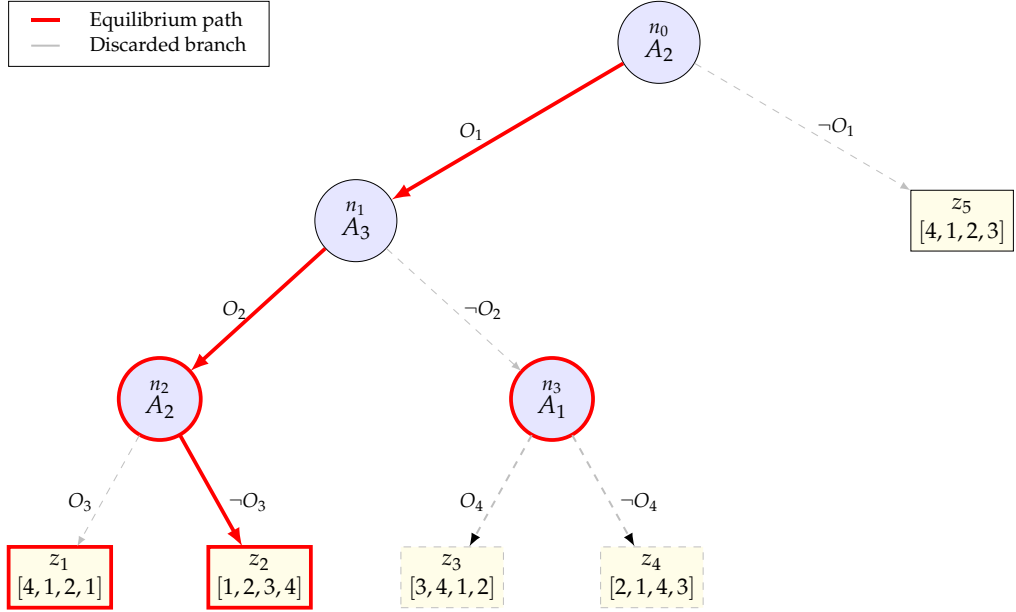
\begin{figure}[htbp]
	\centering
	\resizebox{\textwidth}{!}{%
\begin{tikzpicture}[
		level distance=2.5cm,
		level 1/.style={sibling distance=8.5cm},
		level 2/.style={sibling distance=5.5cm},
		level 3/.style={sibling distance=2.8cm},
		decision/.style={circle, draw, fill=blue!10, inner sep=2pt,
			minimum size=1.15cm, font=\small\bfseries, align=center},
		terminal/.style={rectangle, draw, fill=yellow!10, inner sep=3pt,
			align=center, font=\footnotesize},
		edge from parent/.style={draw, thick, -latex},
		eq/.style={draw=red, line width=1.5pt},
		discard/.style={draw=gray!50, dashed, thin},
		elabel/.style={font=\scriptsize, midway}
		]

		\node[decision] (root) {\shortstack{\scriptsize $n_0$\\[-2pt]$A_2$}}
		child { node[decision] (n1) {\shortstack{\scriptsize $n_1$\\[-2pt]$A_3$}}
			child { node[decision] (n2) {\shortstack{\scriptsize $n_2$\\[-2pt]$A_2$}}
				child { node[terminal] {$z_1$ \\ $[4,1,2,1]$}
					edge from parent[discard] node[elabel, left, pos=0.6] {$O_3$}
				}
				child { node[terminal] (t2) {$z_2$ \\ $[1,2,3,4]$}
					edge from parent[eq] node[elabel, right, pos=0.6] {$\neg O_3$}
				}
				edge from parent[eq] node[elabel, left] {$O_2$}
			}
			child { node[decision] (n3) {\shortstack{\scriptsize $n_3$\\[-2pt]$A_1$}}
				child { node[terminal] {$z_3$ \\ $[3,4,1,2]$}
					edge from parent node[elabel, left, pos=0.6] {$O_4$}
				}
				child { node[terminal] {$z_4$ \\ $[2,1,4,3]$}
					edge from parent node[elabel, right, pos=0.6] {$\neg O_4$}
				}
				edge from parent[discard] node[elabel, right] {$\neg O_2$}
			}
			edge from parent[eq] node[elabel, left, xshift=-5pt] {$O_1$}
		}
		child { node[terminal] {$z_5$ \\ $[4,1,2,3]$}
			edge from parent[discard] node[elabel, right, xshift=2pt] {$\neg O_1$}
		};

		\node[anchor=north west, draw, fill=white, font=\scriptsize, inner sep=3pt]
		at (current bounding box.north west) {
			\begin{tabular}{ll}
				\textcolor{red}{\rule[0.5ex]{0.4cm}{1.5pt}} & Equilibrium path \\
				\textcolor{gray!50}{\rule[0.5ex]{0.4cm}{0.8pt}} & Discarded branch
			\end{tabular}
		};
	\end{tikzpicture}%
}%
	\caption[Backward-induction scenario tree]{Four-actor scenario tree for the backward-induction example.
		Decision nodes are labeled by node identifier \(n_i\) and acting entity \(A_j\).
		Terminal nodes carry preference vectors
		\([u_{a_1}, u_{a_2}, u_{a_3}, u_{a_4}]\).
		The subgame-perfect equilibrium path \(O_1 \to O_2 \to \neg O_3\) is highlighted
		in red.}
	\label{fig:scenario_tree_bi}
\end{figure}

	\textbf{Setup.}
	Four actors \(A_1, A_2, A_3, A_4\) interact sequentially.
	Terminal nodes \(z_i\) carry preference vectors
	\([u_{a_1}, u_{a_2}, u_{a_3}, u_{a_4}]\) in which the \(j\)-th component encodes
	actor \(A_j\)'s ordinal rank over outcomes (higher rank is more preferred).
	Figure~\ref{fig:scenario_tree_bi} shows the tree; red edges mark the equilibrium path
	and dashed gray edges mark discarded branches.

	\textbf{Backward induction steps.}
	The recursion is applied bottom-up from the deepest decision nodes.

	\begin{enumerate}[leftmargin=2em, itemsep=4pt, label=\arabic*.]
		\item \textbf{Node \(n_2\) (actor \(A_2\)):}
		Option \(O_3\) leads to \(z_1\) with \(u_{a_2} = 1\);
		option \(\neg O_3\) leads to \(z_2\) with \(u_{a_2} = 2\).
		Since \(2 > 1\), actor \(A_2\) selects \(\neg O_3\).
		Continuation value at \(n_2\): outcome \(z_2\).

		\item \textbf{Node \(n_3\) (actor \(A_1\)):}
		Option \(O_4\) leads to \(z_3\) with \(u_{a_1} = 3\);
		option \(\neg O_4\) leads to \(z_4\) with \(u_{a_1} = 2\).
		Since \(3 > 2\), actor \(A_1\) selects \(O_4\).
		Continuation value at \(n_3\): outcome \(z_3\).

		\item \textbf{Node \(n_1\) (actor \(A_3\)):}
		Branch \(O_2\) leads (via \(n_2\)'s choice) to \(z_2\) with \(u_{a_3} = 3\);
		branch \(\neg O_2\) leads (via \(n_3\)'s choice) to \(z_3\) with \(u_{a_3} = 1\).
		Since \(3 > 1\), actor \(A_3\) selects \(O_2\).
		Continuation value at \(n_1\): outcome \(z_2\).

		\item \textbf{Root \(n_0\) (actor \(A_2\)):}
		Branch \(O_1\) leads (via \(n_1\)'s choice) to \(z_2\) with \(u_{a_2} = 2\);
		branch \(\neg O_1\) leads directly to \(z_5\) with \(u_{a_2} = 1\).
		Since \(2 > 1\), actor \(A_2\) selects \(O_1\).
	\end{enumerate}

	\textbf{Result.}
	The subgame-perfect equilibrium path is
	\(O_1 \to O_2 \to \neg O_3\), yielding terminal outcome \(z_2 = [1,2,3,4]\).
	Actor \(A_4\) benefits most (\(u_{a_4} = 4\)) and actor \(A_1\) least
	(\(u_{a_1} = 1\)) under this equilibrium.
	Note that \(z_1\) and \(z_5\) -- which would give \(A_1\) rank~4 -- are discarded
	because the actors controlling the relevant decision nodes prefer different continuations.
	This illustrates how strategic interdependence shapes which outcomes are reachable under
	backward induction, independent of individual preferences over terminal nodes.

	\paragraph{Event positions.}
	Event positions are not controlled by an actor, hence no preference maximization applies at \(v\in V_E\). A concrete scenario assumption fixes an event-realization selector
	\[
	\nu:V_E \to \Gamma,
	\qquad
	\nu(v)\in \mathrm{succ}(v),
	\]
	choosing one outgoing edge per event position (e.g.\ ``occurs'' vs.\ ``does not occur'', or a refined realization partition). Backward induction on decision positions remains well-defined because \(\nu\) introduces no preference comparison.
	
	\paragraph{Induced MRP-path.}
	Given a fixed event-realization selector \(\nu\), the pair \((\mu^{\mathrm{MRP}},\nu)\) induces a unique realized trajectory from the root to a terminal position by alternating (i) decision-position choices via \(\mu^{\mathrm{MRP}}\) and (ii) event-position resolutions via \(\nu\). The induced trajectory is referred to as the \emph{MRP-path} for the scenario assumption \(\nu\).
	
	\paragraph{Substituting information sets by epistemic components.}
	Information sets in extensive-form games are replaced by explicit epistemic components in the database: knowledge (\(K_a p\)), belief (\(B_a p\)), and their updates under \(\mathcal{U}\). If an observation resolves uncertainty, the update operator changes the epistemic content of \(DB_t\), and the decision rule applied at the subsequent decision position uses the updated epistemic state as input.
	
	\subsubsection{Refinements of the MRP\index{Tree Analysis!MRP (Most Rational Path)}}
	\label{subsubsec:tree-refinements}
	
	MRP refinements are defined by modifying one of three components: preference aggregation for coalitional decision makers, the temporal structure of evaluation, or the decision rule under uncertainty.
	Coalitional rationality is represented here by an SBA-specific aggregation operator:
	\[
	U_X(s)=\mathrm{Agg}_X\bigl(\{U_a(s):a\in X\}\bigr).
	\]
	This reconstruction is motivated by coalition-formation theory \citep{rayvohra2015},
	but it is not meant as a direct import of a standard cooperative-game solution concept.
	Temporal refinement models time-indexed evaluations and discounting to represent path dependence and dynamic consistency:
	\[
	U_a(t,DB_t)=g_a\bigl(\mathrm{Attr}_a(t),\mathrm{Rel}_t,\mathrm{Att}_a(t)\bigr),
	\qquad
	U_a^{\mathrm{total}}=\sum_t \delta_a^t\,U_a(t,DB_t),
	\]
	with \(\delta_a\in(0,1]\), where \(\mathrm{Attr}_a(t)\) denotes actor \(a\)'s attribute profile at stage \(t\).
	Probabilistic choice rules provide a bounded-rational refinement by mapping utilities into choice probabilities via a quantal-response/logit specification \citep{mckelvey1995}:
	\[
	P_a(o)=\frac{\exp(\lambda U_a(o))}{\sum_{o'}\exp(\lambda U_a(o'))}.
	\]
	
	\subsubsection{Likelihood-maximizing analysis and the Most Likely Path\index{Tree Analysis!MLP (Most Likely Path)} (MLP)}
	\label{subsubsec:tree-mlp}
	
	The MLP is defined as the terminal path that maximizes the product of conditional edge likelihoods. Each outgoing edge \(e\) from a vertex \(v\) is assigned a conditional likelihood value \(\ell(e)\in[0,1]\) satisfying \(\sum_{e\in\mathrm{succ}(v)}\ell(e)=1\).
	
	\paragraph{Global definition.}
	Let \(\mathrm{Paths}(T)\) be the set of root-to-leaf paths in the tree, and let \(P=\langle e_1,\dots,e_n\rangle\) be a path.
	For a vertex \(v\in V\), let \(\mathrm{Paths}(v)\) denote the set of \(v\)-to-leaf paths.
	\[
	L(P)=\prod_{i=1}^{n}\ell(e_i),
	\]
	and the MLP is any maximizer:
	\[
	\mathrm{MLP}\in \arg\max_{P\in\mathrm{Paths}(T)}\ L(P).
	\]
	
	\paragraph{Dynamic-programming policy.}
	Define the maximal continuation likelihood from a vertex \(v\) as
	\[
	J(v)\coloneqq\max_{P\in\mathrm{Paths}(v)}\ \prod_{e\in P}\ell(e).
	\]
	For leaf positions \(v\in L\), set \(J(v)=1\). For non-leaf positions,
	\[
	J(v)=\max_{e\in\mathrm{succ}(v)}\ \ell(e)\,J(\mathrm{head}(e)),
	\]
	where \(\ell(e)\) is the conditional probability assigned to edge \(e\in\Gamma\). The associated MLP policy is
	\[
	\mu^{\mathrm{MLP}}(v)\in \arg\max_{e\in\mathrm{succ}(v)}\ \ell(e)\,J(\mathrm{head}(e)),
	\]
	and iterating \(\mu^{\mathrm{MLP}}\) from the root reconstructs an MLP \citep{bellman1957, bertsekas2005}.
	
\begin{figure}[htbp]
		\centering
		\begin{tikzpicture}[
			scale=0.9,
			transform shape,
			actornode/.style={circle, draw=blue!70!black, very thick, fill=blue!10, minimum size=1cm, font=\small\bfseries},
			eventnode/.style={diamond, draw=purple!70!black, very thick, fill=purple!10, minimum size=1cm, font=\small\bfseries},
			terminal/.style={rectangle, draw=black, fill=yellow!10, rounded corners=2pt, minimum width=1.8cm, minimum height=0.8cm, align=center, font=\tiny},
			mlp/.style={->, line width=2.5pt, >=stealth, draw=green!60!black},
			nonmlp/.style={->, line width=1pt, >=stealth, draw=gray!40, dashed},
			likelihood/.style={font=\tiny\bfseries, fill=white, inner sep=1.5pt, text=green!60!black}
			]
			
			\node[font=\large\bfseries] at (0, 8.5) {Most Likely Path (MLP): Likelihood Maximization};
			
			\node[actornode] (root) at (0, 7) {$A_1$};
			
			\node[actornode] (a1) at (-3, 4.5) {$A_2$};
			\node[eventnode] (e1) at (3, 4.5) {$e_1$};
			
			\node[terminal] (t1) at (-4.5, 2) {$z_1$\\MRP: 3rd\\MLP: 0.14};
			\node[terminal] (t2) at (-1.5, 2) {$z_2$\\MRP: \textbf{1st}\\MLP: \textbf{0.56}};
			\node[terminal] (t3) at (1.5, 2) {$z_3$\\MRP: 2nd\\MLP: 0.18};
			\node[terminal] (t4) at (4.5, 2) {$z_4$\\MRP: 4th\\MLP: 0.12};
			
			\draw[mlp] (root) -- node[likelihood, pos=0.4] {0.7} node[left, font=\tiny, xshift=-2mm] {$o_1$} (a1);
			\draw[nonmlp] (root) -- node[font=\tiny, fill=white, text=gray] {0.3} node[right, font=\tiny, xshift=2mm, text=gray] {$\neg o_1$} (e1);
			
			\draw[mlp] (a1) -- node[likelihood, pos=0.4] {0.8} node[right, font=\tiny] {$o_2$} (t2);
			\draw[nonmlp] (a1) -- node[font=\tiny, fill=white, text=gray] {0.2} (t1);
			
			\draw[nonmlp] (e1) -- node[font=\tiny, fill=white, text=gray] {0.6} (t3);
			\draw[nonmlp] (e1) -- node[font=\tiny, fill=white, text=gray] {0.4} (t4);
			
			\node[draw, green!60!black, fill=green!5, text width=3.2cm, font=\tiny, anchor=north west] at (-6.5, 7) {
				\textbf{MLP calculation:}\\
				$L(z_2)=\ell(o_1)\times\ell(o_2)$\\
				$L(z_2)=0.7\times0.8=\mathbf{0.56}$
			};
			
			\node[draw, red!60!black, fill=red!5, text width=3.2cm, font=\tiny, anchor=north east] at (6.5, 7) {
				\textbf{MRP comparison:}\\
				Here: MLP = MRP.\\
				Divergence indicates tension between\\
				preference-optimality and plausibility.
			};
			
			\node[draw, fill=gray!5, text width=13cm, font=\tiny, align=left] at (0, 0.5) {
				\textbf{Interpretation:} The MLP maximizes path likelihood $L(P)=\prod_{e\in P}\ell(e)$ under local conditional edge likelihoods. Comparing MLP and MRP diagnoses cases where highly plausible evolutions conflict with preference-optimal continuations \citep{selten1999sbm, mckelvey1995}.
			};
			
		\end{tikzpicture}
		\caption{Most Likely Path (MLP) vs.\ preference-based Most Rational Path (MRP).}
		\label{fig:mlp_induction}
	\end{figure}
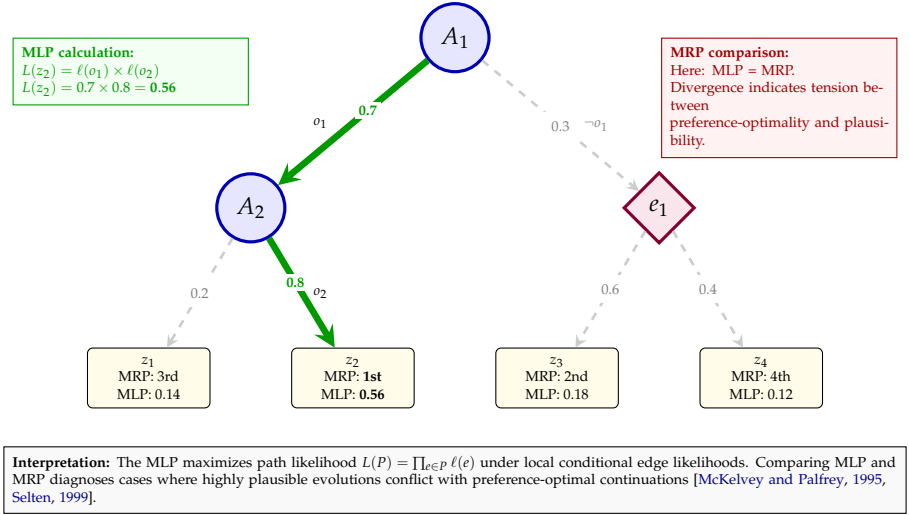

	Figure~\ref{fig:mlp_induction} compares the Most Likely Path against the Most Rational Path for the illustrative two-stage binary tree with four terminal outcomes.
	\paragraph{Tie breaking.}
	Likelihood ties at a vertex induce a set of maximizers rather than a unique edge. Determinacy can be enforced by a tie-break rule applied to the maximizer set:
	\begin{itemize}[leftmargin=1.5em,itemsep=3pt]
		\item \textbf{Lexicographic tie-break:} choose the smallest edge ID among maximizers.
		\item \textbf{Secondary maximization:} among maximizers, maximize a secondary score \(h(e;DB_t)\) (e.g.\ expected impact magnitude or evidential confidence).
		\item \textbf{Set-valued output:} retain all maximizers and treat the MLP as a set of equally most likely paths.
	\end{itemize}
	
	\FloatBarrier
	
	\paragraph{Normative--descriptive duality.}
	MRP expresses normative choice coherence under a fixed decision rule and strict terminal evaluations, while MLP expresses empirical plausibility under conditional likelihood estimates. Divergence between MRP and MLP indicates misperception, commitment problems, or escalation pressure produced by the interaction between incentives and probabilistic constraints.
	
	\subsubsection{Theoretical background}
	\label{subsubsec:tree-theory}
	
	Scenario-tree analysis in SBA integrates extensive-form rationality, epistemic logic, probabilistic inference, and decision theory under uncertainty. Backward-induction reasoning relates to subgame-perfect equilibrium concepts in extensive-form games \citep{selten1975, osborne1994}, while SBA generalizes the node ontology to include coalitions and events and relies on ordinal post hoc evaluations rather than ex ante utility functions.
	
	Epistemic uncertainty is represented semantically by explicit knowledge and belief states in the database and by update operators for epistemic change, drawing on epistemic and dynamic-epistemic perspectives \citep{vanbenthem1996, rescher1971, prior1957}. Decision making under epistemic uncertainty requires an explicit decision rule, with standard options including subjective expected utility \citep{savage1954}, maximin criteria \citep{wald1950}, and ambiguity-sensitive maxmin expected utility \citep{gilboa1989}. Coalitional decision positions employ aggregation rules that connect individual and collective evaluation \citep{rayvohra2015}. Likelihood-based path selection uses conditional probabilities on edges and evaluates complete paths by multiplicative likelihood, with efficient maximization by dynamic programming \citep{bellman1957, bertsekas2005}.
	
	\subsection{Scenario Bundles}
	\label{subsec:from-trees-to-bundles}
	
	A single scenario tree encodes one internally coherent development given an assessment state and fixed modeling assumptions. Multiple coherent trees arise from epistemic variation, alternative root choices, competing event realizations, and different feasibility or salience thresholds, motivating a bundle representation as a model class of trees generated from a database stage.
	
	\subsubsection{Formal structure}
	\label{subsubsec:bundle-formal}
	
	Let the initialized Scenario Database\index{Database Layer} be
	\[
DB_0=\langle A_0, C_0, \mathrm{Attr}_0, \mathrm{Att}_0, \mathrm{Rel}_0, \mathrm{Opt}_0, E_0\rangle
\].
	Let \(J_t\subseteq\mathbb{N}\) be a finite or countable index set. From \(DB_t\), a finite or countable family of Scenario Trees\index{Scenario Trees} is generated:
	\[
	\mathcal{T}(DB_t)=\{\,T_i^{(t)}\mid i\in J_t\,\},
	\]
	When the stage index is irrelevant, \(T_i\) abbreviates \(T_i^{(t)}\). The superscript \((t)\) denotes the assessment stage at which the tree is generated; it does not index within-tree depth.
	
	Each tree \(T_i^{(t)}\) has the structure
	\[
	T_i^{(t)}=\langle V_i^{(t)},\Gamma_i^{(t)},\mathrm{lab}_i^{(t)},\rho_i^{(t)},\Sigma_i^{(t)},\mathrm{Val}_i^{(t)}\rangle,
	\]
	where \(V_i^{(t)}\) is a finite set of positions (vertices),
	\(\Gamma_i^{(t)}\subseteq V_i^{(t)}\times V_i^{(t)}\) the directed edges,
	\(\mathrm{lab}_i^{(t)}:V_i^{(t)}\to A\cup C\cup E\) the label map,
	\(\rho_i^{(t)}\in V_i^{(t)}\) the root,
	\(\Sigma_i^{(t)}:\Gamma_i^{(t)}\to \mathrm{Opt}_t\cup E^\circ\) the edge labels (options or event outcomes), and
	\(\mathrm{Val}_i^{(t)}:(A\cup C)\times L_i^{(t)}\to \mathbb{R}\) assigns numerical leaf scores; unless stated otherwise, only the induced ordering is used (using \(L_i^{(t)}\subseteq V_i^{(t)}\) for the leaf set).
	
	A tree is admissible iff all transitions encoded by its edges are consistent with an assessment state \(DB_t\) reachable from \(DB_0\) under the update family \(\mathcal{U}\), i.e.

	\[
	DB_t = (\mathcal{U}_{t-1}\circ\cdots\circ \mathcal{U}_0)(DB_0).
	\]
	A tree \(T_i^{(t)}\) is admissible at stage \(t\) iff
	\[
	\forall e\in \Gamma_i^{(t)}:\ \mathrm{valid}(e;DB_t).
	\]
	The \emph{admissible tree set} at stage \(t\) is
	\[
	T_t\coloneqq\mathcal{T}(DB_t),
	\]
	the set of all internally coherent scenario trees derivable from the database at that stage. The finite selected \emph{scenario bundle} \(\mathcal{B}_t\coloneqq\mathrm{Sel}_t(T_t)\subseteq T_t\) is the output of a declared selection operator \(\mathrm{Sel}_t\).
	

	\section{Scenario Space and Topology}
	\label{sec:topology}
	
	
	Scenario Space is the topological space induced by a distance on the admissible scenario trees generated from a fixed assessment state \(DB_t\). The construction supports robustness and sensitivity analysis under methodological variation, distinct from temporal evolution of the conflict.
	
\begin{figure}[htbp]
	\centering
	\begin{tikzpicture}[
		x=1cm, y=1cm,
		scale=0.95,
		transform shape,
		sba_db/.style={rectangle, draw=blue!70!black, fill=blue!10, thick,
			rounded corners, minimum width=2.7cm, minimum height=0.95cm,
			font=\small, align=center},
		sba_bundle/.style={rectangle, draw=green!60!black, fill=green!10, thick,
			rounded corners, minimum width=2.1cm, minimum height=0.95cm,
			font=\small, align=center},
		sba_tree/.style={circle, draw=purple!70!black, fill=purple!10, thick,
			minimum size=0.85cm, font=\scriptsize},
		update/.style={-Latex, draw=blue!70!black, very thick},
		process/.style={-Latex, draw=green!60!black, thick},
		method/.style={-Latex, draw=purple!70!black, thick, dashed},
		panel/.style={draw=gray!40, thick, rounded corners},
		ptitle/.style={font=\small\bfseries},
		plabel/.style={font=\scriptsize, align=center},
		inset/.style={draw=gray!45, thick, rounded corners, fill=gray!5,
			font=\scriptsize, align=left, inner sep=4pt},
		legendbox/.style={draw=gray!40, rounded corners, fill=white, inner sep=5pt, font=\scriptsize}
		]
		
		\draw[panel] (-0.3, -0.4) rectangle (5.4, 6.1);
		\node[ptitle, blue!70!black] at (2.55, 5.75) {(i) Assessment-state change};
		
		\node[sba_db] (dbt)  at (2.55, 4.75) {$DB_t$};
		\node[sba_db] (dbt1) at (2.55, 3.35) {$DB_{t+1}$};
		
		\draw[update] (dbt) -- node[right, font=\scriptsize] {$\mathcal{U}_t$} (dbt1);
		
		\node[sba_bundle] (bt)  at (1.1, 1.7) {$\mathcal{B}_t$};
		\node[sba_bundle] (bt1) at (4.0, 1.7) {$\mathcal{B}_{t+1}$};
		
		\draw[process] (dbt1.west) to[out=180, in=90] 
		node[left, font=\scriptsize, align=right, xshift=-2pt] {Gen$_t$,\\Sel$_t$} 
		(bt.north);
		
		\draw[process] (dbt1.east) to[out=0, in=90] 
		node[right, font=\scriptsize, align=left, xshift=2pt] {Gen$_{t+1}$,\\Sel$_{t+1}$} 
		(bt1.north);
		
		\node[plabel] at (2.55, 0.3)
		{Update of $DB$ changes admissibility\\$\Rightarrow$ regenerate $\mathcal{B}$ (no incremental update).};
		
		\draw[panel] (6.0, -0.4) rectangle (11.7, 6.1);
		\node[ptitle, purple!70!black] at (8.85, 5.75) {(ii) Methodological variation};
		
		\node[sba_db, minimum width=3.2cm] (dbfix) at (8.85, 5.0) {$DB_t$ (fixed)};
		
		\draw[gray!45, thick, dashed, rounded corners] (6.55, 2.3) rectangle (11.15, 4.3);
		
		\node[font=\scriptsize, text=gray!90, anchor=north west] at (6.6, 4.25) {$\mathcal{T}(DB_t)$};
		
		\node[sba_tree] (t)   at (7.25, 3.3) {$T$};
		\node[sba_tree] (tp)  at (8.75, 3.7) {$T'$};
		\node[sba_tree] (tpp) at (10.25, 3.3) {$T''$};
		
		\draw[method] (t)  -- node[above, font=\scriptsize, yshift=-1pt] {$\varphi_{\lambda_1}$} (tp);
		\draw[method] (tp) -- node[above, font=\scriptsize, yshift=-1pt] {$\varphi_{\lambda_2}$} (tpp);
		
		\node[sba_bundle, minimum width=1.6cm] (b)  at (7.75, 1.4) {$\mathcal{B}$};
		\node[sba_bundle, minimum width=1.6cm] (bp) at (9.95, 1.4) {$\mathcal{B}'$};
		\draw[method] (b) -- node[above, font=\scriptsize] {$\Phi_{\lambda}$} (bp);
		
		\node[plabel] at (8.85, 0.3)
		{Fixed $DB_t$; vary encodings,\\weights, admissibility, selection.};
		
		\node[legendbox] at (5.7, -1.1) {
			\begin{tabular}{@{}c@{\quad}c@{\quad}c@{}}
				\tikz[baseline=-0.5ex]{\draw[update] (0,0)--(0.6,0);} update $\mathcal{U}_t$ &
				\tikz[baseline=-0.5ex]{\draw[process] (0,0)--(0.6,0);} generate + select &
				\tikz[baseline=-0.5ex]{\draw[method] (0,0)--(0.6,0);} method $\varphi_\lambda,\Phi_\lambda$
			\end{tabular}
		};
		
		\node[inset] at (5.7, -2.1) {In general:\quad $\mathcal{B}_{t+1}\neq \Pi_{t\to t+1}\!\bigl(\Phi_{\lambda}(\mathcal{B}_t)\bigr)$.};
		
	\end{tikzpicture}
	\caption{Assessment-state change versus methodological variation.}
	\label{fig:temporal_methodological_dynamics}
\end{figure}
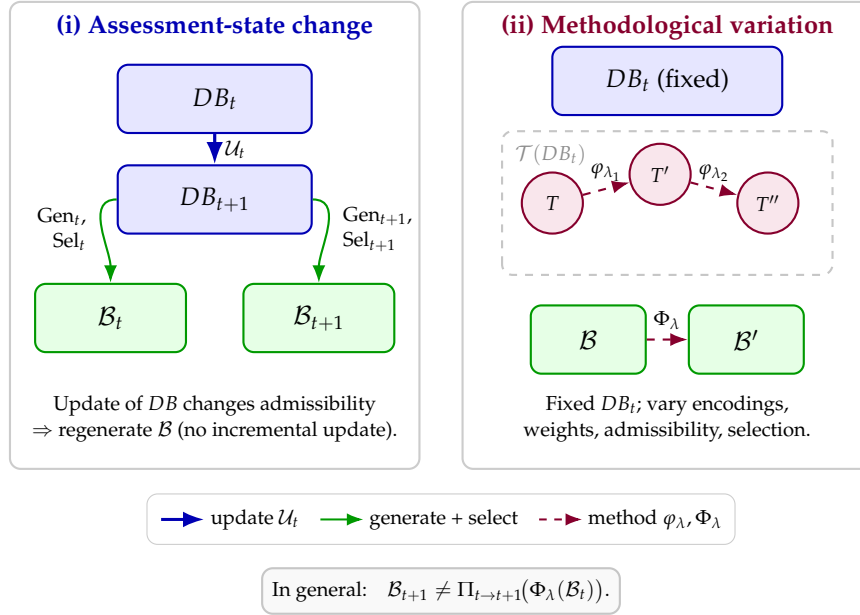

	We distinguish two notions of change:

	Fix \(\lambda\in\Lambda\) as a methodological-choice parameter.
	\begin{enumerate}[label=(\roman*),leftmargin=2em]
		\item \textbf{Temporal evolution (world change):} The conflict evolves over time through assessment-state updates \(\mathcal{U}_t\colon DB_t \to DB_{t+1}\) as specified in Section~\ref{sec:scenario-dynamics}.
		\item \textbf{Methodological variation (model change at fixed assessment state):} For fixed \(DB_t\), different admissibility choices, codings, weights, and modeling conventions yield different admissible trees and bundles. Transformations \(\varphi_{\lambda}\) (tree-level) and \(\Phi_{\lambda}\) (bundle-level) represent these variations (Section~\ref{sec:scenario-dynamics}).
	\end{enumerate}
	The topology defined below conditions on a fixed assessment state \(DB_t\). Figure~\ref{fig:temporal_methodological_dynamics} separates assessment-state change (via \(\mathcal{U}_t\)) from within-state methodological variation (via \(\varphi_\lambda,\Phi_\lambda\)).
	
	\subsection{Assessment-state conditioned scenario space}
	\label{subsec:assessment-conditioned-space}
	
	\paragraph{Assessment state.}
	\(DB_t\) denotes the assessment state at stage \(t\): the instantiated scenario database (actor and coalition sets, attributes, attitudes, relations, options, and events) specified in the Database Layer (Section~\ref{sec:database}). The valuation is fixed throughout.
	
	\paragraph{Admissible scenario trees and bundles.}
	\(T_t\) denotes the set of admissible scenario trees\index{Scenario Trees} generated from \(DB_t\). Admissibility is governed by SBA\index{Scenario Bundle Analysis} constraints (consistency of attitudes and relations, feasibility of options, event plausibility, and related filters). A scenario bundle\index{Scenario Bundles} is a finite subset \(\mathcal{B}\subseteq T_t\), selected by salience, plausibility, or coverage criteria.
	
	\paragraph{Scenario space.}
	The scenario space at assessment stage \(t\) is the topological space
	\[
	\mathcal{S}_t \coloneqq (T_t,\tau_t),
	\]
	where \(\tau_t\) is induced by a pseudo-metric\index{Topology!Pseudo-metric} \(d_t\) defined below.
	
	\paragraph{Fixed signatures, \texorpdfstring{\(DB_t\)}{DB\_t}-dependent valuations.}
	Type signatures remain fixed (attribute symbols, attitude operators, relation symbols). Valuations (attribute values, attitudes held, active relations) are those of \(DB_t\). Temporal change is represented by moving from \((DB_t,T_t)\) to \((DB_{t+1},T_{t+1})\), not by paths inside \(T_t\).
	
	\paragraph{Encoding as comparison-relevant projection.}
	An encoding map
	\[
	\mathrm{Enc}_t \colon T_t \to \mathcal{E}_t
	\]
	represents each admissible tree by a structured descriptor in \(\mathcal{E}_t\). The encoding is purpose-relative: it preserves features relevant for comparison, evaluation, and robustness analysis without aiming at full representation.
	
	Admissible encodings satisfy:
	\begin{enumerate}[label=(\roman*),leftmargin=2em]
		\item \emph{Typing}: Each component of \(\mathrm{Enc}_t(T)\) is defined using the fixed signature of the SBA language and the valuation given by \(DB_t\).
		\item \emph{Isomorphism invariance}: If \(T\cong T'\) as rooted, edge-labeled trees (preserving actor/event labels up to admissible renaming),
		  then \(\mathrm{Enc}_t(T)=\mathrm{Enc}_t(T')\).
		\item \emph{Transparency}: Similarity judgments induced by the encoding are interpretable.
	\end{enumerate}
	Typical encoded components include event-pattern summaries, coalition trajectories, attitude configurations, option realizations, and terminal outcome profiles.
	
	\subsection{Distances and induced topology}
	\label{subsec:distance-topology}
	
	\paragraph{Pseudo-metric.}
	A pseudo-metric
	\[
	d_t \colon T_t \times T_t \to \mathbb{R}_{\ge 0}
	\]
	is defined by aggregating component-wise discrepancies:
	\[
	d_t(T,T') \coloneqq \sum_{k=1}^{K} w_k \cdot d^{(k)}_t\big(\mathrm{Enc}^{(k)}_t(T),\mathrm{Enc}^{(k)}_t(T')\big),
	\]
	where \(w_k \ge 0\) are weights and each \(d^{(k)}_t\) is a discrepancy on the \(k\)-th encoded component.
	
	The function \(d_t\) must be a pseudo-metric on \(T_t\): for all \(T,T',T''\in T_t\),
	\[
	d_t(T,T)=0,\qquad d_t(T,T')=d_t(T',T),\qquad d_t(T,T'')\le d_t(T,T')+d_t(T',T'').
	\]
	If each component distance \(d^{(k)}_t\) is a pseudo-metric on its component space and \(w_k\ge 0\), then \(d_t\) is a pseudo-metric on \(T_t\).
	
	Zero distance between distinct trees occurs when the encoding forgets distinctions or when equivalence is imposed for analysis. The induced equivalence relation
	\[
	T\sim_t T' \iff d_t(T,T')=0
	\]
	supports an optional quotient space \(T_t/{\sim_t}\) when strict separation of points is required.
	
	\paragraph{Topology.}
	Open \(\varepsilon\)-balls
	\[
	N_{t,\varepsilon}(T)\coloneqq \{T'\in T_t \mid d_t(T,T')<\varepsilon\}
	\]
	generate a topology \(\tau_t\) on \(T_t\).
	
	\paragraph{Distance is not time.}
	\(d_t\) carries no temporal interpretation: \(d_t(T,T')\) measures methodological similarity under \(\mathrm{Enc}_t\) and the chosen weights at fixed \(DB_t\), not reachability or succession in conflict time.
	
	\subsection{Bundles as Points of a Derived Space}
	\label{subsec:bundle-space}
	
	Bundle-level robustness is supported by lifting \(d_t\) to finite subsets of \(T_t\). For finite bundles \(\mathcal{B},\mathcal{B}'\subseteq T_t\),
	\[
	d^{\mathcal{B}}_t(\mathcal{B},\mathcal{B}')
	\coloneqq
	\max\Big\{
	\max_{T\in \mathcal{B}}\min_{T'\in \mathcal{B}'} d_t(T,T'),\;
	\max_{T'\in \mathcal{B}'}\min_{T\in \mathcal{B}} d_t(T,T')
	\Big\}.
	\]
	Small values of \(d^{\mathcal{B}}_t(\mathcal{B},\mathcal{B}')\) mean that every tree in one bundle admits a near counterpart in the other, with proximity measured by \(d_t\).
	This supports robustness claims: if bundles computed from the same \(DB_t\) under alternative methodological choices (e.g.\ different \(\lambda\) or selection rules) remain close, then the resulting assessment is stable under that variation.

	Zero distance is possible when \(d_t\) is only a pseudo-metric: distinct trees can satisfy \(d_t(T,T')=0\) when they agree on the declared invariants or encodings.
	In that case \(d^{\mathcal{B}}_t\) is again a pseudo-metric and different bundles can represent the same derived assessment point; analytic statements should be read as statements about the induced equivalence classes on \(\mathcal{P}_{\mathrm{fin}}(T_t)\), not about incidental representational differences.
	Continuity requirements on evaluation maps with respect to \(d^{\mathcal{B}}_t\) then yield sensitivity bounds and clustering diagnostics.

	Here \(\mathcal{P}_{\mathrm{fin}}(X)\) denotes the set of all finite subsets of \(X\). This defines a pseudo-metric on \(\mathcal{P}_{\mathrm{fin}}(T_t)\) when \(d_t\) is a pseudo-metric on \(T_t\).

	\subsection{Worked Example: Distance and Topology in a Border-Incident Scenario}
	\label{subsec:topology-example}

	The following example makes the abstract distance construction concrete by computing
	\(d_t(T, T')\) for two admissible trees generated from the same assessment state.
	It previews the border-incident setup used again in
	Section~\ref{subsec:eval-example}: three actors
	\(a\) (aggressor state), \(b\) (defender), and \(m\) (mediator), with an
	exogenous triggering event \(e_1\) (border incident).

	\paragraph{Assessment state.}
	Fix \(DB_t\) with \(A_t = \{a, b, m\}\), one coalition \(C_t = \{\{a,m\}\}\)
	(potential mediator alignment), and event \(e_1 \in E_t\).
	Actor \(a\) holds \(\mathrm{aim}(a, \mathrm{escalate}) = 3\) (high, ordinal scale 0--3)
	and \(\mathrm{aim}(a, \mathrm{restrain}) = 1\).
	Actor \(b\) holds \(\mathrm{aim}(b, \mathrm{defend}) = 3\).

	\paragraph{Two admissible trees.}
	From \(DB_t\), two trees are generated under different admissibility thresholds
	for coalition activation:

	\medskip
	\noindent
	\textbf{Tree \(T\) --- escalation path.}
	Root \(e_1\); \(a\) executes option \(R\) (retaliate); \(b\) responds with \(E\)
	(escalate); coalition \(\{a,m\}\) does not form.
	Terminal outcome: \(z_{RE} = [u_a{=}3,\, u_b{=}1,\, u_m{=}1]\).

	\medskip
	\noindent
	\textbf{Tree \(T'\) --- de-escalation path.}
	Root \(e_1\); \(a\) executes option \(S\) (show restraint); \(b\) responds with \(W\)
	(withdraw); coalition \(\{a,m\}\) forms as diplomatic channel.
	Terminal outcome: \(z_{SW} = [u_a{=}1,\, u_b{=}3,\, u_m{=}3]\).

	\paragraph{Encoding and component distances.}
	Three encoded components (\(K = 3\), equal weights \(w_k = \tfrac{1}{3}\)):

	\begin{center}
		\renewcommand{\arraystretch}{1.3}
		\small
		\begin{tabular}{>{\RaggedRight\arraybackslash}p{3.2cm}
		                >{\RaggedRight\arraybackslash}p{4.4cm}
		                ccc}
			\toprule
			\textbf{Component} & \textbf{Discrepancy \(d^{(k)}_t\)} &
			\(\mathrm{Enc}^{(k)}_t(T)\) & \(\mathrm{Enc}^{(k)}_t(T')\) & \(d^{(k)}_t\) \\
			\midrule
			\(k{=}1\): Terminal outcome & \(\ell_{\mathrm{norm}}\)-norm on rank vectors (normalised)
			& \([3,1,1]\) & \([1,3,3]\) & \(\tfrac{6}{9}\) \\
			\(k{=}2\): Coalition trajectory & \(0{/}1\) (coalition absent/present)
			& \(0\) & \(1\) & \(1\) \\
			\(k{=}3\): Dominant action of \(a\) & \(0{/}1\) (retaliate/restrain)
			& \(0\) & \(1\) & \(1\) \\
			\bottomrule
		\end{tabular}
	\end{center}

	\paragraph{Pseudo-metric computation.}
	Applying the weighted sum:
	\[
		d_t(T, T')
		= \tfrac{1}{3}\cdot\tfrac{6}{9}
		+ \tfrac{1}{3}\cdot 1
		+ \tfrac{1}{3}\cdot 1
		= \tfrac{2}{9} + \tfrac{1}{3} + \tfrac{1}{3}
		= \tfrac{2}{9} + \tfrac{6}{9}
		= \tfrac{8}{9}
		\approx 0.89.
	\]
	The two trees are nearly maximally distant under this encoding: they differ in
	all three components.

	\paragraph{Topological interpretation.}
	Fix \(\varepsilon = 0.2\).
	The \(\varepsilon\)-ball around \(T\),
	\[
		N_{t,0.2}(T) = \{T'' \in T_t \mid d_t(T, T'') < 0.2\},
	\]
	contains only trees that share both the escalation action and the absent-coalition
	trajectory with \(T\), and whose terminal outcome vector differs from \([3,1,1]\)
	by at most \(0.2\cdot 9 / 3 = 0.6\) ordinal units per component on average.
	Any analysis conclusion that holds for all trees in \(N_{t,0.2}(T)\) is
	\emph{robust} to small methodological perturbations around the escalation scenario.

	\paragraph{Sensitivity to weights.}
	If the weight of the terminal-outcome component is increased to
	\(w_1 = \tfrac{2}{3}\) and \(w_2 = w_3 = \tfrac{1}{6}\), then
	\[
		d_t(T, T')
		= \tfrac{2}{3}\cdot\tfrac{6}{9}
		+ \tfrac{1}{6}\cdot 1
		+ \tfrac{1}{6}\cdot 1
		= \tfrac{4}{9} + \tfrac{3}{9}
		= \tfrac{7}{9}
		\approx 0.78.
	\]
	The distance decreases slightly but the qualitative conclusion---\(T\) and \(T'\) are
	far apart---is robust to this reweighting.
	Systematic weight variation is a methodological transformation \(\Phi_\lambda\)
	in the sense of Section~\ref{subsec:methodological-dynamics}.

	\subsection{Theoretical foundations}
	\label{subsec:space-theory}
	
	The topological construction links scenario comparison to metric structure, epistemic uncertainty, and structuralist model classes.
	
	\paragraph{Metric geometry and structural similarity.}
	Tree-based distances connect SBA to metric geometry and structural comparison of discrete objects \citep{burago2001,sturm2006}. Scenario trees are structured combinatorial objects, and their comparison invokes graph and tree metrics (including edit-type discrepancies). The resulting distances express structural similarity, not temporal succession.
	
	\paragraph{Epistemic and decision-theoretic foundations.}
	Distances reflecting belief distributions, option realizations, and epistemic constraints connect the topology to decision theory under uncertainty and ambiguity. Bayesian networks and probabilistic graphical models formalize dependence structure and probabilistic inference \citep{pearl1988,jensen2007}. SBA uses similarity measures to test robustness of conclusions across admissible scenario representations within a fixed assessment state.
	
	\paragraph{Structuralist philosophy of science.}
	The construction aligns with the structuralist view of theories as structured families of models related by invariants \citep{balzer1987}. The assessment state defines the base structure, each scenario tree is a partial model, and the topology represents structural relations within the admissible model class.
	
	\paragraph{Interpretation.}
	Scenario Space provides an interpretable geometry of admissible futures under fixed assessments: uncertainty, strategic interaction, and modeling assumptions are explored systematically without conflating similarity with temporal evolution.


	\section[Scenario Dynamics]{Scenario Dynamics and\\Transformation Rules}
	\label{sec:dynamics}
	\label{sec:scenario-dynamics}
	
	Scenario dynamics decomposes into two orthogonal mechanisms: assessment-state updates \(\mathcal{U}_t\colon DB_t \to DB_{t+1}\) and within-state methodological variation at fixed \(DB_t\) (Section~\ref{sec:topology}).
The separation is a diagnostic constraint rather than a stylistic preference.
Temporal change updates the evidential base \(DB_t\) and therefore changes what is admissible.
Methodological variation changes how an analyst generates, selects, and compares scenario trees at fixed \(DB_t\), without claiming that the world has changed.
Conflating the two mechanisms collapses error diagnosis: apparent instability may be caused by shifting assumptions rather than shifting evidence, or vice versa.
Robust scenario assessment therefore requires that each change in outputs be attributable either to an explicit update \(DB_t\mapsto DB_{t+1}\) or to a declared methodological parameter choice \(\lambda\).

	\subsection{Temporal dynamics: assessment-state update and bundle regeneration}
	\label{subsec:temporal-dynamics}
	
	\paragraph{Database update operator.}
	Let \(\mathcal{U}_t\) denote an update operator
	\[
	\mathcal{U}_t \colon DB_t \to DB_{t+1},
	\]
	triggered by new evidence, revised assessments, or observed events. Updates modify:
	\begin{itemize}[leftmargin=1.5em,itemsep=2pt]
		\item the actor and coalition sets (\(A_t, C_t\)),
		\item attribute assignments (\(\mathrm{Attr}_t\)),
		\item attitude profiles (\(\mathrm{Att}_t\)),
		\item relations (\(\mathrm{Rel}_t\)),
		\item option feasibility and availability (\(\mathrm{Opt}_t\)),
		\item the event set and event parameters (\(E_t\)).
	\end{itemize}

	\paragraph{Admissible updates (minimal constraints).}
Updates are methodologically meaningful only if they preserve the declared interfaces of the database and make revision choices auditable.
An update \(\mathcal{U}_t\) is called \emph{admissible} if the following conditions hold.
\begin{itemize}[leftmargin=1.5em,itemsep=2pt]
	\item \textbf{Type preservation.} \(\mathcal{U}_t(DB_t)\) is an assessment state of the same product type as \(DB_t\), i.e.\ its components remain well-typed (\(A_{t+1}\subseteq A\), \(C_{t+1}\subseteq C\), and so on).
	\item \textbf{Provenance completeness.} Every updated primitive assertion (attribute value, attitude report, relation, option feasibility claim, event parameter) is accompanied by a provenance record (source, time stamp, extraction or coding method).
	\item \textbf{Declared revision policy.} If new information conflicts with previously stored fragments under the adopted consistency conventions, \(\mathcal{U}_t\) applies a declared priority rule (e.g.\ source quality or recency) or explicitly records the conflict as a disputed item rather than silently overwriting.
	\item \textbf{Event coherence.} If the update is triggered by an observed outcome \(e^\circ\in E^\circ\) of some \(e\in E_t\), then the realized outcome is recorded and downstream constraints (e.g.\ option feasibility or attribute shocks) are changed only through declared event-to-effect links.
	\item \textbf{Feasibility discipline.} Reductions of option feasibility are justified by new constraints or event realizations; expansions of feasibility are justified by newly available capabilities, agreements, or lifted constraints.
\end{itemize}

\paragraph{Contradiction handling.}
Contradictions under update are informative signals about either the quality of sources or the scope convention of the current assessment state.
A minimal operational protocol marks the affected items, assigns them to a revision queue, and regenerates the bundle only after the revision choice has been declared; this keeps \(DB_t\) interpretable as an auditable assessment snapshot rather than a black-box state.

	\paragraph{Tree generation and bundle selection.}
	Let \(\mathsf{Tree}\) be a universe of admissible tree objects, and let \(\mathcal{T}(DB_t)\subseteq \mathsf{Tree}\) denote the set of scenario trees consistent with \(DB_t\).
	Define the admissible-tree generator by
	\[
	\mathrm{Gen}_t \colon DB_t \to \mathcal{P}(\mathsf{Tree}),
	\qquad
	\mathrm{Gen}_t(DB_t)=\mathcal{T}(DB_t).
	\]
	For consistency with Section~\ref{sec:topology}, write
	\[
	T_t \coloneqq \mathcal{T}(DB_t).
	\]
	A bundle selection rule is a map
	\[
	\mathrm{Sel}_t \colon \mathcal{P}(\mathsf{Tree}) \to \mathcal{P}_{\mathrm{fin}}(\mathsf{Tree}),
	\]
	satisfying \(\mathrm{Sel}_t(S)\subseteq S\) for all \(S\subseteq \mathsf{Tree}\). The canonical bundle associated with \(DB_t\) is
	\[
	\mathcal{B}_t \coloneqq \mathrm{Sel}_t(\mathrm{Gen}_t(DB_t)).
	\]
	Assessment-state change regenerates the associated bundle by reapplying generation and selection:
	\[
	DB_t \xrightarrow{\ \mathcal{U}_t\ } DB_{t+1}
	\quad\Longrightarrow\quad
	\mathcal{B}_{t+1}\coloneqq \mathrm{Sel}_{t+1}\bigl(\mathrm{Gen}_{t+1}(DB_{t+1})\bigr).
	\]
	
	\subsection{Methodological dynamics: transformations at fixed \texorpdfstring{\(DB_t\)}{DB\_t}}
	\label{subsec:methodological-dynamics}
	
	\paragraph{Methodological parametrization.}
	Fix \(t\). A methodological choice parameter \(\lambda\) determines generation and selection at fixed \(DB_t\) via \(\lambda\)-indexed maps
	\[
	\mathrm{Gen}_{t,\lambda}\colon DB_t\to \mathcal{P}(\mathsf{Tree}),
	\]
	\[
	\mathrm{Sel}_{t,\lambda}\colon \mathcal{P}(\mathsf{Tree})\to \mathcal{P}_{\mathrm{fin}}(\mathsf{Tree}),
	\]
	and the induced bundle
	\[
	\mathcal{B}_{t,\lambda}\coloneqq \mathrm{Sel}_{t,\lambda}\bigl(\mathrm{Gen}_{t,\lambda}(DB_t)\bigr).
	\]
	Output-level transformations may be represented as endomorphisms on trees and bundles,
	\[
	\varphi_{\lambda}\colon \mathsf{Tree}\to \mathsf{Tree},
	\]
	\[
	\Phi_{\lambda}\colon \mathcal{P}_{\mathrm{fin}}(\mathsf{Tree})\to \mathcal{P}_{\mathrm{fin}}(\mathsf{Tree}),
	\]
	interpreted as recodings, reweightings, or alternative admissibility/selection conventions applied at fixed \(DB_t\).
	
	\paragraph{Examples of methodological transformations.}
	Re-coding changes thresholds or coding schemes for attitudes and attributes while keeping \(DB_t\) fixed. Re-weighting modifies the weights \(w_k\) in the tree/bundle distance \(d_t\), thereby changing the induced topology \(\tau_t\) and the associated robustness notion. Admissibility adjustment tightens or relaxes filters inside \(\mathrm{Gen}_{t,\lambda}\), changing \(\mathcal{T}(DB_t)\) as operationalized under \(\lambda\). Selection adjustment alters \(\mathrm{Sel}_{t,\lambda}\) to prioritize different salience, plausibility, or coverage criteria, including alternative decision rules under epistemic uncertainty when these rules enter admissibility or evaluation.
	
	\paragraph{Robustness under methodological variation.}
	A continuous parameter path \(\lambda\in[0,1]\) induces a deformation family of bundles \(\{\mathcal{B}_{t,\lambda}\}_{\lambda}\). Robust conclusions correspond to statements that are invariant (or approximately invariant) under sufficiently small changes in \(\lambda\) or within sufficiently small \(d_t\)-neighborhoods (Section~\ref{sec:evaluation}).
	
	\subsection{Interaction of temporal and methodological dynamics}
	\label{subsec:interaction-temporal-methodological}
	
	\paragraph{Non-commutation.}
	Temporal updating and methodological transformation do not commute. A comparison across stages requires a transport map \(\Pi_{t\to t+1}\) from time-\(t\) bundles to time-\(t+1\) bundles (or to a common representation space), induced by the chosen encoding of trees. With \(\mathcal{B}_t=\mathrm{Sel}_t(\mathrm{Gen}_t(DB_t))\), non-commutation can be expressed as
	\[
	\mathrm{Sel}_{t+1}\bigl(\mathrm{Gen}_{t+1}(\mathcal{U}_t(DB_t))\bigr)
	\;\neq\;
	\Pi_{t\to t+1}\!\Bigl(\Phi_{\lambda}\bigl(\mathrm{Sel}_{t}(\mathrm{Gen}_{t}(DB_t))\bigr)\Bigr),
	\]
	because \(\mathcal{U}_t\) changes the assessment state and therefore changes the admissible scenario space itself, whereas \(\Phi_{\lambda}\) varies modeling choices within a fixed assessment state.
	
	\paragraph{Approximate commutation as an optional stability criterion.}
	Let \(d_t\) denote the stage-indexed tree distance induced by \(\mathrm{Enc}_t\) and the corresponding weights \(w_k\). For finite bundles \(\mathcal{B},\mathcal{B}'\subseteq \mathsf{Tree}\), define the induced Hausdorff-type distance
	\[
	d^{\mathcal{B}}_t(\mathcal{B},\mathcal{B}')
	\coloneqq
	\max\Bigl\{
	\max_{T\in \mathcal{B}}\min_{T'\in \mathcal{B}'} d_t(T,T'),\;
	\max_{T'\in \mathcal{B}'}\min_{T\in \mathcal{B}} d_t(T,T')
	\Bigr\}.
	\]
	When assessment states change gradually, a stability regime satisfies
	\[
	d^{\mathcal{B}}_{t+1}\!\Bigl(
	\mathrm{Sel}_{t+1}\bigl(\mathrm{Gen}_{t+1}(\mathcal{U}_t(DB_t))\bigr),\;
	\Pi_{t\to t+1}\!\bigl(\Phi_{\lambda}(\mathcal{B}_t)\bigr)
	\Bigr)
	\ \text{small}.
	\]

	\paragraph{Worked example: non-commutativity of \(\mathcal{U}_t\) and \(\Phi_\lambda\).}
	\label{par:noncomm-example}
	The following minimal example makes the non-commutation claim concrete.

	\medskip
	\noindent
	\textbf{Initial state \(DB_t\).}
	Two actors: state \(a\) and opposition \(b\).
	\begin{itemize}[leftmargin=1.5em, itemsep=2pt]
		\item \(\mathrm{aim}(b,\, \text{topple regime}) = 3\) (high).
		\item \(\mathrm{believe}(a,\, b\text{ is moderate}) = \mathrm{True}\).
		\item Option \(o_b\) (mobilise supporters) available to \(b\);
			admissibility threshold for inclusion: salience \(\geq 2\).
		\item Salience of \(o_b\) under \(DB_t\): \(\mathrm{sal}(o_b, DB_t) = 2.5\).
			Hence \(o_b\) is included in trees generated from \(DB_t\).
	\end{itemize}

	The canonical bundle at stage \(t\) is \(\mathcal{B}_t = \mathrm{Sel}_t(\mathrm{Gen}_t(DB_t))\),
	containing trees in which \(b\) may execute \(o_b\).

	\medskip
	\noindent
	\textbf{Update \(\mathcal{U}_t\).}
	New intelligence: \(a\) learns that \(b\) is radical, not moderate.
	The update changes:
	\begin{itemize}[leftmargin=1.5em, itemsep=2pt]
		\item \(\mathrm{believe}(a,\, b\text{ is moderate}) = \mathrm{False}\);\quad
			\(\mathrm{believe}(a,\, b\text{ is radical}) = \mathrm{True}\).
		\item Attribute \(\mathrm{threat}(b)\) increases from level~2 to level~4,
			triggering a new option \(o_a\) (preemptive crackdown) for \(a\),
			previously below the feasibility threshold.
	\end{itemize}

	Result: \(DB_{t+1} = \mathcal{U}_t(DB_t)\) has a strictly larger option set
	(\(o_a\) now feasible), hence \(\mathrm{Gen}_{t+1}(DB_{t+1}) \supsetneq\)
	any transport of \(\mathrm{Gen}_t(DB_t)\) that does not include \(o_a\).

	\medskip
	\noindent
	\textbf{Methodological transformation \(\Phi_\lambda\) at fixed \(DB_t\).}
	Lower the salience threshold from 2 to 1.5.
	Under this relaxed threshold, an additional option \(o_b'\) (online campaign,
	salience 1.8) enters the admissible tree set.
	The transformed bundle \(\Phi_\lambda(\mathcal{B}_t)\) contains trees with \(o_b'\)
	but still does \emph{not} contain \(o_a\), since \(o_a\)'s
	feasibility depends on \(\mathrm{threat}(b) \geq 3\), which holds only in
	\(DB_{t+1}\), not in \(DB_t\).

	\medskip
	\noindent
	\textbf{Non-commutativity.}
	\begin{align*}
		\text{Path 1 (update then recode):} &
		& \mathcal{B}_{t+1}^{(1)} = \Phi_\lambda\bigl(\mathrm{Sel}_{t+1}(\mathrm{Gen}_{t+1}(DB_{t+1}))\bigr) \\
		& & \quad \text{contains } o_a \text{ and } o_b'. \\[4pt]
		\text{Path 2 (recode then update):} &
		& \mathcal{B}_{t+1}^{(2)} = \Pi_{t\to t+1}\!\bigl(\Phi_\lambda(\mathcal{B}_t)\bigr) \\
		& & \quad \text{contains } o_b' \text{ but \emph{not} } o_a.
	\end{align*}
	Since \(o_a \notin \mathcal{B}_{t+1}^{(2)}\) but \(o_a \in \mathcal{B}_{t+1}^{(1)}\), the two paths
	yield structurally different bundles:
	\(d^{\mathcal{B}}_{t+1}(\mathcal{B}_{t+1}^{(1)}, \mathcal{B}_{t+1}^{(2)}) > 0\).

	\medskip
	\noindent
	\textbf{Analytical implication.}
	When assessing robustness across two stages, the order of operations matters.
	Applying a methodological transformation \emph{before} updating the database
	can suppress options that become feasible only after the update, leading to
	systematically incomplete scenario bundles.
	The correct protocol for longitudinal robustness analysis is:
	update \(DB_t \xrightarrow{\mathcal{U}_t} DB_{t+1}\) first, then apply
	\(\Phi_\lambda\) within the updated state.

	\subsection{Theoretical foundations}
	\label{subsec:dynamics-foundation}
	
	Dynamic and evolutionary game theory, nonlinear dynamics, and complex adaptive systems study structural change in interactive systems. SBA's dynamics separates assessment-state updates \(\mathcal{U}_t\) from within-state methodological variation, thereby distinguishing changes in the modeled situation from changes in encoding and inference conventions.
	
	\paragraph{Dynamic and evolutionary game theory.}
	Evolutionary stability \citep{smith1982}, learning dynamics \citep{fudenberg1998}, and adaptive play \citep{young1998} model sequences of strategic adjustment under changing beliefs and information. Assessment-state updates in SBA correspond to revisions of the informational and structural basis of strategic reasoning, while within-state methodological variation corresponds to alternative model instantiations at fixed informational content.
	
	\paragraph{Nonlinear dynamics and attractor theory.}
	Representing scenarios as points in a structured space connects SBA to nonlinear dynamical systems \citep{strogatz2015}. Bifurcations correspond to qualitative shifts in admissible scenario structure, and attractors correspond to self-reinforcing configurations of coalitions, norms, or belief states.
	
	\paragraph{Complex adaptive systems.}
	Feedback between database updates and scenario generation parallels the logic of complex adaptive systems \citep{holland1992, arthur1999}. Macro-level patterns emerge from micro-level updates without requiring equilibrium as a stationary endpoint.
	
	\paragraph{Epistemic and narrative dynamics.}
	Bundle transformations reflect shifts in collective interpretation and in the
	representation of evidence. Narrative evolution \citep{rosenthal1993} and epistemic
	network dynamics \citep{friedkin2011, battiston2020} support a view of scenario
	change as structured belief updating within an evolving interpretive landscape.

	\section{Scenario Evaluation and Comparison}
	\label{sec:evaluation}
	
	Scenario evaluation adds normative orientation to the admissible trees and bundles generated from a fixed assessment state \(DB_t\).
	The database\index{Database Layer} fixes the relevant actors\index{Actors}, coalitions\index{Coalitions}, attributes\index{Attributes}, attitudes\index{Attitudes}, relations\index{Relations}, options\index{Options}, and events\index{Events}; the tree structure fixes the admissible developments.
	Evaluation then makes explicit the aims, preferences, and criteria used by actors, coalitions, and, when required, external observers.
	
	\paragraph{Assessment-state discipline.}
	Evaluation is performed conditional on a fixed assessment state \(DB_t\). Temporal evolution is represented by assessment-state updates \(\mathcal{U}_t\colon DB_t \to DB_{t+1}\) (Section~\ref{sec:scenario-dynamics}). All comparisons range over terminal scenarios within a given admissible tree or bundle generated from \(DB_t\).

	\subsection{Conceptual role}
	\label{subsec:eval-conceptual}
	
	Terminal nodes in a scenario tree\index{Scenario Trees} represent completed configurations of actors\index{Actors}, relations\index{Relations}, attributes\index{Attributes}, and realized events\index{Events}, resulting from a particular sequence of options\index{Options} and exogenous developments. Evaluation induces an ordering over terminal scenarios that supports (i) actor- and coalition-sensitive guidance and (ii) robustness claims under methodological variation (Section~\ref{sec:topology}).
	
	\paragraph{Preference relations.}
	Let \(S\) be a set of terminal scenarios under consideration. For each actor \(a\in A_t\), a weak preference relation \(\succeq_a\) on \(S\) satisfies reflexivity and transitivity. The corresponding strict preference is
	\[
	s \succ_a s' \;\;\Longleftrightarrow\;\; (s \succeq_a s') \wedge \neg(s' \succeq_a s).
	\]
	When a numerical representation is convenient, an ordinal utility \(u_a\colon S\to \mathbb{R}\) represents \(\succeq_a\) via \(s\succeq_a s' \Leftrightarrow u_a(s)\ge u_a(s')\). The representation is optional and depends on measurement commitments.
	
	Beyond actor-level evaluation, SBA admits coalition-level and system-level criteria such as stability, sustainability, or ethical impact, provided the criteria are made explicit as part of the assessment.
	
	\subsection{Evaluation structure}
	\label{subsec:eval-structure}
	
	Let \(T\in T_t\) be an admissible scenario tree generated from \(DB_t\), and let \(S(T)\) denote its set of terminal scenarios. For a scenario bundle \(\mathcal{B}\subseteq T_t\), define the terminal-scenario set
	\[
	S(\mathcal{B}) \coloneqq \bigcup_{T\in \mathcal{B}} S(T),
	\qquad
	S \coloneqq S(\mathcal{B}).
	\]
	
	\paragraph{Actor-level evaluation.}
	Each actor \(a\in A_t\) specifies either (i) a preference relation \(\succeq_a\) on \(S\) or (ii) an ordinal utility \(u_a\colon S\to\mathbb{R}\) representing \(\succeq_a\).
	
	\paragraph{Coalition-level evaluation.}
	For each coalition \(X\in C_t\), a collective evaluation \(U_X\colon S\to\mathbb{R}\) (or an induced preorder \(\succeq_X\)) is derived by an aggregation scheme
	\[
	U_X(s) \;=\; \mathrm{Agg}_X\bigl(\{u_a(s)\}_{a\in X}\bigr),
	\]
	subject to coalition-internal coherence constraints specified in the database layer (e.g., admissible weight vectors, veto rules, or unanimity requirements).
	
	\paragraph{System-level evaluation.}\mbox{}\\
	An external observer may define a meta-evaluation \(\mathcal{V}\) capturing systemic risk, welfare, or institutional desirability. Its type is \(\mathcal{V}\colon S\to\mathbb{R}\). This layer is optional and must be declared as an explicit normative commitment rather than an implicit default.
	
\begin{figure}[htbp]
		\centering
		\resizebox{\textwidth}{!}{%
\begin{tikzpicture}[
			node distance=1.5cm and 2cm,
			sba_level/.style={rectangle, draw=blue!70!black, fill=blue!5, thick,
				rounded corners, minimum width=6cm, minimum height=1.1cm,
				font=\small, align=center},
			sba_arrow/.style={-Latex, blue!70!black, line width=1.2pt},
			sba_in/.style={Latex-, teal!60, dashed, thick},
			sba_box/.style={draw=gray!45, thick, rounded corners, fill=gray!2,
				font=\scriptsize, align=left, inner sep=6pt, text width=4.5cm}
			]
			
			\node[sba_level] (L1) {\textbf{Actor-Level (Micro)}\\ \(u_a\) or \(\succeq_a\) for each \(a \in A_t\)};
			
			\node[sba_level, above=of L1] (L2) {\textbf{Coalition-Level (Meso)}\\ \(U_X\) or \(\succeq_X\) for each \(X \in C_t\)};
			
			\node[sba_level, above=of L2] (L3) {\textbf{System-Level (Macro)}\\ \(\mathcal{V}\) (Institutional / Analyst)};
			
			\draw[sba_arrow] (L1) -- (L2) node[midway, right, font=\tiny] {Aggregation \(\mathrm{Agg}_X\)};
			\draw[sba_arrow] (L2) -- (L3) node[midway, right, font=\tiny] {Meta-Criteria};
			
			\node[sba_box, right=2.5cm of L2, text width=4cm] (Sset) {
				\textbf{Terminal Scenarios (Input)}\\ 
				\(S(\mathcal{B}) = \bigcup_{T \in \mathcal{B}} S(T)\)\\
				\vspace{1mm} \hrule \vspace{1mm}
				Dataset of all admissible outcomes within the bundle \(\mathcal{B}\).
			};
			
			\draw[sba_in] (L1.east) -- ++(1.2,0) |- (Sset.west);
			\draw[sba_in] (L2.east) -- (Sset.west) node[midway, above, font=\tiny, text=teal] {evaluates};
			\draw[sba_in] (L3.east) -- ++(1.2,0) |- (Sset.west);
			
			\node[draw=gray!30, fill=white, rounded corners=2pt, 
			below right=0.5cm and 0cm of Sset.south west, 
			text width=4cm, align=left] {
				\textbf{Legend}\\
				\tikz[baseline=-0.5ex]\draw[sba_arrow, line width=1pt] (0,0)--(0.5,0); \scriptsize Evaluation Flow\\
				\tikz[baseline=-0.5ex]\draw[sba_in, line width=1pt] (0,0)--(0.5,0); \scriptsize Scenario Input Data
			};
			
			\draw[->, gray!40, line width=2pt] ([xshift=-1cm]L1.west) -- ([xshift=-1cm]L3.west) 
			node[midway, above, rotate=90, font=\tiny, text=gray] {Strategic Abstraction};
			
		\end{tikzpicture}%
}%
		\caption{Hierarchical evaluation layers induced by a fixed assessment state \(DB_t\) and a bundle \(\mathcal{B}\).}
		\label{fig:evaluation_structure_new}
	\end{figure}
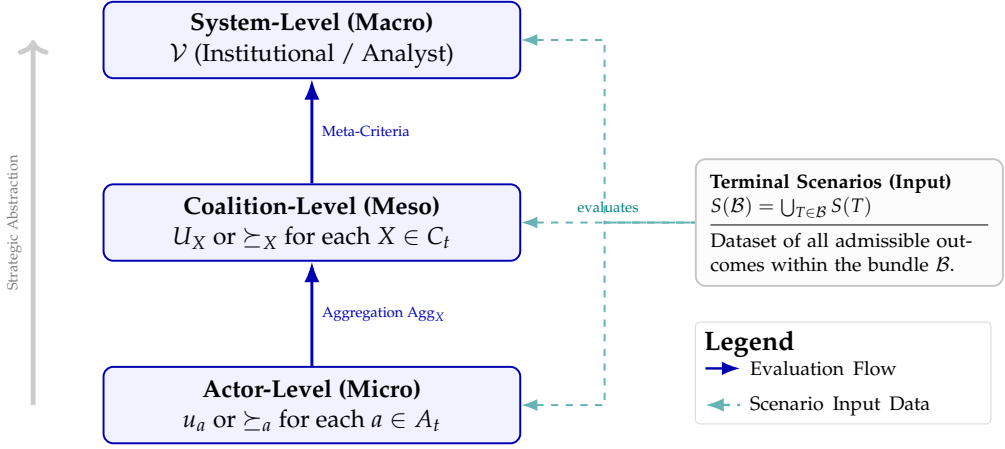

	\paragraph{Evaluation package.}
	Figure~\ref{fig:evaluation_structure_new} shows the hierarchical evaluation layers induced by \(DB_t\) and the bundle \(\mathcal{B}_t\).
	The evaluation state associated with \(\mathcal{B}\) can be summarized as
	\[
	\mathcal{E}_t(\mathcal{B})=
	\big\langle
	S(\mathcal{B}),\,
	\{u_a\}_{a\in A_t},\,
	\{U_X\}_{X\in C_t},\,
	\mathcal{V}
	\big\rangle.
	\]
	If the scenario trees are interpreted as decision/game trees with actor-indexed option nodes, \(\mathcal{E}_t(\mathcal{B})\) supports tree-level reasoning such as backward induction\index{Tree Analysis!MRP (Most Rational Path)} and the extraction of the Most Rational Path (MRP) under the chosen evaluation scheme.
	
	\subsection{Comparison of scenarios}
	\label{subsec:eval-comparison}
	
	Scenario comparison identifies dominance relations, trade-offs, and stable patterns across \(S(\mathcal{B})\).
	
	\paragraph{Actor- and coalition-based comparison.}
	For each \(a\in A_t\) (or \(X\in C_t\)), the evaluation induces an order or preorder on \(S(\mathcal{B})\). Disagreements across \(\{ \succeq_a \}\) indicate conflict potential; alignments indicate coalition plausibility under the fixed \(DB_t\).
	
	\paragraph{System-based comparison.}
	System-level criteria induce a partial order \(\preceq_{\mathrm{sys}}\) on \(S(\mathcal{B})\) when \(\mathcal{V}\) is defined. When multiple system criteria coexist, \(\preceq_{\mathrm{sys}}\) is in standard applications a Pareto-type order on a vector of indicators.
	
\paragraph{Pareto frontier.}\index{Decision Theory!Pareto Frontier}
Assume actor-level utilities \(u_a\) are available (ordinal scales suffice). Pareto dominance on the terminal scenario set \(S(\mathcal{B})\) is defined by:
\begin{equation}
	\begin{aligned}
		s' \succeq_{\mathrm{P}} s \quad &\Longleftrightarrow \quad \forall a \in A_t: u_a(s') \ge u_a(s) \\
		s' \succ_{\mathrm{P}} s \quad &\Longleftrightarrow \quad (s' \succeq_{\mathrm{P}} s) \;\wedge\; (\exists b \in A_t: u_b(s') > u_b(s))
	\end{aligned}
\end{equation}
The Pareto frontier consists of the set of non-dominated scenarios:
\begin{equation}
	S^{\ast} = \{ s \in S(\mathcal{B}) \mid \neg\exists s' \in S(\mathcal{B}): s' \succ_{\mathrm{P}} s \}
\end{equation}
	
	\paragraph{Stability criteria (optional).}
	When terminal scenarios encode strategic profiles or induced normal-form outcomes, stability can be assessed by Nash or coalition-stability conditions at terminal positions. The criterion must be chosen consistently with the tree semantics used in the preceding layers.
	
	\paragraph{Robustness link to topology.}
	Let \(\Psi\) be an evaluation output functional (e.g., \(\Psi(\mathcal{B})=S^\ast\), or \(\Psi(\mathcal{B})\) an ordered list). Robust conclusions correspond to local stability of \(\Psi\) under methodological perturbations: \(\Psi\) remains constant (or changes only within a declared tolerance) for bundles in a sufficiently small neighborhood induced by the pseudo-metric/topology of Section~\ref{sec:topology}.
	
	\subsection{Evaluation heuristics and visualization}
	\label{subsec:eval-visualization}
	
	Practical SBA applications require transparent and interpretable outputs. Common representations include:
	\begin{itemize}[leftmargin=1.5em,itemsep=3pt]
		\item payoff tables or evaluation matrices summarizing \(\{u_a(s)\}_{a,s}\),
		\item multi-criteria profiles (e.g., radar/spider plots) for vector-valued indicators,
		\item dominance graphs visualizing induced partial orders on \(S(\mathcal{B})\),
	\item heatmaps for preference alignment\index{Relations!Alignment / Affinity} and coalition stability.
	\end{itemize}\paragraph{From \(DB_t\) to evaluation matrices.}
Interpretable evaluation outputs arise from an explicit mapping from the assessment state \(DB_t\) and a selected bundle \(\mathcal{B}\) to a finite family of terminal scenarios \(S(\mathcal{B})\) and numerical or ordinal assessments.
A convenient representation is an evaluation matrix
\[
M_t(\mathcal{B})\;=\;\bigl(M_{a,s}\bigr)_{a\in A_t,\; s\in S(\mathcal{B})},
\qquad
M_{a,s}\coloneqq u_a(s),
\]
where \(u_a\) is a declared representation of the actor's terminal order \(\succeq_a\) on \(S(\mathcal{B})\).
When evaluation is ordinal only, \(M_t(\mathcal{B})\) is replaced by a relation table encoding \(\succeq_a\) or its strict part \(\succ_a\); numerical tables then serve only as a visualization device.

\paragraph{Filling payoff tables.}
Operational SBA requires that each entry \(u_a(s)\) be traceable to declared ingredients.
Three constructions cover most practical cases.
\begin{itemize}[leftmargin=1.5em,itemsep=2pt]
	\item \textbf{Attribute-based scoring.} Fix a set of relevant attributes \(J_a\) and nonnegative weights \((w_{a,k})_{k\in J_a}\) with \(\sum_{k\in J_a} w_{a,k}=1\).
	Let \(\ell_s\in L\) be the terminal position corresponding to scenario \(s\), and let \(k(\ell_s)\) be the terminal value of attribute \(k\) under the path leading to \(\ell_s\).
	Define a monotone normalization map \(\mathrm{norm}_{a,k}\) (thresholds and units declared once), and set
	\[
	u_a(s)\coloneqq \sum_{k\in J_a} w_{a,k}\,\mathrm{norm}_{a,k}\!\bigl(k(\ell_s)\bigr).
	\]
	\item \textbf{Rule-based ordinal assessment.} Encode domain rules as constraints on comparisons (e.g.\ ``\(s\) is unacceptable if civilian harm exceeds a threshold'' or ``\(s\) outranks \(s'\) if legitimacy increases and costs do not worsen'').
	The induced \(\succeq_a\) can be visualized by any numerical embedding consistent with the order, without committing to cardinal utilities.
	\item \textbf{Game- or mechanism-induced payoffs.} If terminal scenarios encode induced normal-form outcomes or payoff vectors (e.g.\ from an extensive-form interpretation), then \(u_a(s)\) is read off from the terminal outcome representation, with the semantics fixed by the chosen solution concept.
\end{itemize}
In all three cases, missing information should be represented explicitly (unknown entries or intervals) rather than imputed silently; this keeps dominance and robustness diagnostics meaningful.

\paragraph{Dominance graphs.}
Dominance graphs provide an immediate diagnostic of ``obviously inferior'' terminal scenarios under a declared criterion.
For a fixed dominance relation \(\succ\) on \(S(\mathcal{B})\) (actor-specific, Pareto, or coalition-based), define the directed graph
\[
G^{\mathrm{dom}}(\mathcal{B})\coloneqq \langle S(\mathcal{B}),\,E^{\mathrm{dom}}\rangle,
\qquad
(s\to s')\in E^{\mathrm{dom}}\;\Longleftrightarrow\; s\succ s'.
\]
Acyclicity is expected when \(\succ\) is a strict partial order; cycles indicate either inconsistent evaluations or an encoding error in the dominance predicate and should trigger inspection of the underlying table entries.
Nodes of \(G^{\mathrm{dom}}(\mathcal{B})\) with no incoming edges represent non-dominated scenarios and support rapid screening before more refined comparisons.

\paragraph{Sensitivity and robustness displays.}
Methodological perturbations (e.g.\ alternative weights, alternative feasibility filters, alternative relation aggregation rules) can be summarized by plotting the stability of selected outputs \(\Psi(\mathcal{B})\) as a function of the declared parameter choice \(\lambda\).
A practitioner-relevant report states which conclusions remain invariant across a declared neighborhood in the bundle space and which conclusions change under small perturbations.

\paragraph{Exchange formats.}
Reproducible practice benefits from exporting evaluation outputs in simple machine-readable formats.
A minimal export contains: unique scenario identifiers, a path or leaf identifier, the evaluation table entries \(u_a(s)\) (or the ordinal comparison data), and provenance pointers for each nontrivial entry.

\paragraph{Worked micro-example.}
A minimal bundle already yields the standard artefacts (tables and dominance structure) once a concrete evaluation mapping is declared.
Fix a stage \(t\) and a bundle \(\mathcal{B}_t\) with scenario set \(S(\mathcal{B}_t)=\{s_1,s_2\}\).
Suppose two actors \(a,b\in A_t\) admit numerical representations \(u_a,u_b\colon S(\mathcal{B}_t)\to\mathbb{R}\) of their scenario-wise preorders.
A simple evaluation matrix is:
\begin{table}[htbp]
\centering
\footnotesize
\caption[Scenario evaluation matrix]{Scenario evaluation matrix \(\{u_a(s),u_b(s)\}\) for two scenarios.}
\label{tab:scenario-eval-matrix}
\begin{tabular}{lcc}
\toprule
& \(u_a(\cdot)\) & \(u_b(\cdot)\) \\
\midrule
\(s_1\) & \(4\) & \(3\) \\
\(s_2\) & \(2\) & \(1\) \\
\bottomrule
\end{tabular}
\end{table}
When scenarios correspond to terminal leaves, a declared leaf scoring map (e.g., \(\mathrm{Val}(a,\ell)\)) induces \(u_a(s)\) by composition with the leaf selected by \(s\); otherwise \(u_a(s)\) is a direct assessment attached to \(s\) with provenance and confidence tags.
A dominance graph on scenarios can be generated from a declared dominance criterion.
For instance, Pareto dominance on \(S(\mathcal{B}_t)\) is defined by \(s\succ_{\mathrm{P}} s'\) iff \(u_x(s)\ge u_x(s')\) for all \(x\in\{a,b\}\) and the inequality is strict for at least one actor.
Table~\ref{tab:scenario-eval-matrix} induces \(s_1\succ_{\mathrm{P}} s_2\), hence a directed edge \(s_1\to s_2\).
Dominated scenarios can be deprioritized or removed as a first-pass filter; if the dominance graph has no edges (trade-offs), subsequent analysis requires an explicit aggregation rule (e.g., weighted sums, lexicographic priorities, or coalition-level \(U_X\)).
Cycles (when a graph is built from pairwise ordinal judgements rather than from a transitive numerical representation) are diagnostic of inconsistent elicitation or of mixed criteria treated as commensurable without a declared aggregation rule.

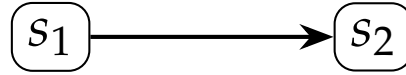
\begin{figure}[htbp]
\centering
\resizebox{0.40\textwidth}{!}{%
\begin{tikzpicture}[>=Stealth, node distance=22mm]
\node (s1) [draw, rounded corners, inner sep=3pt] {\(s_1\)};
\node (s2) [draw, rounded corners, inner sep=3pt, right of=s1] {\(s_2\)};
\draw[->, thick] (s1) -- (s2);
\end{tikzpicture}%
}
\caption[Dominance graph]{Dominance graph induced by Table~\ref{tab:scenario-eval-matrix}: edge \(s_1\to s_2\) denotes Pareto dominance \(s_1\succ_{\mathrm{P}} s_2\).}
\label{fig:dominance-micro}
\end{figure}

	\subsection{Theoretical foundations}
	\label{subsec:eval-foundation}
	
	The evaluation layer connects SBA to decision theory, game theory, and social choice.
	
	\paragraph{Game-theoretic evaluation.}
	When scenario trees are interpreted as extensive-form games, terminal scenarios correspond to terminal histories with induced payoffs \citep{osborne1994,selten1975}. Backward induction and related solution concepts determine rational-path selections, while Pareto and coalition analyses extend evaluation to collective and systemic perspectives.
	
	\paragraph{Decision and value theory.}
	Ordinal and cardinal evaluation rely on preference representation results \citep{arrow1951,sen1977}. Aggregation links to social choice and welfare economics \citep{harsanyi1955,mas1995} under explicit admissibility assumptions on the aggregation rule.
	
	\paragraph{Multi-criteria and scenario evaluation.}
	Cross-scenario comparison uses tools from multi-criteria decision analysis \citep{roy1996,belton2002} and robust scenario ranking \citep{vannotten2003}, combining qualitative judgment with quantitative indicators under declared transparency constraints.

\subsection{Illustrative Example: Border-Incident Evaluation with Transparent Criteria}
\label{subsec:eval-example}

	A minimal conflict scenario illustrates how terminal evaluations are constructed from explicit criteria under a fixed assessment state \(DB_t\). Let \(A_t=\{a,b,m\}\) where \(a\) and \(b\) represent two states and \(m\) denotes an external mediator or international organization. Let \(X=\{a,m\}\in C_t\) denote an alliance-coalition.

\paragraph{Event and Options.}
	An exogenous border incident \(e_1\in E_t\) initiates a two-step tree. State \(a\) chooses an option \(o_a \in \{\mathrm{R}, \mathrm{S}\}\) (retaliate or show restraint), and state \(b\) responds with \(o_b \in \{\mathrm{E}, \mathrm{W}\}\) (escalate or withdraw). The resulting terminal scenarios are defined as:
\begin{equation}
	s_1=(\mathrm{R},\mathrm{E}), \quad s_2=(\mathrm{R},\mathrm{W}), \quad s_3=(\mathrm{S},\mathrm{E}), \quad s_4=(\mathrm{S},\mathrm{W}).
\end{equation}

\paragraph{Transparent Criteria.}
Evaluation utilizes three interpretable indicators, coded at \(DB_t\), and a legitimacy indicator \(p(s)\):
\begin{align*}
	h(s) &\in [0,10] \quad \text{(harm / escalation cost)} \\
	g_i(s) &\in [0,10] \quad \text{(aim attainment for actor \(i \in \{a, b\}\))} \\
	p(s) &\in [0,10] \quad \text{(legitimacy indicator for \(m\))}
\end{align*}
Actor utilities are then derived as linear combinations:
\begin{equation}
	\begin{aligned}
		u_a(s) &= g_a(s) - 0.5 \cdot h(s) \\
		u_b(s) &= g_b(s) - 0.3 \cdot h(s) \\
		u_m(s) &= p(s) - h(s)
	\end{aligned}
\end{equation}
Coalition evaluation follows a declared aggregation rule:
\begin{equation}
	U_X(s) = 0.6 \cdot u_a(s) + 0.4 \cdot u_m(s).
\end{equation}
Table~\ref{tab:border_incident_eval} reports the resulting terminal criteria and induced evaluation values for all four scenarios.

\begin{table}[htbp]
	\centering
	\small
	\begin{tabular}{lcccccccc}
		\toprule
		\textbf{Scenario} & $h$ & $g_a$ & $g_b$ & $p$ & $u_a$ & $u_b$ & $u_m$ & $U_X$ \\
		\midrule
		$s_1=(\mathrm{R},\mathrm{E})$ & 9 & 4 & 4 & 2 & $-0.5$ & $1.3$ & $-7$ & $-3.1$ \\
		$s_2=(\mathrm{R},\mathrm{W})$ & 3 & 8 & 2 & 6 & $6.5$  & $1.1$ & $3$  & $5.1$ \\
		$s_3=(\mathrm{S},\mathrm{E})$ & 6 & 2 & 8 & 3 & $-1$   & $6.2$ & $-3$ & $-1.8$ \\
		$s_4=(\mathrm{S},\mathrm{W})$ & 1 & 6 & 5 & 9 & $5.5$  & $4.7$ & $8$  & $6.5$ \\
		\bottomrule
	\end{tabular}
	\caption{Terminal criteria and induced evaluations at fixed \(DB_t\).}
	\label{tab:border_incident_eval}
\end{table}

\paragraph{Dominance and Conflict Pattern.}
Scenario \(s_1\) is dominated by \(s_4\) for all three actors. The remaining set \(\{s_2, s_3, s_4\}\) forms a non-trivial trade-off region: \(a\) prefers \(s_2\) over \(s_4\), \(b\) prefers \(s_3\) over \(s_4\), while \(m\) and the coalition \(X=\{a,m\}\) prefer \(s_4\). The induced Pareto frontier thus depends on whether dominance is evaluated actor-wise \((a,b,m)\) or coalition-wise \((X,b)\).

\paragraph{Expected-Utility Threshold.}
If \(a\) assigns a probability \(q = \Pr(\mathrm{W} \mid \mathrm{R})\) and \(r = \Pr(\mathrm{W} \mid \mathrm{S})\) to \(b\)'s withdrawal conditional on \(a\)'s choice, the expected utilities are:
\begin{align}
	\mathbb{E}[u_a \mid \mathrm{R}] &= q \cdot u_a(s_2) + (1-q) \cdot u_a(s_1) = 7q - 0.5 \label{eq:exp_R} \\
	\mathbb{E}[u_a \mid \mathrm{S}] &= r \cdot u_a(s_4) + (1-r) \cdot u_a(s_3) = 6.5r - 1 \label{eq:exp_S}
\end{align}
Retaliation is optimal for \(a\) if and only if \(\mathbb{E}[u_a \mid \mathrm{R}] \ge \mathbb{E}[u_a \mid \mathrm{S}]\). This yields the threshold condition:
\begin{equation}
	7q - 0.5 \ge 6.5r - 1 \quad \iff \quad q \ge \frac{6.5r - 0.5}{7}.
\end{equation}
This threshold illustrates how evaluation interacts with tree-level reasoning while maintaining the discipline of the assessment-state.

\section{Empirical Data Sources and Validation}
\label{sec:empirical}

Scenario Bundle Analysis\index{Scenario Bundle Analysis} (SBA) relies on empirical information about actors\index{Actors},
attributes\index{Attributes}, attitudes\index{Attitudes}, relations\index{Relations}, options\index{Options}, and events\index{Events}.
Empirical grounding\index{Empirical grounding} is implemented by mapping heterogeneous sources into an assessment state \(DB_t\)\index{Assessment state}
and by controlling how new evidence triggers updates.
\paragraph{Targets of empirical grounding.}
Empirical work in SBA addresses three targets:
(i) construction of the assessment state \(DB_t\) (data \(\to DB_t\));
(ii) validation of the update operator \(\mathcal{U}_t\)\index{Update operator} (how \(DB_t\) changes under new evidence);
(iii) validation of methodological mappings (how \(DB_t\) induces admissible trees and bundles).

\subsection{Typology of data sources}
\label{subsec:data-typology}

Empirical material is grouped into five categories, each targeting specific database components.

\begin{enumerate}[label=(\alph*),leftmargin=2em,itemsep=4pt]
	\item \textbf{Structured event databases.}
	Repositories such as ICEWS\index{ICEWS} (Integrated Crisis Early Warning System),
	GDELT\index{GDELT} (Global Database of Events, Language, and Tone),
	ACLED\index{ACLED} (Armed Conflict Location and Event Data),
	and EM-DAT\index{EM-DAT} (Emergency Events Database) provide time-stamped, geocoded event\index{Events} data.
	These sources initialize and track the event layer \(E_t\subseteq E\) recorded in \(DB_t\).
	
	\item \textbf{Institutional and economic indicators.}
	Datasets from the World Bank\index{World Bank}, IMF\index{IMF}, OECD\index{OECD}, or national statistical offices quantify macroeconomic
	and institutional attributes (GDP, debt, inflation, governance indices).
	They support the attribute layer \(\mathrm{Attr}_t\)\index{Attributes} and operational constraints relevant for options\index{Options}.
	
	\item \textbf{Expert interviews and elite surveys.}
	Qualitative elicitation\index{Expert elicitation} provides information on attitudes\index{Attitudes} and relations\index{Relations} that are not directly observable.
	Delphi\index{Delphi method} and Q-sort\index{Q-sort} procedures support intersubjective comparability and explicit uncertainty annotation.
	
	\item \textbf{Media and open-source intelligence (OSINT).}
	Open-source intelligence (OSINT)\index{OSINT}\index{Open-source intelligence} and curated media corpora provide information on framing\index{Framing},
	rhetoric, and communication\index{Relations!Information / Communication}.
	These sources complement the attitude\index{Attitudes} and option\index{Options} layers, especially for signaling and discursive actions.
	
	\item \textbf{Social and digital traces.}
	Digital platforms (Twitter/X\index{Twitter/X}, Reddit\index{Reddit}, Telegram\index{Telegram}) provide high-frequency indicators of sentiment\index{Sentiment analysis},
	stance\index{Stance detection}, and mobilization.
	Text-to-variable pipelines map such material to likelihood scores \(\ell\)\index{Events!Likelihood}, preference intensity \(\pi\)\index{Preferences!Intensity}\index{Preferences},
	and temporal horizon \(\vartheta\)\index{Time and Dynamics!Temporal Horizon}.
\end{enumerate}

\subsection{Source valuation and bias handling}
\label{subsec:data-valuation}

Source quality\index{Data quality} and bias\index{Bias} are tracked at the level of both input items and derived database variables, so that downstream scenario generation can distinguish strong evidence from low-confidence or skewed inputs.

\paragraph{Source quality metric.}
Each source \(q_i\) is assigned a reliability score \(r_i\)\index{Reliability}, a coverage score \(c_i\)\index{Coverage}, and a temporal-resolution score \(\chi_i\)\index{Temporal resolution}, summarized as
\[
v(q_i)=\langle r_i,c_i,\chi_i\rangle.
\]
Fix nonnegative weights \(w_r,w_c,w_t\) with \(w_r+w_c+w_t=1\). The aggregate quality score is
\[
Q(q_i)=w_r r_i+w_c c_i+w_t \chi_i.
\]
Low-\(Q(q_i)\) sources are excluded, down-weighted, or represented as latent inputs.

\paragraph{Cross-validation and triangulation.}
Overlap checks across event datasets\index{Cross-validation} (e.g.\ ICEWS vs.\ ACLED) identify coding asymmetries and omissions.
Triangulation\index{Triangulation} with expert judgment is used to adjudicate divergent classifications of events or relations before they enter \(DB_t\).

\paragraph{Bias identification.}
Three bias types are tracked:
(i) coverage bias\index{Bias!Coverage bias} (uneven geographical or actor representation);
(ii) framing bias\index{Bias!Framing bias} (ideological or cultural distortion);
(iii) algorithmic bias\index{Bias!Algorithmic bias} (extraction artifacts).
Mitigation combines diversity weighting in \(Q(q_i)\), explicit uncertainty bookkeeping, and documented provenance\index{Data provenance}.

\subsection{Coding and transformation procedures}
\label{subsec:data-coding}

Empirical information is converted into SBA entities by auditable coding rules\index{Coding}.

\paragraph{Entity extraction and alignment.}
Named-entity recognition\index{Named-entity recognition (NER)} (NER) and coreference resolution\index{Coreference resolution} identify actors, coalitions, and locations,
aligning them to unique identifiers in \(A_t\subseteq A\) (the active actor slice recorded in \(DB_t\)).

\paragraph{Verb and action extraction.}
Dependency parsing\index{Dependency parsing} and predicate analysis extract verbal expressions representing attitudes or options.
Results are classified via action-family typologies (Section~\ref{subsubsec:opt-theory}; \citealp{ballmer1981}).

\paragraph{Sentiment and stance coding.}
Polarity and modality markers are mapped to \(\ell,\pi,\vartheta\) under explicit scale conventions and coder guidelines\index{Inter-rater reliability}.

\paragraph{Hybrid strategies.}
Automated extraction (statistical or LLM-based)\index{Large language models}
is combined with expert correction to preserve interpretability and reduce silent drift
\citep{ballmer1981}.

\subsection{Automated extraction and machine assistance}
\label{subsec:data-automation}

Machine assistance is used as a coverage amplifier under explicit provenance control\index{Data provenance}\index{Provenance metadata}.
Each automatically produced record is stored with metadata (source span, model identifier, timestamp, confidence score)\index{Confidence score}.
Such outputs remain defeasible and require expert acceptance criteria.

\paragraph{Target object and interface.}
Automated extraction targets \emph{typed database fragments} rather than prose.
Let \(\widehat{DB}_t\) denote a candidate assessment state constructed from a collection of source items \(\{q_i\}_{i\in I}\).
A record is accepted only if it can be represented as a well-typed entry of one of the declared components of \(DB_t\)
(actors or coalitions, attributes, attitudes, relations, options, or events).
Each record carries a provenance pointer to an explicit source span (document identifier and character offsets) and a trace identifier for the extraction run.

\paragraph{Record template (minimal).}
A practical exchange format is a schema-validated JSON/CSV representation.
The following minimal fields suffice to keep the database auditable while remaining tool-agnostic:
\begin{verbatim}
	{ 
		"kind": "att" | "rel" | "event" | "attr" | "option",
		"subject": "a" | "X", 
		"predicate": "B" | "K" | "Rel" | "Attr" | "Opt",
		"object": "p" | "b" | "e" | "value", 
		"qualifiers": {"ell":..., "pi":..., "vartheta":...},
		"time": "t" | "interval", 
		"confidence": c in [0,1],
		"provenance": {"source": "q_i", "span": "start..end"} 
	}
\end{verbatim}
The field \texttt{time} records empirical time (or an interval) stated or implied by the sources; the database stage index \(t\) is internal versioning of the assessment snapshot.
The field \texttt{predicate} stores the concrete relation/attribute/option symbol name used in the database signature; the placeholders \texttt{Rel/Attr/Opt} indicate the slot rather than introducing new operators.
The template is not a commitment to a particular implementation; it fixes the \emph{minimum} information required for validation and revision.

\paragraph{Reference pipeline (coverage-to-acceptance).}
A reliable extraction protocol decomposes the task into passes with explicit outputs.
\begin{enumerate}[leftmargin=1.5em,itemsep=2pt]
  \item \emph{Segmentation and citation anchoring.} Split each \(q_i\) into extractable spans and attach stable identifiers to all spans.
  \item \emph{Entity inventory.} Propose actors and coalitions and align them to stable identifiers (merge/split decisions are logged).
  \item \emph{Predicate extraction.} Extract candidate attributes, relations, attitudes, options, and events as typed records.
  \item \emph{Constraint checking.} Reject ill-typed records and flag constraint violations (e.g. polarity conflicts, impossible coalition commitments, temporal incoherence).
  \item \emph{Expert acceptance.} Route only flagged items to human review; accept all remaining items subject to random spot checks\index{Spot checking}.
  \item \emph{Commit as update.} Accepted records induce a change set \(\Delta DB_t\); the new state is obtained by applying the declared revision operator to \(DB_t\) with \(\Delta DB_t\) under the declared revision policy\index{Belief revision}. This commit step instantiates \(\mathcal{U}_t\).

\end{enumerate}

\paragraph{Prompt discipline and decomposition (LLM-assisted extraction).}
If large-language models are used\index{Large language models}, prompts should be schema-first.
A stable pattern is: (i) present the allowed record kinds and field types, (ii) provide the relevant local context span, (iii) demand output \emph{only} in the record template.
Coverage increases when extraction is decomposed into narrow passes (entity linking, then attitudes, then relations, then events), because the main error modes differ by kind.
A second pass can be dedicated to contradiction mining: given \(\widehat{DB}_t\), request only (typed) pairs that violate explicit constraints.

\paragraph{Acceptance gates.}
A candidate record is admitted to \(\widehat{DB}_t\) only if all of the following conditions are satisfied:
\begin{itemize}[leftmargin=1.5em,itemsep=2pt]
  \item \emph{Schema validity:} required fields are present and type-correct (no free-form text in predicate slots).
  \item \emph{Entity alignment:} all referenced actors/coalitions/events resolve to identifiers already declared in \(DB_t\) or introduced by the entity inventory pass.
  \item \emph{Provenance completeness:} every record points to an explicit source span; derived records list their parent records.
  \item \emph{Constraint satisfaction:} internal checks (consistency, closure, temporal coherence) do not fail, or else the record is routed to review.
  \item \emph{Calibration tag:} a confidence score is stored, but it never substitutes for provenance or constraints.
\end{itemize}

\paragraph{Common failure modes.}
Automated extraction is predictable in its mistakes; recording error types improves both supervision and prompt design.
Frequent failures include: entity merge/split errors (two actors conflated or one actor duplicated), role drift (actor treated as coalition or event), polarity flips in attitudes, temporal leakage (future information attributed to \(t\)), spurious relations from narrative adjacency, and silent omission of negation or modality markers.

\paragraph{Human supervision and audit queues.}
Human time is reserved for high-impact or high-uncertainty items.
A minimal queue prioritization uses: (i) constraint violations, (ii) records affecting high-centrality actors or coalitions, (iii) records that change feasibility of options, and (iv) disagreements between independent extraction runs.
Accepted corrections are fed back as explicit coding rules and gold fragments for spot-check calibration\index{Inter-rater reliability}.

\paragraph{Reproducibility and drift control.}
Each extraction run is identified by a versioned configuration (prompt set, model identifier, decoding parameters, and constraint set).
Storing these parameters with \(\Delta DB_t\) prevents silent drift: differences in \(DB_t\) can be attributed to either new evidence or to a declared methodological change.
\subsection{Validation, calibration, and updating}
\label{subsec:validation-merged}

Validation\index{Validation} constrains both the internal coherence of \(DB_t\) and its empirical correspondence.

\paragraph{Internal validation.}
Logical and structural integrity checks include:
\begin{itemize}[leftmargin=1.5em,itemsep=2pt]
	\item \emph{Relational coherence}\index{Relations!Coherence}: no cyclic contradictions in authority or dependence graphs;
	\item \emph{Attitudinal consistency}\index{Attitudes!Consistency}: explicit contradictions are absent (e.g.\ no simultaneous \(B_a p\) and \(B_a \neg p\);
	equivalently, no simultaneous \(B_a(p)\) and \(B_a(\neg p)\))\index{Belief consistency};
	\item \emph{Coalitional closure}\index{Coalitions!Coherence}: aggregation assumptions match coalition commitments;
	\item \emph{Tree determinacy}\index{Tree Analysis!Determinacy}: well-defined MRP\index{Tree Analysis!MRP (Most Rational Path)} and MLP\index{Tree Analysis!MLP (Most Likely Path)} for the selected trees.
\end{itemize}

\paragraph{External validation and calibration.}
Calibration\index{Calibration}\index{Model calibration} compares scenario outputs to historical analogues, out-of-sample episodes, or independent expert constructions.
Quantitative parameters \(\theta\) (e.g.\ thresholds or weights) can be adjusted by
\[
\theta_{t+1}=\theta_t+\eta(\hat{\theta}_t-\theta_t),
\]
with learning rate \(\eta\)\index{Learning rate}. The MRP corresponds to subgame-perfect reasoning \citep{selten1975,osborne1994};
the MLP connects to probabilistic inference under subjective likelihood assessments \citep{pearl1988,jensen2007}.

\subsection{Dynamic updating of the database}
\label{subsec:validation-dynamic}

Temporal evolution is represented by controlled updates of the assessment state.

\paragraph{Event-driven update.}
When an event or action at node \(v_i\) is observed, affected entities are updated by an operator
\[
\mathcal{U}_{t,v_i}\colon DB_t\to DB_{t+1},
\qquad
DB_{t+1}=\mathcal{U}_{t,v_i}(DB_t),
\]
with explicit change logs\index{Audit trail} (what changed, why, and on which evidence).

\paragraph{Feedback and learning.}
Likelihood scores \(\ell_o\in[0,1]\) can be updated by an adaptive rule
\[
\ell_o^{\,t+1}=(1-\beta)\,\ell_o^{\,t}+\beta\,f(o),
\]
where \(f(o)\) is an empirical frequency estimate and \(\beta\) is a learning rate\index{Learning rate}.

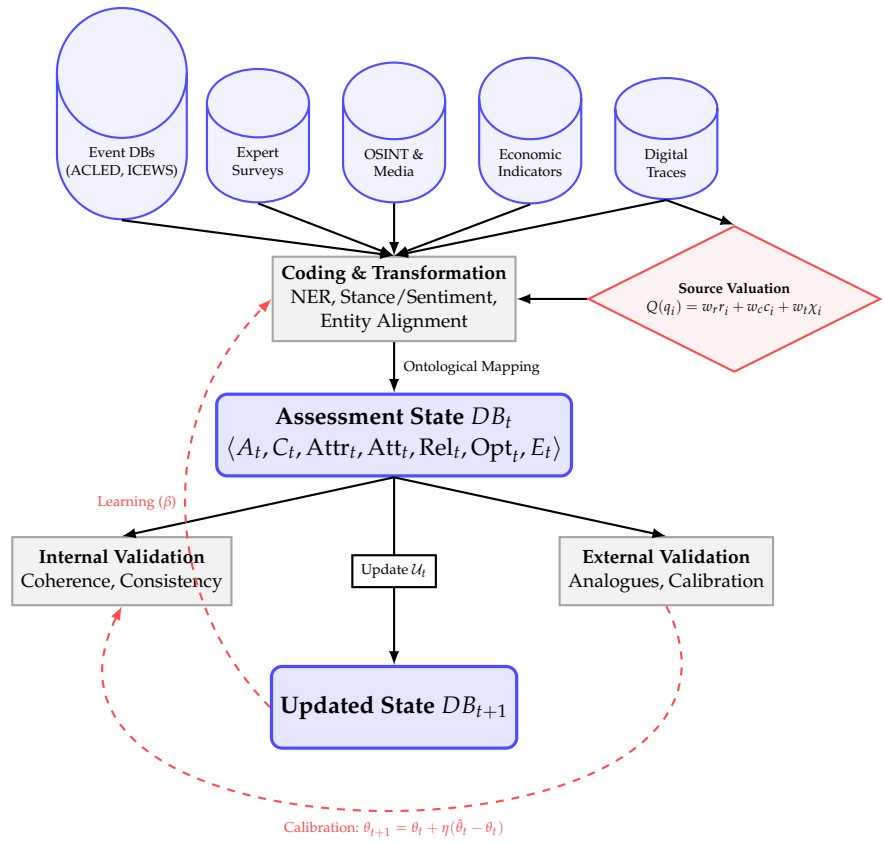
\begin{figure}[htbp]
	\centering
	\begin{tikzpicture}[
		scale=0.9,
		transform shape,
		sba_source/.style={cylinder, draw=blue!70, fill=blue!5, thick,
			shape border rotate=90, minimum width=1.5cm, minimum height=1cm,
			font=\tiny, align=center},
		sba_proc/.style={rectangle, draw=gray!70, fill=gray!10, thick,
			minimum width=2.5cm, minimum height=1cm, font=\scriptsize, align=center},
		sba_db/.style={rectangle, draw=blue!70, fill=blue!10, very thick,
			rounded corners, minimum width=3cm, minimum height=1.2cm,
			font=\small\bfseries, align=center},
		sba_val/.style={diamond, draw=red!70, fill=red!5, thick,
			aspect=2, minimum width=2.5cm, font=\tiny, align=center},
		sba_arrow/.style={-latex, thick},
		sba_feedback/.style={-latex, thick, red!70, dashed}
		]
		
		\node[sba_source] (s1) at (-4, 4) {Event DBs\\(ACLED, ICEWS)};
		\node[sba_source] (s2) at (-2, 4) {Expert\\Surveys};
		\node[sba_source] (s3) at (0, 4) {OSINT \&\\Media};
		\node[sba_source] (s4) at (2, 4) {Economic\\Indicators};
		\node[sba_source] (s5) at (4, 4) {Digital\\Traces};
		
		\node[sba_proc] (coding) at (0, 2) {
			\textbf{Coding \& Transformation}\\
			NER, Stance/Sentiment,\\
			Entity Alignment
		};
		
		\node[sba_val] (quality) at (5.0, 2) {
			\textbf{Source Valuation}\\
			$Q(q_i)=w_r r_i+w_c c_i+w_t \chi_i$
		};
		
		\foreach \n in {s1,s2,s3,s4,s5} {
			\draw[sba_arrow] (\n.south) -- (coding.north);
		}
		\draw[sba_arrow] (s5.south) -- (quality.north);
		\draw[sba_arrow] (quality.west) -- (coding.east);
		
		\node[sba_db] (db) at (0, 0) {
			Assessment State $DB_t$\\
			\small $\langle A_t, C_t, \mathrm{Attr}_t, \mathrm{Att}_t, \mathrm{Rel}_t, \mathrm{Opt}_t, E_t \rangle$
		};
		
		\draw[sba_arrow] (coding.south) -- (db.north) node[midway, right, font=\tiny] {Ontological Mapping};
		
		\node[sba_proc] (internal) at (-4, -2) {
			\textbf{Internal Validation}\\
			Coherence, Consistency
		};
		
		\node[sba_proc] (external) at (4, -2) {
			\textbf{External Validation}\\
			Analogues, Calibration
		};
		
		\draw[sba_arrow] (db.south) -- (internal.north);
		\draw[sba_arrow] (db.south) -- (external.north);
		
		\node[sba_db] (db_next) at (0, -4) {Updated State $DB_{t+1}$};
		
		\draw[sba_arrow] (db.south) -- (db_next.north) node[midway, fill=white, font=\tiny, draw] {Update $\mathcal{U}_t$};
		
		\draw[sba_feedback] (external.south) .. controls (6,-6.5) and (-6,-6.5) .. (internal.south)
		node[midway, below, font=\tiny] {Calibration: $\theta_{t+1}=\theta_t+\eta(\hat{\theta}_t-\theta_t)$};
		
		\draw[sba_feedback] (db_next.west) to[bend left=45] node[midway, left, font=\tiny] {Learning ($\beta$)} (coding.west);
		
	\end{tikzpicture}
	\caption{Empirical grounding and validation workflow in SBA.}
	\label{fig:data_workflow}
\end{figure}

\paragraph{Temporal integration.}
A temporal trace is recorded as
\[
DB_0 \xrightarrow{\ \mathcal{U}_0\ } DB_1 \xrightarrow{\ \mathcal{U}_1\ } \dots \xrightarrow{\ \mathcal{U}_{T-1}\ } DB_T.
\]
Figure~\ref{fig:data_workflow} shows the full empirical grounding and validation workflow.
This supports retrodictive validation\index{Validation!Retrodictive} (realized developments remain within admissible scenario sets).

\subsection{Theoretical foundations}
\label{subsec:validation-theory}

The validation and update mechanisms connect SBA to decision theory, learning, and belief revision\index{Belief revision}.

\paragraph{Equilibrium refinement and subgame perfection.}
MRP reasoning aligns with subgame-perfect equilibrium analysis \citep{selten1975,osborne1994}.
MLP selection aligns with probabilistic graphical reasoning \citep{pearl1988,jensen2007}.

\paragraph{Adaptive rationality.}
Parameter updating and database learning relate to adaptive expectations and bounded rational learning\index{Learning} \citep{simon1982,mckelvey1995}.

\paragraph{Dynamic epistemic logic and belief revision.}
Updates of beliefs and attitudes follow dynamic epistemic logic\index{Dynamic epistemic logic} and AGM belief revision\index{Belief revision!AGM}
\citep{baltag2008,vanbenthem2011,halpern2003}, constraining how new information changes epistemic states.

\paragraph{Iterative model calibration.}
Feedback-based adjustment connects to system dynamics\index{System dynamics} and simulation-based calibration\index{Calibration}
\citep{sterman2000,railsback2019}.

\section{Applications, Limitations\index{Applications}, and Outlook}
\label{sec:reflection}

	Scenario Bundle Analysis (SBA) was designed to support structured
	reasoning under uncertainty in settings where quantitative modeling
	is infeasible or uninformative.
	Its formal apparatus---database layer, scenario trees, topology, and
	evaluation structure---is domain-independent: the same representational
	machinery applies to geopolitical crises, economic disruptions, organizational
	conflicts, and hybrid scenarios.
	The present section situates extended SBA in applied and research contexts,
	identifies its principal limitations, and outlines three structurally natural
	development lines.
\subsection{Fields of Application}
\label{subsec:reflection-applications}

The modular architecture supports diverse policy and research environments:

\begin{itemize}[leftmargin=1.5em,itemsep=4pt]
	\item \textbf{Crisis foresight and early warning\index{Early warning}\index{Crisis foresight}.}
	Topology-aware analysis of attitude\index{Attitudes} and relation\index{Relations}
	configurations can identify branching points in escalation pathways.
	
	\item \textbf{Negotiation and mediation support\index{Negotiation}\index{Mediation}.}
	SBA externalizes belief\index{Attitudes!Belief} and dependency structures as a shared
	reference object, improving traceability of disagreement sources.
	
	\item \textbf{Policy design and intervention planning\index{Policy design}\index{Intervention}.}
	Scenario trees\index{Scenario Trees} function as an ex-ante evaluation laboratory,
	exposing trade-offs and unintended consequences.
	
	\item \textbf{Integrative hybrid modeling\index{Hybrid modeling}.}
	SBA can constrain and interpret outputs from agent-based
	models\index{Agent-based modeling} and LLM-supported
	extraction\index{Large language models}, ensuring that machine-assisted
	inputs remain human-auditable.
\end{itemize}

\paragraph{Domain independence.}
The ontology\index{Logical Foundations!Ontology} abstracts from domain content while preserving action, uncertainty, and interaction structure.
Attributes\index{Attributes} may represent financial capacities, trust, legitimacy, or technological constraints without changing the formal role they play in admissibility and evaluation.

\subsection{Limitations and Epistemic Challenges}
\label{subsec:reflection-limitations}

The framework inherits methodological and epistemic constraints\index{Modal Logics!Doxastic / Epistemic} that affect interpretation:

\begin{enumerate}[leftmargin=2em,itemsep=4pt, label=\protect\small\arabic*.]
	\item \textbf{Bias propagation\index{Bias!Propagation}.}
	Database-level sampling and coding biases (e.g.\ actor underrepresentation\index{Actors}) can amplify during generation and selection.
	
	\item \textbf{Subjective variance\index{Subjectivity}\index{Inter-rater reliability}.}
	Qualitative coding remains defeasible; stable results require explicit protocols, coder training, and adjudication rules.
	
	\item \textbf{Combinatorial complexity\index{Complexity}\index{Scenario space!Complexity}.}
	The scenario space grows rapidly with actors and options; pruning heuristics may miss low-probability/high-impact events\index{Black swan}.
	
	\item \textbf{Temporal lag\index{Time and Dynamics!Temporal Horizon}.}
	Without systematic updating, the model becomes stale; validity depends on
	disciplined transitions \(DB_t \xrightarrow{\ \mathcal{U}_t\ } DB_{t+1}\)\index{Update operator}.
\end{enumerate}

	\paragraph{Methodological responses.}
	Limitations~1 and~2 are addressable through explicit provenance tracking:
	each database entry should record its source, coder, and timestamp, supporting
	post-hoc bias audits and inter-rater reliability checks.
	Limitation~3 calls for configurable pruning thresholds and explicit
	tree-size budgets as parameters of the generation procedure.
	Limitation~4 is addressed structurally by the dynamic update operator
	\(\mathcal{U}_t\) and the requirement of versioned assessment states.

\subsection{Research Outlook}
\label{subsec:reflection-outlook}

Three development lines follow naturally from the extended framework.

\begin{itemize}[leftmargin=1.5em,itemsep=3pt]
	\item \textbf{Adaptive SBA\index{Learning}.}
	Bayesian and learning-based updates can connect bundles to near-real-time tracking while preserving explicit audit trails.
	
	\item \textbf{LLM--SBA integration\index{Large language models}\index{NLP}.}
	LLMs can support high-throughput extraction of entities, relations, and stances under provenance\index{Data provenance} and consistency constraints.
	
	\item \textbf{Formal refinements\index{Topology}\index{Category theory}.}
	Topological and categorical methods can sharpen invariance and stability claims for belief change and intersubjective coordination.
\end{itemize}

\section*{Conclusion}
\addcontentsline{toc}{section}{Conclusion}

	This paper has presented a formal refinement and extension of Scenario Bundle
	Analysis (SBA), the game-theoretic framework for structured representation of
	strategic interaction under uncertainty originally developed by Amos Perlmutter
	and Reinhard Selten.

	\paragraph{Main contributions.}
	The extended framework introduces three principal advances over the original SBA.
	First, the two-layer architecture
	\(\mathcal{M}_{\mathrm{SBA}} = \langle \mathrm{DB}_{\mathrm{SBA}}, \mathbb{T} \rangle\)
	makes the separation between the static database and the dynamic scenario tree
	system explicit, enabling controlled updates and versioned model states.
	Second, the extended attitude vocabulary---beliefs, desires, intentions,
	fears, and coalitional commitments, with expectations treated as doxastic
	attitudes---provides a richer
	representational basis for epistemic and motivational states, grounded in modal
	and doxastic logic.
	Third, the domain typology \(\mathcal{D} \times \mathcal{M}\) and the
	topological scenario space supply a formal language for comparative analysis and
	robustness assessment that was absent from the original framework.

	\paragraph{Methodological position.}
	Extended SBA occupies a semiformal position between classical game theory and
	qualitative scenario planning.
	It preserves the interpretability and elicitation-friendliness of the original
	framework while adding checkable structural constraints, explicit consistency
	conditions, and formal update semantics.
	The result is a method that can be applied by domain experts without formal
	training in game theory, while remaining auditable and computationally tractable.

	\paragraph{Formal coverage.}
	Taken together, the database, tree, topology, update, evaluation, and validation layers define a single audit trail from evidence intake to scenario comparison. This is the paper's central methodological claim: SBA can remain qualitative in elicitation while becoming formally inspectable at every modeling interface.
	The formal definitions matter because they turn SBA from a workshop-centered elicitation method into a typed modeling interface whose assumptions, updates, and evaluation choices can be inspected and reproduced.

	\paragraph{Outlook.}
	The development lines sketched in Section~\ref{subsec:reflection-outlook} matter because they test whether SBA can remain auditable when its inputs become faster, noisier, and more computationally mediated.
	The central constraint for future work is therefore not automation alone, but preserving explicit provenance, typed update rules, and interpretable evaluation under scale.

\appendix
\section{List of Symbols}
\label{app:symbols}

\paragraph{Indices and parameters.}
\begin{description}[leftmargin=3.6cm,labelwidth=3.3cm,style=nextline]
\item[\(t\)] Database stage index (assessment/update stage).
\item[\(h\)] Within-tree depth index (independent of \(t\)).
\item[\(i\in J_t\)] Index of an admissible tree \(T_i^{(t)}\) generated from \(DB_t\) (finite or countable).
\item[\(\lambda\)] Methodological-variation parameter (modeling choice), used in \((\varphi_\lambda,\Phi_\lambda)\).
\item[\(\vartheta\)] Temporal-horizon parameter, distinct from the topology \(\tau_t\).
\item[\(k\)]
Index of a feature component in weighted pseudo-metrics; weights \(w_k\ge 0\).
\end{description}

\paragraph{Core sets and database objects.}
\begin{description}[leftmargin=3.6cm,labelwidth=3.3cm,style=nextline]
\item[\(A\)] Set of actors.
\item[\(C\subseteq \Pow(A)\)] Set of coalitions.
\item[\(\Omega \coloneqq A\cup C\)] Unified endpoint domain (actors and coalitions); used as the relational domain in $R\subseteq\Omega\times\Omega$.
\item[\(D_{\mathrm{Attr}}\)] Generic value codomain for the attribute effect map \(\Delta_e\) and the stage valuation \(\mathrm{val}_t\).
\item[\(E_{\mathrm{Attr}}\)] Generic effect codomain for the event-effect map \(\Delta_e\); per-attribute effect type \(E_k\).
\item[\(\mathrm{Upd}_k\)] Typed update rule for attribute \(k\), \(\mathrm{Upd}_k:D_k\times E_k\to D_k\), composing prior value and event effect into a post-event value.
\item[\(P\)] Set of propositional contents (attitude contents).
\item[\(E\)] Universe of event labels; \(E_t\subseteq E\) denotes active events in \(DB_t\).
\item[\(\mathrm{Attr}_t\)] Attribute layer at stage \(t\).
\item[\(D_k\)] Domain of admissible values for attribute type \(k\).
\item[\(\mathrm{Sc}_k\)] Score function for aggregative attribute \(k\), used before ordinalization by \(Q_k\).
\item[\(Q_k\)] Ordinalization map returning aggregated scores for attribute \(k\) to \(D_k\).
\item[\(\mathrm{Rel}_t\)] Relation layer at stage \(t\).
\item[\(\Xi_{X,Y}^R\)] Member-level tie set generating the aggregate relation between \(X\) and \(Y\) for relation type \(R\).
\item[\(\operatorname{vis}_R\)] Visibility attribute for relation type \(R\) on \(\Omega\times\Omega\).
\item[\(\mathrm{Att}_t\)] Attitude layer at stage \(t\).
\item[\(\mathrm{Opt}_t\)] Option vocabulary at stage \(t\) (including designated non-execution labels in binary encodings).
\item[\(DB_t\)] Scenario assessment state at stage \(t\):
\[
DB_t=\langle A_t,C_t,\mathrm{Attr}_t,\mathrm{Att}_t,\mathrm{Rel}_t,\mathrm{Opt}_t,E_t\rangle .
\]
\item[\(\widehat{DB}_t\)] Candidate assessment state constructed by automated extraction (prior to acceptance).
\item[\(\Delta DB_t\)] Change set (accepted records) applied to \(DB_t\) under the declared revision policy.
\item[\(q_i\)] Source item used in empirical grounding and provenance tracking.
\item[\(\chi_i\)] Temporal-resolution score assigned to source item \(q_i\).
\item[\(M_t(\mathcal{B})\)] Evaluation matrix with entries \(M_{a,s}=u_a(s)\) for \(a\in A_t\), \(s\in S(\mathcal{B})\).
\item[\(G^{\mathrm{dom}}(\mathcal{B})\)] Dominance graph on \(S(\mathcal{B})\): edge \(s\to s'\) iff \(s\succ s'\) under a declared dominance criterion.
\item[\(\mathcal{M}_{\mathrm{SBA}}\)] Two-layer SBA architecture: \(\mathcal{M}_{\mathrm{SBA}}=\langle DB_{\mathrm{SBA}},\mathbb{T}\rangle\).
\item[\(\mathbb{T}\)] Dynamic scenario tree system (tree generation, selection, and update interaction).
\item[\(\Pow(X)\)] Power set of \(X\); \(\mathcal{P}_{\mathrm{fin}}(X)\) denotes the family of all finite subsets of \(X\).
\end{description}

\paragraph{Attitude operators.}
\begin{description}[leftmargin=3.6cm,labelwidth=3.3cm,style=nextline]
\item[\(K_a p\)] Actor \(a\) knows that \(p\).
\item[\(B_a p\)] Actor \(a\) believes that \(p\).
\item[\(W_a p\)] Actor \(a\) desires that \(p\).
\item[\(I_a p\)] Actor \(a\) intends that \(p\).
\item[\(F_a p\)] Actor \(a\) fears (negatively anticipates) that \(p\).
\item[\(\mathrm{Com}_X p\)] Coalition \(X\in C\) is (weakly) committed to \(p\).
\item[\(\mathrm{aim}(a,p)\)] Defined aim predicate: \(\mathrm{aim}(a,p)\Leftrightarrow W_a p\wedge \neg B_a\neg p\).
\end{description}

\paragraph{Scenario trees and branching.}
\begin{description}[leftmargin=3.6cm,labelwidth=3.3cm,style=nextline]
\item[\(T=\langle V,\Gamma,\rho,\mathrm{lab},\Sigma\rangle\)] Scenario tree structure (positions, edges, root, labels, edge labels).
\item[\(V\)] Set of positions (vertices).
\item[\(\Gamma\subseteq V\times V\)] Set of directed edges (transitions).
\item[\(\rho\in V\)] Root position (with root label \(x_0=\mathrm{lab}(\rho)\)).
\item[\(\mathrm{lab}:V\to A\cup C\cup E\)] Position label map (acting entity/event at a position).
\item[\(\mathrm{tail}(e),\mathrm{head}(e)\)] Tail/head projections for edges \(e\in\Gamma\).
\item[\(\mathrm{succ}(v)\)] Successor-edge set of \(v\): \(\mathrm{succ}(v)=\{e\in\Gamma:\mathrm{tail}(e)=v\}\).
\item[\(L\subseteq V\)] Leaf positions: \(L=\{v\in V:\mathrm{succ}(v)=\varnothing\}\).
\item[\(z_i\)] Terminal-node label for the \(i\)-th terminal outcome in worked tree examples.
\item[\(\Sigma:\Gamma\to \mathrm{Opt}_t\cup E^\circ\)] Edge-label map (option labels on decision edges; outcome labels on event edges).
\item[\(\mathrm{Avail}(x,DB_t)\subseteq \mathrm{Opt}_t\)] Options available to entity \(x\in A\cup C\) at stage \(t\).
\item[\(\Pi_{\mathrm{lab}(v)}(DB_t)\)] Non-empty set of priors used in ambiguity-averse evaluation at decision position \(v\) (maxmin expected utility).
\item[\(E^\circ\)] Event-realization labels induced by \(E_t\) (binary and multivalued realizations).
\item[\(\operatorname{obs}_e\)] Observability marker for event \(e\) (unobserved, partial, or public).
\item[\(\mathrm{valid}(e;DB_t)\)] Admissibility predicate for transitions (edge \(e\) is consistent with constraints encoded in \(DB_t\)).
\item[\(\pi_o,\ell_o,\vartheta_o\)] Option parameters: preference intensity, success likelihood, temporal horizon.
\item[\(\ell_e,\eta_e,\vartheta_e,\delta_e\)] Event parameters: likelihood, impact, temporal horizon, duration/delay (as used in event branching).
\end{description}

\paragraph{Bundles, topology, dynamics, and evaluation.}
\begin{description}[leftmargin=3.6cm,labelwidth=3.3cm,style=nextline]
\item[\(T_t=\mathcal{T}(DB_t)\)] Set of admissible scenario trees generated from \(DB_t\).
\item[\(\mathcal{B}_t=\mathrm{Sel}_t(T_t)\subseteq T_t\)] Finite selected scenario bundle at stage \(t\).
\item[\(\mathrm{Gen}_t\)] Tree-generation operator (maps \(DB_t\) to \(\mathcal{T}(DB_t)\)).
\item[\(\mathrm{Sel}_t\)] Selection operator (maps admissible trees to a finite bundle).
\item[\(\mathcal{U}_t:DB_t\to DB_{t+1}\)] Database update operator (assessment-state update).
\item[\((\varphi_\lambda,\Phi_\lambda)\)] Methodological-variation maps; controlled modeling transformations.
\item[\(d_t\)] Stage-dependent pseudo-metric on admissible trees and bundles.
\item[\(\tau_t\)] Topology induced by \(d_t\).
\item[\(\mathcal{S}_t=(T_t,\tau_t)\)] Topological scenario space at stage \(t\).
\item[\(S(\mathcal{B})\)] Set of scenarios (terminal paths) encoded by a bundle \(\mathcal{B}\).
\item[\(\succeq_a\)] Actor-relative preorder on scenarios (or outcomes).
\item[\(u_a\)] Numerical representation of \(\succeq_a\) (if assumed), unique up to strictly increasing transforms.
\item[\(U_X\)] Coalition-level evaluation functional for \(X\in C_t\) (aggregation of member evaluations).
\item[\(\mathrm{Agg}_X\)] Aggregation operator used in MRP refinements: \(U_X(s)=\mathrm{Agg}_X(\{U_a(s):a\in X\})\).
\item[\(\Psi\)] Evaluation output functional on bundles (e.g.\ \(\Psi(\mathcal{B})=S^\ast\)); robustness diagnostic in the topology section.
\item[\(\mathcal{V}\)] Optional observer/meta-evaluation functional on scenarios/bundles.
\end{description}

\bibliographystyle{plainnat}
\bibliography{sba_references_v37}

@misc{acled2024,
	title        = {Armed Conflict Location \& Event Data Project (ACLED)},
	author       = {{ACLED Project}},
	year         = {2024},
	url          = {https://acleddata.com/},
	note         = {Accessed: 2024-12-01}
}

@article{altafini2013,
	title        = {Consensus Problems on Networks with Antagonistic Interactions},
	author       = {Altafini, Claudio},
	year         = {2013},
	journal      = {IEEE Transactions on Automatic Control},
	volume       = {58},
	number       = {4},
	pages        = {935--946},
	doi          = {10.1109/TAC.2012.2224251}
}

@book{anscombe1957,
	title        = {Intention},
	author       = {Anscombe, G. E. M.},
	year         = {1957},
	publisher    = {Blackwell},
	address      = {Oxford}
}

@book{arrow1951,
	title        = {Social Choice and Individual Values},
	author       = {Arrow, Kenneth J.},
	year         = {1951},
	publisher    = {Wiley},
	address      = {New York}
}

@article{arthur1999,
	title        = {Complexity and the economy},
	author       = {Arthur, W. Brian},
	year         = {1999},
	journal      = {Science},
	volume       = {284},
	number       = {5411},
	pages        = {107--109}
}

@book{austin1962,
	title        = {How to Do Things with Words},
	author       = {Austin, J. L.},
	year         = {1962},
	publisher    = {Oxford University Press},
	address      = {Oxford}
}

@book{axelrod1984,
	title        = {The Evolution of Cooperation},
	author       = {Axelrod, Robert},
	year         = {1984},
	publisher    = {Basic Books},
	address      = {New York}
}

@book{ballmer1981,
	title        = {Speech Act Classification: A Study in the Lexical Analysis of English Speech Activity Verbs},
	author       = {Ballmer, Thomas T. and Brennenstuhl, Waltraud},
	year         = {1981},
	publisher    = {Springer},
	address      = {Berlin}
}

@incollection{baltag2008,
	title        = {The Logic of Public Announcements, Common Knowledge, and Private Suspicions},
	author       = {Baltag, Alexandru and Moss, Lawrence S. and Solecki, Slawomir},
	year         = {2008},
	booktitle    = {Handbook of Logic and Rationality},
	publisher    = {Springer},
	address      = {Dordrecht},
	pages        = {49--146},
	editor       = {Gabbay, D. and Guenthner, F.}
}

@book{balzer1987,
	title        = {An Architectonic for Science: The Structuralist Program},
	author       = {Balzer, Wolfgang and Moulines, C. Ulises and Sneed, Joseph D.},
	year         = {1987},
	publisher    = {Reidel},
	address      = {Dordrecht},
	series       = {Synthese Library},
	volume       = {186},
	doi          = {10.1007/978-94-009-3765-9}
}

@article{battiston2020,
	title        = {Networks beyond Pairwise Interactions: Structure and Dynamics of Higher-Order Systems},
	author       = {Battiston, Federico and Cencetti, Giulia and Iacopini, Iacopo and Latora, Vito and Lucas, Maxime and Patania, Alice and Young, Jean-Gabriel and Petri, Giovanni},
	year         = {2020},
	journal      = {Physics Reports},
	volume       = {874},
	pages        = {1--92},
	doi          = {10.1016/j.physrep.2020.05.004}
}

@book{beck1992,
	title        = {Risk Society: Towards a New Modernity},
	author       = {Beck, Ulrich},
	year         = {1992},
	publisher    = {SAGE Publications},
	address      = {London}
}

@book{bellman1957,
	author       = {Bellman, Richard},
	title        = {Dynamic Programming},
	publisher    = {Princeton University Press},
	year         = {1957}
}

@book{belton2002,
	title        = {Multiple Criteria Decision Analysis: An Integrated Approach},
	author       = {Belton, Valerie and Stewart, Theodor J.},
	year         = {2002},
	publisher    = {Springer}
}

@article{benson2016,
	title        = {Higher-Order Organization of Complex Networks},
	author       = {Benson, Austin R. and Gleich, David F. and Leskovec, Jure},
	year         = {2016},
	journal      = {Science},
	volume       = {353},
	number       = {6295},
	pages        = {163--166},
	doi          = {10.1126/science.aad9029}
}

@book{bercovitch1992,
	title        = {Social Conflicts and Third Parties: Strategies of Conflict Resolution},
	author       = {Bercovitch, Jacob},
	year         = {1992},
	publisher    = {Westview Press},
	address      = {Boulder, CO}
}

@book{bertsekas2005,
	author       = {Bertsekas, Dimitri P.},
	title        = {Dynamic Programming and Optimal Control},
	publisher    = {Athena Scientific},
	year         = {2005},
	edition      = {3}
}

@article{boccaletti2014,
	title        = {The Structure and Dynamics of Multilayer Networks},
	author       = {Boccaletti, Stefano and Bianconi, Ginestra and Criado, Regino and del Genio, Charo I. and G{\'o}mez-Garde{\~n}es, Jes{\'u}s and Romance, Miguel and Sendi{\~n}a-Nadal, Irene and Wang, Zhen and Zanin, Massimiliano},
	year         = {2014},
	journal      = {Physics Reports},
	volume       = {544},
	number       = {1},
	pages        = {1--122},
	doi          = {10.1016/j.physrep.2014.07.001}
}

@book{bratman1987,
	title        = {Intention, Plans, and Practical Reason},
	author       = {Bratman, Michael E.},
	year         = {1987},
	publisher    = {Harvard University Press},
	address      = {Cambridge, MA}
}

@book{burago2001,
	title        = {A Course in Metric Geometry},
	author       = {Burago, Dmitri and Burago, Yuri and Ivanov, Sergei},
	year         = {2001},
	publisher    = {American Mathematical Society}
}

@book{buzan1998,
	title        = {Security: A New Framework for Analysis},
	author       = {Buzan, Barry and W{\ae}ver, Ole and de Wilde, Jaap},
	year         = {1998},
	publisher    = {Lynne Rienner Publishers},
	address      = {Boulder}
}

@book{carnap1947,
	title        = {Meaning and Necessity: A Study in Semantics and Modal Logic},
	author       = {Carnap, Rudolf},
	year         = {1947},
	publisher    = {University of Chicago Press},
	address      = {Chicago}
}

@article{cartwright1956,
	title        = {Structural Balance: A Generalization of Heider's Theory},
	author       = {Cartwright, Dorwin and Harary, Frank},
	year         = {1956},
	journal      = {Psychological Review},
	volume       = {63},
	number       = {5},
	pages        = {277--293},
	doi          = {10.1037/h0046049}
}

@book{castells1996,
	title        = {The Rise of the Network Society},
	author       = {Castells, Manuel},
	year         = {1996},
	publisher    = {Blackwell},
	address      = {Oxford}
}

@misc{cepii2024,
	title        = {BACI Trade Data and Geographical Datasets},
	author       = {CEPII},
	year         = {2024},
	url          = {https://www.cepii.fr/},
	note         = {Accessed: 2024-12-01}
}

@book{cloreortonysmith1988,
	title        = {The Cognitive Structure of Emotions},
	author       = {Ortony, Andrew and Clore, Gerald L. and Collins, Allan},
	year         = {1988},
	publisher    = {Cambridge University Press},
	address      = {Cambridge}
}

@book{coleman1990,
	title        = {Foundations of Social Theory},
	author       = {Coleman, James S.},
	year         = {1990},
	publisher    = {Harvard University Press},
	address      = {Cambridge, MA}
}

@misc{cowalliance2024,
	title        = {Formal Interstate Alliance Dataset},
	author       = {{Correlates of War Project}},
	year         = {2024},
	url          = {https://correlatesofwar.org/data-sets/formal-alliances},
	note         = {Accessed: 2024-12-01}
}

@book{cyert1963,
	title        = {A Behavioral Theory of the Firm},
	author       = {Cyert, Richard M. and March, James G.},
	year         = {1963},
	publisher    = {Prentice Hall},
	address      = {Englewood Cliffs, NJ}
}

@incollection{davidson1963,
	title        = {Actions, Reasons, and Causes},
	author       = {Davidson, Donald},
	year         = {1980},
	booktitle    = {Essays on Actions and Events},
	publisher    = {Oxford University Press},
	address      = {Oxford},
	pages        = {3--20},
	note         = {Originally published 1963}
}

@incollection{davidson1967,
	title        = {The Logical Form of Action Sentences},
	author       = {Davidson, Donald},
	year         = {1967},
	booktitle    = {The Logic of Decision and Action},
	publisher    = {University of Pittsburgh Press},
	pages        = {81--95},
	editor       = {Rescher, Nicholas}
}

@misc{desinventar2024,
	title        = {DesInventar Disaster Information System},
	author       = {{UNDRR}},
	year         = {2024},
	url          = {https://www.desinventar.net/},
	note         = {Accessed: 2024-12-01}
}

@book{easley2010,
	title        = {Networks, Crowds, and Markets: Reasoning About a Highly Connected World},
	author       = {Easley, David and Kleinberg, Jon},
	year         = {2010},
	publisher    = {Cambridge University Press},
	address      = {Cambridge}
}

@misc{ecmwf2024,
	title        = {ERA5 Global Climate Reanalysis},
	author       = {{ECMWF}},
	year         = {2024},
	url          = {https://www.ecmwf.int/en/forecasts/datasets/reanalysis-datasets/era5},
	note         = {Accessed: 2024-12-01}
}

@misc{emdat2024,
	title        = {The International Disaster Database (EM-DAT)},
	author       = {{CRED/EM-DAT}},
	year         = {2024},
	url          = {https://www.emdat.be/},
	note         = {Accessed: 2024-12-01}
}

@article{emerson1962,
	title        = {Power-Dependence Relations},
	author       = {Emerson, Richard M.},
	year         = {1962},
	journal      = {American Sociological Review},
	volume       = {27},
	number       = {1},
	pages        = {31--41},
	doi          = {10.2307/2089716}
}

@book{fagin1995,
	title        = {Reasoning About Knowledge},
	author       = {Fagin, Ronald and Halpern, Joseph Y. and Moses, Yoram and Vardi, Moshe Y.},
	year         = {1995},
	publisher    = {MIT Press},
	address      = {Cambridge, MA},
	note         = {Foundational text on epistemic logic and knowledge in multi-agent systems.}
}

@misc{faostat2024,
	title        = {FAOSTAT Statistical Database},
	author       = {{Food and Agriculture Organization of the United Nations}},
	year         = {2024},
	url          = {https://www.fao.org/faostat/},
	note         = {Accessed: 2024-12-01}
}

@book{fidler2004,
	title        = {SARS, Governance and the Globalization of Disease},
	author       = {Fidler, David P.},
	year         = {2004},
	publisher    = {Palgrave Macmillan},
	address      = {Basingstoke}
}

@book{floridi2019,
	title        = {The Logic of Information: A Theory of Philosophy as Conceptual Design},
	author       = {Floridi, Luciano},
	year         = {2019},
	publisher    = {Oxford University Press},
	address      = {Oxford}
}

@article{frege1892,
	title        = {{\"U}ber Sinn und Bedeutung},
	author       = {Frege, Gottlob},
	year         = {1892},
	journal      = {Zeitschrift f{\"u}r Philosophie und philosophische Kritik},
	volume       = {100},
	pages        = {25--50},
	note         = {English translation: ``On Sense and Reference,'' in \emph{Philosophical Writings}, ed. Geach \& Black, 1952.}
}

@book{friedkin2011,
	title        = {Social Influence Network Theory: A Sociological Examination of Small Group Dynamics},
	author       = {Friedkin, Noah E. and Johnsen, Eugene C.},
	year         = {2011},
	publisher    = {Cambridge University Press},
	address      = {Cambridge}
}

@book{fudenberg1998,
	title        = {The Theory of Learning in Games},
	author       = {Fudenberg, Drew and Levine, David K.},
	year         = {1998},
	publisher    = {MIT Press}
}

@book{galtung1996,
	title        = {Peace by Peaceful Means: Peace and Conflict, Development and Civilization},
	author       = {Galtung, Johan},
	year         = {1996},
	publisher    = {SAGE Publications},
	address      = {London}
}

@misc{gdelt2023,
	title        = {Global Database of Events, Language, and Tone},
	author       = {{GDELT Project}},
	year         = {2023},
	howpublished = {\url{https://www.gdeltproject.org/}}
}

@book{gilbert1989,
	title        = {On Social Facts},
	author       = {Gilbert, Margaret},
	year         = {1989},
	publisher    = {Routledge},
	address      = {London}
}

@article{gilboa1989,
	author       = {Gilboa, Itzhak and Schmeidler, David},
	title        = {Maxmin Expected Utility with Non-Unique Prior},
	journal      = {Journal of Mathematical Economics},
	year         = {1989},
	volume       = {18},
	number       = {2},
	pages        = {141--153}
}

@incollection{gillies1959,
	author       = {Gillies, Donald B.},
	title        = {Solutions to General Non-Zero-Sum Games},
	booktitle    = {Contributions to the Theory of Games {IV}},
	editor       = {Tucker, Albert W. and Luce, Robert Duncan},
	series       = {Annals of Mathematics Studies},
	number       = {40},
	pages        = {47--85},
	year         = {1959},
	publisher    = {Princeton University Press},
	address      = {Princeton, NJ}
}

@article{greenberg1993,
	title        = {Stable Coalition Structures with a Unidimensional Set of Alternatives},
	author       = {Greenberg, Joseph and Weber, Shlomo},
	year         = {1993},
	journal      = {Journal of Economic Theory},
	volume       = {60},
	number       = {1},
	pages        = {62--82}
}

@book{greig2012,
	title        = {International Mediation},
	author       = {Greig, J. Michael and Diehl, Paul F.},
	year         = {2012},
	publisher    = {Polity Press},
	address      = {Cambridge},
	note         = {Discusses effectiveness of mediation and peacekeeping in conflict resolution.}
}

@book{halpern2003,
	title        = {Reasoning about Uncertainty},
	author       = {Halpern, Joseph Y.},
	year         = {2003},
	publisher    = {MIT Press},
	address      = {Cambridge, MA}
}

@article{harsanyi1955,
	title        = {Cardinal Welfare, Individualistic Ethics, and Interpersonal Comparisons of Utility},
	author       = {Harsanyi, John C.},
	year         = {1955},
	journal      = {Journal of Political Economy},
	volume       = {63},
	number       = {4},
	pages        = {309--321}
}

@article{heider1946,
	title        = {Attitudes and Cognitive Organization},
	author       = {Heider, Fritz},
	year         = {1946},
	journal      = {The Journal of Psychology},
	volume       = {21},
	number       = {1},
	pages        = {107--112},
	doi          = {10.1080/00223980.1946.9917275}
}

@article{helbing2013,
	title        = {Globally Networked Risks and How to Respond},
	author       = {Helbing, Dirk},
	year         = {2013},
	journal      = {Nature},
	volume       = {497},
	pages        = {51--59}
}

@book{heuer1999,
	title        = {Psychology of Intelligence Analysis},
	author       = {Heuer, Richards J.},
	year         = {1999},
	publisher    = {Center for the Study of Intelligence, Central Intelligence Agency},
	address      = {Washington, DC},
	note         = {Available online at cia.gov}
}

@book{heuer2010,
	title        = {Structured Analytic Techniques for Intelligence Analysis},
	author       = {Heuer, Richards J. and Pherson, Randolph H.},
	year         = {2010},
	publisher    = {CQ Press},
	address      = {Washington, DC},
	edition      = {1st}
}

@book{hintikka1962,
	title        = {Knowledge and Belief: An Introduction to the Logic of the Two Notions},
	author       = {Hintikka, Jaakko},
	year         = {1962},
	publisher    = {Cornell University Press},
	address      = {Ithaca, NY}
}

@book{holland1992,
	title        = {Adaptation in Natural and Artificial Systems},
	author       = {Holland, John H.},
	year         = {1992},
	publisher    = {MIT Press}
}

@article{holling1973,
	title        = {Resilience and stability of ecological systems},
	author       = {Holling, C. S.},
	year         = {1973},
	journal      = {Annual Review of Ecology and Systematics},
	volume       = {4},
	pages        = {1--23}
}

@book{homer1999,
	title        = {Environment, Scarcity, and Violence},
	author       = {Homer-Dixon, Thomas F.},
	year         = {1999},
	publisher    = {Princeton University Press},
	address      = {Princeton}
}

@misc{icews2024,
	title        = {Integrated Crisis Early Warning System Event Data},
	author       = {{ICEWS Program}},
	year         = {2024},
	url          = {https://www.lockheedmartin.com/en-us/capabilities/research-labs/advanced-technology-labs/icews.html},
	note         = {Harvard Dataverse}
}

@misc{icow2024,
	title        = {Issue Correlates of War Dataset},
	author       = {{ICOW Project}},
	year         = {2024},
	url          = {https://www.paulhensel.org/icow.html},
	note         = {Accessed: 2024-12-01}
}

@misc{imfifs2024,
	title        = {International Financial Statistics},
	author       = {{International Monetary Fund}},
	year         = {2024},
	url          = {https://data.imf.org/},
	note         = {Accessed: 2024-12-01}
}

@book{jackendoff1983,
	title        = {Semantics and Cognition},
	author       = {Jackendoff, Ray},
	year         = {1983},
	publisher    = {MIT Press},
	address      = {Cambridge, MA},
	series       = {Current Studies in Linguistics}
}

@book{jackson2008,
	title        = {Social and Economic Networks},
	author       = {Jackson, Matthew O.},
	year         = {2008},
	publisher    = {Princeton University Press},
	address      = {Princeton, NJ}
}

@book{jensen2007,
	title        = {Bayesian Networks and Decision Graphs},
	author       = {Jensen, Finn V. and Nielsen, Thomas D.},
	year         = {2007},
	publisher    = {Springer},
	address      = {New York},
	edition      = {2nd}
}

@book{kahn1967,
	title        = {The Year 2000: A Framework for Speculation on the Next Thirty-Three Years},
	author       = {Kahn, Herman and Wiener, Anthony J.},
	year         = {1967},
	publisher    = {Macmillan},
	address      = {New York}
}

@article{kivela2014,
	title        = {Multilayer Networks},
	author       = {Kivel{\"a}, Mikko and Arenas, Alex and Barthelemy, Marc and Gleeson, James P. and Moreno, Yamir and Porter, Mason A.},
	year         = {2014},
	journal      = {Journal of Complex Networks},
	volume       = {2},
	number       = {3},
	pages        = {203--271},
	doi          = {10.1093/comnet/cnu016}
}

@book{kreps1988,
	author    = {Kreps, David M.},
	title     = {Notes on the Theory of Choice},
	publisher = {Westview Press},
	address   = {Boulder, CO},
	year      = {1988}
}

@book{latour2005,
	title        = {Reassembling the Social: An Introduction to Actor--Network Theory},
	author       = {Latour, Bruno},
	year         = {2005},
	publisher    = {Oxford University Press},
	address      = {Oxford}
}

@article{leeds2003,
	title        = {Alliance Reliability in Times of War: Explaining State Decisions to Violate Treaties},
	author       = {Leeds, Brett Ashley},
	year         = {2003},
	journal      = {International Organization},
	volume       = {57},
	number       = {4},
	pages        = {801--827},
	doi          = {10.1017/S0020818303574057}
}

@book{listpettit2011,
	title        = {Group Agency: The Possibility, Design, and Status of Corporate Agents},
	author       = {List, Christian and Pettit, Philip},
	year         = {2011},
	publisher    = {Oxford University Press},
	address      = {Oxford},
	note         = {Comprehensive account of group agency and collective decision-making}
}

@article{march1984,
	title        = {The New Institutionalism: Organizational Factors in Political Life},
	author       = {March, James G. and Olsen, Johan P.},
	year         = {1984},
	journal      = {American Political Science Review},
	volume       = {78},
	number       = {3},
	pages        = {734--749}
}

@book{mas1995,
	title        = {Microeconomic Theory},
	author       = {Mas-Colell, Andreu and Whinston, Michael D. and Green, Jerry R.},
	year         = {1995},
	publisher    = {Oxford University Press}
}

@article{mckelvey1995,
	title        = {Quantal Response Equilibria for Normal Form Games},
	author       = {McKelvey, Richard D. and Palfrey, Thomas R.},
	year         = {1995},
	journal      = {Games and Economic Behavior},
	volume       = {10},
	number       = {1},
	pages        = {6--38},
	doi          = {10.1006/game.1995.1023}
}

@book{meyer1999,
	title        = {Epistemic Logic for AI and Computer Science},
	author       = {Meyer, John-Jules Ch. and van der Hoek, Wiebe},
	year         = {1999},
	publisher    = {Cambridge University Press},
	address      = {Cambridge},
	note         = {Comprehensive treatment of epistemic and doxastic logics for artificial agents.}
}

@article{montague1970,
	title        = {Universal Grammar},
	author       = {Montague, Richard},
	year         = {1970},
	journal      = {Theoria},
	volume       = {36},
	number       = {3},
	pages        = {373--398},
	doi          = {10.1111/j.1755-2567.1970.tb00434.x}
}

@book{morrow1994,
	title        = {Game Theory for Political Scientists},
	author       = {Morrow, James D.},
	year         = {1994},
	publisher    = {Princeton University Press},
	address      = {Princeton, NJ}
}

@book{newman2010,
	title        = {Networks: An Introduction},
	author       = {Newman, Mark},
	year         = {2010},
	publisher    = {Oxford University Press},
	address      = {Oxford}
}

@misc{noaa2024,
	title        = {NOAA Climate Data Records},
	author       = {{NOAA National Centers for Environmental Information}},
	year         = {2024},
	url          = {https://www.ncei.noaa.gov/},
	note         = {Accessed: 2024-12-01}
}

@book{north1991,
	title        = {Institutions, Institutional Change and Economic Performance},
	author       = {North, Douglass C.},
	year         = {1991},
	publisher    = {Cambridge University Press},
	address      = {Cambridge}
}

@book{olson1965,
	title        = {The Logic of Collective Action: Public Goods and the Theory of Groups},
	author       = {Olson, Mancur},
	year         = {1965},
	publisher    = {Harvard University Press},
	address      = {Cambridge, MA}
}

@book{osborne1994,
	title        = {A Course in Game Theory},
	author       = {Osborne, Martin J. and Rubinstein, Ariel},
	year         = {1994},
	publisher    = {MIT Press},
	address      = {Cambridge, MA}
}

@article{patania2017shape,
	title        = {The shape of collaborations},
	author       = {Patania, Alice and Petri, Giovanni and Vaccarino, Francesca},
	year         = {2017},
	journal      = {EPJ Data Science},
	volume       = {6},
	number       = {1},
	pages        = {18},
	doi          = {10.1140/epjds/s13688-017-0114-8}
}

@book{pearl1988,
	title        = {Probabilistic Reasoning in Intelligent Systems: Networks of Plausible Inference},
	author       = {Pearl, Judea},
	year         = {1988},
	publisher    = {Morgan Kaufmann},
	address      = {San Mateo, CA}
}

@book{phersonheuer2021sat,
	title        = {Structured Analytic Techniques for Intelligence Analysis},
	author       = {Pherson, Randolph H. and Heuer Jr., Richards J.},
	year         = {2021},
	publisher    = {CQ Press},
	address      = {Thousand Oaks, CA},
	edition      = {3rd}
}

@book{prior1957,
	title        = {Time and Modality},
	author       = {Prior, Arthur N.},
	year         = {1957},
	publisher    = {Oxford University Press},
	address      = {Oxford}
}

@article{quine1956,
	title        = {Quantifiers and Propositional Attitudes},
	author       = {Quine, Willard Van Orman},
	year         = {1956},
	journal      = {The Journal of Philosophy},
	volume       = {53},
	number       = {5},
	pages        = {177--187}
}

@book{quine1960,
	title        = {Word and Object},
	author       = {Quine, Willard V. O.},
	year         = {1960},
	publisher    = {MIT Press},
	address      = {Cambridge, MA}
}

@book{railsback2019,
	title        = {Agent-Based and Individual-Based Modeling: A Practical Introduction},
	author       = {Railsback, Steven F. and Grimm, Volker},
	year         = {2019},
	publisher    = {Princeton University Press},
	address      = {Princeton},
	edition      = {2nd}
}

@incollection{rayvohra2015,
	title        = {Coalition Formation},
	author       = {Ray, Debraj and Vohra, Rajiv},
	year         = {2015},
	booktitle    = {Handbook of Game Theory with Economic Applications},
	volume       = {4},
	pages        = {239--326},
	publisher    = {Elsevier},
	doi          = {10.1016/B978-0-444-53766-9.00005-7}
}

@book{rescher1971,
	title        = {Temporal Logic},
	author       = {Rescher, Nicholas and Urquhart, Alasdair},
	year         = {1971},
	publisher    = {Springer},
	address      = {Vienna}
}

@book{riker1962,
	title        = {The Theory of Political Coalitions},
	author       = {Riker, William H.},
	year         = {1962},
	publisher    = {Yale University Press},
	address      = {New Haven}
}

@book{rosenthal1993,
	title        = {Reconstruction of Life Stories: Principles of Selection in Generating Stories for Narrative Biographical Interviews},
	author       = {Rosenthal, Gabriele},
	year         = {1993},
	publisher    = {SAGE}
}

@book{roy1996,
	title        = {Multicriteria Methodology for Decision Aiding},
	author       = {Roy, Bernard},
	year         = {1996},
	publisher    = {Springer}
}

@book{savage1954,
	author       = {Savage, Leonard J.},
	title        = {The Foundations of Statistics},
	publisher    = {John Wiley \& Sons},
	address      = {New York},
	year         = {1954}
}

@book{schelling1966,
	title        = {Arms and Influence},
	author       = {Schelling, Thomas C.},
	year         = {1966},
	publisher    = {Yale University Press},
	address      = {New Haven, CT}
}

@article{schoemaker1995,
	title        = {Scenario Planning: A Tool for Strategic Thinking},
	author       = {Schoemaker, Paul J. H.},
	year         = {1995},
	journal      = {Sloan Management Review},
	volume       = {36},
	number       = {2},
	pages        = {25--40}
}

@book{schwartz1991,
	title        = {The Art of the Long View: Planning for the Future in an Uncertain World},
	author       = {Schwartz, Peter},
	year         = {1991},
	publisher    = {Currency Doubleday},
	address      = {New York}
}

@book{searle1969,
	title        = {Speech Acts: An Essay in the Philosophy of Language},
	author       = {Searle, John R.},
	year         = {1969},
	publisher    = {Cambridge University Press},
	address      = {Cambridge}
}

@incollection{searle1975,
	title        = {A Taxonomy of Illocutionary Acts},
	author       = {Searle, John R.},
	year         = {1975},
	booktitle    = {Language, Mind, and Knowledge},
	editor       = {Gunderson, Keith},
	pages        = {344--369},
	publisher    = {University of Minnesota Press},
	address      = {Minneapolis}
}

@book{searle1983,
	title        = {Intentionality: An Essay in the Philosophy of Mind},
	author       = {Searle, John R.},
	year         = {1983},
	publisher    = {Cambridge University Press},
	address      = {Cambridge}
}

@book{searle1995,
	title        = {The Construction of Social Reality},
	author       = {Searle, John R.},
	year         = {1995},
	publisher    = {Free Press},
	address      = {New York},
	note         = {Explores institutional facts and collective intentionality, relevant to coalitions.}
}

@article{selten1975,
	title        = {Reexamination of the Perfectness Concept for Equilibrium Points in Extensive Games},
	author       = {Selten, Reinhard},
	year         = {1975},
	journal      = {International Journal of Game Theory},
	volume       = {4},
	number       = {1},
	pages        = {25--55},
	doi          = {10.1007/BF01766400}
}

@misc{selten1976,
	title        = {Scenario Bundle Analysis: A Strategic Perspective on the Persian Gulf},
	author       = {Selten, Reinhard and Perlmutter, Amos},
	year         = {1976},
	howpublished = {Unpublished manuscript},
	note         = {Often cited as an early SBA application; archival availability is limited.}
}

@incollection{selten1999sbm,
	title        = {The Scenario Bundle Method},
	author       = {Selten, Reinhard},
	year         = {1999},
	booktitle    = {Game Theory and Economic Behaviour: Selected Essays, Volume I},
	publisher    = {Edward Elgar},
	address      = {Cheltenham},
	pages        = {291--325},
	note         = {Reprinted essay outlining the SBA methodology. (Originally appeared in Zeitschrift f{\"u}r Betriebswirtschaft, 69, Erg{\"a}nzungsheft.)}
}

@incollection{selten2004kosovo,
	title        = {Die Szenariob{\"u}ndelmethode -- Am Beispiel Kosovo},
	author       = {Selten, Reinhard and Chmura, Thorsten and Pitz, Thomas},
	year         = {2004},
	booktitle    = {Zur L{\"o}sung des Kosovokonflikts},
	publisher    = {Nomos Verlagsgesellschaft},
	address      = {Baden-Baden},
	pages        = {11--44},
	editor       = {Reiter, Erich and Selten, Reinhard}
}

@incollection{selten2004bosnien,
	title        = {Die Szenariob{\"u}ndelmethode -- Am Beispiel Bosnien und Herzegowina},
	author       = {Selten, Reinhard and Chmura, Thorsten and Pitz, Thomas},
	year         = {2004},
	booktitle    = {Bosnien und Herzegowina: Europas Balkanpolitik auf dem Pr{\"u}fstand},
	publisher    = {Nomos Verlag},
	address      = {Baden-Baden},
	pages        = {163--179},
	editor       = {Reiter, Erich and Jurekovi{\'c}, Predrag}
}

@misc{selten2004interview,
	title        = {Laureates Interview},
	author       = {Selten, Reinhard},
	year         = {2004},
	note         = {Nobel Prize Organization.},
	howpublished = {\url{https://www.nobelprize.org/prizes/economic-sciences/1994/selten/interview/}}
}

@article{sen1977,
	title        = {Rational Fools: A Critique of the Behavioral Foundations of Economic Theory},
	author       = {Sen, Amartya},
	year         = {1977},
	journal      = {Philosophy \& Public Affairs},
	volume       = {6},
	number       = {4},
	pages        = {317--344}
}

@incollection{shapley1953,
	author    = {Shapley, Lloyd S.},
	title     = {A Value for n-Person Games},
	booktitle = {Contributions to the Theory of Games {II}},
	editor    = {Kuhn, Harold W. and Tucker, Albert W.},
	publisher = {Princeton University Press},
	address   = {Princeton, NJ},
	year      = {1953},
	pages     = {307--317}
}

@book{simon1957,
	title        = {Models of Man: Social and Rational},
	author       = {Simon, Herbert A.},
	year         = {1957},
	publisher    = {Wiley},
	address      = {New York}
}

@book{simon1982,
	title        = {Models of Bounded Rationality, Volume 2: Behavioral Economics and Business Organization},
	author       = {Simon, Herbert A.},
	year         = {1982},
	publisher    = {The MIT Press},
	address      = {Cambridge, MA},
	volume       = {2},
	isbn         = {9780262192064}
}

@book{smith1982,
	title        = {Evolution and the Theory of Games},
	author       = {Maynard Smith, John},
	year         = {1982},
	publisher    = {Cambridge University Press}
}

@book{sterman2000,
	title        = {Business Dynamics: Systems Thinking and Modeling for a Complex World},
	author       = {Sterman, John D.},
	year         = {2000},
	publisher    = {Irwin/McGraw-Hill},
	address      = {Boston}
}

@book{strawson1959,
	title        = {Individuals: An Essay in Descriptive Metaphysics},
	author       = {Strawson, Peter F.},
	year         = {1959},
	publisher    = {Methuen},
	address      = {London}
}

@book{strogatz2015,
	title        = {Nonlinear Dynamics and Chaos: With Applications to Physics, Biology, Chemistry, and Engineering},
	author       = {Strogatz, Steven H.},
	year         = {2015},
	publisher    = {Westview Press},
	edition      = {2nd}
}

@article{sturm2006,
	title        = {On the geometry of metric measure spaces},
	author       = {Sturm, Karl-Theodor},
	year         = {2006},
	journal      = {Acta Mathematica},
	volume       = {196},
	number       = {1},
	pages        = {65--131}
}

@article{szell2010,
	title        = {Multirelational Organization of Large-Scale Social Networks in an Online World},
	author       = {Szell, Michael and Lambiotte, Renaud and Thurner, Stefan},
	year         = {2010},
	journal      = {Proceedings of the National Academy of Sciences (PNAS)},
	volume       = {107},
	number       = {31},
	pages        = {13636--13641},
	doi          = {10.1073/pnas.1004008107}
}

@article{taeihagh2021,
	title        = {Governance of Artificial Intelligence},
	author       = {Taeihagh, Araz},
	year         = {2021},
	journal      = {Policy and Society},
	volume       = {40},
	number       = {2},
	pages        = {137--157},
	doi          = {10.1080/14494035.2021.1928377}
}

@book{tuomela2002,
	title        = {The Philosophy of Social Practices: A Collective Acceptance View},
	author       = {Tuomela, Raimo},
	year         = {2002},
	publisher    = {Cambridge University Press},
	address      = {Cambridge}
}

@article{tverskykahneman1992,
	title        = {Advances in Prospect Theory: Cumulative Representation of Uncertainty},
	author       = {Tversky, Amos and Kahneman, Daniel},
	year         = {1992},
	journal      = {Journal of Risk and Uncertainty},
	volume       = {5},
	number       = {4},
	pages        = {297--323},
	doi          = {10.1007/BF00122574}
}

@misc{ucdp2024,
	title        = {Uppsala Conflict Data Program -- Georeferenced Event Data},
	author       = {{UCDP/PRIO}},
	year         = {2024},
	url          = {https://ucdp.uu.se/},
	note         = {Accessed: 2024-12-01}
}

@misc{undesa2022,
	title        = {World Population Prospects},
	author       = {{UN DESA}},
	year         = {2022},
	url          = {https://population.un.org/wpp/},
	note         = {Accessed: 2024-12-01}
}

@book{vanbenthem1996,
	title        = {Exploring Logical Dynamics},
	author       = {van Benthem, Johan},
	year         = {1996},
	publisher    = {CSLI Publications},
	address      = {Stanford}
}

@book{vanbenthem2011,
	title        = {Logical Dynamics of Information and Interaction},
	author       = {van Benthem, Johan},
	year         = {2011},
	publisher    = {Cambridge University Press},
	address      = {Cambridge}
}

@article{vanderhoek2003,
	title        = {Towards a Logic of Rational Agency},
	author       = {van der Hoek, Wiebe and Wooldridge, Michael J.},
	year         = {2003},
	journal      = {Logic Journal of the IGPL},
	volume       = {11},
	number       = {2},
	pages        = {135--159},
	doi          = {10.1093/jigpal/11.2.135}
}

@article{vannotten2003,
	title        = {An updated scenario typology},
	author       = {van Notten, Philip W. F. and Rotmans, Jan and van Asselt, Marjolein B. A. and Rothman, Dale S.},
	year         = {2003},
	journal      = {Futures},
	volume       = {35},
	number       = {5},
	pages        = {423--443}
}

@book{vonneumann1944,
	title        = {Theory of Games and Economic Behavior},
	author       = {von Neumann, John and Morgenstern, Oskar},
	year         = {1944},
	publisher    = {Princeton University Press},
	address      = {Princeton},
	edition      = {1st}
}

@book{vonwright1963,
	title        = {Norm and Action: A Logical Enquiry},
	author       = {von Wright, Georg Henrik},
	year         = {1963},
	publisher    = {Routledge \& Kegan Paul},
	address      = {London}
}

@book{wald1950,
	author       = {Wald, Abraham},
	title        = {Statistical Decision Functions},
	publisher    = {John Wiley \& Sons},
	address      = {New York},
	year         = {1950}
}

@book{waltz1959,
	title        = {Man, the State, and War: A Theoretical Analysis},
	author       = {Waltz, Kenneth N.},
	year         = {1959},
	publisher    = {Columbia University Press},
	address      = {New York}
}

@book{wasserman1994,
	title        = {Social Network Analysis: Methods and Applications},
	author       = {Wasserman, Stanley and Faust, Katherine},
	year         = {1994},
	publisher    = {Cambridge University Press},
	address      = {Cambridge}
}

@misc{wdi2024,
	title        = {World Development Indicators},
	author       = {{World Bank}},
	year         = {2024},
	url          = {https://data.worldbank.org/},
	note         = {Accessed: 2024-12-01}
}

@misc{who2022,
	title        = {WHO's response to health emergencies: annual report 2022},
	author       = {{World Health Organization}},
	year         = {2023},
	isbn         = {978-92-4-007464-4 (electronic), 978-92-4-007465-1 (print)},
	url          = {https://www.who.int/publications/i/item/9789240074644},
	note         = {Geneva: World Health Organization}
}

@book{wooldridge2000,
	title        = {Reasoning About Rational Agents},
	author       = {Wooldridge, Michael},
	year         = {2000},
	publisher    = {MIT Press},
	address      = {Cambridge, MA},
	note         = {Formalization of agent logics combining belief, desire, and intention (BDI).}
}

@book{young1998,
	title        = {Individual Strategy and Social Structure: An Evolutionary Theory of Institutions},
	author       = {Young, H. Peyton},
	year         = {1998},
	publisher    = {Princeton University Press}
}

\printindex

\end{document}